\documentclass[aps,prx,twocolumn, notitlepage,longbibliography,superscriptaddress,nofootinbib]{revtex4-2}
\usepackage{times}
\usepackage{amsmath}
\usepackage{amsfonts}
\usepackage{amssymb}
\usepackage{dsfont}
\usepackage{graphicx}
\usepackage{physics}
\usepackage{tikz}
\usepackage{hyperref}
\usepackage{mathtools}
\usepackage{comment}
\usepackage{multirow}
\usepackage{ulem}

\usepackage[english]{babel}

\newtheorem{theorem}{Theorem}

\newtheorem{definition}{Definition}
\newtheorem{lemma}{Lemma}

\usepackage{pst-all}
\usepackage{leftidx}

\def\H{\mathcal{H}}
\def\M{\mathcal{M}}
\def\Z{\mathbb{Z}}
\def\C{\mathcal{C}}
\def\D{\mathcal{D}}
\def\B{\mathcal{B}}
\def\R{\mathcal{R}}

\def\L{\mathcal{L}}
\def\SS{\mathcal{S}}
\DeclareMathOperator*{\Motimes}{\text{\raisebox{0.25ex}{\scalebox{0.8}{$\bigotimes$}}}}

\newcommand{\pd}{\partial}

\newcommand{\ZZ}{\mathbb{Z}}
\newcommand{\I}{\mathbb{I}}

\newcommand{\boket}[3]{\langle\, #1 \,|\, #2 \,|\, #3 \,\rangle}

\newcommand{\be}{\begin{equation}}
\newcommand{\ee}{\end{equation}}
\newcommand{\bc}{\begin{center}}
\newcommand{\ec}{\end{center}}
\newcommand{\nin}{\noindent}

\newcommand{\non}{\nonumber}
\newcommand{\lo}{\overline}

\definecolor{dualblue}{RGB}{3,101,192}

\DeclarePairedDelimiter{\ceil}{\lceil}{\rceil}

\usepackage[bottom]{footmisc}

\begin{document}

\title{Topological Order, Quantum Codes and Quantum Computation on Fractal Geometries}

\author{Guanyu Zhu}
\email{guanyu.zhu@ibm.com}
\affiliation{IBM Quantum, IBM T.J. Watson Research Center, Yorktown Heights, NY 10598 USA}
\affiliation{IBM Almaden Research Center, San Jose, CA 95120 USA}

\author{Tomas Jochym-O'Connor}
\affiliation{IBM Quantum, IBM T.J. Watson Research Center, Yorktown Heights, NY 10598 USA}
\affiliation{IBM Almaden Research Center, San Jose, CA 95120 USA}

\author{Arpit Dua}
\affiliation{IBM Quantum, IBM T.J. Watson Research Center, Yorktown Heights, NY 10598 USA}
\affiliation{Department of Physics and Institute for Quantum Information and Matter, \mbox{California Institute of Technology, Pasadena, CA 91125 USA}}

\begin{abstract}
We investigate  topological order on fractal geometries embedded in $n$~dimensions. We consider the  $n$-dimensional lattice with holes at all length scales whose corresponding fractal (Hausdorff) dimension is $D_H=n-\delta$.  In particular, we diagnose the existence of the topological order through the lens of quantum information and geometry, i.e., via its equivalence to a quantum error-correcting code with a macroscopic code distance or the presence of macroscopic systoles in systolic geometry.  We first prove a no-go theorem that $\ZZ_N$ topological order cannot survive on any fractal embedded in two spatial dimensions and with $D_H=2-\delta$. For fractal lattice models embedded in 3D or higher spatial dimensions, $\ZZ_N$ topological order survives if the boundaries on the holes condense only loop or more generally $k$-dimensional membrane excitations ($k \ge 2$), thus predicting the existence of  fractal topological quantum memories (at zero temperature) or topological codes that are embeddable in 3D. Moreover, for a class of models that contain only loop or membrane excitations, and are hence self-correcting on an $n$-dimensional manifold, we prove that $\ZZ_N$ topological order survives on a large class of fractal geometries independent of the type of hole  boundaries and is hence extremely robust.  We further construct fault-tolerant logical gates in the $\ZZ_2$ version of  these fractal models, which we name as \textit{fractal surface codes}, using their connection to global and higher-form topological symmetries equivalent to sweeping the corresponding gapped domain walls.  In particular, we have discovered a logical CCZ gate corresponding to a global symmetry in a class of fractal codes embedded in 3D with  Hausdorff dimension asymptotically approaching  $D_H=2+\epsilon$ for arbitrarily small $\epsilon$, which hence only requires a space-overhead $\Omega(d^{2+\epsilon})$ with $d$ being the code distance. This in turn leads to the surprising discovery of certain exotic gapped boundaries that only condense the combination of loop excitations and certain gapped domain walls. We  further obtain logical $\text{C}^{p}\text{Z}$ gates with $p\le n-1$ on fractal codes embedded in $n$ dimensions. In particular, for the logical $\text{C}^{n-1}\text{Z}$ in the $n^\text{th}$~level of Clifford hierarchy, we can reduce the space overhead to $\Omega(d^{n-1+\epsilon})$.  On the mathematical side, our findings in this paper also lead to the discovery of macroscopic relative systoles in a class of fractal geometries. 

\end{abstract}

\maketitle

\tableofcontents

\section{Introduction}
\textit{Topology, geometry}, and \textit{symmetry} play crucial roles in the progress of modern physics. Concerning topology, in particular, the rapid development of condensed matter physics and the theory of quantum computing and information in the past decades have established a deep connection between seemingly distant fields including topological order, quantum error correction, and fault-tolerant quantum computation~\cite{Wen:1990tm, Wen:1990wk, kitaev2003, kitaev2006, Dennis:2002ds,  nayak2008, Terhal:2015ks,  campbell2017}.  
%The roles of geometry and symmetry will also be explored in this paper.

At the heart of the above connection is quantum entanglement.  A many-body quantum system having \textit{topological order}  possesses \textit{long-range entanglement}, which means that it has a sharp distinction from a product state and cannot be prepared from it via a local constant-depth circuit~\cite{Chen:2010gb}. Such long-range entanglement ensures that local noise cannot decohere the logical information stored non-locally across the system.  In this sense, one can view a system with topological order as a \textit{quantum error-correcting code}, which is protected against any local noise with support up to the order of the code distance.  Thus, such a system can serve as a \textit{topological quantum memory}.  Well-known candidates of such topological memories include exotic materials with passive topological protection like fractional quantum Hall states \cite{Wen:1990wk, nayak2008}, quantum spin liquids \cite{savary2017, Lukin:2021_spin_liquid}, and Majorana wires \cite{Kitaev_2001}, as well as active quantum error correcting codes using topological codes including toric (surface) codes \cite{kitaev2003, Bravyi:1998uq, fowler2012}, color codes \cite{Bombin:2006hw, Bombin:2015hia, Kubica:2015, Kubica:2015br}, and non-Abelian codes \cite{levin2005, Koenig:2010do, schotte2020quantum, Zhu:2020_constant_depth, Lavasani2019universal} supported on conventional qubits such as superconducting qubits and trapped ions. 

As another manifestation of this deep connection, the code distance of a topological memory must be macroscopic to ensure topological degeneracies \cite{Bravyi:2010jn, Bravyi:2011wg}, another key signature of topological order. More concretely, it means the distance needs to grow with system size, such that the splitting of ground-state degeneracies and logical error rate induced by the perturbation of the environment can be exponentially suppressed and vanishes in the thermodynamic limit. 
%Physical systems exhibiting such topological degeneracies and topological order

To this point, the majority of the studies of topological order are typically associated with a manifold, including the situation of a lattice forming the cellulation of a manifold. In this context, the main physical properties hence only depend on the topological properties of the manifold rather than the geometry. For example, in a 2D topological order, the topological degeneracies depend only on the  genus of a closed manifold \cite{Wen:1990wk}. This is related to the fact that topological orders can typically be described by a \textit{topological quantum field theory} (TQFT) \cite{nayak2008}, which is usually  defined on an $n$-dimensional manifold. As an example, the path integral of the Turaev-Viro topological quantum field theory is a 3-manifold topological invariant \cite{turaev1992}. A natural question arises: whether or not there exists topological order and long-range entanglement on \textit{exotic geometries} beyond manifolds and what role does geometry, rather than topology, play in this context.  

An obvious candidate of such geometries is fractal geometry, which is nowhere differentiable and hence is in sharp contrast to a continuum model. There are several other fundamental or practical motivations to consider topological orders on fractal geometries besides the conceptual extension of TQFT.  First of all, from the information-theoretical perspective, it has been realized that an encoded classical memory with local interactions supported on a fractal geometry can exist in nature. In particular, it can be formed by an Ising model supported on a fractal lattice such as a Sierpinski carpet \cite{Gefen:1980_fractal, Gefen_1984_fractal, Bonnier:1987_fractal, Monceau:1998_fractal}.  Therefore, a natural question to ask is that whether or not a topological quantum memory supported on a fractal geometry could also exist in nature, i.e., embeddable in three dimensions.  In this paper, we will answer this question affirmatively.   

More interestingly, by punching holes in a topological quantum memory to form a fractal geometry, one could  significantly reduce the space-time overhead of fault-tolerant quantum computation. This is essentially a form of the \textit{code puncturing} idea in error correction.  
%This is because the space cost, or equivalently the total number of physical qubits in the memory scales as $L^{D_H}$ where $L$ is the linear size of the system and $D_H$ is the Hausdorff dimension of the fractal.  For example, one can consider replacing a  a 2D and 3D topological memory with a memory defined on fractals with Hausdorff dimensions $D_H=2-\delta$ or $D_H=3-\delta$. In this paper, we show that the former is impossible while the later is possible. 
In particular, our motivation here is not only to reduce the resource cost for memory storage but mainly the cost for performing fault-tolerant logical gates. Currently, the major approaches in fault-tolerant quantum computation utilize topological stabilizer codes such as 2D surface codes and color codes. However, topological stabilizer codes in 2D can only perform fault-tolerant logical gates within the Clifford group~\cite{Bravyi:2013dx}, and to have a universal logical gate set one hence needs to perform magic-state distillation which leads to a huge space-time overhead~\cite{bravyi2005, Fowler:2012fi}. An alternative approach is performing code-switching to a higher-dimensional topological stabilizer code equipped with fault-tolerant non-Clifford logical gates~\cite{Paetznick:2013fu,Bombin:2015jk}, yet such switching methods also incur high space-time overheads~\cite{Kubica:2021}. This requirement for higher-dimensional codes to obtain non-Clifford gates arises due to the Bravyi-K\"onig bound~\cite{Bravyi:2013dx}. This bound states that a topological stabilizer code supported in an $n$-dimensional lattice can only have local constant-depth logical gates within the $n^\text{th}$~level of the Clifford hierarchy. Such a result would suggest a lower bound in the space overhead $\Omega(L^n)$ for gates in the $n^\text{th}$~level but outside the $(n-1)^\text{th}$~level \cite{Paetznick:2013fu, JochymOConnor:2014, Bombin:2015jk, Kubica:2015, Bombin:2016dq, Kubica:2015br, JochymOConnor:2018is, Vasmer2019, JochymOConnor:2021ih}. Yet, by trading space for time, there exist methods to simulate the action of higher-dimensional codes using lower-dimensional space resources, such as simulating the action of a $3^{\text{rd}}$~level gate using only two-dimensional spatial resources~\cite{Bombin:2018wj, Browneaay4929}. However, these methods are intrinsically based on underlying 3D code models, which is in contrast to the code we present here whose space overhead has Hausdorff dimension $D_H = 3-\delta$ or more generally $D_H = n-\delta$.

Next, one may consider the stability of topological order under large spatial disorder (e.g., in the presence of islands of trivial phases). The spatial disorder can contain holes (trivial phases) at all length scales and is hence similar to the situation of a fractal lattice. This problem is also related to the issue of fabrication errors in a topological quantum error-correcting code~\cite{Auger_fabrication_error_2017}, for example, the presence of islands of corrupted or unusable qubits or couplers at all length scales.

\begin{figure}[t]
  \includegraphics[width=1\columnwidth]{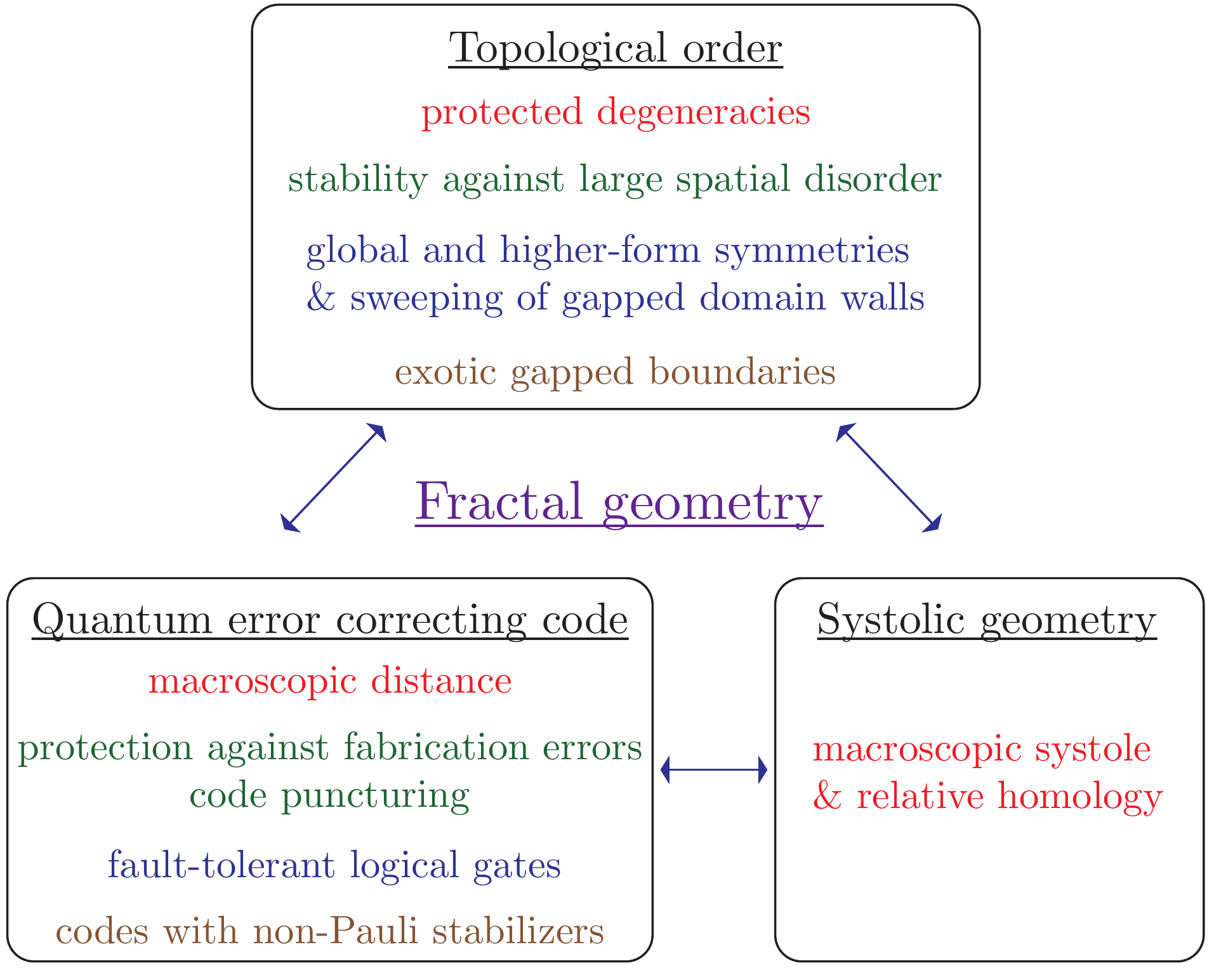}
  \caption{Summary of the motivation and scope of this paper, and the correspondence among the three fields at the interface of fractal geometry. Items with the same color indicate the equivalence of concepts in different fields.}
  \label{fig:venn_diagram}
\end{figure}

The mathematical essence behind the above questions and motivations is the \textit{systolic geometry} (e.g., see Ref.~\cite{Gromov:1996}).  In particular, the connection between code distance and the systole (i.e., the shortest length of a non-contractible cycle) of a cell complex was first discovered in the context of quantum codes defined on the cellulation of manifolds by Freedman-Meyer-Luo~\cite{Freedman_systole_2002, Freedman:1999_Z2_sysole}. In other words, the existence of macroscopic code distance, i.e., topological order, is equivalent to the existence of  macroscopic systole which should also grow with system size. This connection has been further extended to recent studies of certain quantum low-density parity-check (LDPC) codes defined on hyperbolic manifolds \cite{Guth:2014cj}, or even cell complexes built from expander graphs~\cite{Zemor:2020_higher_dimensional_expander} which also provide examples of nontrivial quantum many-body states supported on complexes beyond manifold. These nontrivial states can be considered as a generalization of the toric code states, and cannot be prepared from a product state via a local constant-depth circuit applied on the cell complexes and hence also possess long-range entanglement. In particular, they also exhibit macroscopic combinatorial systoles and code distances.  Therefore, geometrically speaking, searching for topological order and long-range entanglement on fractal geometries is also equivalent to searching for the existence of macroscopic systoles.  The potential practical advantage of fractal geometries, compared with hyperbolic manifold or expander graphs approaches, is that they can still be realized with geometrically local physical systems embedded in low dimensions (e.g., in 3D), and thus can more easily be implemented with topological materials existing in nature or near-term qubit technology. 

The main motivation and the scope of this paper, as well as the correspondence between different fields mentioned above is summarized in Fig.~\ref{fig:venn_diagram}.

 \textit{Main contributions.---} In this paper, we use two complementary descriptions to study the existence of topological order on fractal geometries: (1) A TQFT description, in particular, the physical pictures of gapped boundaries and condensation of anyons, strings, and membranes. (2) The homology theory, especially the \textit{relative homology} \cite{Hatcher:2001ut}, providing a mathematical description of gapped boundaries in Abelian topological orders \cite{Bravyi:1998uq}, and the corresponding relative systoles \cite{Babenko_2002} which are the conceptual extension of systoles to the shortest length of non-contractible cycles connecting boundaries.  
 
 Quite surprisingly, the TQFT description, which was commonly considered to be suitable mainly for the manifold case,  can be well extended to the case of fractal geometries.  The key is to start with an $n$D topological order or an $n$+1D TQFT and then punch holes with gapped boundaries in all length scales. The fractal geometries along with the types of gapped boundaries determine the preservation of long-range entanglement and hence the fate of topological order.  First of all, we prove a no-go theorem for the existence of $\ZZ_N$ topological order on a fractal embedded in two dimension and with Hausdorff dimension $D_H=2-\delta$ by showing the absence of macroscopic distance. This attributes to the presence of short logical strings (i.e., with $O(1)$ length) connecting nearby gapped boundaries due to the condensation of anyons on these boundaries. This also shows the absence of a macroscopic relative systole. 
 
 When considering fractals embedded in three dimension and with Hausdorff dimension $D_H=3-\delta$, or equivalently puncturing holes in a 3D $\ZZ_N$ topological order,  the situation becomes more subtle and interesting. A 3D $\ZZ_N$ topological order supports two types of excitations: the particle-like excitations called $e$ and the loop-like excitations called $m$. Now we summarize the results in two different scenarios: (1) We show that if the boundaries of the holes inside the fractal are so-called $e$-boundaries on which $e$-particles can condense, there are always short logical strings connecting nearby gapped boundaries leading to the absence of macroscopic code distance or relative systoles. Therefore, topological order does not exist in this type of models.  (2) In contrast, if the hole boundaries are so-called $m$-boundaries on which $m$-loops  can condense, no short logical string or membrane can connect such boundaries for a particular class of fractals called \textit{simple fractals} such that all the holes inside are equivalent to 3D balls. Hence, in this case, the code distance or relative systole is macroscopic and topological order can exist.  This provides an answer to a major question posed above: a topological quantum memory supported on a fractal geometry can indeed exist in nature, i.e., embeddable in 3D, either in the form of a passive memory using $\ZZ_N$ spin liquid materials at a temperature much smaller than the gap ($T \ll \Delta$), or in the form of an active error correcting code consisting of conventional qubits. We further construct a class of models supported on such simple fractals with Hausdorff dimension $D_H=3-\delta$, where $0<\delta<1$, which can asymptotically approach $D_H=2+\epsilon$ for arbitrary small~$\epsilon$. The code distance in the whole class of models remains invariant as the shortest length of the logical string operator, which is the order of the linear size: $d=O(L)$. We call these class of models 3D \textit{fractal surface codes} for simplicity.

Even more interestingly, these fractal codes still preserve the capability of performing a fault-tolerant logical 
non-Clifford gate, i.e., the logical CCZ gate,  as in the case of 3D surface codes \cite{Kubica:2015br,  Vasmer2019}. The logical gates in both the fractal codes embedded in $n$~dimensions and the usual $n$D surface codes can be understood via a TQFT picture. In general, a large class of fault-tolerant logical gates, including transversal logical gates and more generally local constant-depth circuits, is associated with an onsite  \textit{topological symmetry} associated with multiple copies of topological orders or more generally a symmetry enriched topological order \cite{barkeshli2014SDG}, as has been pointed out in Ref.~\cite{Zhu:2017tr}. These symmetries can either be global symmetries which act on the entire system, or higher-form ($q$-form) symmetries which act on a codimension-$q$ sub-manifold $\M^{n-q}$ \cite{Yoshida_gate_SPT_2015, Yoshida_global_symmetry_2016, Yoshida2017387, Webster_gates_2018, Zhu:2017tr}. Moreover, applying these logical gates, or topological  symmetries, is equivalent to sweeping certain gapped domain walls which are one dimension lower than the corresponding symmetries \cite{Yoshida_gate_SPT_2015, Yoshida_global_symmetry_2016, Yoshida2017387, Webster_gates_2018, Zhu:2017tr}. When considering three copies of 3D $\ZZ_2$ fractal surface codes, one can first apply a transversal CCZ gate, equivalent to a global \textit{topological symmetry} on three copies of $\ZZ_2$ topological orders.  This is equivalent to sweeping the CCZ domain wall across the system and it remaining attached on all the hole boundaries, which leads to the discovery of new types of exotic gapped boundaries and maps the original codes to new codes. We hence call this operation a \textit{transversal logical map}, in contrast to the usual transversal logical gate which keeps the code space invariant. These exotic boundaries are interesting in their own right since they only allow the combination of $m$-loops and CCZ domain walls to condense on them and correspond to a parent Hamiltonian with non-Pauli stabilizers. A subsequent lattice surgery method with additional ancilla code blocks can map the new composite code back to the original code space and complete the logical CCZ gate.  The space overhead in the presented scheme can be lowered to $\Omega(L^{2+\epsilon}) = \Omega(d^{2+\epsilon})$, which is surprising given all known methods for implementing a non-Clifford constant-depth logical gate are based on codes whose space overhead are intrinsically~$\Omega(L^3)$ \cite{Bombin:2015hia, Kubica:2015br, Vasmer2019, JochymOConnor:2018is, JochymOConnor:2021ih, Bombin:2018wj, Browneaay4929}, as suggested by the Bravyi-K\"onig bound.

We can further extend our classification of topological order and logical gates to those supported on fractals embedded in $n$ dimensions, or equivalently puncturing holes in $n$-dimensional topological order. Note that one can classify $n$-dimensional $\ZZ_N$ topological orders with the dimensions of its two types of excitations using the bi-label $(i, n-i)$ (with $i\le n-i$), which states that the world-volume of  $e$-excitation and the corresponding logical operator are $i$-dimensional and those of the $m$-excitation is $(n-i)$-dimensional. We summarize our classification in two different scenarios: 
\begin{enumerate}
\item 
$(i=1)$---The $e$-excitation is particle-like and the corresponding logical operator is string-like. In this case, topological order is still absent if the hole boundaries are $e$-boundaries which allow $e$-particles to condense on them. On the other hand, we prove that for a simple fractal (with holes being equivalent to $n$-balls) embedded in $n$D with $m$-boundaries, topological order does exist. We hence construct a class of $n$D fractal surface codes with Hausdorff dimension $n-\delta$ ($0<\delta<1$) which can asymptotically approach $D_H=n-1+\epsilon$.  Interestingly, in a similar way to the 3D case, the $n$D fractal surface codes support fault-tolerant logical $\text{C}^p\text{Z}$ gates, with $p\le n-1$. In particular, the logical $\text{C}^n\text{Z}$ gate involves applying the global symmetry in $n$ copies of $\ZZ_2$ topological orders, while other logical $\text{C}^p\text{Z}$ gates for $p<n$ are single-shot transversal logical gates corresponding to $p$-form symmetries in $n-p$ copies of  $\ZZ_2$ topological orders. The logical gate $\text{C}^n\text{Z}$  belongs to the $n^\text{th}$~level Clifford hierarchy and requires minimal space overhead $\Omega(L^{n-1+\epsilon})=\Omega(d^{n-1+\epsilon})$, again improving on the space overhead of~$\Omega(L^{n})$ as suggested by the Bravyi-K\"onig bound.

\item ($i\ge 2$)---No particle-like excitation or string-like logical operator exists and as such the topological memory is expected to be self-correcting when supported on a manifold. Quite interestingly, we show that in this case topological order can exist in simple fractals independent of the boundary type ($e$ or $m$) of each hole inside.  This reveals that topological order in a self-correcting model is not only protected against thermal noise and hence stable  at finite temperature (in contrast to the 2D topological order which is unstable at finite temperature), it can also be extremely robust under large spatial disorder or fabrication errors.  We have also constructed a family of such fractal codes with Hausdorff dimension $n-\delta$ ($0<\delta<1$). We have proven that the transversal logical gate or the logical gate composed of constant-depth circuit in multiple copies of $(i,n-i)$ surface codes or color codes in $n$ dimensions is preserved in the corresponding fractal codes and lies within the $n^\text{th}$~level Clifford hierarchy.  The gain in our case is that the Hausdorff dimension can be lowered to $D_H=n-1+\epsilon$. We note that a fractal topological model embedded in 4D has been proposed before,  which uses the specific construction of a  homological product of two classical codes \cite{Brell:2016co}.  The fractal models we study in this paper is more general, especially in the context of capturing the essence of the spatial disorder. 
\end{enumerate}

Besides the physics discovery in this paper, our main mathematical contributions are the extension of systolic geometry to the context of fractals, and proof of the existence of macroscopic relative systoles in simple fractals embedded in three and higher dimensions. 

\textit{Outline of the paper.---}
In Sec.~\ref{sec:TO_def}, we introduce the precise definition of topological order and its relation to the code distance of a quantum error correcting code as well as the systolic geometry. In Sec.~\ref{sec:2D}, we introduce the definition of quantum models defined on fractals and prove the no-go theorem of topological order existing in a fractal embedded in 2D.  In Sec.~\ref{sec:3D}, we first introduce the TQFT, stabilizer, and algebraic topology descriptions of 3D topological order.  We then prove the no-go theorem of topological order on arbitrary fractals embedded in 3D with hol$e$-boundaries being $e$-boundaries. We then use different ways, i.e., with the TQFT and the algebraic topology descriptions respectively, to prove the existence of topological order on a simple fractal embedded in 3D with all the holes having $m$-boundaries. The former way is more inclined towards condensed matter physicists while the latter is more suitable for quantum information theorists or mathematical physicists. In addition, we construct the family of 3D fractal surface codes with Hausdorff dimension $2<D_H<3$.  In Sec.~\ref{sec:nD}, we extend the theory to $n$ dimensions, discuss both the topological order with and without string-like logical operators. In Sec.~\ref{sec:gates}, we discuss fault-tolerant logical gates in 3D and $n$D fractal surface codes. 

\section{Definition of topological order and its relation to quantum codes and systolic geometry}\label{sec:TO_def}

We consider a \textit{local physical system} of qudits occupying the sites of an  $n$-dimensional lattice $\mathcal{L}$ with linear size $L$ and governed by a geometrically local Hamiltonian $H$.  The associated physical Hilbert space is a tensor product of local Hilbert spaces,  $\H=\Motimes_j \H_j$, where $j$ is the site label of $\L$.  
We provide the following definition of \textit{topological order} (T.O.) which is commonly used in the literature \cite{ Bravyi:2006kt, Bravyi:2010jn, Bravyi:2011wg, Terhal:2015ks}:
\begin{definition}\label{def:TQO}
Topological order is a local physical system whose  ground-state subspace forms a quantum error correcting code with a macroscopic code distance.
\end{definition}

\nin One can also rephrase the above definition more formally into the following mathematical condition for topological order \cite{Bravyi:2010jn, Bravyi:2011wg}: 
\textit{There exist $d = O(L^a)$ for some constant $a>0$ and sufficiently large $L$ such that}
\be\label{eq:TQO1}
   P_C\mathcal{O}_AP_C = cP_C.
\ee
\textit{Here, $P_C$ is the projector onto the ground-state subspace $\H_C \subset \H$ or equivalently the code space, $\mathcal{O}_A$ represents any local operator supported in a region $A$ involving at most $d$ sites, and $c$ is some complex number. `Sufficiently large L' also requires that the system size $L$ is much larger than the correlation length $\xi$, i.e., $L \gg \xi$}.

In the above condition, the code distance $d$ scales as $L^a$, i.e., a power law of the linear size of the system, and is hence macroscopic\footnote{One could in principle relax the definition of macroscopic distance to include $polylog(L)$. In this paper, we stick to the more restrictive power-law definition of a macroscopic distance to ensure the exponential suppression of the topological degeneracy in Eq.~\eqref{eq:degeneracy_splitting} according to the traditional definition of T.O. and the stability of gap \cite{Bravyi:2010jn, Bravyi:2011wg}. Since none of the situations considered in this paper will have a $polylog(L)$ scaling, it does not matter.}. Definition \ref{def:TQO} and condition Eq.~\eqref{eq:TQO1} ensure that any local operator cannot induce transition between or distinguish two orthogonal ground states, and hence coincide with the traditional definition of topological order in the literature of condensed-matter physics \cite{Wen:1990wk, kitaev2003, nayak2008, freedman2003topological, Bravyi:2006kt}\footnote{We note that Def.~\ref{def:TQO} and condition Eq.~\eqref{eq:TQO1} is also referred to as the  TQO1 condition in Ref.~\cite{Bravyi:2010jn, Bravyi:2011wg}. Meanwhile, there is an additional TQO2 condition to ensure the stability of the gap under perturbation, which is not included in our current definition and will be discussed in future works.},  
i.e., 
\be\label{eq:degeneracy_splitting}
\boket{\lo{a}}{\mathcal{O}_A}{\lo{b}} \propto \delta_{a,b} +   O(e^{-L/\xi}),
\ee
where $\ket{\lo{a}}$ and $\ket{\lo{b}}$ label the ground states or equivalently the logical states.  
This hence leads to the topologically protected degeneracies: a local perturbation can only lift the ground-state degeneracy only in the $k$th order perturbation theory, where $k$ increases with the system size $L$. 
The generalization of the above definition to a continuous system is also straightforward.

% \textbf{TQO-2}: Local ground-state subspaces are consistent with the global one.   More concretely, let $B \in \text{Cube}(r+1)$ be a region containing A and all nearest neighbors of A (A is defined the same as in TQO-1).  If $\mathcal{O}_A$ is any operator satisfying $\mathcal{O}_A P =0$, then one has
% \be\label{eq:TQO2}
% \mathcal{O}_A P_B =0. 
% \ee
 
\begin{figure}[hbt]
  \includegraphics[width=0.6\columnwidth]{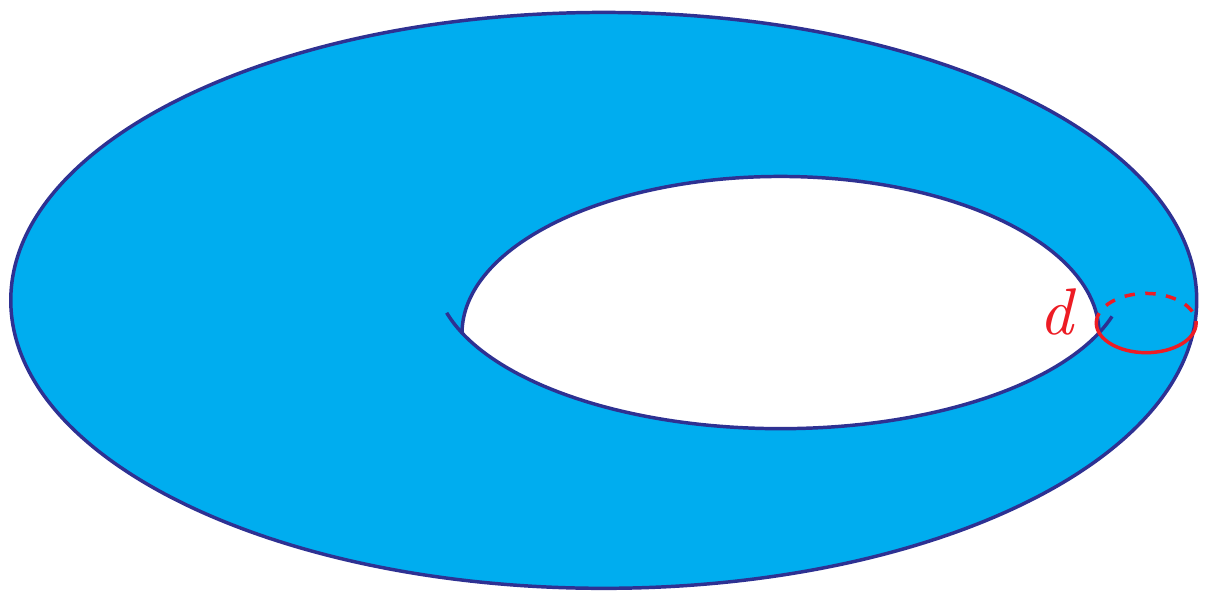}
  \caption{The 1-systole on a torus  $sys_1(T^2)=d$ is equivalent to the code distance of a 2D quantum error correcting code or topological order defined on the torus, such as the toric code model.}
  \label{fig:systole}
\end{figure}

 As we see, the definitions of topological order and quantum error correcting codes are deeply connected, and the key is the presence of a macroscopic code distance. On the other hand, the connection between a class of quantum codes and systolic geometry has been established in Ref.~\cite{Freedman_systole_2002}. One can define a quantum code on the cellulation of a manifold or more generally a cell complex. The code distance corresponds to the \textit{systole}, i.e., the shortest cycle, in a manifold or a cell complex. A simple example is illustrated in Fig.~\ref{fig:systole} for the case of a torus, where the 1-systole is the shortest  1-cycle around the narrow handle.  
 The formal definition including the generalization to $i$-systole will be given in Sec.~\ref{sec:chain_complex}.  In this paper, we need to further extend this connection to include the case of a manifold with boundaries, where the code distance corresponds to the \textit{relative systole} \cite{Babenko_2002}, i.e., the shortest \textit{relative cycles} which connects two boundaries.   
 
 Throughout this paper, we consider the possible existence of topological order on various fractal geometries. A fractal embedded in an  $n$-dimensional manifold $\M^n$  will have Hausdorff dimension $D_H=n-\delta$, with $0<\delta < n$.  Some well-known fractals are:  Sierpi\'nski carpet ($D_H=1.8928$), Sierpi\'nski triangle ($D_H=1.585$), and Apollonian gasket ($D_H=1.3057$) are embedded in $\mathbb{R}_2$, while Menger sponge ($D_H=2.7268$) and Sierpi\'nski tetrahedron ($D_H=2$) are embedded in $\mathbb{R}_3$.  As one can see, in contrast to the torus case,  it is very easy to have short cycles or relative cycles almost everywhere in these fractal geometries. Therefore, determining the existence of topological order in this context is equivalent to determining the existence of macroscopic systoles and equivalently a macroscopic code distance.

\begin{figure*}[t]
  \includegraphics[width=1.8\columnwidth]{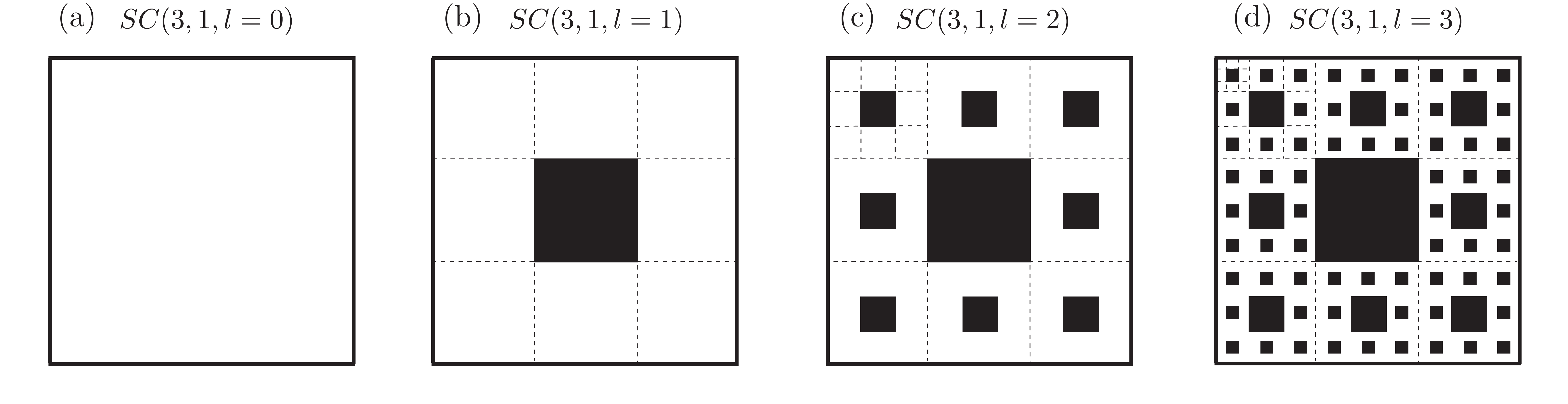}
  \caption{Recursive definition of the Sierpi\'nski carpet $SC(3,1)$. In the $l^\text{th}$ iteration, we punch level-$l$ holes with linear size $\frac{1}{3^l}$ and get the Sierpski carpet at level $l$. One approaches the Sierpi\'nski carpet as $l \rightarrow \infty$.}
  \label{fig:fractal_2D}
\end{figure*}

\section{Absence of $\ZZ_N$ Topological Order on fractals embedded in 2D}\label{sec:2D}

In this section, we show that there is no (intrinsic) $\ZZ_N$ topological order which can survive on a fractal embedded in a 2D surface.   

We start by introducing a simple example of fractal embedded in 2D, the Sierpi\'nski carpet, as shown in Fig.~\ref{fig:fractal_2D}.   The Sierpi\'nski carpet, as the case of most fractals, can be defined recursively.  We start with a square patch with linear size $L$ in $\mathbb{R}_2$,  as shown in Fig.~\ref{fig:fractal_2D}(a). In the first iteration, we punch a square-shaped hole in the center with the linear size being 1/3 of the entire square patch, i.e., $L/3$, as shown in Fig.~\ref{fig:fractal_2D}(b).  One can then consider the division of the largest square patch into 9 square patches with the same size as the hole, i.e., with linear size $L/3$.  In the second iteration, we punch a hole in the center of the 8 square patches without holes, while the linear size of this hole is 1/3 that of the hole in the previous iteration, i.e., $L/9$. We then repeat the above iteration and dig holes with linear size $L/27$ in the third iteration as shown in Fig.~\ref{fig:fractal_2D}(d).  We denote the shape in each iteration as $SC(3,1,l)$, where `$3,1$' represents the relative linear size of each square patch and that of the hole inside the square patch which is punched in each iteration.  Here, $l$ refers to the $l^\text{th}$ iteration, which is also called the \textit{level} of the Sierpi\'nski carpet.  We hence can call $SC(3,1,l)$ as the (3,1) Sierpi\'nski carpet at level $l$. The fractal $SC(3,1)$ is defined as the limit of the sequence $SC(3,1,l)$ at infinite level, i.e., $SC(3,1) \equiv \text{lim}_{l\rightarrow \infty} SC(3,1,l)$. We also call the new holes introduced at the $l^\text{th}$ iteration as level-$l$ holes. We can now calculate the Hausdorff dimension of $SC(3,1)$ in the following way: when we increase the linear size of the fractal by a factor of 3, the volume (area in 2D case) increases by a factor of 8.  Therefore the Hausdorff dimension is $D_H = \frac{\text{ln}(8)}{\text{ln}(3)} \approx 1.8927$. We can generalize the above Sierpi\'nski carpet sequence to more general cases as $SC(p,q,l)$, where $p$ and $q$ are both integers.

\begin{figure}[hbt]
  \includegraphics[width=1\columnwidth]{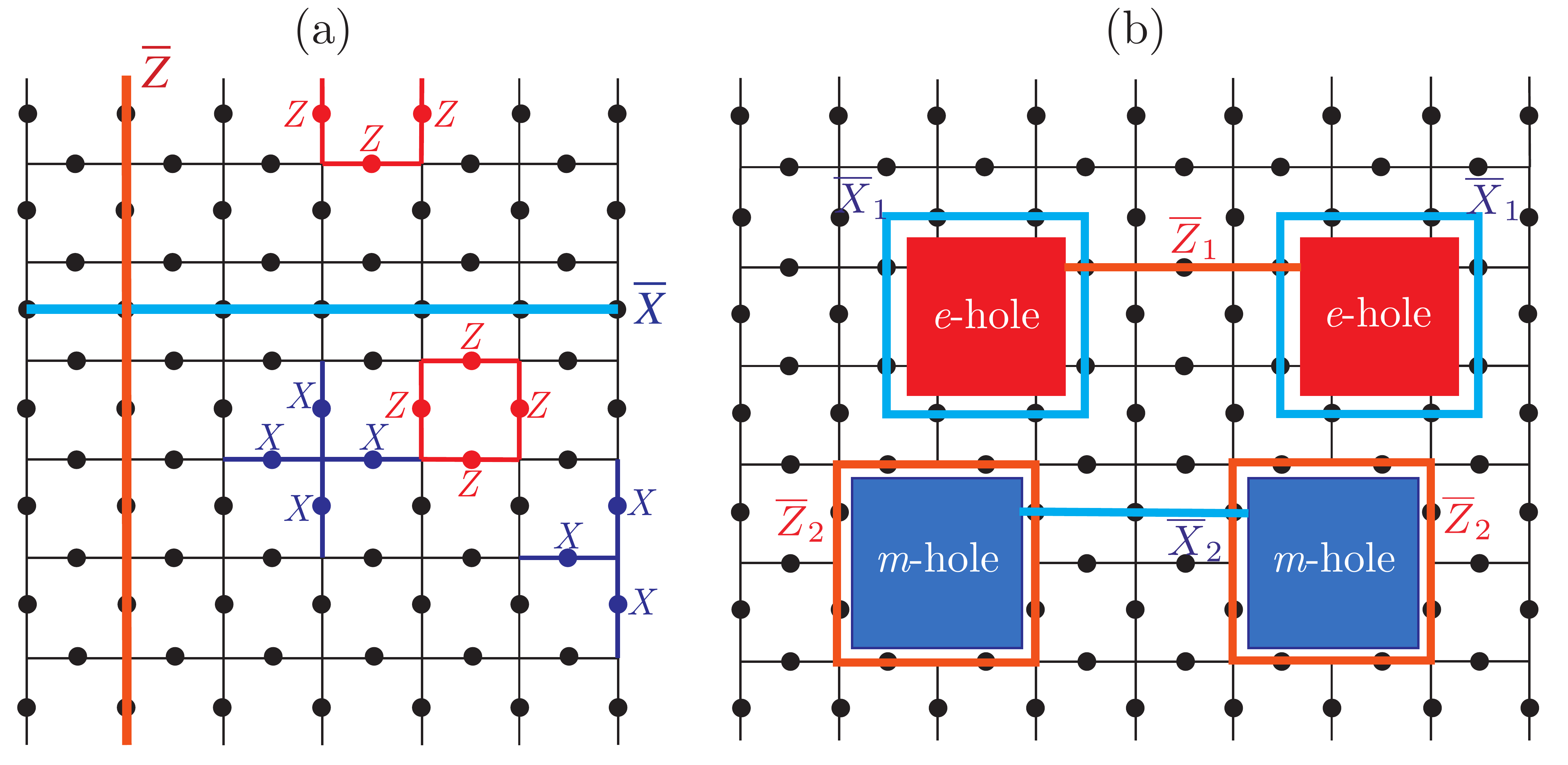}
  \caption{(a) The surface code and two types of gapped boundaries: the horizontal (vertical) $e$-boundaries ($m$-boundaries) condense $e$-anyons ($m$-anyons) such that the logical $Z$-string ($X$-string) can terminate on them. (b) Two types of holes. The $e$-hole ($m$-hole) has $e$-boundaries ($m$-boundaries), which trap a logical $X$-loop  ($Z$-loop) around it, and allows a logical $Z$-string ($X$-string) to terminate on it. }
  \label{fig:surface_code}
\end{figure}

\begin{figure*}[hbt]
  \includegraphics[width=2\columnwidth]{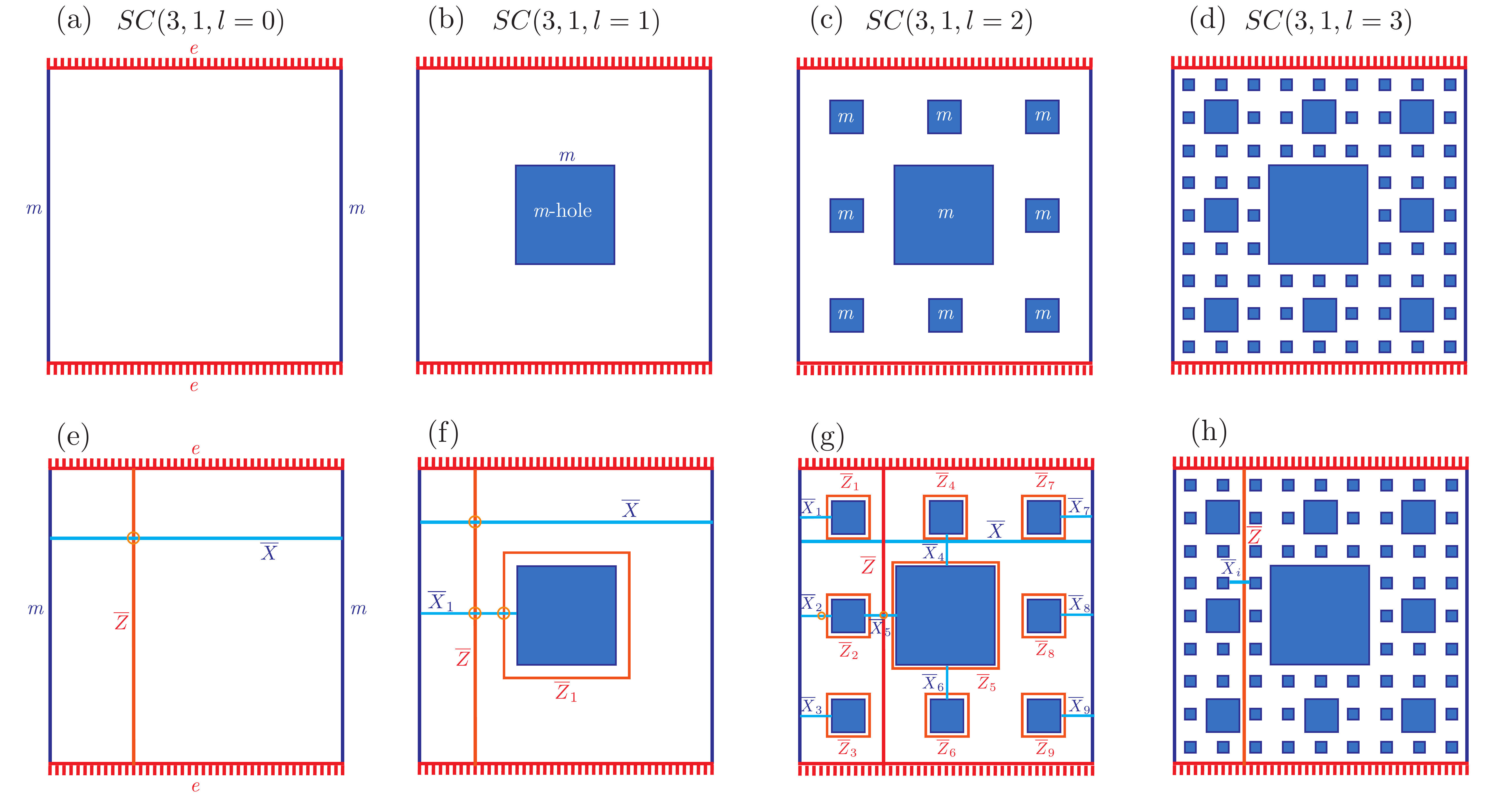}
  \caption{(a-d) The recursive definition of the fractal models with the external boundaries of the surface code and $m$-holes inside. (e-h) Logical string operators in each iteration. In panels (f-h), any representative of the macroscopic logical-$Z$ string, such as $\overline{Z}$ illustrated in the figure, always intersects with a short logical-$X$ string, which will have $O(1)$ length in the final iteration, as illustrated in (h) in this example.  This leads to an $O(1)$ code distance. }
  \label{fig:fractal_surface_code}
\end{figure*}

We now investigate the possible existence of topological orders on fractals embedded in 2D.  We first consider the simplest type of topological order, i.e., $\ZZ_N$ topological order.  We can describe it via the $\ZZ_N$ gauge theory with the Chern-Simons action being:
\be\label{eq:CS}
S_\text{CS} = \int  \frac{N}{2\pi} b \wedge da.
\ee
Here, $a$ and $b$ are 1-form compact U(1) gauge fields describing the electric and magnetic degree of freedoms.  A $\ZZ_{N}$ fusion rule exists for both $e$ (electric charge) and $m$ (magnetic flux) anyonic excitations, with the corresponding gauge group being $G=\ZZ_{N}$. Note that the $\ZZ_N$ gauge theory discussed here is also a topological quantum field theory (TQFT).

In the case of $N$$=$$2$, the above action describes the $\ZZ_2$ gauge theory.  One can construct an exactly-solvable microscopic lattice model for this gauge theory, i.e., the $\ZZ_2$ toric code model in 2D, with the Hamiltonian being:
\begin{align}\label{eq:2DTC}
\nonumber H_\text{2DTC}=&-J\sum_v A_v - J\sum_p B_p, \\
\text{with} \quad  A_v=&\Motimes_{j\in \{e_v\}}X_j, \quad B_p=\Motimes_{j\in \{e_p\}}Z_j,
\end{align}
where $A_v$ is the vertex stabilizer with all the Pauli-$X$ operators supported on the edges connected to the vertex $v$ (denoted by $\{e_v\}$) and $B_p$ is the plaquette/face stabilizer with all the Pauli-$Z$ operator supported on the edges surrounding the plaquette $p$ (denoted by $\{e_p\}$). Violation of vertex and plaquette stabilizers creates $e$ and $m$ anyons respectively. When defining such a model on the 2D square lattice, both $A_v$ and $B_p$ are 4-body operators, as illustrated in Fig.~\ref{fig:surface_code}(a).  The generalization to the $\ZZ_N$ toric code model in any spatial dimension is discussed in App.~\ref{append:ZN}. Note that in the case of exactly solvable topological stabilizer models, such as the toric code, one has zero correlation length, i.e.,  $\xi=0$,  while in the more general case the topological order has finite correlation length $\xi$. 

Now we consider the properties of gapped boundaries. There are two types of gapped boundaries in a single copy of 2D toric code model:  the \textit{$e$-boundaries} and \textit{$m$-boundaries} which condense $e$-anyons and $m$-anyons respectively. In terms of microscopic realization, one can implement the  $e$-boundaries with the so-called \textit{rough boundaries} having dangling edges sticking out, as shown in the upper and lower boundaries in Fig.~\ref{fig:surface_code}(a).  The rough boundaries in this example have 3-body boundary $Z$ stabilizers, while $X$ stabilizers are taken away from the rough boundaries.  Similarly, one can implement the $m$-boundaries with the so-called \textit{smooth boundaries} having no dangling edges, as shown in the left and right boundaries in Fig.~\ref{fig:surface_code}(a).  The smooth boundaries in this example have 3-body boundary $X$ stabilizers, while the $Z$ stabilizers are taken away from the smooth boundaries.  A square patch of toric code with both $e$- and $m$-boundaries on opposite sides form the so-called \textit{surface code}, which encodes a single logical qubit. The logical $Z$ and $X$ string (denoted by $\overline{Z}$ and $\overline{X}$) connect the $e$-boundaries and $m$-boundaries respectively and are illustrated in Fig.~\ref{fig:surface_code}(a). Since the $e$ and $m$ charges condense on the $e$ and $m$ boundaries respectively, these logical strings are Wilson lines (worldlines) of the $e$ and $m$ anyons.    

With these two types of boundaries, one can also make two types of holes, which we call \textit{$e$-holes} and \textit{$m$-holes}, with their boundaries being the $e$ (rough) and $m$ (smooth) boundaries respectively, as illustrated in Fig.~\ref{fig:surface_code}(b).  These holes can also encode logical information.  For example, as shown in  Fig.~\ref{fig:surface_code}(b), a pair of $e$-holes and a pair of $m$-holes can both store one logical qubit.  As one can see, logical string $\overline{X}_1$ ($\overline{Z}_1$) encircles  $e$-holes ($m$-holes), while the $\overline{Z}_1$ ($\overline{X}_1$) strings connect the $e$-holes ($m$-holes).  One might consider the hole encoding as a `blessing' for quantum information storage and processing.  However, in the context of fractal topological orders, the fact that these holes can encode logical information becomes a `curse'. 

With the understanding of the basic properties stated above, we now put the surface code on a fractal.  A necessary question to consider is how to associate the boundary conditions to the holes.  Without losing generality, we begin with putting the surface code introduced above on the level-0 Sierpi\'nski carpet $SC(3,1,l=0)$, i.e., a square patch with linear size $L$, with $e$ (rough) boundaries on the upper and lower sides, and $m$ (smooth) boundaries on the left and right sides, as shown in Fig.~\ref{fig:fractal_surface_code}(a). A pair of \textit{macroscopic} logical operators $\overline{Z}$ and $\overline{X}$ is shown in Fig.~\ref{fig:fractal_surface_code}(e). We say $\overline{Z}$ and $\overline{X}$ are  \textit{macroscopic} since they have size $O(L)$, and hence are macroscopic in the thermodynamic limit $L \rightarrow \infty$. This initial surface code has macroscopic code distance, i.e., $d=O(L)$. On top of the level-1 Sierpi\'nski carpet $SC(3,1,l=1)$, one makes a level-1 $m$-hole with linear size $L/3$ in the center, i.e., hole with $m$ (smooth) boundaries, as shown in Fig.~\ref{fig:fractal_surface_code}(b).  The introduction of the $m$-hole leads to the encoding of an additional logical qubit, with the corresponding logical operators being the $\overline{Z_1}$ string circulating the $m$-hole, and the $\overline{X_1}$ string connecting the hole boundary and the outer $m$-boundary of the surface code, as shown in Fig.~\ref{fig:fractal_surface_code}(f).  Note that the logical string $\lo{X}_1$ intersects with the macroscopic logical string $\lo{Z}$ once or odd number of times (if bending the strings). Hence, the algebraic intersection of the operator support can be  written as 
\be
\mathbf{supp}(\lo{Z}) \cap \mathbf{supp}(\lo{X}_1 )=1, 
\ee
where `$\mathbf{supp}$' represents the support of an operator, and `$\cap$' denotes the algebraic intersection. 
This means, in the case of $\ZZ_2$ topological order, the logical operators $\lo{X}_1$ and $\lo{Z}$ anti-commute, i.e., $\{\lo{X}_1, \lo{Z}\}=0$.   This anti-commutation relation is replaced by the group commutator $\lo{Z} \ \lo{X}_1 \lo{Z}^\dag \lo{X}_1^\dag $$=$$e^{2\pi i/N}$ in the general case of $\ZZ_N$ topological order. In any case, the two operators $\lo{X}_1$ and $\lo{Z}$ fail to commute, i.e., 
$[\lo{X}_1, \lo{Z}] \neq 0$.  The consequence is that, an error in the form of $\overline{X}_1$ changes the logical information stored in the macroscopic logical operator $\lo{Z}$ and hence splits the corresponding topological degeneracy,  which also means the $X$ code distance $d_X$ corresponds to the original macroscopic logical qubit is decreased to the length of $\lo{X}_1$ instead of the original length of $\lo{X}$, i.e., $d_X=L/3$.  Note that the overall code distance of the macroscopic logical qubit is the minimum of the $X$-distance and $Z$-distance, i.e., $d=\text{min}(d_X,d_Z)=L/3$.

We then go to the next iteration, i.e., $SC(3,1, l=2)$, by introducing another eight level-2 $m$-holes with linear size $L/9$ as shown in Fig.~\ref{fig:fractal_surface_code}(c).  This leads to the introduction of another eight logical qubits with the representatives of the logical operators shown in Fig.~\ref{fig:fractal_surface_code}(g), where the logical string $\lo{Z}_i$ encircles the $i^\text{th}$-hole, and the logical string $\lo{X}_i$ connects $m$-holes either to the outer boundaries or to the neighboring $m$-holes. 
% Note that due to the presence of the holes, there are several non-equivalent representatives of the macroscopic logical-$Z$ string.  As shown in Fig.~\ref{fig:fractal_surface_code}(g), we label the leftmost macroscopic string as $\overline{Z}$, while the one next to it is labeled by $\overline{Z}'$.  One can see these two macroscopic logical $Z$-string  differ by three $Z$-strings circulating on the level-2 holes 1, 2, and 3.  Therefore, they can be related to each other as $\lo{Z}'= \lo{Z} \ (\lo{Z}_1 \lo{Z}_2 \lo{Z}_3)$. On the other hand, there is only one equivalent representative of the macroscopic logical string $\lo{X}$. This is because different macroscopic $X$-strings separated by multiple holes can be made equal to each other by multiplying $X$-strings around the $m$-holes, which are equal to  $X$ stabilizers, as illustrated in Fig.~\ref{fig:fractal_surface_code}(g). 
As we can see, any representative of the macroscopic logical-$Z$ strings, such as $\lo{Z}$ in (g), always intersects with some short logical-$X$ strings, such as $\lo{X}_5$ in (g).  In the case of both $\ZZ_2$ and more generally $\ZZ_N$ topological order, we have the following non-commutation relation,  $[\lo{Z}, \lo{X}_5] \neq 0$. This further reduces the $X$-distance $d_X$ of the original macroscopic logical operator in the surface code at level 0 to the length of $\lo{X}_5$, i.e., $d_X=L/9$.

For any actual physical system on a lattice, the iteration continues until the size of the smallest hole reaches the lattice constant of the underlying spin model, i.e., $O(1)$, or more generally the order of the correlation length, i.e.,  $O(\xi)$, if we have nonzero $\xi$ either in a lattice model or continuum topological quantum matter.  Note that in the case of exact solvable models such as the toric code, one has zero correlation length, i.e., $\xi=0$. Since for any topological order, $\xi$ is a constant independent of system size, we can always replace $O(\xi)$ just with $O(1)$ in the case of a lattice model. For the continuum case, if we represent lengths in the unit of correlation length $\xi$, then we can again replace $O(\xi)$ with $O(1)$.  For simplicity, we consider the case that the level-3 hole  [illustrated in Fig.~\ref{fig:fractal_surface_code}(d)] has reached the size of the lattice constant or correlation length, i.e., $L/27 = O(1)$.  As shown in Fig.~\ref{fig:fractal_surface_code}(h), we can pick any representative of the macroscopic logical $Z$ string, denoted by $\lo{Z}$, which will always intersect with a size $O$(1) logical string $\lo{X}_i$.  Therefore, the code distance of the originally  `macroscopic' logical qubit is determined by its $X$-distance, i.e.,  $d=d_X=O(1)$. For the similar reason, any logical qubit in this code has distance $O(1)$.  Due to the absence of a macroscopic code distance, we can say that there is no topological order in this particular fractal geometry.

\begin{figure}[hbt]
  \includegraphics[width=0.6\columnwidth]{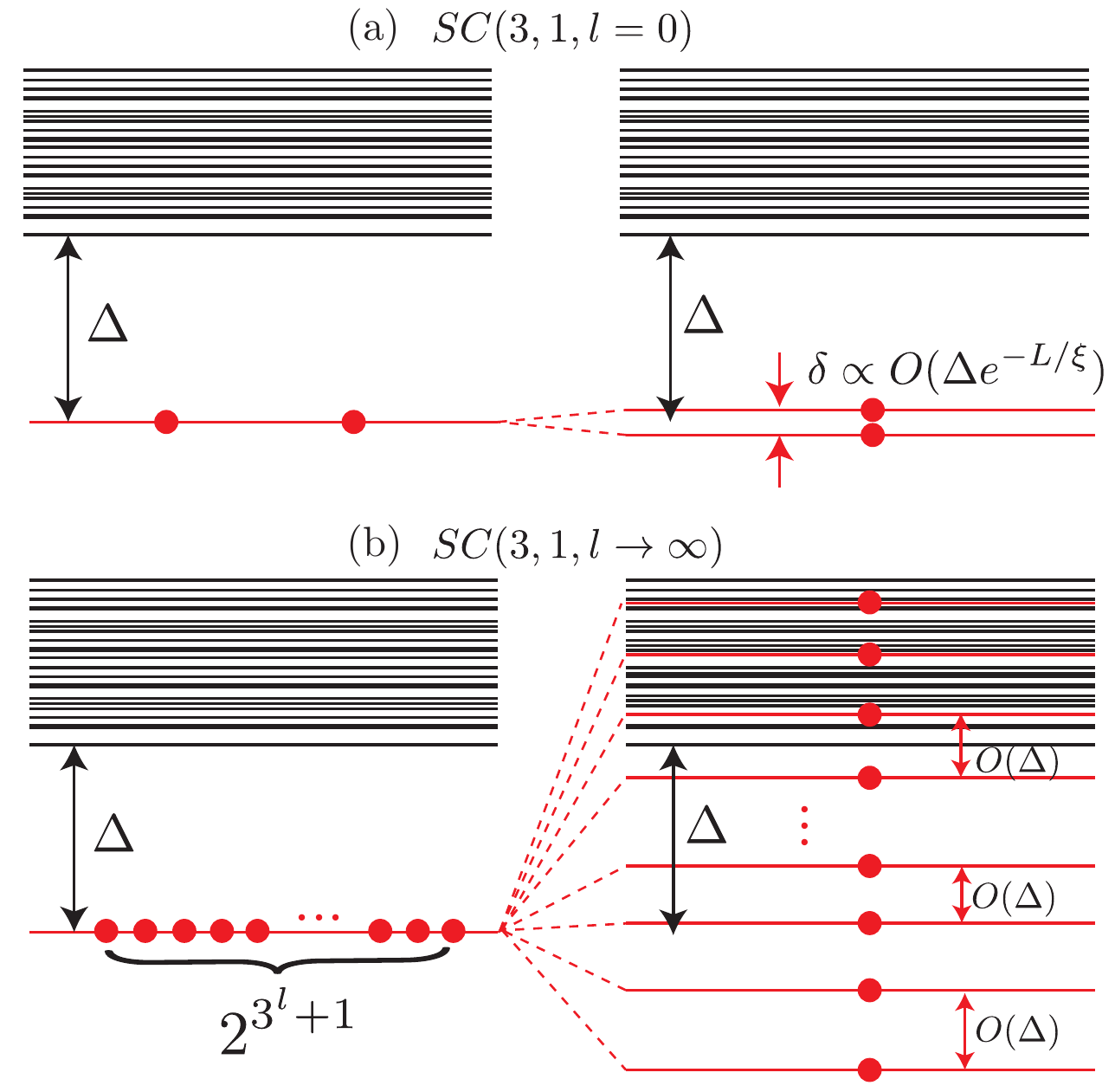}
  \caption{(a) The exponential suppression of the ground-state degeneracy under a perturbation in the Hamiltonian of the conventional surface code defined on $SC(3,1,l=0)$.   (b) On the Sierspiki carpet model defined on $SC(3,1,l \rightarrow \infty)$, a perturbation in the Hamiltonian can lead to the splitting of the extensive ground-state degeneracy at a scale comparable to the gap.}
  \label{fig:degeneracy_splitting}
\end{figure}

We can also comprehend this result in terms of super-selection rules and the robustness of  topological degeneracies. The absence of macroscopic code distance, i.e., $d=O(1)$, tells us that for a local operator  $\mathcal{O}_A$ supported in region $A$ such that $|\textbf{supp}(A)|$$=$$O(1)$, one can get a nonzero off-diagonal matrix element $\boket{\lo{a}}{\mathcal{O}_A}{\lo{b}} =c$ when $\mathcal{O}_A$ is a short-distance logical operator $\lo{X}_i$ connecting neighboring boundaries.  Here, $c$ is a constant independent of the overall system size $L$ and $a \neq b$ label different degenerate ground-state sectors. This is in contrast to the super-selection rule in the presence of topological order in 2D such that $\boket{\lo{a}}{\mathcal{O}_A}{\lo{b}} \propto \delta_{a,b} +   O(e^{-d/\xi})$ with $d=O(L)$,  meaning that the coupling between different ground-state sectors ($a \neq b$) is either exponentially suppressed with system size $L$ when the correlation length $\xi$ is non-zero or always stays zero in the case of zero correlation length, i.e., $\xi=0$. 
% This super-selection rule breaks down when $d=O(1)$. 
This also leads to the exponential suppression of the ground-state degeneracy splitting $O(\Delta e^{-L/\xi})$ in the presence of topological order \cite{Wen:1990wk, kitaev2003, Kitaev:2009ut}, where $\Delta$ is the many-body gap. On the other hand, the degeneracy splitting in the case of Sierpi\'nski carpet model is instead $O(\Delta)$ and hence independent of the linear system size $L$, as illustrated in Fig.~\ref{fig:degeneracy_splitting} and explained in details in App.~\ref{app:degeneracy_splitting}.

\begin{figure}[hbt]
  \includegraphics[width=0.8\columnwidth]{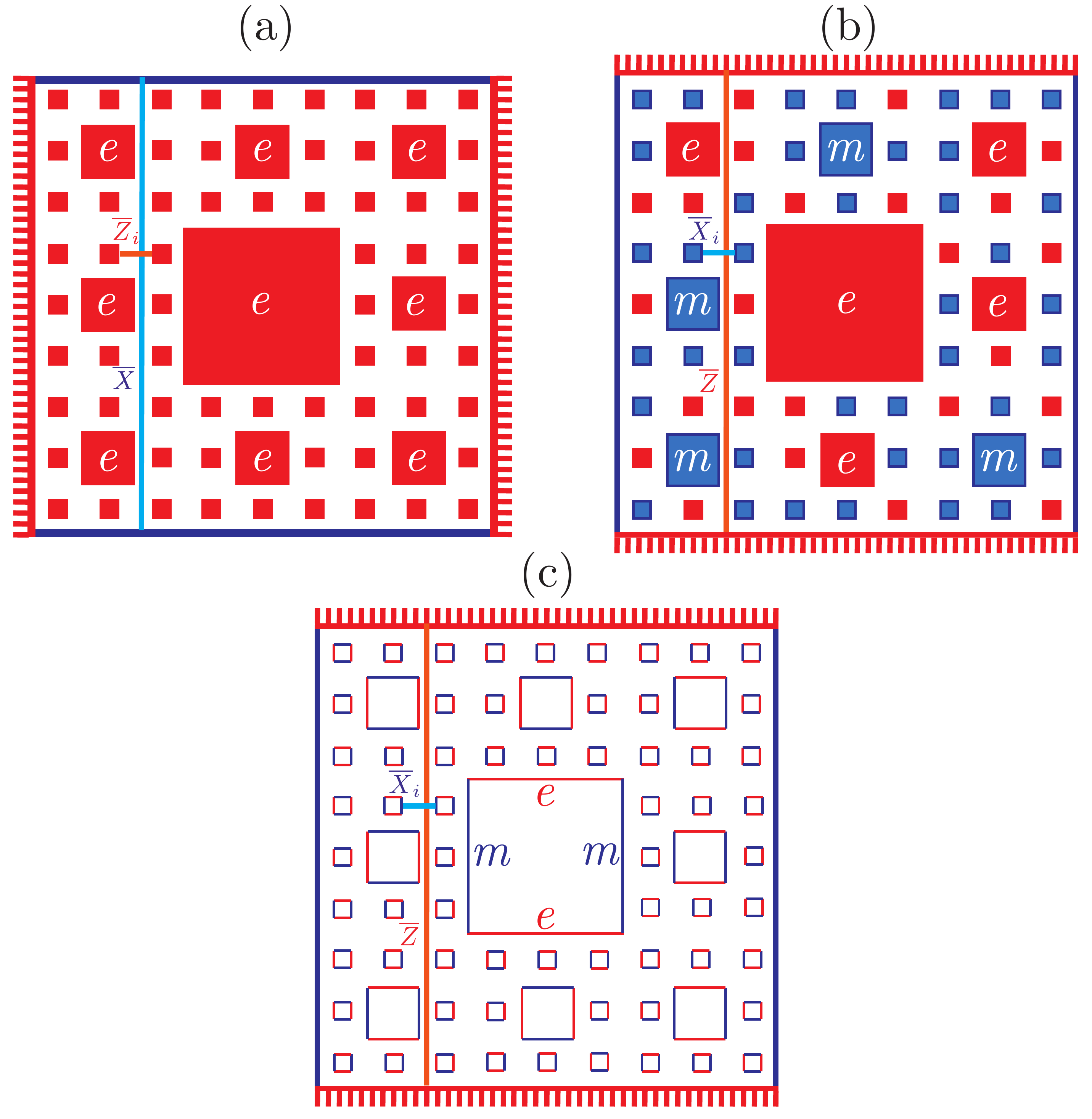}
  \caption{ (a-c)  Three fractal models with only $e$-holes, both $e$ and $m$ holes,  and  holes with alternating $e$- and $m$-boundaries respectively. In all three cases, a macroscopic logical string $\lo{X}$ or $\lo{Z}$ is intersected by a short dual logical string with $O(1)$ length, leading to an $O(1)$ code distance.}
  \label{fig:mixed_holes}
\end{figure}

In all the above discussion, we have shown the absence of topological order on a Sierpi\'nski carpet with $m$-holes.  Due to the $e$-$m$ duality symmetry of the toric code, topological order is also absent on a Sierpi\'nski carpet with $e$-holes. Furthermore, topological order is also absent in the general case that there are holes with both types in the Sierpi\'nski carpet, and even more generally, in the case where a single hole can have multiple types of boundaries.  All of the above situations are illustrated in Fig.~\ref{fig:mixed_holes}.

\begin{figure}[t]
  \includegraphics[width=1\columnwidth]{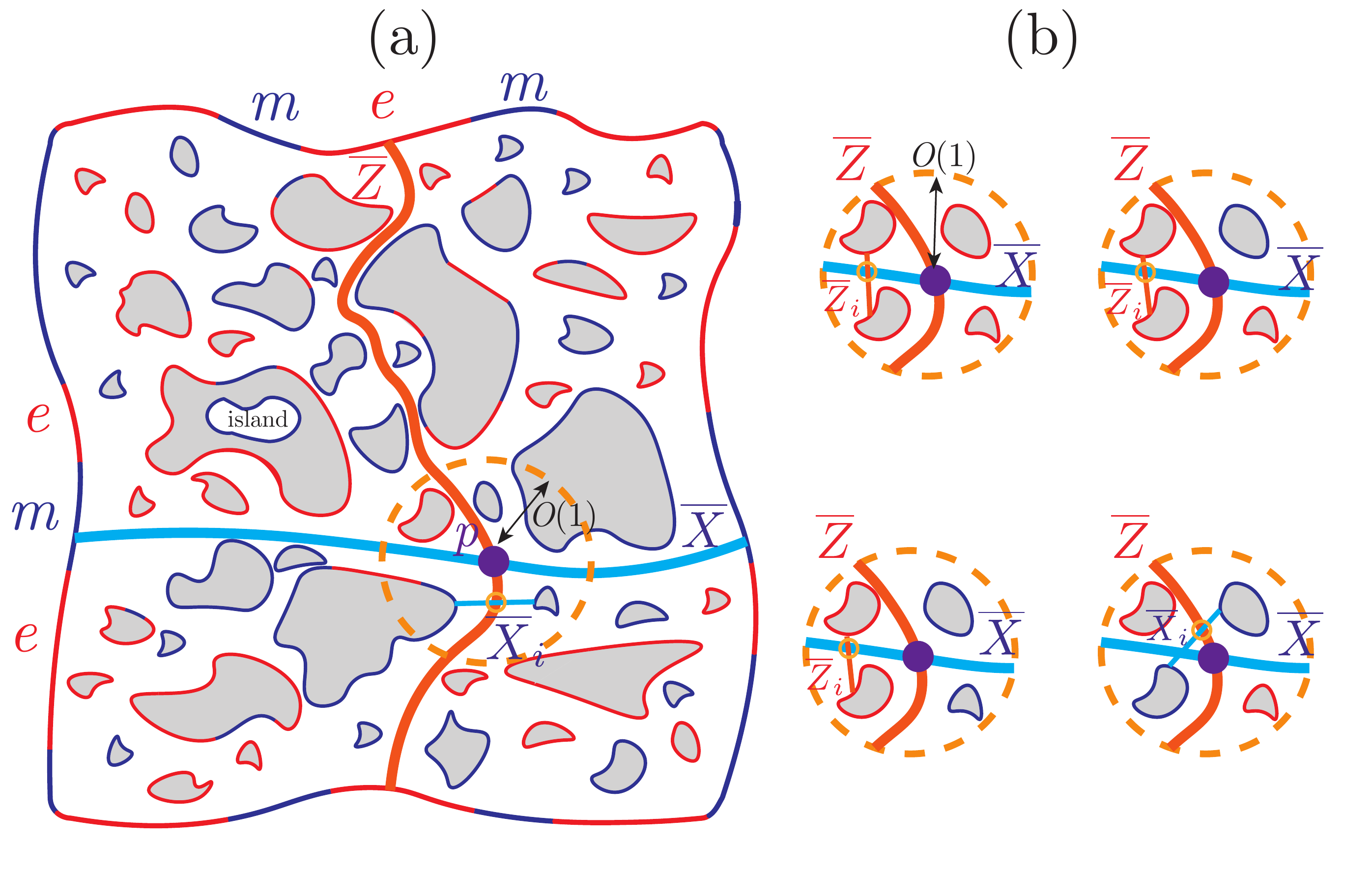}
  \caption{Illustration for the proof of Theorem \ref{theorem:no_go_2D} with a generic random  fractal embedded in $\mathbb{R}_2$.  (a) The holes are colored in grey and with a possible alternating types of   boundaries in general. 
  The holes can also contain `islands' (white) within it which can be simply ignored.  A pair of macroscopic logical string $\lo{X}$ and $\lo{Z}$ intersect at a single point $p$. %According to self-similarity, within $O(1)$ distance from $p$ along any direction (indicated by the dashed circle), there must exists a hole with $O(1)$ linear size or larger. 
  Panel (b) enumerates different scenarios within the $O(1)$ radius from the intersection point $p$ with different configurations of hole boundaries.
  %Either the macroscopic string $\lo{X}$ or $\lo{Z}$ is intersected by an $O(1)$-length short logical string of the dual type leading to an $O(1)$ code distance. Other scenarios can be obtained from the listed ones by symmetry.
  }
  \label{fig:no_go_2D}
\end{figure}

It is obvious that such a no-go result is not limited to a particular type of fractal, in this case, the Sierpi\'nski carpet. We hence state and then prove the following no-go theorem: 
\begin{theorem}\label{theorem:no_go_2D}
	 $\mathbb{Z}_N$ topological order cannot survive on a fractal embedded in a 2D Euclidean space $\mathbb{R}_2$ \footnote{More specifically, we refer to the $(1,1)$-$\ZZ_N$ topological order. This notation will be introduced in Sec.~\ref{sec:nD} for the general theory in $n$D.}.
\end{theorem}
\textbf{Proof:} We consider generic fractals embedded in 2D, including the case of random fractals as illustrated in Fig.~\ref{fig:no_go_2D}.  The holes in the fractal can be simply connected, such as the holes in the Sierpi\'nski carpet, or more generally not simply connected, i.e., circulating some `islands' inside as indicated in Fig.~\ref{fig:no_go_2D}(a). However, one can simplify the situation of non-simply-connected holes to the situation of simply-connected holes just by ignoring the islands inside since they do not affect the states outside.  A generic hole can either have a single $e$- or $m$-boundary or alternating $e$ and $m$ boundaries, as illustrated in Fig.~\ref{fig:no_go_2D}. 

A key property of a fractal is self-similarity.  Therefore, holes at a characteristic linear scale $\lambda$ are separated by a characteristic distance $c\lambda$, where $c$ is a constant independent of the scale $\lambda$. Therefore, holes at $O(1)$ characteristic linear scale are also separated with $O(1)$ distance. We can hence conclude that for any given point in the fractal geometry, when moving radially inside a finite angle $\Delta \theta$ and  within $O(1)$ radius, one must encounter a hole or an outer boundary. This hole could either be a hole at $O(1)$ scale (based on the $O(1)$ characteristic distance between holes at $O(1)$ scale), or a hole at larger scale if it happens to be there. 

Now we consider a pair of dual macroscopic logical strings $\overline{X}$ and $\overline{Z}$ crossing at a single point $p$ as shown in Fig.~\ref{fig:no_go_2D}(a), which is a prerequisite for topological order to exist in this setup.   Then the point $p$ must have a neighboring hole or outer boundary at $O(1)$ distance on each side of string $\lo{X}$ and string $\lo{Z}$. Now we can assign one or multiple boundaries with either $e$- or $m$-type to each of the holes or outer boundaries under consideration.  With any assignment, either $\lo{X}$ will have one hole or outer boundary on each side of the string containing $m$-boundaries within a $O(1)$ distance to $p$,   or $\lo{Z}$ will have one neighboring hole or outer boundary on each side of the string containing $e$-boundaries within a $O(1)$ distance, as illustrated in Fig.~\ref{fig:no_go_2D}(b). This implies either $\lo{X}$ will cross a short logical operator $\lo{Z}_i$ with $O(1)$ length connecting two $e$-boundaries, or $\lo{Z}$ will cross a short logical operator $\lo{X}_i$ with $O(1)$ length connecting two $m$-boundaries. 

Note that in the above discussion we have considered any possible pair of macroscopic logic operators, and the conclusion holds for arbitrary choice of logical basis.   We have also used the fact that the logical operators always need to be a string-like operator, which is guaranteed in the $\ZZ_N$ topological order in 2D.  This is because in this theory, the logical operators are described by the 1st $\ZZ_N$-homology group and any logical operator corresponds to a non-trivial 1-cycle (string-like), which will be explained in details in Sec.~\ref{sec:chain_complex}. 
 
Therefore, we can conclude that the code distance of the corresponding code defined on the fractal is $d = O(1)$, and hence no $\ZZ_N$ topological order exists on a fractal embedded in 2D.      

\nin Q.E.D.

Although our current proof is limited to the $\ZZ_N$ topological order,  the generalization of the no-go theorem to arbitrary types of existing 2D topological orders in the literature supported on fractals with Hausdorff dimension $D_H=2-\delta$ ($\delta>0$) should be straightforward and will be discussed in future. The generalization can be done via extending the Chern-Simons action in Eq.~\eqref{eq:CS} to include cases such as the Non-Abelian TQFT, or more generally via an extension to all 2D T.O.~captured by modular tensor categories \cite{Wen:2015_TO_review}. The essence of the no-go results is that all these 2D T.O.~have string-like logical operators.

\section{Topological order on fractals embedded in 3D}\label{sec:3D}
\subsection{$\ZZ_N$ topological order  in 3D}\label{sec:3Dmodel}
% \arpit{we could specify we are just considering the bosonic 3d toric code even though some ideas could extend to the fermionic toric code}
In three dimensions, we consider a class of topological orders described by a particular type of discrete gauge theory called the BF theory, with the Chern-Simons action being:
\be\label{eq:BF}
S_\text{BF} = \int  \frac{N}{2\pi} b^{(2)} \wedge da^{(1)}.
\ee
Here, the 1-form $a^{(1)}$ and 2-form $b^{(2)}$ are both compact U(1) gauge fields describing the $e$-particle and  $m$-loop (closed string) degree of freedom respectively.   A $\mathbb{Z}_{N}$ fusion rule exists for both $e$-particle and $m$-string respectively. The corresponding gauge group is $G=\mathbb{Z}_{N}$. This theory can describe the deconfined phase of 3D type-II superconductor with a charge-$N$ condensate, or a 3D $\ZZ_N$ spin liquid.  A particle-loop braiding corresponding to a process that carries a particle $e$ in a closed path $l_e$ around a loop $m$ induces a quantized Aharanov-Bohm phase $\frac{2\pi}{N} \cdot \text{Hopf}(m, l_e)$.   Here,  $\text{Hopf}(m, l_e)$ is the Hopf invariant corresponding to the particle worldline $l_e$ and the loop $m$.  This type of topological order is the  $\ZZ_N$ topological order. 

\begin{figure}[hbt]
  \includegraphics[width=1\columnwidth]{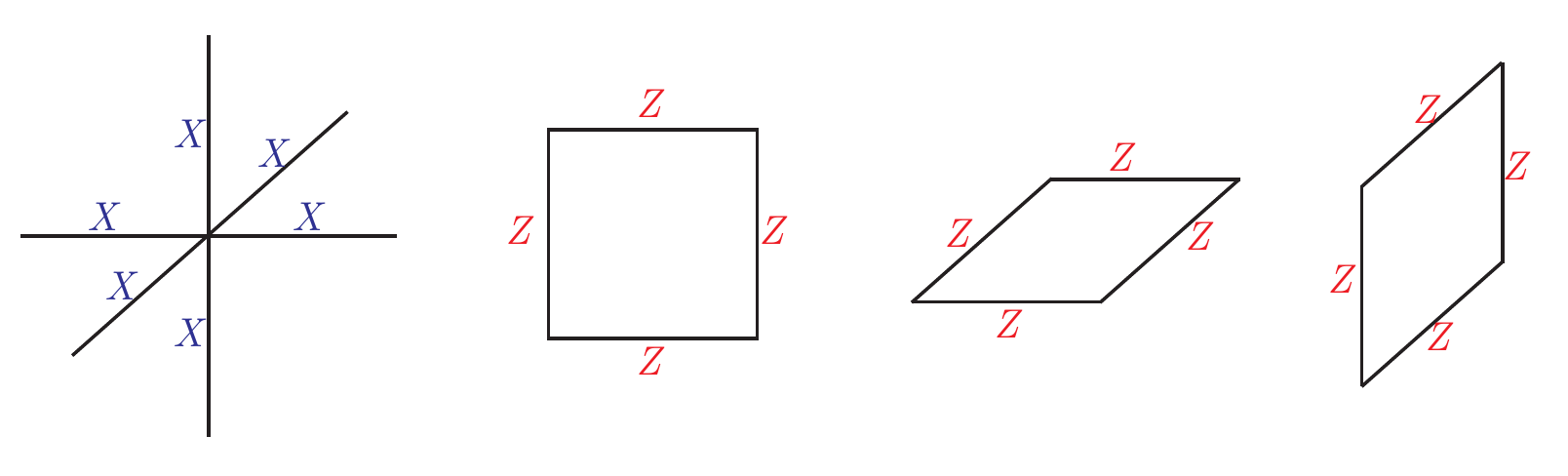}
  \caption{The 6-body vertex operators and three types of 4-body plaquette stabilizers lying in the $yz$, $xy$, and $xz$ planes respectively in the 3D toric code model defined on a cubic lattice.}
  \label{fig:stabilizers_3D}
\end{figure}

In the $N=2$ case, one can construct the following exact-solvable microscopic model, i.e., the $\mathbb{Z}_2$ toric code model on a 3D lattice $\L$, with the following Hamiltonian:
\begin{align}\label{eq:3DTC}
\nonumber H_\text{3DTC}=&-J\sum_v A_v - J\sum_p B_p, \\
\text{with} \quad  A_v=&\Motimes_{j\in \{e_v\}}X_j, \quad B_p=\Motimes_{j\in \{e_p\}}Z_j.
\end{align}
Here, $A_v$ is the vertex stabilizer with all the Pauli-$X$ operators supported on the edges connected to the vertex $v$ (denoted by $\{e_v\}$), while $B_p$ is the plaquette/face stabilizer with all the Pauli-$Z$ operator supported on the edges surrounding the plaquette $p$ (denoted by $\{e_p\}$). 
Violation of $A_v$ corresponds to $e$-particle excitations, while violation of $B_p$ corresponds to $m$-loop excitations as will be explained later in details. 
When defining such a model on the cubic lattice, the vertex stabilizers are 6-body operators and the plaquette stabilizers are 4-body operators on faces lying in $xy$, $yz$, and $xz$ planes respectively, as shown in Fig.~\ref{fig:stabilizers_3D}. Generalization of the above Hamiltonian to the $\ZZ_N$ toric code model is presented in Appendix~\ref{append:ZN}.

\subsection{The TQFT and stabilizer descriptions}

%We start our discussion by considering the excitation picture on the exact-solvable lattice model Eq.~\eqref{eq:3DTC}. As illustrated in  Fig.~\ref{fig:toric_3D_illustration}(a), a pair of $e$-particles can be created by an $Z$-string operator along the string $s_e$ formed by sequence of connected edges, where $e$-particles' locations $(v_1, v_2)$ are on the two boundaries of the string: $(v_1, v_2) = \partial  s_e$.  A $m$-loop excitation along a loop of plaquettes/faces can be created by a membrane operator $X^\otimes$ acting on the edges.  In the TQFT picture, one can consider the membrane operator $X^\otimes$ being  supported on a surface $A$, where the loop excitation $l_m$ forms the boundary of this surface: $A=\partial l_m$. We emphasize that when studying the lattice model Eq.~\eqref{eq:3DTC}, we always consider the lattice analog of continuous strings, loops and membranes (abbreviated as branes from now on). 

We now start with an axiomatic presentation of the TQFT, accompanied with illustrations with the corresponding exact-solvable 3D toric-code model in Eq.~\eqref{eq:3DTC} using the stabilizer language.  We consider a TQFT Hilbert space $\H_\M$ defined on a closed 3-manifold $\M$, which provides a low-energy effective theory of the corresponding topological phase of matter.  The TQFT Hilbert space $\H_\M$ is also isomorphic to the code space $\H_C$ of the quantum error correcting code corresponding to the exact-solvable toric-code model, i.e., $\H_\M  \cong \H_C$.  In the case when the manifold has a boundary $\partial M$, the TQFT Hilbert space (code space) is generalized to $\H_{\M, \partial \M} \cong \H_C$.  

\begin{figure}[hbt]
  \includegraphics[width=1\columnwidth]{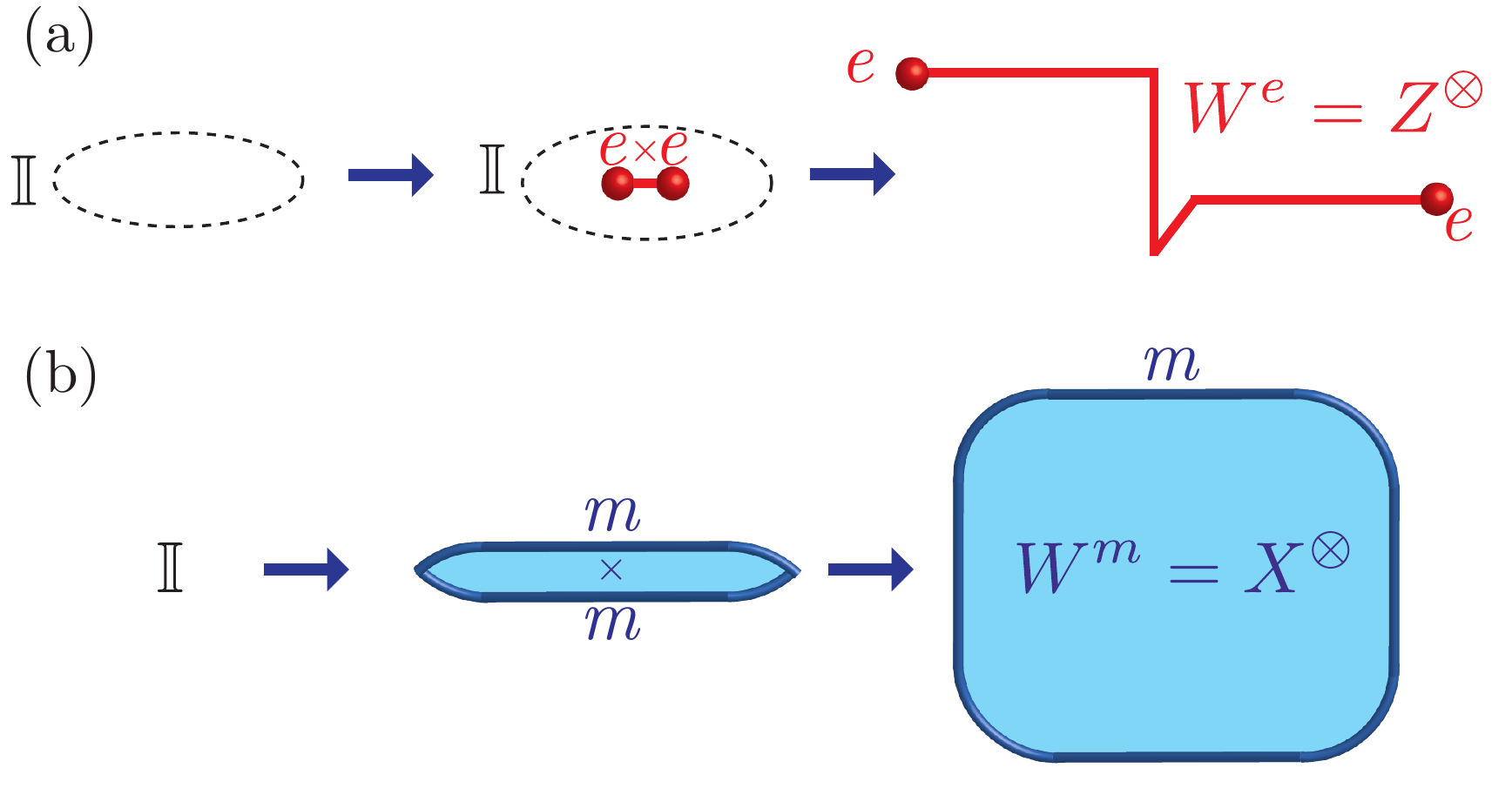}
  \caption{Illustration of the $\ZZ_2$ fusion rules of the $e$-particles and $m$-strings respectively. (a) A pair $e$-particles can be created out of the vacuum $\I$ via a Wilson-line $W^e$ in between, which corresponds to a $Z$-string operator $Z^\otimes $ in the 3D toric code model. (b) A pair of $m$-strings forming an $m$-loop can be created out of the vacuum $\I$ via a Wilson-sheet $W^m$ in between, which corresponds to an $X$-brane operator $X^\otimes$ in the toric code.}
  \label{fig:fusion_rule}
\end{figure}

\begin{figure}[hbt]
  \includegraphics[width=0.6\columnwidth]{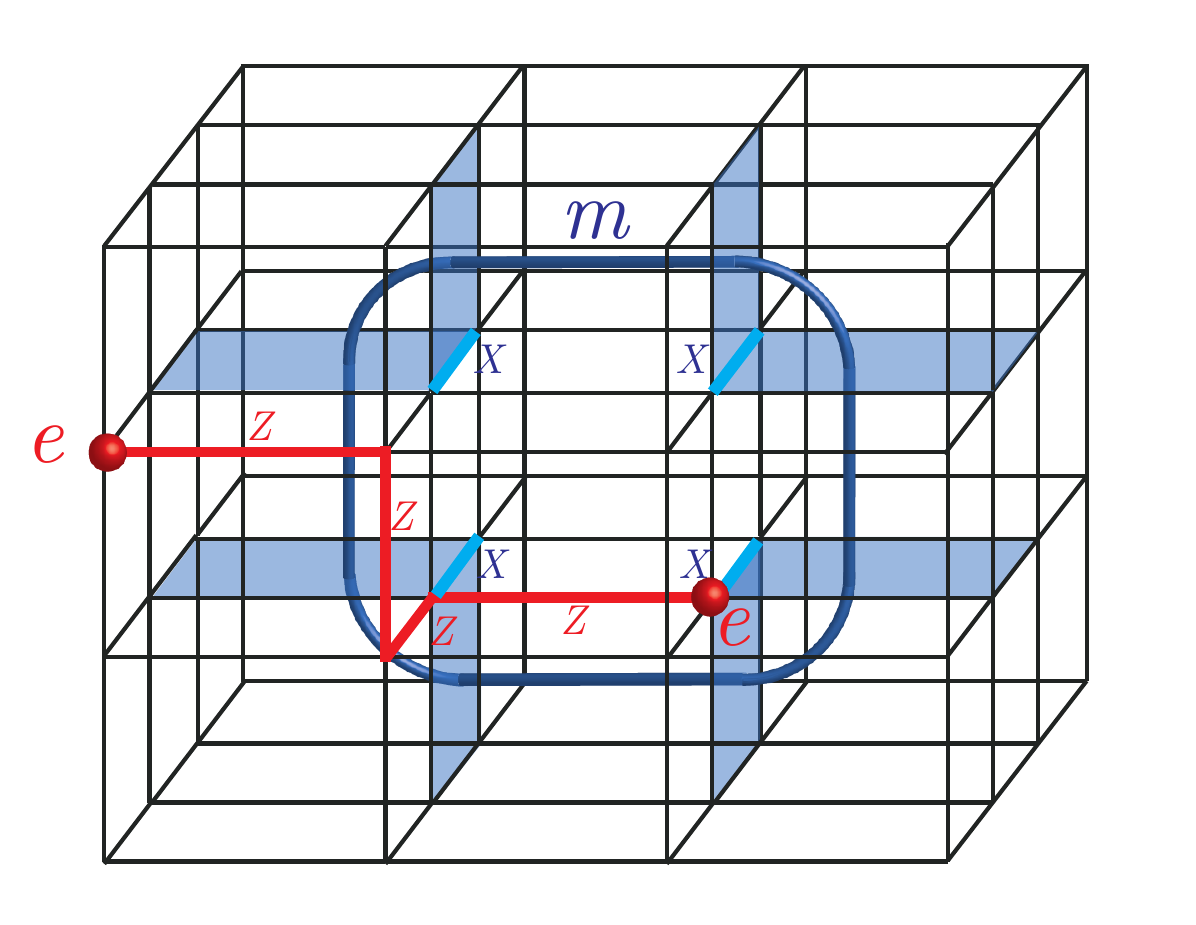}
  \caption{The lattice realization of Fig.~\ref{fig:fusion_rule} with the 3D toric code model. A pair of $e$-particle excitations (red) are created from the ground state by a $Z$-string operator (four Pauli-$Z$'s on the red edges), and the vertex stabilizers occupied by the $e$-particles are violated. An $m$-loop excitations (blue) are created from the ground state by an $X$-brane operator (four Pauli-$X$'s on the blue edges), and the plaquettes (blue) penetrated by the $m$-loop are violated.  }
  \label{fig:toric_3D_illustration}
\end{figure}

In the TQFT and gauge theory picture, one has a $\ZZ_2$ fusion rule in the $N=2$ case for both the $e$-particle and $m$-string:
	\begin{align}
		\raisebox{-0.15cm}{\includegraphics[scale=.60]{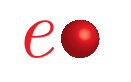}} \times &
		\raisebox{-0.15cm}{\includegraphics[scale=.60]{e-particle.pdf}}
		 = \quad \mathbb{I}  \label{eq:fusion_rule_e} \\
		\raisebox{-0.1cm}{\includegraphics[scale=.60]{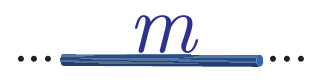}} \times &
		\raisebox{-0.1cm}{\includegraphics[scale=.60]{m-string.pdf}}
		 = \quad \mathbb{I}.
		 \label{eq:fusion_rule_m}
	\end{align}
Here, the $e$-fusion rule  Eq.~\eqref{eq:fusion_rule_e} states that two $e$-particles fuse into the \textit{vacuum} sector $\I$.  This suggests that one can create (annihilate)  a pair of $e$-particles out of (into) the particle vacuum, i.e., the ground state sector, via applying a string operator $W^e$ connecting the two particles which corresponds to the Wilson line (worldline) of particle $e$ as shown in Fig.~\ref{fig:fusion_rule}(a).  This property can also be illustrated with the exact-solvable lattice model Eq.~\eqref{eq:3DTC}. As shown in  Fig.~\ref{fig:toric_3D_illustration}, a pair of vertex excitations on vertices $(v_1, v_2)$ corresponding to two  $e$-particles can be created (annihilated) by a $Z$-string operator $W^e$$=$$Z^\otimes$$:=$$  \Motimes_{j \in s_e} Z_j$ along the string $s_e$ formed by sequence of connected edges, where $e$-particles' locations $(v_1, v_2)$ are on the two boundaries of the string: $(v_1, v_2) = \partial  s_e$.

The $m$-fusion rule Eq.~\eqref{eq:fusion_rule_m} states that two  $m$-string segments can locally fuse into the vacuum sector $\I$. One should consider $m$ and $\I$ as the local charge of this particular string segment.  This fusion rule also implies one can create two semi-circles with $m$ charges and shared ending points (equivalent to a thin $m$-loop) out of the vacuum $\I$ via a membrane operator $W^m$, which corresponds to the Wilson sheet (world-sheet) describing the expansion history of the $m$-loop, as shown in Fig.~\ref{fig:fusion_rule}(b).  Now we can again illustrate this property via the exact-solvable lattice model Eq.~\eqref{eq:3DTC}. As shown in Fig.~\ref{fig:toric_3D_illustration}, a collection of plaquette excitation along a loop of plaquettes/faces (denoted by $l_m$) corresponding to a $m$-loop excitation can be created by a membrane operator $W^m$$=$$X^\otimes$$:=$$\Motimes_{j\in A_m} X_j$ acting on the edges $j \in A_m$. Consistent with the TQFT picture, one can consider the membrane operator $X^\otimes$ being supported on a surface $A_m$, where the loop $l_m$ (location of the $m$-loop excitation) forms the boundary of this membrane: $l_m =\partial A_m$\footnote{In the language of chain complex discussed later in Sec.~\ref{sec:chain_complex}, it is more precise to say $l_m$ is the co-boundary of the surface $A$, i.e., $l_m = \delta A_m$ when considering the model supported on the lattice $\L$ shown in Fig.~\ref{fig:toric_3D_illustration}, while $l_m$  can still be described as the boundary of $A_m$ in the dual lattice $\L^*$.}.

We emphasize that when studying the lattice model Eq.~\eqref{eq:3DTC}, we always consider the lattice analog of continuous strings, loops and membranes (abbreviated as branes from now on). We also note that the Wilson operators $W^e$ and $W^m$ only take the forms of $Z$-string $Z^\otimes$ and $X$-brane $X^\otimes$ in the case of the exact-solvable toric code model Eq.~\eqref{eq:3DTC} but not in the more general case of other models in the same phase described by the same  $\ZZ_N$ gauge theory (TQFT). For brevity, we sometimes use $Z^\otimes$ and $X^\otimes$ to represent the Wilson operators even when discussing the more general case, but the readers should be aware of their limitation to the exact solvable model and replace them with $W^e$ and $W^m$ in the general case of gauge theory and TQFT. 

\begin{figure*}[hbt]
  \includegraphics[width=1.7\columnwidth]{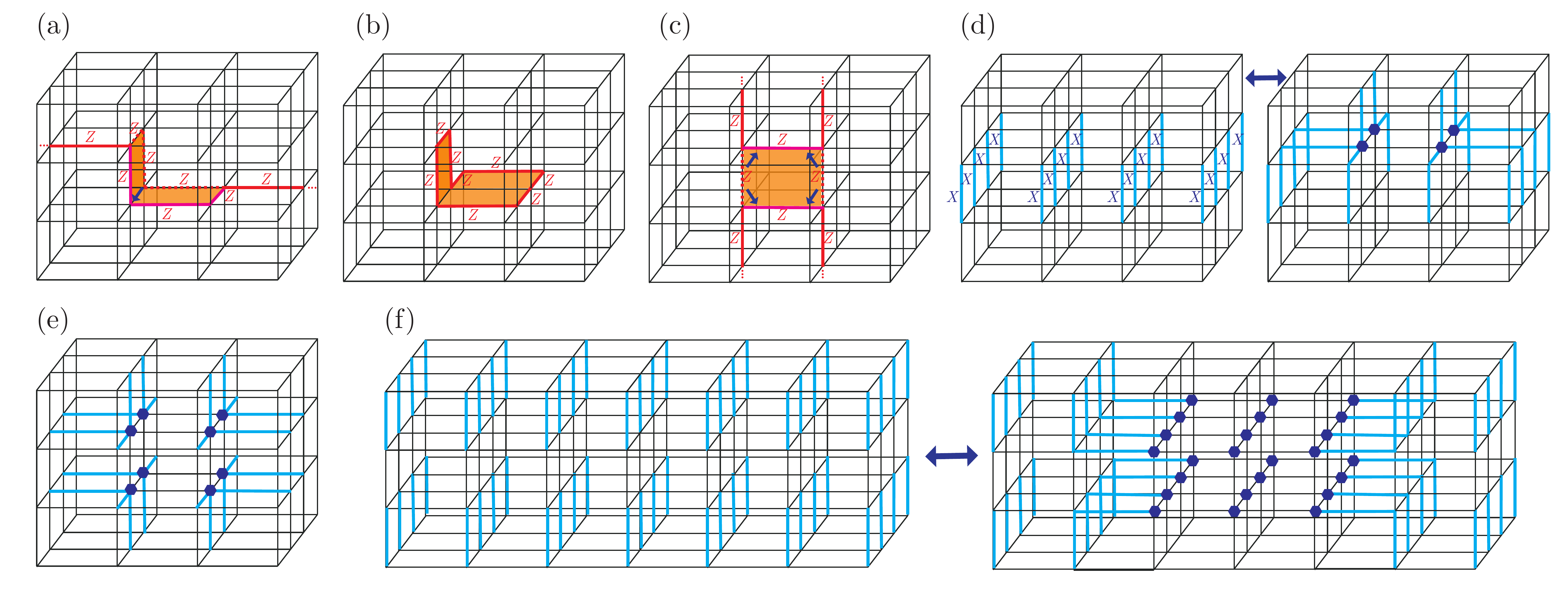}
  \caption{Illustrating the TQFT relations Eqs.~(\ref{eq:string_bending}-\ref{eq:sheet_recoupling}) with the 3D toric code model Eq.~\eqref{eq:3DTC}. (a) Bending of a $Z$-string by multiplying two plaquette $Z$ stabilizers (orange) corresponding to Eq.~\eqref{eq:string_bending}. The dashed lines indicating the previous $Z$-string location before bending.  (b) A closed $Z$-loop can be shrunk to logical identity by multiplying three plaquette $Z$ stabilizers (orange) enclosed by the loop corresponding to Eq.~\eqref{eq:circle}. (c) Recoupling two $Z$-strings via multiplying a plaqette $Z$ stabilizers in between corresponding to Eq.~\eqref{eq:string_recoupling}. (d) Locally bending the $X$-brane by multiplying four vertex $X$ stabilizers (blue hexagons) corresponding to Eq.~\eqref{eq:sheet_recoupling}. (e) A closed $X$-brane topologically equivalent to a sphere can be shrunk into a logical identity by multiplying four vertex $X$ stabilizers corresponding to Eq.~\eqref{eq:sphere}.  (f) Recoupling two $X$-branes by multiplying vertex $X$ stabilizers corresponding to Eq.~\eqref{eq:sheet_recoupling}.  }
  \label{fig:stabilizer_relations_3D}
\end{figure*}

Now we continue with other axioms in the TQFT. The worldlines of $e$-particles $W^e$ and world-sheets of $m$-loops $W^m$ in the 3+1D $\ZZ_2$ gauge theory obey the following diagrammatic relations:
\begin{align}
\raisebox{-0.15cm}{\includegraphics[scale=.60]{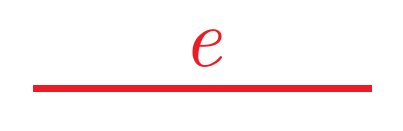}} =&\raisebox{-0.15cm}{\includegraphics[scale=.60]{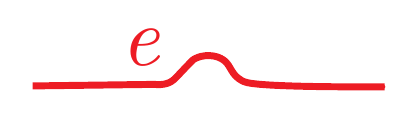}} \label{eq:string_bending}  \\
\raisebox{-0.3cm}{\includegraphics[scale=.60]{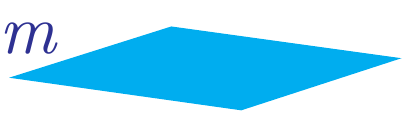}} =& \ \raisebox{-0.3cm}{\includegraphics[scale=.60]{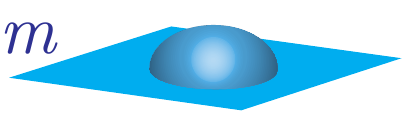}}  \label{eq:sheet_bending}  \\
\raisebox{-0.3cm}{\includegraphics[scale=.60]{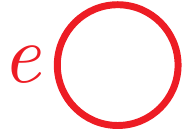}} \quad =& \quad \quad \mathbb{I}  \label{eq:circle}  \\
\raisebox{-0.3cm}{\includegraphics[scale=.60]{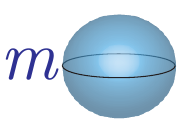}} \ \ \ \ =& \quad \quad \mathbb{I} \label{eq:sphere}  \\
\raisebox{-0.45cm}{\includegraphics[scale=.60]{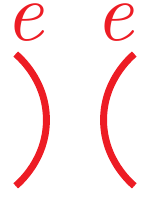}} \quad =& \quad \raisebox{-0.45cm}{\includegraphics[scale=.60]{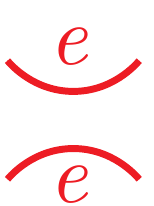}}  \label{eq:string_recoupling} \\
\raisebox{-0.45cm}{\includegraphics[scale=.60]{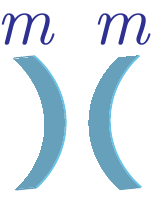}} \quad =& \quad \raisebox{-0.45cm}{\includegraphics[scale=.60]{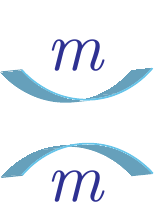}}.   \label{eq:sheet_recoupling}  
\end{align}
Here, Eqs.~(\ref{eq:string_bending},\ref{eq:sheet_bending}) state that the worldline of $e$-particle and the world-sheet of $m$-loop are equivalent under local continuous deformation, i.e., correspond to the same physical observable in the TQFT Hilbert space $\H_\M$ or equivalently the code space $\H_C$.  Eqs.~(\ref{eq:circle},\ref{eq:sphere}) state that a worldline of $e$-anyon and a world-sheet of $m$-loop topologically equivalent to a circle $S^1$ and a sphere $S^2$ respectively are equivalent to the vacuum sector $\I$ in the TQFT Hilbert space $\H_\M$, or equivalently logical identity $I$ in the code space $\mathcal{H}_C$. Note that both the loop and the sphere can be shrunk into nothing under continuous local deformation. 
% The fact that we get a logical identity with coefficient $1$ is a manifestation that both the particle and loop excitations have quantum dimensions $\mathbf{d}_e=\mathbf{d}_m=1$.
This property can also be interpreted in the following relations in the context of exact-solvable model:  $P_C Z^{\otimes_c}  P_C = I$   and  $P_C X^{\otimes_s}  P_C = I$, where $P_C$ represents the projector onto the code space or equivalently the ground-state subspace $\H_C$, $Z^{\otimes_c}$ represents the $Z$-string along a contractible circle, and $X^{\otimes_s}$ represents the $X$-brane wrapping around a sphere.  We note that Eqs.~(\ref{eq:string_bending}, \ref{eq:sheet_bending}) can be derived from the fusion rules Eqs.~(\ref{eq:fusion_rule_e}, \ref{eq:fusion_rule_m}) along with  Eqs.~(\ref{eq:circle}, \ref{eq:sphere}).  Similarly, we can also derive Eqs.~(\ref{eq:circle},\ref{eq:sphere}) from the fusion rules along with the relations Eqs.~(\ref{eq:string_bending}, \ref{eq:sheet_bending}), or equivalently derive the fusion rules from Eqs.~(\ref{eq:string_bending}-\ref{eq:sphere}).

The relations in Eqs.~(\ref{eq:string_recoupling}, \ref{eq:sheet_recoupling}) show that how two worldlines of $e$-particle or two world-sheets of $m$-loop  can be recoupled, which can also be derived from the fusion rules Eqs.~(\ref{eq:fusion_rule_e}, \ref{eq:fusion_rule_m}) along with the relations Eqs.~(\ref{eq:circle}, \ref{eq:sphere}). One can think of these relations as 3+1D generalization  of the 2-2 Pachner moves (F-moves) in 2+1D TQFT, i.e.,  higher-dimensional Pachner moves. 
% In particular, since this is an Abelian TQFT, the fusion of particles and string only have deterministic outcome.  Therefore, the recoupling relations do not involve superposition as in the more general case of non-Abelian TQFTs.
Due to the Abelian nature of this theory, all the relations Eqs.~(\ref{eq:string_bending}-\ref{eq:sheet_recoupling}) can also be understood as equivalence relations in the $\ZZ_2$ homology group, which will be discussed in Sec.~\ref{sec:chain_complex}.  In particular, relations of the worldlines of $e$-particle correspond to the equivalence relations of the 1st-homology, and the relations of world-sheets of the $m$-loop correspond to the equivalence relations of the 2nd-homology. 

Now we show that the relations Eqs.~(\ref{eq:string_bending}-\ref{eq:sheet_recoupling}) can all be understood in the context of the exact-solvable 3D toric-code model Eq.~\eqref{eq:3DTC} in terms of the stabilizer properties.  In particular, the left hand side of these relations differ from the right hand side by multiplying a stabilizer $S_i \in \mathcal{S}$, as illustrated in Fig.~\ref{fig:stabilizer_relations_3D}.  In particular, we can see that the bending of a $Z$-string corresponding to Eq.~\eqref{eq:string_bending} is achieved by multiplying two plaquette $Z$ stabilizers in the example shown in Fig.~\ref{fig:stabilizer_relations_3D}(a).  Similarly, a closed $Z$-loop corresponding to Eq.~\eqref{eq:circle} is equivalent to the multiplication of three plaquette $Z$ stabilizers in the example shown in Fig.~\ref{fig:stabilizer_relations_3D}(b), and hence equals to logical identity, i.e., $P_C Z^\otimes P_C = I $. Furthermore, as an illustration for Eq.~\eqref{eq:string_recoupling}, two parallel vertical $Z$-strings can be recoupled into two U-shaped strings by multiplying a vertical plaquette $Z$ stabilizers in the example shown in Fig.~\ref{fig:stabilizer_relations_3D}(c).  For the worldsheet of $m$-string, we see that the local deformation of $X$-brane corresponding to Eq.~\eqref{eq:sheet_bending} is achieved by multiplying four vertex $X$ stabilizers (highlighted by blue hexagons on the vertices) in the example illustrated in Fig.~\ref{fig:stabilizer_relations_3D}(d).  Furthermore, a closed $X$-brane (sphere) corresponding to Eq.~\eqref{eq:sphere}  is equivalent to the multiplication of four $X$ stabilizers in the example illustrated in Fig.~\ref{fig:stabilizer_relations_3D}(e), and is hence a logical identity, i.e.,  $P_C X^\otimes P_C = I $. Finally, as an illustration for Eq.~\eqref{eq:sheet_recoupling}, two parallel horizontal $X$-branes can be recoupled into two U-shaped branes by multiplying 24 X~stabilizers in between in the example shown in Fig.~\ref{fig:stabilizer_relations_3D}(f).

Besides the basic relations  Eqs.~(\ref{eq:string_bending}-\ref{eq:sheet_recoupling}), we can also derive the following relation for a general configuration of a Wilson brane (world-sheet) operator $W^m$ supported on a genus-$g$ surface $\mathcal{M}^2_g$:
\begin{align}\label{eq:genus-g}
\nonumber & \raisebox{-0.5cm}{\includegraphics[scale=1.6]{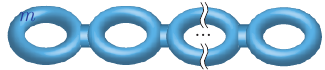}} \\
\nonumber =&  \raisebox{-0.4cm}{\includegraphics[scale=1.6]{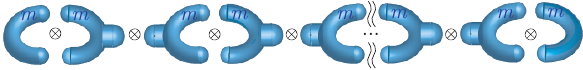}} \\ 
=& \raisebox{-0.5cm}{\includegraphics[scale=1.6]{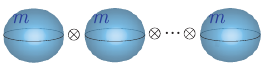}}= \I.  
\end{align}
The first equality essentially utilizes the idea of the pants decomposition of a generic closed surface, which decomposes $\mathcal{M}^2_g$ into a disjoint union of pair of pants in most region and two cylinders on the left and right ends respectively.  A pair of pants is homeomorphic to a 3-punctured sphere, i.e., a sphere with three disks being removed:  $S^2 \setminus (D^2\cup D^2\cup D^2)$, and a cylinder is homeomorphic to a 2-punctured sphere. The Wilson brane operator $W^m$
can hence decomposed as a direct product of the branes acting on the pair of pants and cylinders with their punctures being filled by putting additional `caps' on them.  To reach the decomposition, we have also used the following recoupling relation:
\be\label{eq:cylinder_recoupled}
\raisebox{-0.4cm}{\includegraphics[scale=4]{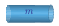}} = \raisebox{-0.4cm}{\includegraphics[scale=4]{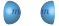}}, 
\ee
which is essentially a variant of the brane recoupling relation Eq.~\eqref{eq:sheet_recoupling} with a rotation of the central axis by $2\pi$.
Now the pair of pants and cylinders with filled punctures are just all equivalent to 2-spheres $S^2$. Therefore, the Wilson brane operator $W^m$ supported on a genus-$g$ surface can just be decomposed into a tensor product of spherical branes, which then equal to the vacuum sector $\mathbb{I}$ or logical identity according to Eq.~\eqref{eq:sphere}.
In the case of exact-solvable toric code model, one can show that this $X$-brane equals the logical identity by multiplying a product stabilizers supported on the volume enclosed by the genus-$g$ surface $\mathcal{M}^2_g$.

\begin{figure}[t]
  \includegraphics[width=0.6\columnwidth]{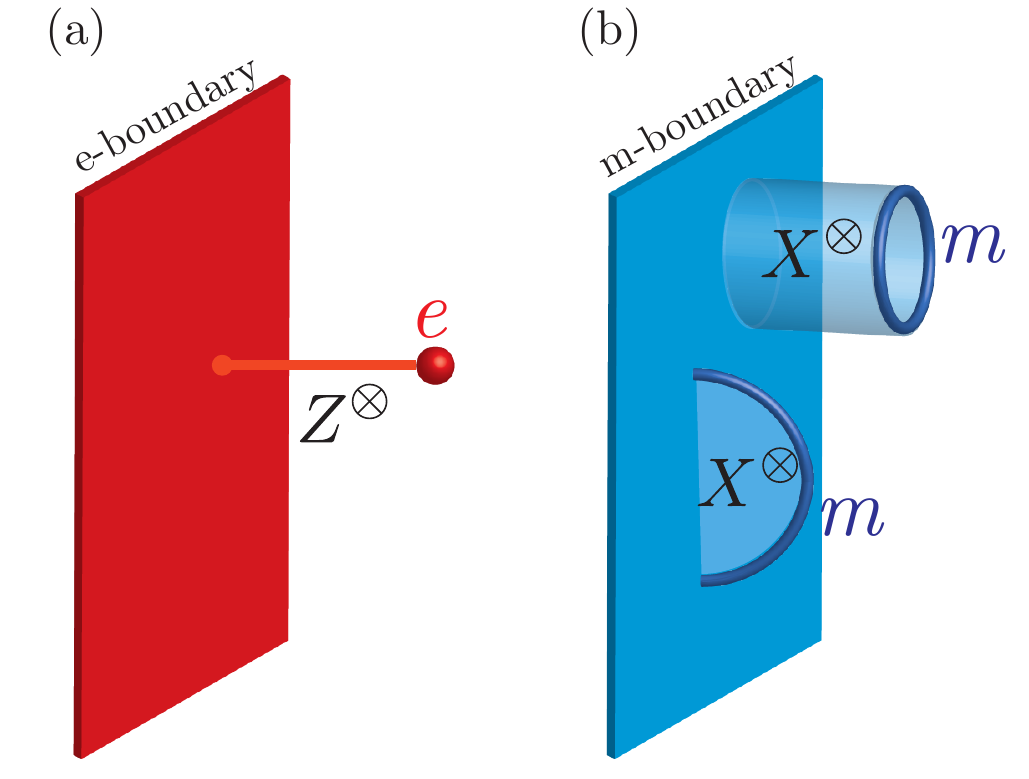}
  \caption{(a) An $e$-particle excitation can condense on the $e$-boundary when applying a Wilson line operator $W^e$, which corresponds to a $Z$-string operator $Z^\otimes$ in the 3D toric code. (b) An $m$-string excitation can condense on the $m$-boundary when applying a Wilson sheet operator $W^m$, which corresponds to a $X$-brane operator $X\otimes$ in the 3D toric code.}
  \label{fig:condensation_picture}
\end{figure}

As a crucial ingredient for understanding the situations in the presence of fractals, we now consider the gapped boundaries in this TQFT.  Similar to the 2+1D case, there are also two types of gapped boundaries in the 3+1D $\Z_N$ gauge theory: the $e$-boundary and $m$-boundary which condense $e$-particle and $m$-string/loop  respectively, as illustrated in Fig.~\ref{fig:condensation_picture}.  In particular, as shown in Fig.~\ref{fig:condensation_picture}(b), a partial $m$-loop can terminate at the $m$-boundary before fully condensing onto the boundary.  The condensation property also means the Wilson line operator (worldline) of the $e$-particle $W^e$ can terminate at the $e$-boundary, while the Wilson brane operator (world sheet) of the $m$-string $W^m$ can terminate at the $m$-boundary. In the case of the exact-solvable toric-code model, these terminated Wilson line and brane operators corresponding to terminated $Z$-strings and $X$-brane respectively, i.e., $W^e=Z^\otimes$ and $W^m=X^\otimes$.  In addition to the relations of Wilson operators in the bulk Eqs.~(\ref{eq:string_bending}-\ref{eq:sheet_recoupling}), we also have the following relations for the Wilson operators on the boundaries:
\begin{align}
\raisebox{-0.8cm}{\includegraphics[scale=.24]{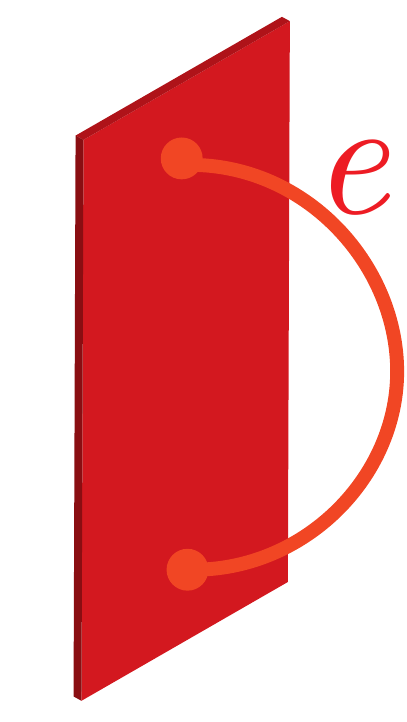}} \quad =& \quad \raisebox{-0.8cm}{\includegraphics[scale=.24]{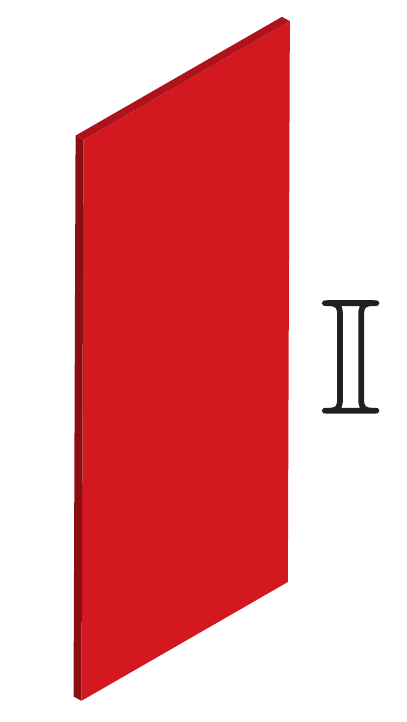}}, \label{eq:relation_boundary_e} \\
\raisebox{-0.8cm}{\includegraphics[scale=.24]{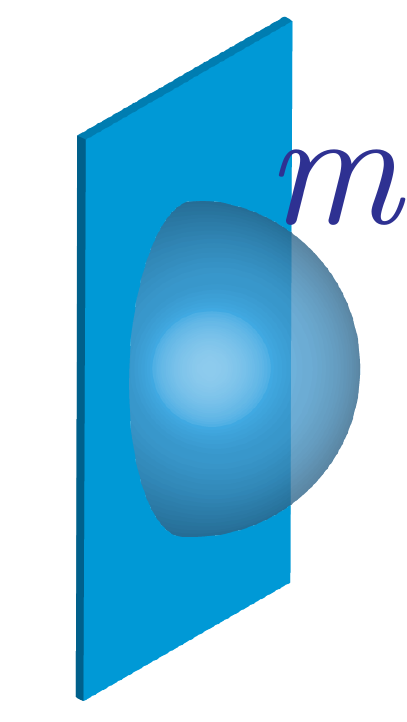}} \quad =& \quad \raisebox{-0.8cm}{\includegraphics[scale=.24]{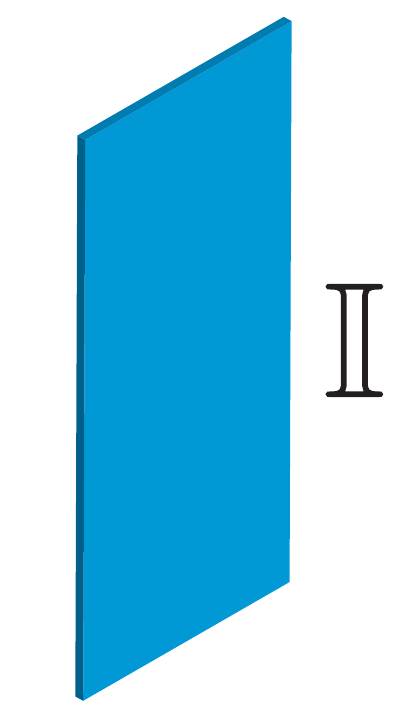}}, \label{eq:relation_boundary_m}
\end{align}
which states that the worldline of $e$-particle $W^e$ and the worldsheet of $m$-string $W^m$ attached to the $e$- and $m$-boundaries respectively can be completely absorbed into the boundary and are hence equivalent to the vacuum sector or logical identity.

\begin{figure}[t]
  \includegraphics[width=1\columnwidth]{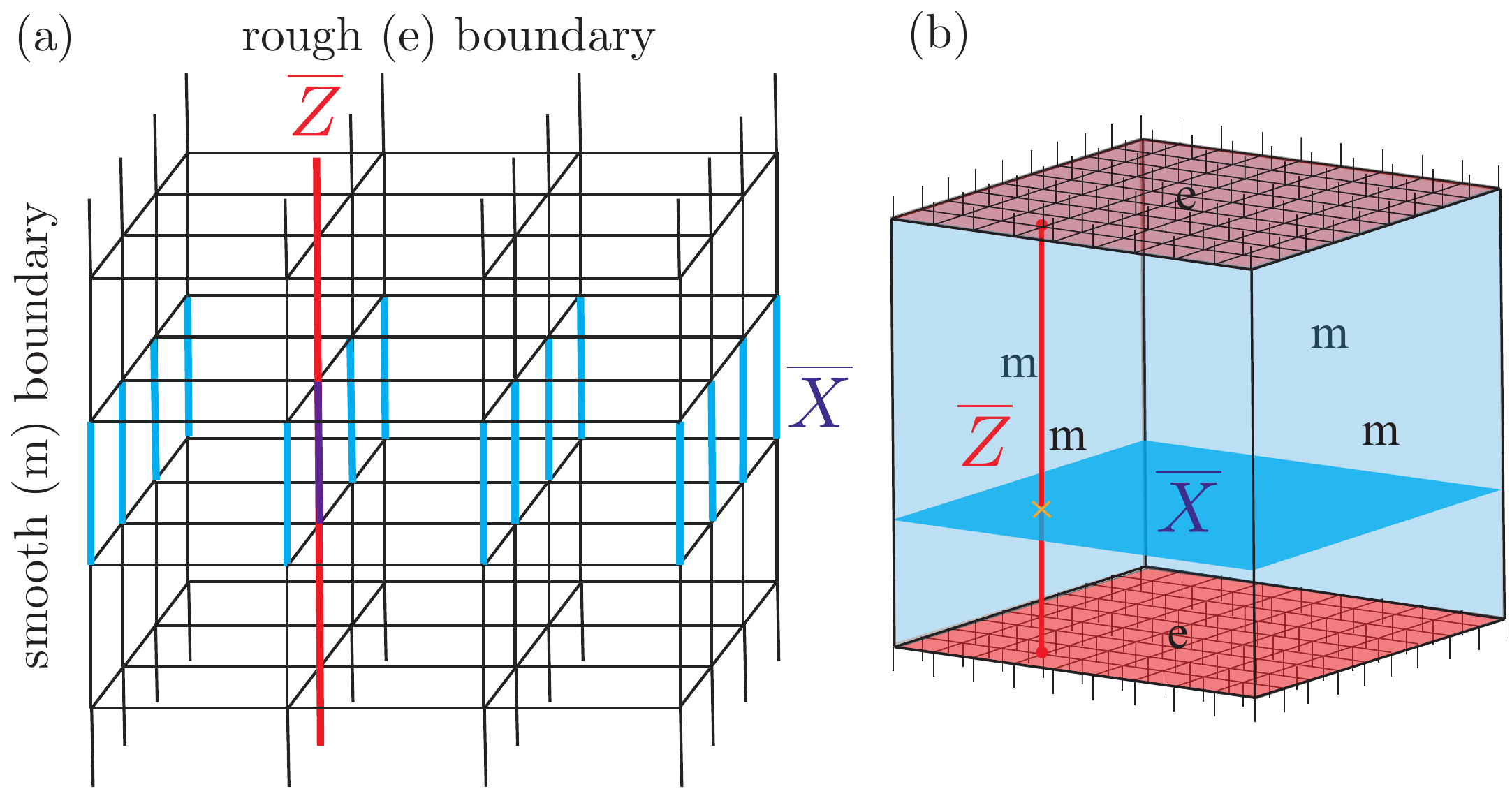}
  \caption{The logical $Z$-string and the logical $X$-brane operators are illustrated on the discrete  lattice (a) and continuous TQFT (b) pictures respectively.  They intersect on a single edge (purple) in the lattice picture or a single point (yellow cross) in the TQFT picture. The dangling edges on the top and bottom faces (red) in (b) symbolize the rough boundaries, and will sometimes be omitted in the later figures for simplicity.}
  \label{fig:surface_code_3D_logical}
\end{figure}

Now we consider the microscopic realizations of these gapped boundaries in the 3D toric code model. As illustrated in Fig.~\ref{fig:surface_code_3D_logical}(a), we choose the top and bottom faces to be $e$-boundaries.  This can be achieved by removing all the $X$-type boundary stabilizers such that they will not be violated when an $e$-particle disappear on the boundary.  These boundaries are also called \textit{rough boundaries} since dangling edges are sticking out of the boundaries.  The boundary stabilizers are 3-body $Z$-type stabilizers, formed by two dangling edges perpendicular to the boundary and one edge parallel to the boundary.   The four side faces are chosen to be $m$-boundaries, which can be achieved by removing all the $Z$-type boundary stabilizers such that they will not be violated when an $m$-string disappear on the boundary.  These boundaries are also called \textit{smooth boundaries} since the boundaries are formed by smooth surfaces without dangling edges. The boundary stabilizers are 5-body $X$-type stabilizers, formed by four edges parallel to the boundary and one edge going inward.  

The setup with the particular boundary choice in Fig.~\ref{fig:surface_code_3D_logical} is often called the 3D surface code.  This code contains a logical-$Z$ string operator $\lo{Z}=\lo{W}^e$  corresponding to the Wilson line of $e$-particle in the TQFT picture. In the corresponding exact-solvable lattice model,  this logical-$Z$ operator corresponds to Pauli-$Z$'s supported on a string $s_e$, i.e.,  $\lo{Z}:=\Motimes_{j\in s_e}X_j$.  The model also contains a logical-$X$ brane operator $\lo{X}=\lo{W}^m$ corresponding to the Wilson brane of the $m$-string $W^m$. In the exact-solvable model, this logical-$X$ operator corresponds to Pauli-$X$'s supported on the brane $A_m$, i.e., $\lo{X}:=\Motimes_{j\in A_m}X_j$. Both the lattice and TQFT pictures of these logical operators are shown in Fig.~\ref{fig:surface_code_3D_logical}(a) and (b) respectively.

% In the following, we focus on the situation of the 3D toric code with boundaries, i.e., the 3D surface code. The generalization of the case on the 3-torus (with periodic boundary conditions) is straightforward.  In particular we consider the models defined on a cubic lattice. 

\subsection{The algebraic topology description: introduction of chain complexes and homology theory}\label{sec:chain_complex}

To describe the more general situations of three- and higher-dimensional topological order in a more concise way, here we introduce the mathematical framework of chain complex.  We focus on the explanation of the $\Z_2$ chain complexes, while the generalization to $\Z_N$ chain complex is straightforward and also discussed below.  

The lattice $\mathcal{L}$ considered in this work is, mathematically speaking, an $n$-dimensional cell complex. We call the $i$-dimensional elements of the cell complex \textit{$i$-cells}, denoted by $e^{(i)}_\alpha$: 0-cells are vertices, 1-cells are edges, 2-cells are faces, 3-cells are volumes and so on. We now define a vector space over $\Z_2$:  $C_i$$=$$C_i(\mathcal{L}; \Z_2)$, which is an Abelian group under component-wise addition modulo 2 and is hence called a \textit{chain group}. The element of this vector space $c_i \in C_i$ is called \textit{$i$-chain}, defined as $c_i=\sum_\alpha z_\alpha e^{(i)}_\alpha$, where $z_\alpha \in \{0, 1\}$ is a $\Z_2$ coefficient. The generalization to the $\Z_N$ and $\Z$ cell complex is straightforward by promoting $z_\alpha$ to $\Z_N$ and $\Z$ (integer) coefficients respectively, with the corresponding chain group denoted by $C_i(\mathcal{L}; \Z_N)$ and $C_i(\mathcal{L}; \Z)$. Note that the usual homology group considered in the general cases corresponds to the homology group over $\Z$, i.e., $C_i(\mathcal{L}) \equiv C_i(\mathcal{L}; \Z)$. In the following discussion, the notation $C_i$ apply to all the three cases ($\ZZ_2$, $\ZZ_N$ and $\ZZ$) mentioned above. 

We now define the \textit{boundary map} $\partial_i$ as:
\be
\partial_i C_i \rightarrow C_{i-1}, 
\ee 
which can also be abbreviated  as $\partial$ for simplicity. Here, $\partial_i c_i$ is an $(i-1)$-chain which is the boundary of $c_i$.  The boundary map satisfies the following identity
\be\label{eq:partial_square}
\partial_{i} \circ  \partial_{i+1} = 0,
\ee
which can also be abbreviated as $\partial^2=0$, i.e., the boundary of a boundary is zero.  

To define the homology group which describes the logical operators, we introduce the following \textit{chain complex}: 
\be\label{eq:exact_sequence_nD}
C_{i+1} \xrightarrow[]{\partial_{i+1}} C_i \xrightarrow[]{\partial_i} C_{i-1}.
\ee
A chain $c_i$ is called a \textit{cycle} if it is in the  kernel of the boundary map $\partial_i$, namely $\partial_i c_i =0$.  We define the subgroup of $i$-cycles as $Z_i:=\text{Ker} (\partial_i) \subset C_i$.  We call an $i$-chain $c_i$ \textit{boundary}, if there exists an $(i+1)$-chain $c_{i+1}$ such that $c_i$$=$$\partial_{i+1}c_{i+1}$. This means that $c_i$ is in the image of the boundary map $\partial_{i+1}$ i.e., $c_i \in \text{Img}(\partial_{i+1})$. We define the subgroup of $i$-boundaries as $B_i := \text{Img}(\partial_{i+1})\subset C_i$. One can see that the boundary $c_i$ is also an $i$-cycle, since $\partial_i c_i=(\partial_i \circ \partial_{i+1})c_{i+1} =0$ due to Eq.~\eqref{eq:partial_square}. We hence have $B_i \subset Z_i$. We also call the boundary $c_i$ a \textit{trivial cycle} since it is contractible.  The trivial cycles generate continuous deformation of a particular non-contractible cycle which forms a homology class.   We hence define the $i^\text{th}$-\textit{homology group} $H_i$ with its elements the corresponding homology class:
\be\label{eq:homology_def}
H_i:=Z_i/B_i =\text{Ker} (\partial_i)/\text{Img}(\partial_{i+1}),
\ee    
i.e., formed by the cycles quotient the boundaries.  

To represent the $X$-type logical operators later, we also introduce a dual description: the \textit{cochain complex}.  
We define the dual vector space of $C_i$ as the abelian group $C^i$.  The elements in this group $c^i \in C^i$ is called \textit{cochain}. Throughout this paper, the chain and co-chain are considered isomorphic to each other, i.e., $C_i \cong C^i$.
We also define the dual of the boundary map, i.e., the \textit{coboundary map} $\delta^i$ as:
\be
\delta^{i} C^i \rightarrow C^{i+1},
\ee 
which can also be abbreviated  as $\delta$ for simplicity. The co-boundary map satisfies the following identity:
\be\label{eq:delta_square}
\delta^{i} \circ  \delta^{i-1} = 0.
\ee
Since both $\delta$ and $\partial$ are linear maps between vector spaces, one can define their transpose as the induced maps on the dual vector spaces, and we get the following relation:
\be
\delta^i = \partial^T_{i+1},
\ee
which can also be interpreted as the transpose relation of their matrix representation.  This relation also implies that the rank of these two linear maps are the same:
\be\label{eq:rank_equality}
\rank(\delta^i) = \rank(\partial_{i+1}).
\ee
Now we introduce the \textit{cochain complex} as the dual of the chain complex in Eq.~\eqref{eq:exact_sequence_3D}:
\be\label{eq:dual_exact_sequence_nD}
C^{i+1} \xleftarrow[]{\delta_{i}} C^i \xleftarrow[]{\delta_{i-1}} C^{i-1}.
\ee
We then define the subgroup of \textit{$i$-cocycles} as $Z^i$$:=$$\text{Ker} (\delta^i) $$\subset$$C^i$. We also define the subgroup of \textit{$i$-coboundaries} as $B^i$$:=$$\text{Img}(\delta^{i-1})$$\subset$$C^i$.  We can hence define the $i^\text{th}$-\textit{cohomology group} as 
\be\label{eq:cohomology_def}
H^i:=Z^i/B^i =\text{Ker} (\delta^i)/\text{Img}(\delta^{i-1}).
\ee

Now we apply the homology theory to the description of topological orders and codes. As the example studied in this section, we consider the $\ZZ_2$ (or in general $\ZZ_N$) topological order in 3D is supported on a 3-manifold $\M^3$ with a  corresponding cellulation $\L$. We hence have the following chain complex associated with $\L$:
\be\label{eq:exact_sequence_3D}
 C_2 \xrightarrow[]{\partial_2} C_1 \xrightarrow[]{\partial_1} C_0,
\ee
as well as its dual cochain complex:
\be\label{eq:dual_exact_sequence_3D}
C^2 \xleftarrow[]{\delta^1} C^1 \xleftarrow[]{\delta^0} C^0,
\ee
with the identification $C_i \cong C^i$. 
We put qubits or more generally $N$-level qudits on the edges (1-cells) of $\L$. The physical Hilbert space is hence $\H$$=$$\mathbb{C}^{C_1(\L; \ZZ_N)}$$=$$(\mathbb{C}^N)^{\otimes |E|}$, where one sets $N=2$ for the qubit ($\ZZ_2$) case. The total number of qubits (qudits), i.e., the total number of edges $|E|$, equals the vector-space dimension or equivalently the rank of the 1-chain group, i.e.,  $|E|=\dim(C_1(\L))=\rank(C_1(\L))$ \footnote{Note that the chain group is just a vector space over $\ZZ_2$ (or more generally $\ZZ_N$) as introduced before.}.  

For the corresponding $\ZZ_2$ toric code model in 3D, we associate the $X$ stabilizers with the 0-cells (vertices), and the $Z$ stabilizers with the 2-cells (faces/plaquettes). More concretely, each $Z$ stabilizer is supported on the boundary of a face/plaquette: $B_p=\Motimes_{j\in \partial_1 p}  Z_j$ (where $p$ labels the face/plaquette), while each $X$ stabilizers is supported on the coboundary of a vertex: $A_v=\Motimes_{j\in \delta^0  v}X_j$. 

% We can hence express the $X$- and $Z$ stabilizer subgroups as:
% \begin{align}
% \non \SS_Z=&\{\Motimes_{j\in c_1} Z_j | c_1 \in B_1 = \text{Img}(\partial_{2}) \},  \\ \SS_X=&\{\Motimes_{j\in c^1} X_j | c^1 \in B^1 = \text{Img}(\delta^{0}) = \text{Img}(\partial^T_{1}) \},  
% \end{align}
% which are isomorphic to the 1-boundary and 1-coboundary groups $B_1$ and $B^1$.  We can also define the $X$- and $Z$-type parity matrices $S_X$ and $S_Z$ \footnote{The parity matrix corresponds to the $\ZZ_2$ case, while for the $\ZZ_N$ case the entries of the matrices are defined modulo $N$ and have three possible values ${0, 1, -1}$.} to encode the support of the $X$- and $Z$ stabilizers, which have the following correspondence with the matrix representation of the (co)boundary operators: $S_Z$$=$$ \partial_2, S_X$$=$$\delta^0$$=$$ \partial^T_1$.   Now we can see that the condition in Eq.~\eqref{eq:partial_square} leads to the following matrix  relation:
% \be
% \partial_1 \partial_2 = S_X^T S_Z =0,
% \ee
% which implies the commutation relation Eq.~\eqref{eq:commutation} between the two types of stabilizers.  

For a system defined on a closed 3-manifold such as a 3D torus $T^3$, the logical-$Z$ string operators $\lo{Z}([c_1])$$=$$ \Motimes_{j\in [c_1]} Z_j  $ are  supported on a class of non-trivial 1-cycles belonging to the 1st-homology group: $[c_1] \in H_1(\L)=\text{Ker} (\partial_1)/\text{Img}(\partial_{2})$.  The logical-$X$ brane operators $\lo{X}([c^1])$$=$$ \Motimes_{j\in [c^1]} X_j$ are  supported on a class of non-trivial 1-cocycles belonging to the 1st-cohomology group:  $[c^1] \in H^1(\L) = \text{Ker} (\delta^1)/\text{Img}(\delta^{0})$. 

As an alternative way to describe logical-$X$ brane operators, we can also consider using the dual cell complex (lattice) of $\L$.  The dual cell complex $\L^*$, which is a dual cellulation of an $n$-dimensional manifold $\mathcal{M}^n$, is obtained by replacing any $i$-cell in $\L$ with a $(n-i)$-cell, leading to the following identification of the chain groups: $C_i \cong C^*_{n-i} \cong C^i$. One hence get the following chain complex on $\L^*$ in 3D as a dual description of Eq.~\eqref{eq:exact_sequence_3D}:
\be
C^*_1 \xleftarrow[]{\partial^*_2} C^*_2 \xleftarrow[]{\partial^*_3} C^*_3.
\ee
The logical-$X$ brane operators $\lo{X}([c^*_2])$$=$$ \Motimes_{j\in [c^*_2]} X_j$ are supported on a class of non-trivial 2-cycles on the dual complex (lattice) $\L^*$ belonging to the 2nd-homology group: $[c^*_2]$$\in$$ H_2(\L^*)$$=$$\text{Ker} (\partial^*_2)/\text{Img}(\partial^*_{3})$. 

% The ground-state subspace (code space) of the $\mathbb{Z}_2$ topological order in 3D can hence be expressed as 
% \be\label{eq:ground_space}
% \mathcal{H}_C= \mathbb{C}^{H_1(\mathcal{L}; \mathbb{Z}_2)} = \mathbb{C}^{H^1(\mathcal{L}; \mathbb{Z}_2)} = \mathbb{C}^{H_2(\mathcal{L^*}; \mathbb{Z}_2)}.
% \ee
% Note that the three equalities here correspond to different logical basis choices for the code space: the first one chooses the logical-Z basis, while both the second and the third choose the logical-X basis. The second one uses the cohomology representation in the original complex (lattice) $\L$, while the third one uses the homology representation in the dual complex (lattice) $\L^*$.
% These equivalent basis choices imply the following isomorphism between the corresponding (co)homology groups:
% \be\label{homology_isomorphism}
% H_1(\mathcal{L}; \mathbb{Z}_2) \cong H^1(\mathcal{L}; \mathbb{Z}_2) \cong H_2(\mathcal{L^*}; \mathbb{Z}_2).
% \ee
 
In the presence of gapped boundaries, the logical operators are associated with the relative homology groups $H_i(\L, \B_a)$, where $\B_a$ stands for a subcomplex corresponding to the gapped boundary of type $a$ \cite{Bravyi:1998uq}. Simply speaking, the relative homology can be considered as the absolute (ordinary) homology of a modified complex ($\L'$) of the original complex  $\L$ such that the corresponding boundary subcomplex $\B_a$ is identified into a single point.  Mathematically this can be expressed as:
\be
H_i(\L, \B_a)  \cong H_i(\L / \B_a)  \quad (\text{for} \  i>0),
\ee
where $\L / \B_a$ stands for the quotient space equivalent to the modified complex $\L'$. More precise definition and detailed discussion of the relative homology and its correspondence with gapped boundaries can be found in App.~\ref{append:relative_homology}. In the example of a 3D surface code, the logical-$Z$ string operator $\lo{Z}([c_1])$ is associated with the non-trivial relative 1-cycle  belonging to the 1st relative homology group, i.e., $[c_1]$$\in$$ H_1(\L, \B_e)$, where $\B_e$ stands for the $e$-boundary. Note that relative $i$-cycles  refer to cycles terminated on the boundaries (see explanation in App.~\ref{append:relative_homology}), in contrast to the absolute (ordinary) $i$-cycles in the bulk as introduced above in the case of closed manifolds.  On the other hand, the logical-$X$ brane operator $\lo{X}([c^*_2])$ is associated with non-trivial  relative 2-cycles belonging to the 2nd relative homology group on the dual complex (lattice) $\L^*$, i.e., $[c^*_2]$$\in$$ H_2(\L^*, \B_m^*)$, where $\B_m^*$ stands for the dual  subcomplex corresponding to the $m$-boundaries.

\begin{figure*}[t]
  \includegraphics[width=1.8\columnwidth]{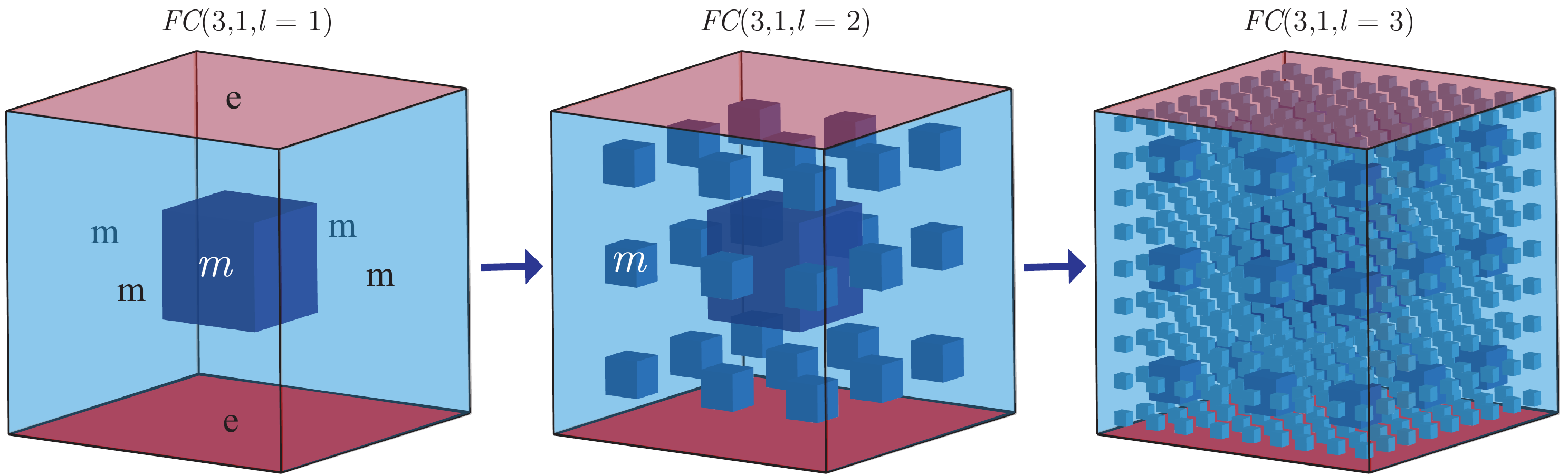}
  \caption{A 3D fractal surface code defined on the  fractal cube geometry $FC(3,1)$  (at level $l=$ 1,2 and 3). The illustrated model contains only  $m$-holes (blue). The outer boundaries are the same as those of the 3D surface codes, with the top and bottom faces being $e$-boundaries, while the rest are $m$-boundaries.}
  \label{fig:fractal_cube_m_hole}
\end{figure*}

Now we can also relate the code distance $d$ with the combinatorial  systoles of a cell complex $\L$ \cite{Freedman_systole_2002}.  Given an $n$-dimensional cellulated manifold $\M^n$, together with its cellulation $\L$ and dual cellulation $\L^*$,  we may define its combinatorial $i$-systole $sys_i(\L)$ as the fewest number of $i$-cells of $\L$ that form a non-trivial cycle in the homology group $H_i(\L)$, and the dual combinatorial $(n-i)$-systole $sys_{n-i}(\L^*)$ as the fewest number of $(n-i)$-cells of $\L^*$ that form a non-trivial cycle in $H_{n-i}(\L^*)$.  The continuous version of this is just the geometric systole of the manifold $\M^n$: the $i$-systole is defined as $sys_i(\M^n):= \inf_{\alpha \neq \mathbf{0}} i\text{-area}(\alpha)$, where $\alpha$ is a smooth non-trivial oriented $i$-cycle whose class $[\alpha] \neq \mathbf{0} \in H_i(\M^n)$. The dual systole of the manifold is just $sys_{n-i}(\M^n)$.  

Similarly, we can define the combinatorial relative $i$-systole $sys_i(\L, \B)$ as the fewest number of $i$-cells of $\L$ that form a non-trivial relative cycle terminated at the boundaries $\B$ in the relative homology groups $H_i(\L, \B)$. The continuous version, i.e., the geometric relative $i$-systole can be defined accordingly:  $sys_i(\M^n, \B) := \inf_{\alpha \neq \mathbf{0}} i\text{-area}(\alpha)$, where $\alpha$ is a smooth non-trivial oriented relative $i$-cycle whose class $[\alpha] \neq \mathbf{0} \in H_i(\M^n, \B)$. 

In terms of fractals $\mathcal{F}$ considered in this paper, one can define the absolute and relative systoles in a similar fashion. In particular, as mentioned previously, we have defined the fractal $\mathcal{F}$ iteratively by punching holes in a manifold $\M^n$ with decreasing length scales at each iteration $l$.  Therefore, at the $l^\text{th}$ iteration, the considered geometry $\mathcal{F}(l)$ is just a punctured manifold with boundaries $\B(l)$, and one can hence define the absolute and relative geometric systoles of the corresponding manifold as $sys_i(\mathcal{F}(l))$ and  $sys_i(\mathcal{F}(l), \B(l))$. The geometric systoles of the fractal are hence defined in the limit of infinite iteration, i.e., $sys_i(\mathcal{F})=\lim_{l\rightarrow \infty} sys_i(\mathcal{F}(l))$ and $sys_i(\mathcal{F}, \B)=\lim_{l\rightarrow \infty} sys_i(\mathcal{F}(l),\B(l))$. One can also simply define the combinatorial systole of $\mathcal{F}$ through a finite cellulation $\L$ of $\mathcal{F}$. %Finer cellulation with increasing number of cells can capture holes in $\mathcal{F}$ at smaller scales.   

Note that in the above discussions we are defining the absolute and relative homology groups and the corresponding systoles in the general situations, i.e., over $\ZZ$-coefficients. When associating them with the quantum codes defined on qubits or qudits, we can simply define them over $\ZZ_2$- or $\ZZ_N$-coefficients. For example, we can denote the corresponding absolute or relative $\ZZ_2$-homology groups and $\ZZ_2$-systoles as $H_i(\L; \ZZ_2)$, $H_i(\L, \B; \ZZ_2)$, $sys_i(\L; \ZZ_2)$ and $sys_i(\L, \B; \ZZ_2)$ etc.

In the above discussion, we have made a connection between the exact-solvable stabilizer models (toric codes) with the chain-complex description.  However, we note that this connection exists even beyond the exact-solvable models, i.e., we can associate the TQFT description, namely the $\ZZ_2$ gauge theory with the mathematical structure of $\ZZ_2$ chain complex and $\ZZ_2$-homology theory.  In short, we have the following isomorphism between the three descriptions:
\be
\ZZ_2 \text{ toric code} \cong \ZZ_2 \text{ gauge theory}  \cong \ZZ_2\text{-homology},
\ee
which obviously generalizes also to the $\ZZ_N$ case.

\subsection{The fractal cube geometry and simple fractals}

We now start considering the possible existence of topological orders on fractals embedded in 3D. We begin with a simple construction which serves an important example in the following discussions: the \textit{fractal cube geometry} $FC(3,1)$ defined recursively in Fig.~\ref{fig:fractal_cube_m_hole}. Here, `3,1' represents the relative linear size of each cubic region and the hole punched in each iteration, which can be considered as a 3D generalization of the Sierpi\'nski carpet $SC(3,1)$.  In the illustration in Fig.~\ref{fig:fractal_cube_m_hole}, the top and bottom faces (horizontal) are $e$-boundaries and the rest of the faces (vertical) being $m$-boundaries, similar to the case of a 3D surface code shown in Fig.~\ref{fig:surface_code_3D_logical}. In general, one can also consider the fractal cube defined on a 3D torus $T^3$ or 3D sphere $S^3$ as will also be discussed below. In the specific example shown in Fig.~\ref{fig:fractal_cube_m_hole}, we have assigned  $m$-boundaries on all the holes inside the bulk, and we call these holes as \textit{$m$-holes} from now on.  In a general setup, these holes can also have $e$-boundaries, and will be called \textit{$e$-holes}  instead. 

In the fractal cube geometry, all the holes in the system are topologically equivalent (homeomorphic) to a 3-ball $D^3$ and hence simply-connected. We call such a fractal a \textit{simple fractal}:

\begin{definition}\label{def:simple_fractal}
A simple fractal embedded in $n$ dimensions is a fractal obtained by punching $n$-dimensional  holes homeomorphic to  $n$-balls $D^n$ in an $n$-dimensional manifold.	
\end{definition}

\subsection{Fractals formed by $e$-holes (no-go theorem)}
We first consider fractals where all the holes inside are $e$-holes, i.e.,  assigned with  $e$-boundaries. In this case, the $e$-particles can condense on the $e$-boundaries of these holes. Therefore, there exist logical-$Z$ string operators connecting the boundaries of two $e$-holes, which are Wilson-line  (worldline) operators of $e$-particles.

\begin{figure}[hbt]
  \includegraphics[width=0.6\columnwidth]{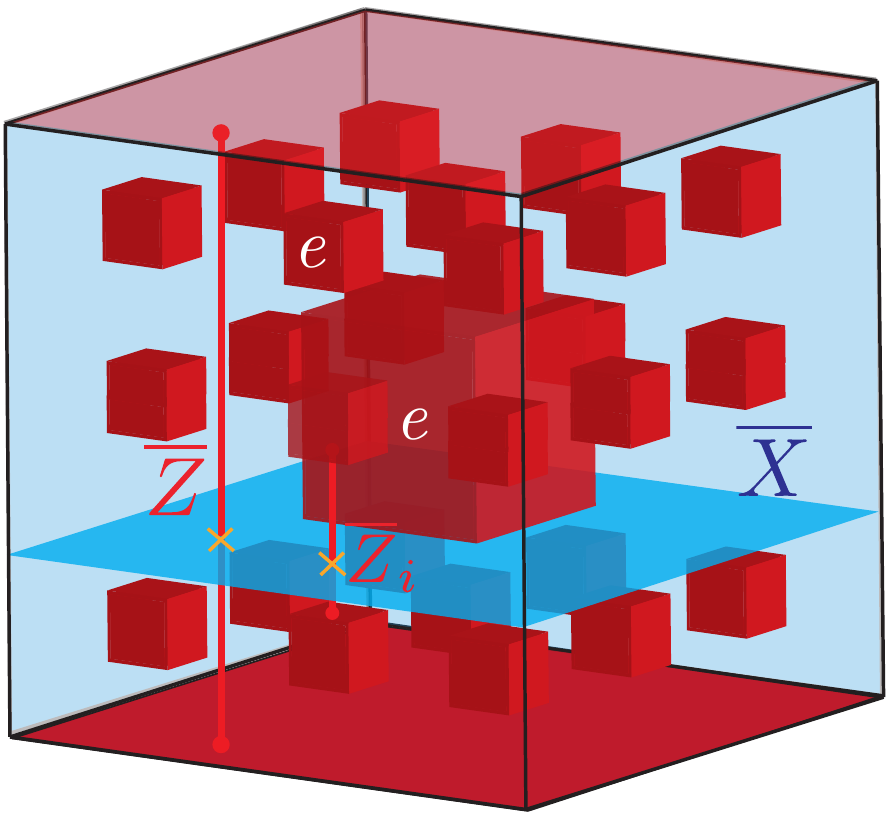}
  \caption{A 3D fractal model defined on the fractal cube geometry $FC(3,1)$ with only $e$-holes (red). A pair of macroscopic logical operators $\lo{X}$ and $\lo{Z}$ intersects at a single point (yellow cross).  The macroscopic logical $\lo{X}$ is intersected by an $O(1)$-length dual logical string $\lo{Z}_i$, leading to an $O(1)$ code distance. }
  \label{fig:no-go_theorem_e-holes}
\end{figure}

\begin{figure}[hbt]
  \includegraphics[width=0.8\columnwidth]{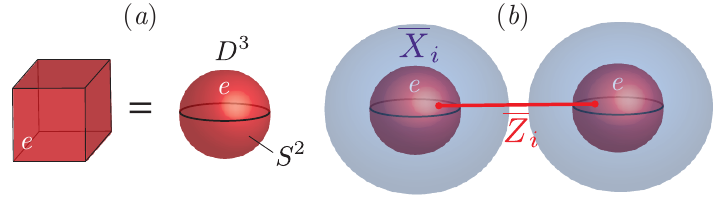}
  \caption{(a) A cubic hole is topologically equivalent to a 3-ball $D^3$ with its boundary being 2-sphere $S^2$. (b) An $e$-hole can trap a logical $X$-brane and let a logical $Z$-string terminated on it due to condensation of $e$-anyons on its $e$-boundary.  A pair of $e$-holes can encode a single logical qubit, similar to the situation of 2D surface code in Fig.~\ref{fig:surface_code}(b).}
  \label{fig:hole_trapping}
\end{figure}

We start with the special case of a simple fractal, i.e., all  $e$-holes in the system ($N_h$ in total) are homeomorphic to a 3-ball $D^3$ and hence simply-connected. One example of such a simple fractal is the fractal cube geometry with $e$-holes, as illustrated in Fig.~\ref{fig:no-go_theorem_e-holes}. Indeed, each cubic $e$-hole is homeomorphic to a 3-ball $D^3$ as shown in Fig.~\ref{fig:hole_trapping}(a). In this case, a similar `curse' to the 2D case occurs: each $e$-hole (homeomorphic to $D^3$) can trap a  $X$-brane logical operator $\lo{X}_i$ (associated with the $i^\text{th}$ $e$-hole). This is because the closed  $X$-brane is homeomorphic to a 2-sphere $S^2$, which is equivalent to the boundary of the $e$-hole (3-ball), i.e., $S^2= \partial D^3$ and can hence enclose the $e$-hole, as shown in Fig.~\ref{fig:hole_trapping}(b). Meanwhile, there is a $Z$-string logical operator $\lo{Z}_i$ connecting the $i^\text{th}$ and $(i+1)^\text{th}$ $e$-holes, as illustrated in Fig.~\ref{fig:no-go_theorem_e-holes} and  Fig.~\ref{fig:hole_trapping}(b). In addition, the logical operator $\lo{Z}_0$ corresponds to the $Z$-string connecting the top $e$-boundary on the surface to the $1^\text{st}$ $e$-hole, while $\lo{Z}_{N_h}$ corresponds to the $Z$-string connecting the $N_h^\text{th}$ $e$-hole to the bottom $e$-boundary on the surface.  Now we denote the macroscopic logical $Z$-string connecting the top and bottom boundaries as $\lo{Z}$. It has a length of $O(L)$. We note that  the logical $Z$-strings mentioned above are not all  independent, due to the following constraint:  $\lo{Z} = \prod_{i=0}^{N_h} \lo{Z}_i$. For the case of $\ZZ_2$ gauge theory or toric code, pairs of logical operators satisfying the anti-commutation relations $\{\lo{Z}_i, \lo{X}_i  \}=0$ (for $1 \le i \le N_h$) correspond to $N_h$ logical qubits. We also denote one of the macroscopic X-logical branes as $\lo{X}$ [with a size of $O(L^2)$]. The logical pair satisfying the anti-commutation relation $\{\lo{Z}, \lo{X}\}=0$ defines the main logical qubit which is already present in the absence of the holes.   However, due to the anti-commutation relation $\{\lo{Z}_i, \lo{X} \}=0$ for certain microscopic logical string $\lo{Z}_i$ with length $O(1)$,  we know that the distance (for the $Z$-type error/noise) of any of the logical qubits (independent of different choice of logical basis) is $d_Z \sim O(1)$.  Therefore, none of the degeneracies in the system is topologically protected similar to the 2D Sierpinski carpet case and topological order does not exist.     

In general, the absence of topological order can be extended  well beyond the case of simple fractals, and we can have the following generic no-go theorem: 
\begin{theorem}\label{theorem:no_go_3D}
	 $\mathbb{Z}_N$ topological order does not exist on a fractal embedded in 3D with only $e$-holes \footnote{More specifically, we refer to the $(1,2)$-$\ZZ_N$ topological order. This notation  will be introduced in Sec.~\ref{sec:nD} for the general theory in $n$D.}.
\end{theorem}
\textbf{Proof:} In the generic situation, the $e$-holes are not necessary simply connected as in the case of the simple fractal. However, this issue does not necessarily complicates our proof.  We start with an arbitrary macroscopic logical brane $\lo{X}$. As discussed in the proof for the 2D case, due to self-similarity of a fractal, for any point on the fractal geometry, when moving along any direction $\hat{n}$ within $O(1)$ distance, one must encounter either a hole or  the outer boundary.  Therefore, for any point $p$ on $\lo{X}$, there must exist one $e$-boundary on each side of $\lo{X}$.  Thus, there also exists a short logical string $\lo{Z}_i$ with $O(1)$ length connecting these two nearby $e$-boundaries having $O(1)$ separation, as can be illustrated in Fig.~\ref{fig:no-go_theorem_e-holes} without loss of generality. This short string necessarily intersects odd number of times with the macroscopic  logical brane $\lo{X}$ and has a single algebraic intersection, i.e.,  
\be
\mathbf{supp}(\lo{Z}_i) \cap \mathbf{supp}(\lo{X} )=1, 
\ee
where `$\cap$' denotes the algebraic intersection.  
There is hence an 
anti-commuting relation in the case of $\mathbb{Z}_2$ gauge theory: $\{\lo{Z}_i, \lo{X}\}$$=$$0$. This anti-commutation relation is replaced by the group commutator $\lo{Z}_i \ \lo{X} \  \lo{Z}_i^\dag \lo{X}^\dag $$=$$e^{2\pi i/N}$ in the general case of  $\mathbb{Z}_N$ gauge theory (with $N=2$ being the special case giving the above anti-commutator), where $2\pi /N$ is just the Aharanov-Bohm phase in this theory as mentioned above in Sec.~\ref{sec:3Dmodel}.  In either case, one gets a non-commuting relation : 
\be
[\lo{Z}_i, \lo{X} ] \neq 0.
\ee
This means a logical-$Z$ error can be caused by an $O(1)$ Z-type error (noise). Therefore, the distance corresponding to the logical-$Z$ error is only $d_Z \sim O(1)$, which also leads to the overall code distance being $d = \min(d_X, d_Z)  \sim O(1)$. The above statement is independent of the choice of logical basis. Note that in the proof above we have used the fact that the logical-$X$ operator has to be brane-like. This is a key property of $\ZZ_2$ gauge theory, since the logical-$X$ operators are associated with the second $\ZZ_N$-homology group and hence corresponds to a nontrivial 2-cycles.  We can hence conclude that no topological order exists on such a fractal geometry. 
\nin Q.E.D.

\subsection{Fractals formed by $m$-holes: existence of topological order}

Now we switch to discuss the case of fractals with $m$-holes. In particular, we focus on the setup where the $m$-holes in the fractal are homeomorphic to 3-balls $D^3$, i.e., the simple 3D fractal as defined above in Def.~\ref{def:simple_fractal}.  One example of such simple 3D fractal is the fractal cube geometry $FC(3,1)$ introduced above in Fig.~\ref{fig:fractal_cube_m_hole}.  
% In this example, the $e$-particle cannot condense on the $m$-holes, so $Z$-strings cannot terminate on these holes. Therefore, the only nontrivial logical string operators $\lo{Z}$ must connect the top and bottom $e$-boundaries, where the $e$-particles can condense.  This string is hence macroscopic and corresponds to a distance $d_Z \sim O(L)$. 

In this scenario, we can claim the following theorem:    

\begin{theorem}
	$\mathbb{Z}_N$ topological order exists on a simple fractal embedded in a 3D manifold with only $m$-holes.
\end{theorem}
In the following, we provide three alternative proofs of the above theorem. The former two are based on the TQFT and stabilizer descriptions, while the latter based on homology theory and will be extend to the higher-dimensional theories in Sec.~\ref{sec:nD}. The purpose of introducing different proofs is to explicitly show different aspects of the underlying physical meaning and pave the way for different types of generalization of the current results in future. We focus on proving the case of $\mathbb{Z}_2$ topological order (a special case being the $\mathbb{Z}_2$ toric code)  with a gapped boundary configuration shown in Fig.~\ref{fig:fractal_cube_m_hole}, i.e., with $e$-boundaries on the top and bottom faces of the cube and $m$-boundaries on the rest of the faces. The generalization to the cases of $\mathbb{Z}_N$ topological order with the corresponding exact-solvable model being the  $\mathbb{Z}_N$ toric code (see App.~\ref{append:ZN}) and situations of a periodic boundary condition, i.e., a 3-torus $T^3$,  is straightforward and will also be discussed below. For the convenience of discussion, we just call the Wilson-line (worldline) operators of $e$-particle $W^e$ as $Z$-string $Z^\otimes$, and the Wilson-brane (world-sheet) operators of $m$-string $W^m$ as $X$-brane $X^\otimes$, as in the case of the exact solvable 3D toric code model.  The reader should keep in mind that our proof also applies to the more general case of the TQFT description in the absence of an exact-solvable model.

\begin{figure}[t]
  \includegraphics[width=1\columnwidth]{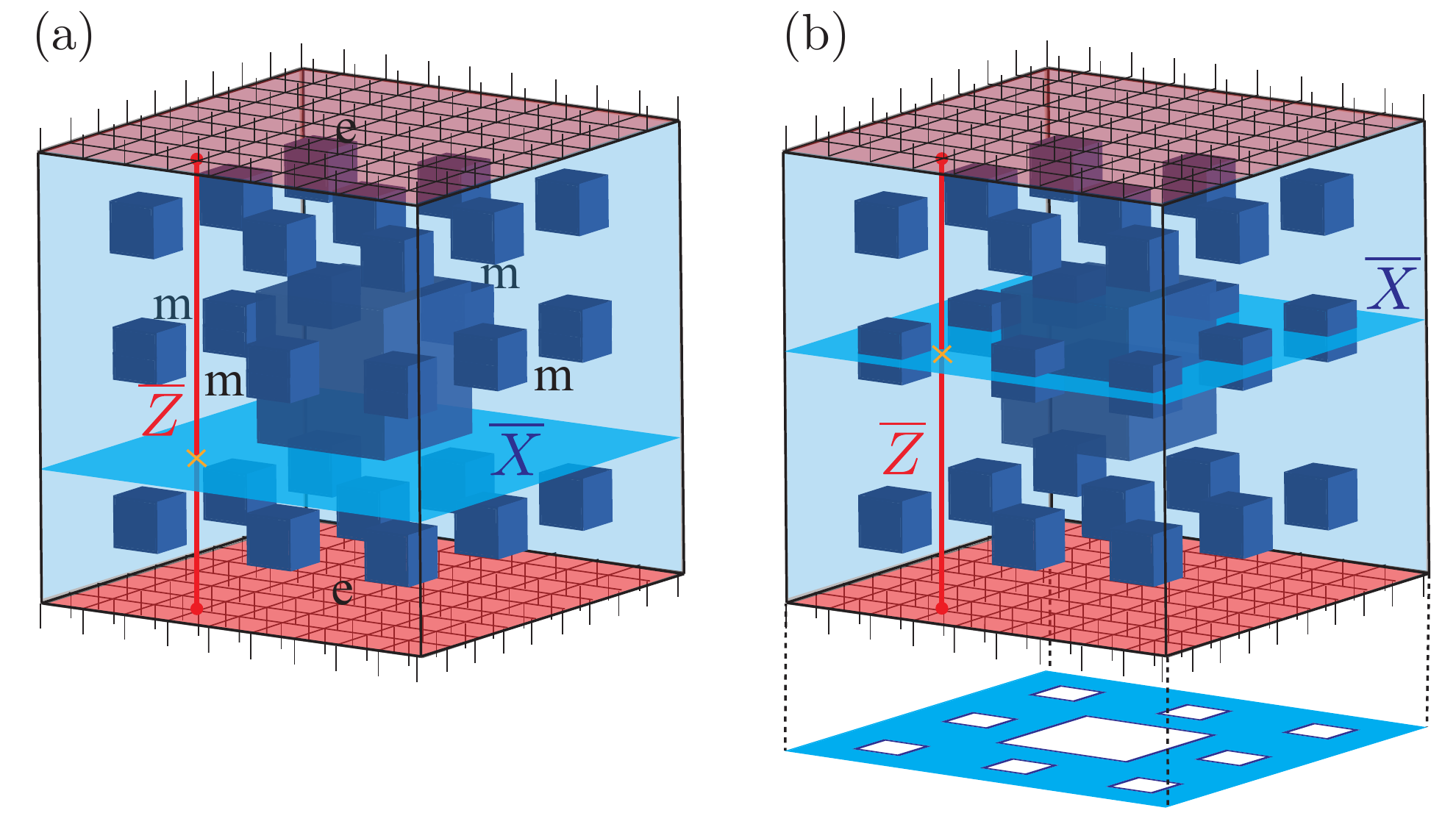}
  \caption{(a) A pair of macroscopic logical operator representatives on the 3D fractal surface code defined on the fractal cube geometry $FC(3,1)$ with $m$-holes: the logical string $\lo{Z}$ and the logical brane $\lo{X}$. (b) The minimal-area representative of the logical brane $\lo{X}$ is interrupted by the $m$-holes and hence has the shape of a Sierpi\'nski carpet with area scaling $O(L^{D_H})$, where $D_H=1.893$ is the Hausdorff dimension of the 2D Sierpi\'nski carpet.}
  \label{fig:fractal_logical}
\end{figure}

\nin \textbf{Proof 3a}: In this setup, the $e$-particle cannot condense on the $m$-holes, so $Z$-strings (Wilson-lines of $e$-particle $W^e$) cannot terminate on these holes. On the other hand, a $Z$-loop  also cannot enclose any $m$-holes due to simple topological constraints. A $Z$-loop $Z^\otimes$ (supported on $S^1$) orbiting around an $m$-hole ($D^3$) can always be shrunk into a trivial point via a gauge transformation (local deformation) according to Eq.~\eqref{eq:circle}, also corresponding to multiplying $Z$-plaquette stabilizers in the exact-solvable model as shown in Fig.~\ref{fig:toric_3D_illustration}(b).  This property is shown by the following relation:
\be
\raisebox{-0.5cm}{\includegraphics[scale=1]{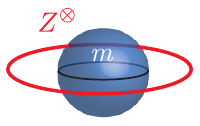}} \quad  =  \raisebox{-0.5cm}{\includegraphics[scale=1]{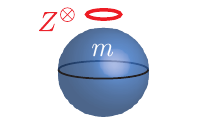}}  = \raisebox{-0.5cm}{\includegraphics[scale=1]{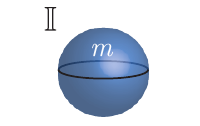}}, 
\ee
which claims that the $Z$-loop is an logical identity in the code (ground) space, i.e., $P_C Z^\otimes P_C= I$. Mathematically, this can also be interpreted by the following property of the homology groups of a 2-sphere, i.e., the boundary of the $m$-hole as a 3-ball ($S^2=\partial D^3$):
\begin{align}\label{eq:sphere_homology}
     H_i(S^2) = \begin{cases} \ZZ  \quad  \ &i=0 \ \text{or} \ 2; \\ 
     \mathsf{0} \quad &i=1. 
     \end{cases}
\end{align}
Here, `$\mathsf{0}$' represents the trivial group containing a single identity element. Here, $H_1(S^2)=\mathsf{0}$ corresponds to a loop $S^1$ on the sphere $S^2$ and is trivial (contractible),  meaning that a loop $S^1$ cannot enclose a sphere $S^2$ (the boundary of $m$-holes).

Therefore, in the case of the fractal cube with top and bottom $e$-boundaries (Fig.~\ref{fig:fractal_cube_m_hole}) or any simple fractal with the same boundary conditions, the only nontrivial logical string operator $\lo{Z}$$=$$ \lo{W}^e$ must connect the top and bottom $e$-boundaries, where the $e$-particles can condense.  This string is hence macroscopic and corresponds to a distance $d_Z \sim O(L)$. In the mean time, there is a macroscopic brane operator $\lo{X}$$=$$\lo{W}^m$ terminated at the $m$-boundaries on the four vertical faces, with a potential distance $d_X \sim O(L^{D_H})$ (if no other shorter logical $X$-brane exists).  Here $D_H$ is the Hausdorff dimension of a representative of the brane operator with the smallest operator support, which is itself a 2D  fractal.   For example, in the case of the fractal cube shown in Fig.~\ref{fig:fractal_cube_m_hole}, the smallest brane is a 2D Sierpi\'nski carpet with Hausdorff dimension $D_H= \log 8/ \log 3 \approx 1.893$.

\begin{figure*}[hbt]
  \includegraphics[width=2\columnwidth]{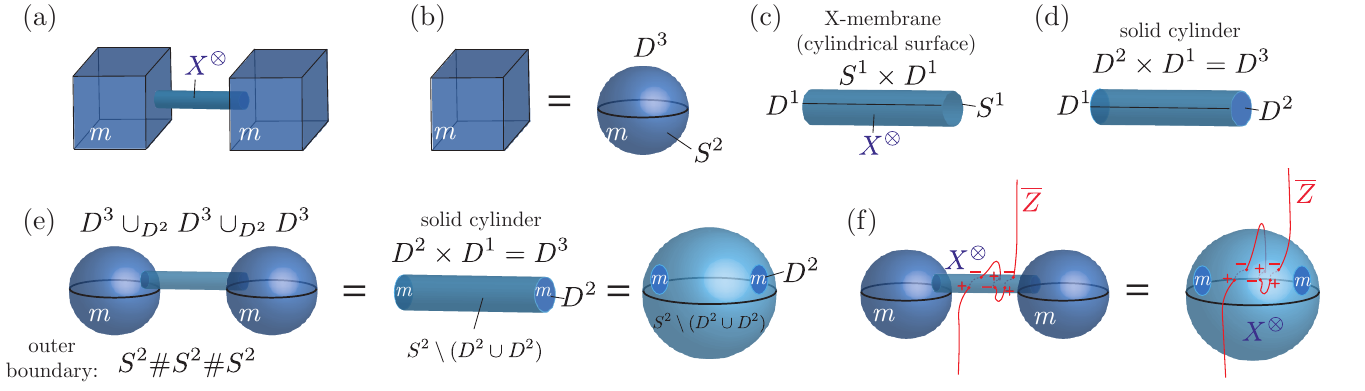}
  \caption{Illustration for Proof 3a. (a) A local $X$-brane connecting two $m$-holes. (b) Equivalence of the $m$-hole to a 3-ball $D^3$ with boundary being a 2-sphere $S^2$. (c) Equivalence of the $X$-brane to a cylinder $S^1\times D^1$. (d) The filled (solid) cylinder is equivalent to a 3-ball $D^3$. (e) Gluing two 3-balls and a solid cylinder in between is equivalent to a solid cylinder, while its boundary is equivalent to a connected sum of three 2-spheres. Excluding the two $m$-boundaries leads to a 2-punctured sphere with two disks $D^2$ being removed, i.e., $S^2\setminus(D^2 \cup D^2)$. (f) The intersection of the macroscopic logical string $\lo{Z}$ and the $X$-brane is equivalent to the intersection of the string with a 2-punctured sphere. The algebraic intersection is always \textit{zero} since the string enters and leaves the interior of the punctured sphere same number of times.}
  \label{fig:surgery}
\end{figure*}

\begin{figure*}[hbt]
  \includegraphics[width=2\columnwidth]{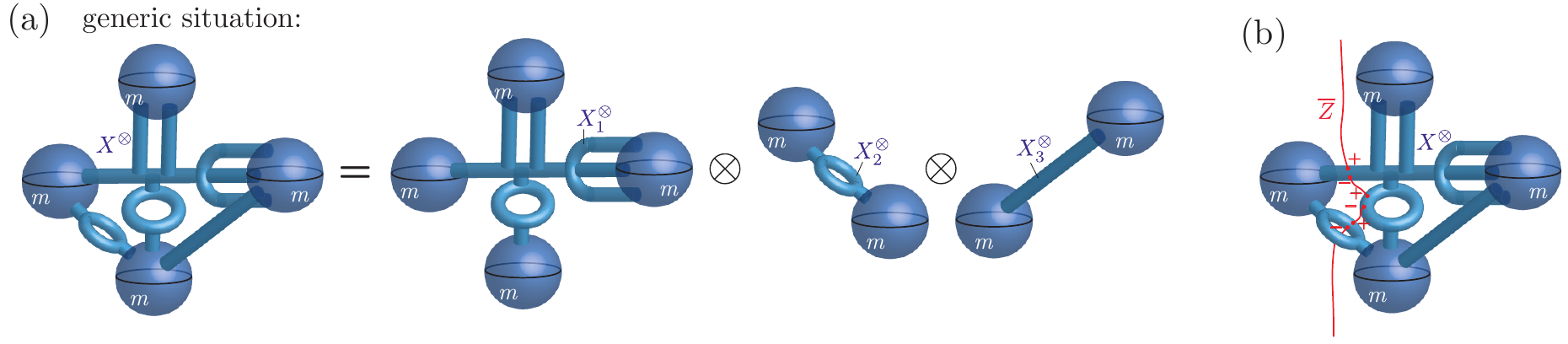}
  \caption{(a) The generic situation considered in Proof 3a and 3b, where a local $X$-brane operator connects multiple $m$-holes and can be decomposed as a tensor product of several connected components. (b) The macroscopic logical string $\lo{Z}$ has \textit{zero} algebraic intersection with the local $X$-brane, since the string enters and leaves the interior of the punctured closed 2-manifolds formed by the $X$-brane operator the same number of times.}
  \label{fig:surgery_generic}
\end{figure*}

The macroscopic logical string operator $\lo{Z}$ connecting the top and bottom $e$-boundaries has a single algebraic intersection with the macroscopic logical brane operator $\lo{X}$ (hence with odd number of geometric intersections), i.e., 
\be
\mathbf{supp}(\lo{Z}) \cap \mathbf{supp}(\lo{X} )=1. 
\ee
 In the case of $\mathbb{Z}_2$ topological order, we get an anti-commutation relation $\{\lo{Z}, \lo{X}\}$$=$$0$. This anti-commutation relation is replaced by the group commutator $\lo{Z} \ \lo{X} \  \lo{Z}^\dag \lo{X}^\dag $$=$$e^{2\pi i/N}$ in the general case of  $\mathbb{Z}_N$ topological order, where $2\pi/N$ comes from the Aharanov-Bohm phase introduced in Sec.~\ref{sec:3Dmodel}.

 We can also see from Eq.~\eqref{eq:sphere_homology} that $H_2(S^2)$ corresponding to a non-contractible 2-brane can enclose the sphere $S^2$ by integer number of times.  However, although the $X$-brane $X^\otimes$ (Wilson brane of $m$-string $W^m$) can enclose the $m$-hole, it is still trivial since it can be absorbed onto the boundary of the $m$-hole due to the condensation of $m$-string and hence disappear.  This can be achieved via a gauge transformation (local deformation) as follows:
\be\label{eq:membrane_trapping}
\raisebox{-0.9cm}{\includegraphics[scale=1]{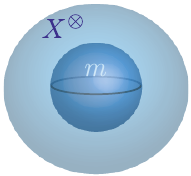}} \quad  =  \raisebox{-0.9cm}{\includegraphics[scale=1]{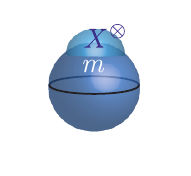}}  = \raisebox{-0.5cm}{\includegraphics[scale=1]{loop_shrinking_vac.pdf}}, 
\ee
 which corresponds to multiplying  X~stabilizers in the exact-solvable model. The first equality in Eq.~\eqref{eq:membrane_trapping} is due to the multiplication of a partial spherical $X$-membrane attached to the $m$-hole, which is itself a logical identity in the code space $\H_C$ according to Eq.~\eqref{eq:relation_boundary_m} and corresponds to a stabilizer in the exact-solvable model.  The second equality is due to the absorption of the attached partial spherical X-membrane into the $m$-boundary which is hence equivalent to the vacuum sector $\I$, again due to  Eq.~\eqref{eq:relation_boundary_m}. The entire Eq.~\eqref{eq:membrane_trapping} states that the X-brane $X^\otimes$ wrapping around the $m$-hole is trivial and equivalent to logical identity in the code (ground) space, i.e., $P_C X^\otimes P_C= I$.

Now the only remaining question is whether there exists any microscopic logical brane operator $\lo{X}_j$ with support of $O(1)$ or more generally less then $O(d_X)$, which does not commute with the macroscopic string operator $\lo{Z}$, i.e., $[\lo{Z}, \lo{X}_j]$$\neq$$0$. 

% having an anti-commutation relation $\{\lo{Z}, \lo{X}_j\}$$=$$0$ for the $\mathcal{D}(\mathbb{Z}_2)$ topological order or more generally the group commutator  $\lo{Z} \ \lo{X}_j \lo{Z}^\dag \lo{X}_j^\dag $$=$$e^{2\pi i/N}$ for the $\mathcal{D}(\mathbb{Z}_N)$ topological order.

We know that a logical $X$-brane operator has to be a closed brane, i.e., contains no boundary corresponding to the string excitation.  It can only terminates at $m$-boundaries due to the condensation of the  $m$-string.  Since the $X$-brane $X^\otimes$ does not have any boundary, its intersection on the $m$-boundaries must also be closed curves, i.e., a union of closed loops: $S^1 \cup S^1 \cup S^1 \cup S^1 \cdots \equiv \cup_j S^1_{(j)}$. Note that a closed loop is homeomorphic (topologically equivalent) to a circle $S^1$.  In general, these closed loops can either be non-overlapping, i.e., $S^1_{(i)}  \cap S^1_{(j)} = \emptyset$ (for any $i$ and $j$),  or overlapping, i.e., crossed at certain discrete set of points and hence $S^1_{(i)}  \cap S^1_{(j)} \neq \emptyset$.  There is no essential difference between these two cases, since one can always remove the intersection between the loops via local deformation, equivalent to gauge transformation or stabilizer multiplication.

The simplest configuration satisfying the above constraints is a cylindrical $X$-brane operator connecting two $m$-holes, as shown in Fig.~\ref{fig:surgery}(a).  As mentioned earlier and illustrated in Fig.~\ref{fig:surgery}(b), the $m$-holes in these simple fractal are homeomorphic (topologically equivalent) to 3-balls ($D^3$), with the boundaries being 2-spheres ($S^2$). The support of the $X$-brane is a cylindrical surface $S^1 \times D^1$, i.e., decomposed by the circle $S^1$ and a line segment $D^1$ (1-dimensional disk), as illustrated in Fig.~\ref{fig:surgery}(c). If including the inner region as well, we obtain a solid cylinder $D^2 \times D^1$ (with the cross section being a 2-disk $D^2$), which is homeomorphic to a 3-ball $D^3$.  Now the configuration in Fig.~\ref{fig:surgery}(a) is equivalent to two 3-balls connected by a solid cylinder, i.e., $D^3\cup_{D^2} D^3 \cup_{D^2} D^3$, as shown Fig.~\ref{fig:surgery}(e).  Here, we have used the notation from the surgery theory, where $D^3\cup_{D^2} D^3$ means that two 3-balls ($D^3$) are glued together along a common (identified) disk ($D^2$) region. In Fig.~\ref{fig:surgery}(e), we have a ball glued with a solid cylinder and then glued with another ball, along two identified disks. The boundary of this configuration is a connected sum of three 2-spheres, i.e., $S^2 \# S^2 \# S^2$.  Here, the connected sum ($\#$) between two 2-manifolds means cutting out two disks $D^2$ on the two manifolds and glue the two manifolds by identifying the boundaries of the two disks, i.e., two circles $S^1$. As we can see from the first equality in Fig.~\ref{fig:surgery}(e), the gluing of the two balls with the solid cylinder is equivalent to a single solid cylinder $D^2 \times D^1$ which is also equivalent to a 3-ball $D^3$, with the boundary being 2-sphere $S^2$, as shown by the second equality. Therefore, as shown by the rightmost image in Fig.~\ref{fig:surgery}(e), the $X$-brane operator $X^\otimes$ has the support of a sphere with two disk regions (corresponding to $m$-boundaries) being cut out:  $\mathbf{supp}(X^\otimes)=S^2 \setminus (D^2 \cup D^2)$. 

Now we consider the intersection of the $X$-brane operator $X^\otimes$ and the macroscopic logical string $\lo{Z}$, as illustrated in Fig.~\ref{fig:surgery}(f).  As one can see, the geometric intersection number between the $Z$-string and the $X$-cylindrical surface is always even, and the algebraic intersection number is always 0, i.e.,
\be\label{eq:intersection2}
\mathbf{supp}(\lo{Z}) \cap \mathbf{supp}(X^\otimes)=0.
\ee   
The zero comes from the cancellation of positive intersection and negative intersection numbers. When choosing a direction of the string, one has a positive (negative) intersection number $+1$ ($-1$) when the string goes into (out of) the surface, labeled as $+$ ($-$) in Fig.~\ref{fig:surgery}(f).  One can further understand this by deforming the boundary of this configuration to an $X$-brane with the support of a 2-sphere with two disks corresponding to the $m$-boundary being cut out. The $\lo{Z}$-string can intersect with the 2-sphere except the two $m$-disk regions, i.e., the region $S^2 \setminus (D^2 \cup D^2)$. However, we know that the algebraic intersection number is a topological invariant in the sense that one can perturb the intersection point of the $\lo{Z}$-string with the 2-sphere without affecting the algebraic intersection number.  This means that the algebraic intersection of the string with the punctured sphere $S^2 \setminus (D^2 \cup D^2)$ is the same as the algebraic intersection with the sphere $S^2$ since one can perturb the intersection point away from the two disks. The algebraic intersection number is hence zero if there is no $e$-particle source inside the sphere.  This is because that the $\lo{Z}$-string has to go into and come out of the sphere the same number of times, leading to the cancellation of algebraic intersection numbers. A source of $e$-particle is only possible if there are $e$-holes enclosed by the $X$-brane, which is excluded in the considered scenario.  Alternatively, one can also consider the intersection of the string with the punctured sphere as a more restrictive case (excluding the two disk regions) of the general situation of the intersection with the entire sphere, therefore the algebraic intersection number must be the same.      

In the above example, the intersection of the $X$-brane with the $m$-holes is a single circle $S^1$.  In more general cases, the intersection can be a union of non-overlapping or crossed circles $\cup_j S^1_{(j)}$.  In these cases, the $X$-brane can be decomposed as a tensor product of several connected components $\prod_j X^\otimes_j$. Each $X^\otimes_j$ intersects with the $m$-hole on a circle $S^1$.  We can hence analyze the intersection of the string $\lo{Z}$ and each brane $X^\otimes_j$ separately.

Now we consider the most generic situation, as illustrated in Fig.~\ref{fig:surgery_generic}. The $X$-brane connects multiple $m$-holes with a complicated topology.  As mentioned above, we can first decompose the $X$-brane into a tensor product of multiple connected components $\prod_j X^\otimes_j$, as shown in Fig.~\ref{fig:surgery_generic}(a).  Each connected component $X^\otimes_j$ along with the surface of the $m$-holes ($S^2$) form a generic orientable closed surface (2-manifold) $\mathcal{M}^2$ as illustrated by the first connected component in Fig.~\ref{fig:surgery_generic}(a). Such generic surface $\mathcal{M}^2$ can also have non-zero genus as illustrated in the figure, which is more general than the cylinder case considered above. The macroscopic logical string $\lo{Z}$ are excluded from the boundary of the $m$-holes $\mathcal{B}_j$ and hence only intersects in the rest of the region on the $X$-brane, i.e., $\mathcal{M}^2 \setminus (\cup_j \mathcal{B}_j)$. Note that different from the simple case of cylindrical $X$-brane, each excluded boundary $\mathcal{B}_j$ is not necessarily a disk $D^2$, but in general a punctured sphere with several disk regions being cut out, i.e., $\mathcal{B}_j=S^2 \setminus (\cup_i D^2_{(i)})$, as illustrated in Fig.~\ref{fig:surgery_generic}(a). In the previously discussed special case of a punctured sphere with a single disk removed, this punctured sphere is just a disk, i.e., $S^2 \setminus D^2 = D^2$.  

However, as mentioned before, such a reduced configuration is just a more restrictive case of the generic situation of the intersection between the string $\lo{Z}$ and the entire closed surface $\mathcal{M}^2$, so the algebraic intersection number in the reduced configuration $\mathcal{M}^2 \setminus (\cup_j \mathcal{B}_j)$ should just be the same as the one in the case of the entire generic closed surface $\mathcal{M}^2$. We note that the $Z$-string has to go into and come out of the interior of the  closed surface the same number of times, since there is no $e$-hole and a source of $Z$-string inside the closed surface does not exist, as illustrated in Fig.~\ref{fig:surgery}(f). Therefore, in the most generic situation, the intersection property Eq.~\eqref{eq:intersection2} is still satisfied: the geometric intersection number is even and the algebraic intersection number is 0.  

In sum, these local $X$-branes will always commute with the macroscopic logical string, i.e., $[\lo{Z}, X^\otimes_j]=0$, and hence can not lead to any logical error.  Therefore, the $X$-distance of the code is macroscopic, i.e., $d_X =  O(L^{D_H})$, which  is determined by the size of the minimum support of the macroscopic logical brane $\lo{X}$, with $D_H$ being the Hausdorff dimension of such brane as mentioned before. The overall code distance is hence $d=\min(d_X,d_Z)=O(L)$ and remains macroscopic.
% In addition, we also note that any of such local $X$-branes must be trivial, namely be equal to logical identity in the code (ground) space: \be\label{eq:identity}
% P_C X^\otimes_j P_C= I,
% \ee 
% since they do not anti-commute (or in general non-commute) with any logical-$Z$ operator. Note that in the case with gapped boundary configuration shown in Fig.~\ref{fig:fractal_logical}, there is only one such operator denoted by  $\lo{Z}$, while there are three macroscopic logical-$Z$ strings if the system is defined on a 3-torus $T^3$.

\nin Q.E.D.

\nin \textbf{Proof 3b}:
In the above proof, we have focused on the intersection property between the local closed $X$-branes attached to the $m$-holes and the macroscopic logical $Z$-string.  In this alternative proof, we focus on proving that any of such local closed $X$-branes is equivalent to logical identity in the code (ground) space, i.e.,   
\be\label{eq:identity}
P_C X^\otimes_j P_C= I.
\ee 
This property can be easily shown either using the basic TQFT relations stated above or through the stabilizer properties in the exactly-solvable toric-code model. 

We first consider the simple case of a cylindrical $X$-brane connecting two $m$-holes as shown in Fig.~\ref{fig:surgery}.  We have the following relation between equivalent configurations:
\begin{align}
\nonumber & \raisebox{-0.65cm}{\includegraphics[scale=1.4]{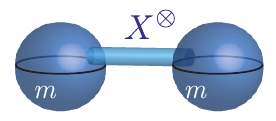}}  =  \raisebox{-0.65cm}{\includegraphics[scale=1.4]{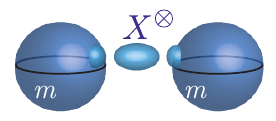}} \\ 
&= \raisebox{-0.65cm}{\includegraphics[scale=1.4]{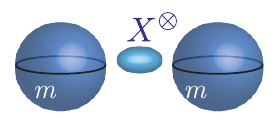}}  
= \raisebox{-0.65cm}{\includegraphics[scale=1.4]{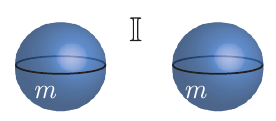}}. \label{eq:membrane_shrinking}
\end{align}
The first equality applies the recoupling relation in Eq.~\eqref{eq:cylinder_recoupled} twice on the left and right end of the cylinder respectively.
 With this recoupling, one can split the cylindrical $X$-brane into a piece equivalent to a sphere $S^2$ in the center and two partial spheres attached to the $m$-holes.  The two attached partial spheres are then shrunk into logical identity in the next equality due to the $m$-string condensation relation Eq.~\eqref{eq:relation_boundary_m}. Now the central spherical $X$-brane can again be shrunk into logical identity in the last equality due to the relation Eq.~\eqref{eq:sphere}.

We can also derive the above equivalence relation in a slightly different way. As we mentioned before, any local deformation of the string and brane operator is equivalent to a gauge transformation, also corresponding to multiplying stabilizers in the case of exact solvable model.  The cylindrical $X$-brane intersects with the $m$-hole on a circle $S^1$. We can then  continuously shrink this intersection circle to a single point $p$, and then detach the $X$-brane form the $m$-hole boundary followed by further shrinking it into the vacuum sector, as shown by the following relation:
\begin{align}
\nonumber & \raisebox{-0.65cm}{\includegraphics[scale=1.4]{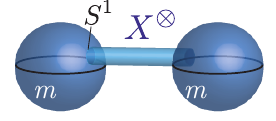}}  =  \raisebox{-0.65cm}{\includegraphics[scale=1.4]{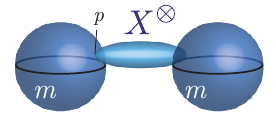}} \\ 
&= \raisebox{-0.65cm}{\includegraphics[scale=1.4]{membrane_shrinking3.pdf}}  
= \raisebox{-0.65cm}{\includegraphics[scale=1.4]{membrane_shrinking4.pdf}}. \label{eq:membrane_shrinking2}
\end{align}

Now we can consider the most general case where the $X$-brane along with the $m$-hole boundaries together form a generic closed surface (2-manifold) $\mathcal{M}^2$ as shown in Fig.~\ref{fig:surgery_generic}. The whole $X$-brane can be locally deformed into vacuum as shown below:
\begin{align}
\raisebox{-1.7cm}{\includegraphics[scale=1]{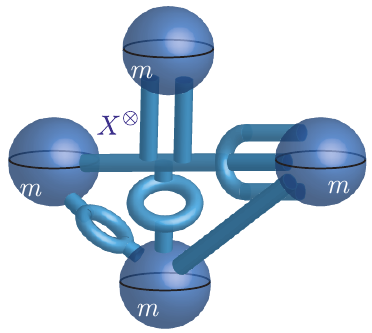}} = \raisebox{-1.7cm}{\includegraphics[scale=1]{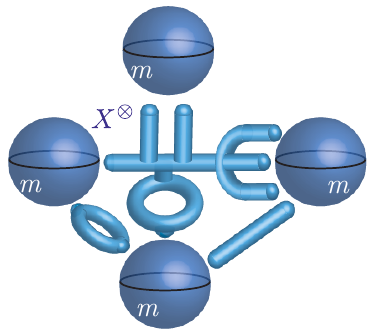}} = \I \label{eq:membrane_shrinking_general}
\end{align}
The first equality uses the detaching move in either Eq.~\eqref{eq:membrane_shrinking} or Eq.~\eqref{eq:membrane_shrinking2}. The second equality uses the relation in  Eq.~\eqref{eq:genus-g} that the $X$-brane supported in any closed genus-$g$ surface is equal to the vacuum (logical identity). Therefore, we have proved the property stated in Eq.~\eqref{eq:identity}, i.e., any local  $X$-membrane connecting different $m$-holes is a logical identity.

In the case of exactly solvable model, one can multiply all the $X$ stabilizers inside the closed surface $\mathcal{M}^2$, which cancels the whole $X$-brane.   Therefore, we can see such local $X$-brane is just a stabilizer and hence equals the logical identity.

\nin Q.E.D.

\nin \textbf{Proof 3c}: There is an even more concise proof based on algebraic topology, without the need to investigate the detailed shape of the local branes as in the previous two proofs.   Since we are focusing only on Abelian topological orders in this paper, the mathematical structure of our theory can be completely captured by the homology and relative homology groups.  The ground-state subspace (code space) of the $\mathbb{Z}_2$ topological order and codes in 3D can be expressed as
\be\label{eq:homology_group}
\mathcal{H}_C= \mathbb{C}^{H_1(\mathcal{L}, \mathcal{B}_e; \mathbb{Z}_2)},
\ee
where $H_1(\mathcal{L}, \mathcal{B}_e; \mathbb{Z}_2)$ represents the 1st relative homology group.  Here,  the argument $\mathcal{L}$ represents the lattice, $\mathcal{B}_e$ represents all the $e$-boundaries and $\mathbb{Z}_2$ means we consider the homology over the $\mathbb{Z}_2$ coefficients. Note that Eq.~\eqref{eq:homology_group} also includes the situation without $e$-boundaries, i.e., $\B_e=0$, and in that case we are investigating the absolute 1st homology group $H_1(\L; \ZZ_2)$. One can easily generalize Eq.~\eqref{eq:homology_group} to the case of $\mathbb{Z}_N$ topological order by replacing the $\mathbb{Z}_2$-homology group with the $\mathbb{Z}_N$-homology group: $H_1(\mathcal{L}, \mathcal{B}_e; \mathbb{Z}_N)$.  To deepen our understanding, in the following we investigate three variants for the setup of $\L$ which have the same fractal structure in the bulk but different choices of external boundary conditions (or in other words background manifolds): (1) The fractal surface code $\L_\text{FSF}$ introduced earlier in Fig.~\ref{fig:fractal_surface_code} with the background manifold being the usual 3D surface code and the corresponding relative homology group being  $H_1(\mathcal{L}_\text{FSF}, \mathcal{B}_e; \mathbb{Z}_2)$; (2) The punctured 3-torus fractal  geometry $\L(T^3)$, with the fractal supported on a background manifold corresponding to a 3D torus $T^3$ with a periodic boundary condition (a 3D cube with opposite faces being identified and hence $\B_e=\emptyset$) and the corresponding absolute homology group being $H_1(\L(T^3); \mathbb{Z}_2)$; (3) The punctured 3-sphere fractal geometry $\L(S^3)$ , with the fractal supported on a background manifold corresponding to a 3D sphere $S^3$ (a 3D cube with all its boundary being identified to a single point and hence $\B_e=\emptyset$) and the corresponding absolute homology group being $H_1(\L(S^3); \mathbb{Z}_2)$.  

As has been mentioned in the previous proofs, all the holes in the bulk are $m$-holes, so there are no $e$-boundaries in the bulk which the $e$-particle could condense onto.  This means there is no non-trivial (non-contractible) relative 1-cycle (which could end on the $e$-boundaries) in the bulk region connecting the holes. According to Eq.~\eqref{eq:sphere_homology}, one has $H_1(S^2)=\mathsf{0}$, suggesting the absolute (ordinary) 1-cycle around any $m$-hole (with boundary being $S^2$) in the bulk is also trivial (contractible).

\begin{figure}[hbt]
  \includegraphics[width=0.7\columnwidth]{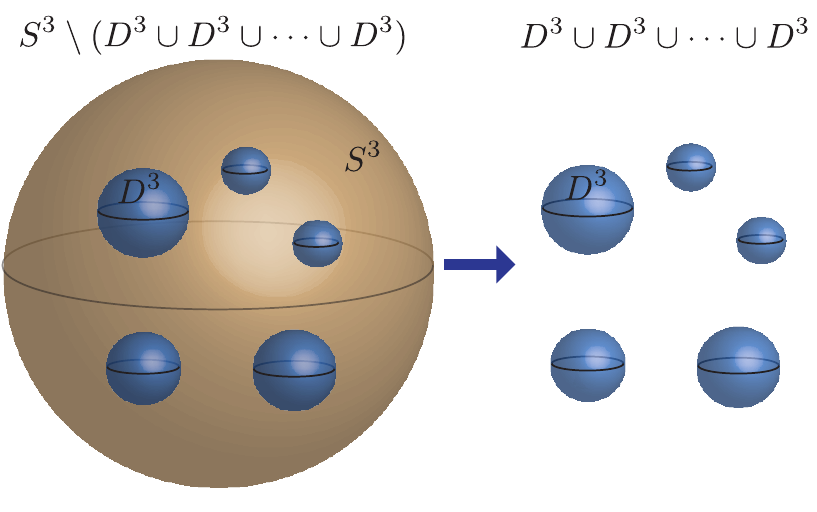}
  \caption{Illustration for the application of  the Alexander duality: the 1st-homology of the punctured 3-sphere (with interior holes homeomorphic to 3-balls $D^3$) is isomorphic to the 1st-homology of the union of the 3-balls $D^3$. }
  \label{fig:Alexander_duality}
\end{figure}

A rigorous way to prove the triviality of any local relative or absolute 1-cycle in the bulk is to directly compute the first relative homology group of the entire complex $\L$: $H_1(\L, \B_e ; \ZZ_2)$. To begin with, we first consider the setup of a punctured 3-sphere fractal geometry $\L(S^3)$ mentioned above such that there is no external boundary, i.e., $\B_e~=~\emptyset$. We then aim to compute the absolute homology group  $H_1(\L(S^3) ; \ZZ_2)$. Note that topologically speaking, this punctured 3-sphere fractal geometry is just a 3-sphere with a set of 3-balls $D^3$ (the $m$-holes) inside being cut out, i.e., $\L(S^3) = S^3 \setminus (D^3 \cup D^3 \cup \cdots \cup D^3)=  S^3 \setminus \cup_j D^3_{(j)}$. One can hence use a `divide and conquer' strategy to compute the homology of the entire complex from the homology of the individual piece.  In particular, this can be achieved via the \textit{Alexander duality} \cite{Hatcher:2001ut}:
\be\label{eq:Alexander_duality}
\tilde{H}_i(S^n \setminus X) \cong \tilde{H}^{n-i-1}(X),
\ee
where $X$ stands for arbitrary sub-manifold. Here, $\tilde{H}$ stands for reduced (co)homology, which is just a slightly modified version of the usual (co)homology: for $i>0$,  these two are the same, i.e.,  $H_{i}=\tilde{H}_{i}$ and $H^{i}=\tilde{H}^{i}$;  for $i=0$, one has $H_{i}=\tilde{H}_{i} \oplus \ZZ$ and $H^{i}=\tilde{H}^{i}\oplus \ZZ$ in the general case for the $\ZZ$-coefficients. The additional contribution $\ZZ$ is replaced by $\ZZ_2$ (or  $\ZZ_N$) for the $\ZZ_2$ (or $\ZZ_N$)  (co)homology. The essence of the Alexander duality is that it  converts the computation of homology of the manifold $S^n\setminus X$ to the computation of the cohomology of its complement $X$ in $S^3$. When applying the Alexander duality to our case, we can convert the homology of $\L(S^3)$ into the cohomologies of individual 3-balls $D^3$:
\begin{align}\label{eq:punctured_sphere_homology}
\nonumber &  H_1(S^3 \setminus \cup_j D^3_{(j)}; \ZZ_2)
\cong H^1(\cup_j D^3_{(j)}; \ZZ_2) \\
\nonumber \cong& H^1(D^3 \cup D^3 \cup \cdots \cup D^3; \ZZ_2)  \\
\nonumber \cong& H^1(D^3; \ZZ_2) \oplus H^1(D^3; \ZZ_2) \oplus \cdots \oplus H^1(D^3; \ZZ_2) \\ 
=& \mathsf{0} \oplus \mathsf{0} \oplus \cdots \oplus \mathsf{0} = \mathsf{0}, 
\end{align}
as illustrated in Fig.~\ref{fig:Alexander_duality}. Here we use only the usual homology instead of the reduced homology since we have $i>0$ and $n-i-1>0$. The third line follows from the fact that the cohomology of a union of disjoint sub-manifold can be decomposed into the direct sum of the homology of the individual sub-manifold  \cite{Hatcher:2001ut}.  The final line follows from the fact that $H^1(D^3)=H_1(D^3)=\mathsf{0}$, since any 1-cocycle or 1-cycle is contractible and hence trivial in a 3-ball.  We have hence proved that any absolute 1-cycle in the punctured 3-sphere and the corresponding 1st-homology group are trivial, i.e., $H_1(\L(S^3); \ZZ_2)=\mathsf{0}$. 

\begin{figure*}[hbt]
  \includegraphics[width=2\columnwidth]{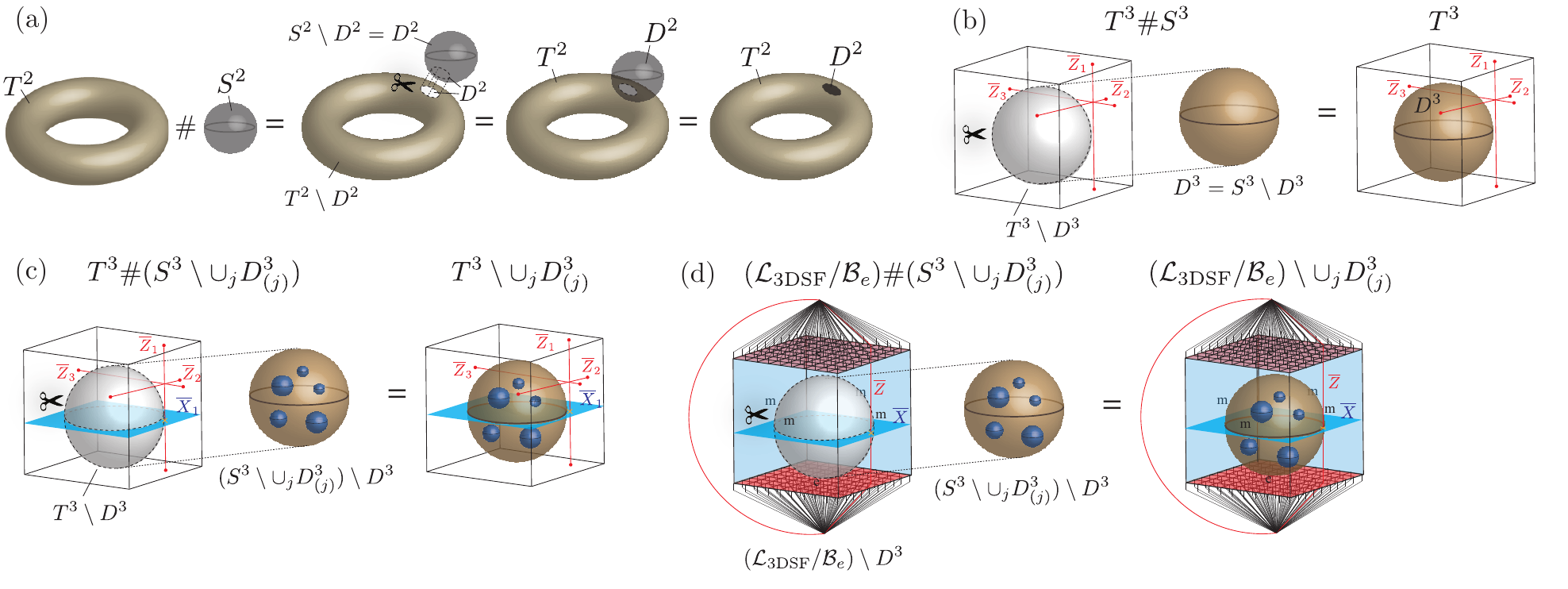}
  \caption{Illustration of the homology decomposition using a connected sum. (a) Illustrating Eq.~\eqref{eq:connected_sum_presentation} in the 2D case. The connected sum of a 2-torus $T^2$ and 2-sphere $S^2$ is obtained by cutting out a 2-ball (disk) $D^2$ on each piece and glue them together by identifying the boundaries of the 2-ball on each piece, i.e., a circle $S^1$.  Note that the 2-sphere with a disk being cut out is just a disk: $S^2 \setminus D^2=D^2$. (b) Illustrating Eq.~\eqref{eq:connected_sum_presentation} in the context of a 3D toric code, where the opposite sides of the cube is identified and hence represent a 3-torus. The connected sum of a 3-torus $T^3$ and a 3-sphere $S^3$ is obtained by cutting out a 3-ball $D^3$ on each piece and gluing them together by identifying the  boundaries of the 3-ball on each piece, i.e., a 2-sphere $S^2$.  Note that the 3-sphere with a 3-ball being cut out is just a 3-ball: $D^3=S^3\setminus D^3$, which gets glued into the 3-torus. (c) Illustrating Eq.~\eqref{eq:torus_homology}, i.e., decomposing  the homology of the punctured 3-torus $T^3 \setminus  \cup_j D^3_{(j)}$ to that  of a connected sum of a punctured sphere $S^3 \setminus  \cup_j D^3_{(j)}$ and a 3-torus. 
  %Note that the macroscopic logical brane $\lo{X}_1$ gets intersected by the removed 3-ball, such that the remaining part has a size $O(L)$, giving a lower bound on the $X$-distance. 
  (d) Illustrating Eq.~\eqref{eq:FSF_homology}, i.e., decomposing the relative homology of the punctured fractal surface code to that of a connected sum of the 3D surface code and a punctured sphere. The relative 1st-homology of the  3D surface code is equivalent to the absolute 1st-homology of the 3D surface code geometry with its $e$-boundaries being identified into a single point: i.e., $\L_\text{3DSF}/\B_e$. The logical string $\lo{Z}$ (red) hence becomes a nontrivial absolute 1-cycle traveling around the entire complex.}
  \label{fig:connected_sum_3D}
\end{figure*}

We next consider the setup of a  punctured 3-torus fractal geometry $\L(T^3)$, which is homeomorphic to a 3-torus cutting out a set of 3-balls, i.e., $\L(T^3)=T^3 \setminus \cup_j D^3_{(j)}$. Note that any $n$-dimensional manifold $\M^n$ can be presented as a connected sum of itself and a $n$-sphere: 
\be\label{eq:connected_sum_presentation}
\M^n = \M^n \# S^n,
\ee
where $S^n$ acts like a zero element in the connected sum. In $n$D, the connected sum ($\#$) between two $n$-manifolds means cutting out two $n$-balls $D^n$ on the two manifolds and glue the two manifolds by identifying the boundaries of the two balls, i.e, two $(n-1)$-spheres $S^{n-1}$, where a 2D example with 2-torus, i.e., $\M^2 = T^2$, is illustrated in Fig.~\ref{fig:connected_sum_3D}(a).   In our context, a 3-torus can also be presented as a connected sum of a 3-torus and a 3-sphere, i.e., $T^3=T^3 \# S^3$, as illustrated in Fig.~\ref{fig:connected_sum_3D}(b). Here, the two pieces glued together are a torus with a 3-ball being cut out, i.e., $T^3\setminus D^3$, and a 3-sphere with a  3-ball being cut out which is equivalent to a 3-ball, i.e., $S^3\setminus D^3= D^3$. Similarly, the punctured 3-torus can also be expressed as a connected sum of a 3-torus and a large  punctured 3-sphere with radius of $O(L)$, i.e., $\L(T^3)=T^3 \setminus \cup_j D^3_{(j)}= T^3 \# S^3 \setminus \cup_j D^3_{(j)}$, as illustrated in Fig.~\ref{fig:connected_sum_3D}(c). Using the Mayer–Vietoris sequence \cite{Hatcher:2001ut}, one can express the 1st-homology of the connected sum of two 3-manifolds as a direct sum of the 1st-homology of each 3-manifold, i.e.,
\begin{align}\label{eq:torus_homology}
\non & H_1(\L(T^3); \ZZ_2) = H_1(T^3 \# S^3 \setminus \cup_j D^3_{(j)}; \ZZ_2) \\
\non \cong & H_1(T^3; \ZZ_2) \oplus H_1(S^3 \setminus \cup_j D^3_{(j)}; \ZZ_2)  \\
= &  (\ZZ_2 \oplus \ZZ_2 \oplus \ZZ_2)  \oplus \mathsf{0} = \ZZ_2 \oplus \ZZ_2 \oplus \ZZ_2.
\end{align}
The last line in the above equation uses the homology of a 3-torus, i.e., $H_1(T^3; \ZZ_2)=\ZZ_2 \oplus \ZZ_2 \oplus \ZZ_2$, where each copy of $\ZZ_2$ comes from the contribution of one of the non-contractible 1-cycles in the 3-torus where the macroscopic logical-$Z$ strings (worldlines of $e$-particle)   can travel through. These macroscopic logical strings have minimal length corresponding to  $Z$-distance $d_z = O(L)$, since they need to circumvent the $m$-holes along the way without terminating on them.

Now we consider the setup of a 3D fractal surface code geometry $\L_\text{FSF}$. Topologically speaking, it is equivalent to a usual 3D surface code geometry $\L_\text{3DSF}$ with a set of 
3-balls being cut out, i.e., $\L_\text{FSF} = \L_\text{3DSF} \setminus  \cup_j D^3_{(j)}$. By using the fact in Eq.~\eqref{eq:connected_sum_presentation}, we can present  $\L_\text{3DSF}$ as a connected sum of itself and a 3-sphere, i.e., $\L_\text{3DSF} = \L_\text{3DSF} \# S^3$.  We can then present the 3D fractal surface code geometry as $\L_\text{FSF} = (\L_\text{3DSF} \# S^3) \setminus  \cup_j D^3_{(j)}= \L_\text{3DSF} \# ( S^3 \setminus  \cup_j D^3_{(j)})$, namely the connected sum of the 3D surface code and a large punctured sphere with radius of $O(L)$. Now we can compute the 1st relative homology of this connected sum with the following homology decomposition: 
\begin{align}\label{eq:FSF_homology}
\non & H_1(\L_\text{FSF}, \B_e; \ZZ_2) \cong  H_1(\L_\text{FSF}/\B_e; \ZZ_2) \\
\non \cong &H_1((\L_\text{3DSF}/\B_e) \# ( S^3 \setminus  \cup_j D^3_{(j)}) ; \ZZ_2) \\
\non \cong & H_1(\L_\text{3DSF}/\B_e; \ZZ_2) \oplus  H_1( S^3 \setminus  \cup_j D^3_{(j)} ; \ZZ_2)\\
 \cong &
H_1(\L_\text{3DSF}, \B_e; \ZZ_2) \oplus  \mathsf{0} = \ZZ_2. 
\end{align}
Note that $\L'=\L_\text{FSF}/\B_e$ and $\L''=\L_\text{3DSF}/\B_e$ are modified complexes obtained from identifying  the $e$-boundaries of the fractal surface code and 3D surface code into a single point, and the corresponding connected sum $(\L_\text{3DSF}/\B_e) \# ( S^3 \setminus  \cup_j D^3_{(j)})$ is illustrated in Fig.~\ref{fig:connected_sum_3D}(d).  The last equality have used the homology of the 3D surface code, i.e., 
$H_1(\L_\text{3DSF}, \B_e; \ZZ_2) = \ZZ_2$.  For both the fractal and usual 3D surface codes, the single $\ZZ_2$ comes from the
the relative 1-cycle connecting the $e$-boundaries on the top and bottom faces of the cube, which is also equivalent to the absolute 1-cycle when the rough boundaries of the fractal and usual 3D surface codes are all identified to a single point, i.e., corresponding to the quotient complex $\L'=\L_\text{FSF}/\B_e$ and $\L''=\L_\text{3DSF}/\B_e$ as illustrated in Fig.~\ref{fig:connected_sum_3D}(d).  This 1-cycle is the support of the macroscopic logical string $\lo{Z}$ which has minimal length corresponding to the  $Z$-distance $d_Z = O(L)$.  

% Therefore, we have $H_1(\mathcal{L}, \mathcal{B}_e; \mathbb{Z}_2)=\mathbb{Z}_2$, same as the case of the 3D surface code.  Thus, the ground-state subspace (code space) of the 3D fractal cubic surface code is $\mathcal{H}_C=\mathbb{C}^{\mathbb{Z}_2}=\mathbb{C}^2$, corresponding to a single logical qubit and 2-fold ground-state degeneracy.  The only non-trivial logical-$Z$ operator (denoted by $\lo{Z}$) is supported on the string connecting the top and bottom $e$-boundaries, corresponding to the only non-contractible relative 1-cycle.  Note that we have obtained the number of logical qubits and ground-state degeneracy by only investigating the 1st relative homology group corresponding to the logical-$Z$ operators, without even investigating the details of the logical-$X$ operators associated with the 2nd  relative cohomology group $H^2(\mathcal{L}, \mathcal{B}_m; \mathbb{Z}_2)=H_2(\mathcal{L}^*, \mathcal{B}_m; \mathbb{Z}_2)$ (equivalent to the 2nd relative homology group on the dual lattice $\mathcal{L}^*$), which was a central part in proofs 1a and 1b.  The essence of this simplification is due to a generalized version of the \textit{Poincaré duality} for an open manifold, i.e., the \textit{Poincaré-Lefschetz duality}. 

Note that we have computed the 1st (relative) homology groups on $\L$ in all three setups.   It would seem that we need to also compute all the 2nd relative homology groups on the dual cellulation $\L^*$, i.e., $H_2(\L^*, B_m, \ZZ_2)$, which is equivalent to the central parts in proofs 3a and 3b.  However, due to the intrinsic symmetries in homology, we do not really need to directly compute the 2nd relative homology.  The essence of this simplification is due to a generalized version of the \textit{Poincaré duality} for a manifold with boundary, i.e., the \textit{Poincaré-Lefschetz duality}. 

The Poincaré duality states that, for any  $n$-dimensional orientable closed manifold $\mathcal{M}$ (compact and without boundary), the $i^\text{th}$ cohomology group is isomorphic to the $(n-i)^\text{th}$ homology group:
\be
 H^i(\mathcal{M}) \cong H_{n-i}(\mathcal{M}).   
\ee
When formulating this theorem in terms of the cellulation $\L$ of the manifold $\M$, it becomes
\be
 H^i(\mathcal{L}) \cong H_{n-i}(\mathcal{L}^*),  
\ee
where $\L^*$ represents the dual cell complex of $\L$. The above two formulations are equivalent since both $\L$ and $\L^*$ are cellulations of the same manifold $\M$, and their homologies are equal to the homology of the manifold. As a generalization to the case with boundaries, the \textit{Poincaré-Lefschetz duality} states that, for any $n$-dimensional orientable compact manifold $\mathcal{M}$ with boundary $\mathcal{B}$, the $i^\text{th}$ relative cohomology (homology) group is isomorphic to the $(n-i)^\text{th}$ absolute homology (cohomology) group:
 \be\label{eq:Poincare_duality}
 H^i(\mathcal{M}) \cong H_{n-i}(\mathcal{M}, \mathcal{B}), \quad  H_i(\mathcal{M}) \cong H^{n-i}(\mathcal{M}, \mathcal{B}),  
 \ee
with the equivalent formalism for the cellulation being:
 \be\label{eq:Poincare_duality_cellulation}
 H^i(\mathcal{L}) \cong H_{n-i}(\mathcal{L^*}, \mathcal{B}^*), \quad  H_i(\L^*) \cong H^{n-i}(\L, \mathcal{B}).  
 \ee

In our case, we consider our fractal lattice $\mathcal{L}$ as a cellulation of an orientable compact punctured 3-manifold $\mathcal{M}^3$ ($n=3$) and the corresponding cellular homology over $\mathbb{Z}_2$ or more generally $\mathbb{Z}_N$.  We start with the punctured 3-sphere fractal geometry $\L(S^3)$. We have shown above in Eq.~\eqref{eq:punctured_sphere_homology} that the absolute 1st homology (corresponding to logical $Z$-string) is trivial, i.e.,  $H_1(\mathcal{L}(S^3); \mathbb{Z}_2) = \mathsf{0}$. The Poincaré-Lefschetz duality hence leads to
\be
H^1(\mathcal{L}(S^3); \mathbb{Z}_2) \cong H_2(\mathcal{L}^*(S^3), \mathcal{B}^*_m; \mathbb{Z}_2).
\ee
Now we use the property that the $i^\text{th}$-homology and cohomology of the same complex being isomorphic to each other \cite{Hatcher:2001ut}, i.e., 
\be\label{eq:cohomology_eqaul_homology}
H^i(\L) \cong H_i(\L). 
\ee
This isomorphism is also reflected in the ranks (dimensions) of the (co)homology groups and can be simply explained as follows.  According to Eqs.~\eqref{eq:homology_def} and \eqref{eq:cohomology_def}, we have
\begin{align}\label{eq:universal_coefficient}
\non  \dim H_i(\L) =& \dim \text{Ker}\partial_i - \dim \text{Img}\partial_{i+1}   \\
\non =& (\dim C_i -\rank \partial_i)-\rank \partial_{i+1} \\
\non =&(\dim C^i - \rank \delta^i) - \rank \delta^{i-1}  \\ 
 =& \dim \text{Ker}\delta^i - \dim \text{Img}\delta^{i-1} = \dim H^i(\L).
\end{align}
The second line of the above equation comes from the \textit{rank-nullity theorem} for the boundary map, i.e., $\rank\partial_i$$+$$\dim \text{Ker}\partial_i$$=$$\dim C_i$, where $\rank \partial_i$$=$$ \dim 
\text{Img}\partial_i$.  The third line uses the isomorphism $C_i \cong C^i$ and Eq.~\eqref{eq:rank_equality}. The fourth line uses the rank-nullity theorem for the coboundary map.
We hence also get the equality of the ranks (equals the vector-space dimensions), i.e., $\rank H_i(\L)$$=$$\rank H^i(\L)$. For the $\ZZ_2$ case, the equality of the ranks alone already ensures the isomorphism of the homology groups. Therefore, we get the following property:
\be\label{eq:sphere_homology_2}
H_2(\mathcal{L}^*(S^3), \mathcal{B}^*_m; \mathbb{Z}_2)  \cong H_1(\mathcal{L}(S^3); \mathbb{Z}_2)=\mathsf{0}.
\ee
This means that the 2nd relative homology in the dual complex (lattice) $\L^*$, which corresponds to the logical $X$-brane,  is also trivial.  Indeed, in the case of punctured 3-sphere fractal geometry,  the ground (code) space can be expressed as
\be\label{eq:ground_space_sphere}
\mathcal{H}_C= \mathbb{C}^{H_1(\mathcal{L}(S^3); \mathbb{Z}_2)} \cong \mathbb{C}^{H_{2}(\mathcal{L}^*(S^3), \mathcal{B}^*_m; \mathbb{Z}_2)}=\mathbb{C}^1,
\ee
which has only a single ground state and hence no degeneracy. Still, it is expected that there is long-range entanglement in the ground-state wavefunction which cannot be prepared by a constant-depth local circuit from a trivial product state.   The number of logical qubit $k$ is determined by the rank or dimension of the homology groups, i.e.,
\be
k=\dim H_1(\L(S^3); \ZZ_2)=  \dim H_{2}(\mathcal{L}^*(S^3), \mathcal{B}^*_m; \mathbb{Z}_2)=0.
\ee
The essence of the Poincaré-Lefschetz duality in this context is revealed in Eq.~\eqref{eq:ground_space_sphere}
as the equivalence of two logical basis choices, i.e.,  logical-$Z$ (electric) basis and logical-$X$ (magnetic) basis, to represent the same ground space $\H_C$, which is essentially a Fourier transform. The physical interpretation of this duality is just the electro-magnetic ($e$-$m$) duality in the $\ZZ_2$ gauge theory. 
It is hence not surprising that computing only the 1st-homology is enough to know the absence of logical degrees of freedom in the bulk, including logical-$Z$ strings and logical-$X$ branes.  

Now we switch to the case of the punctured 3-torus fractal geometry $\L(T^3)$.  According to the Poincaré-Lefschetz duality,  Eq.~\eqref{eq:torus_homology} and Eq.~\eqref{eq:cohomology_eqaul_homology}, we hence get
\be\label{eq:torus_homology2}
H_2(\mathcal{L}^*(T^3), \mathcal{B}^*_m; \mathbb{Z}_2)
\cong H_1(\L(T^3); \ZZ_2)  =
\ZZ_2 \oplus \ZZ_2 \oplus \ZZ_2.
\ee
The ground (code) space can hence be expressed as
\be
\H_C =\mathbb{C}^{H_1(\mathcal{L}(T^3); \mathbb{Z}_2)} \cong \mathbb{C}^{H_{2}(\mathcal{L}^*(T^3), \mathcal{B}^*_m; \mathbb{Z}_2)}= \mathbb{C}^{\mathbb{Z}_2^{\oplus 3}}=\mathbb{C}^{2^3}, 
\ee
with the number of logical qubit being
\be
k=\dim H_1(\L(T^3); \ZZ_2)=  \dim H_{2}(\mathcal{L}^*(T^3), \mathcal{B}^*_m; \mathbb{Z}_2)=3.
\ee
Note that all the nontrivial contribution $\mathbb{Z}_2^{\oplus 3}$ in  $H_2(\mathcal{L}^*(T^3), \mathcal{B}^*_m; \mathbb{Z}_2)$ from Eq.~\eqref{eq:torus_homology2} comes from the three   homology classes associated with macroscopic brane operators $\lo{X}_{1,2,3}$ travelling through the 3-torus. The three homology classes contain representatives parallel to the $xy$-, $xz$-, and $yz$- planes respectively. This can be seen from the homology decomposition  dual to Eq.~\eqref{eq:torus_homology}, i.e.,  $H_2(\L^*(T^3); \ZZ_2)$$=$$ H_2(T^3; \ZZ_2) \oplus H_2(\L^*(S^3), \B^*_m; \ZZ_2)= H_2(T^3; \ZZ_2) \oplus \mathsf{0}$, where the last equality uses the trivial relative 2nd-homology of the punctured sphere from Eq.~\eqref{eq:sphere_homology_2}. In combination with Eq.~\eqref{eq:torus_homology}, it shows that the nontrivial homology classes only come from those of the usual 3-torus, i.e., and can hence only be macroscopic. From Fig.~\ref{fig:connected_sum_3D}(c), we can see that the minimal support of these macroscopic branes, e.g. $\lo{X}_1$, is lower bounded by the minimal size of the membrane outside the ball region being cut out, which is at least proportional to the linear system size $L$. We can hence obtain the lower bound of the $X$-distance as $d_X \sim  \Omega(L)$. One can also get an exact scaling of  $X$-distance as $d_X =   O(L^{D_H})$, where $D_H$ is the Hausdorff dimension of the minimal-area fractal brane operator as we have calculated in the case of the fractal cube geometry. In sum, there is neither any nontrivial absolute 1-cycle or relative 2-cycle with $O(1)$ distance in the presence of $m$-holes in the fractal.

Finally, we come back to the fractal surface code geometry $\L_\text{FSF}$, where we have both the $e$-  and $m$-boundaries. 
% As we have stated above, we have the 1st relative homology group being: $H_1(\mathcal{L}, \mathcal{B}_e; \mathbb{Z}_2)$$=$$\mathbb{Z}_2$.
% Note that we have intentionally divided the boundaries inside the lattice $\mathcal{L}$ into two classes, i.e., the $e$-boundary $\mathcal{B}_e$ associated with the 1st relative homology group and $m$-boundary $\mathcal{B}_m$ associated with the 2nd relative homology group. 
We now use a more general  version of the Poincaré-Lefschetz  duality: let $\M$ be an  orientable compact $n$-dimensional manifold and $\B=\mathcal{B}_1 \cup \mathcal{B}_2$ be its boundary decomposed as a union of two orientable compact $(n-1)$-dimensional manifolds  $\mathcal{B}_1$ and $\mathcal{B}_2$ with a common boundary  $\partial \B_1 = \partial \B_2 =  \mathcal{B}_1 \cap \mathcal{B}_2$, then the following isomorphism holds:
\be
H^i(\mathcal{M}, \B_1) \cong H_{n-i}(\mathcal{M}, \B_2).
\ee
% Recall that absolute homology should just be considered as a special situation of relative homology where the boundary $\mathcal{B}$ is absent. 
In the context of the fractal surface code, the Poincaré-Lefschetz duality for the cellulation translates to
\be
H^1(\L, \B_e; \mathbb{Z}_2) \cong H_2(\L^*, \B^*_m; \mathbb{Z}_2).
\ee
Now by using the isomorphism in Eq.~\eqref{eq:cohomology_eqaul_homology}, we can get $H^1(\L, \B_e) \cong H^1(\L/ \B_e) \cong H_1(\L/ \B_e) \cong H_1(\L, \B_e)$.  Combined with Eq.~\eqref{eq:FSF_homology}, this leads to the following expression
\be
 H_2(\mathcal{L}^*_\text{FSF}, \mathcal{B}^*_m; \mathbb{Z}_2) \cong H_1(\L_\text{FSF}, \B_e; \ZZ_2) = \ZZ_2.
\ee
The ground (code) space can then be expressed as
\be
\H_C =\mathbb{C}^{H_1(\mathcal{L}_\text{FSF}, \B_e; \mathbb{Z}_2)} \cong \mathbb{C}^{H_{2}(\mathcal{L}^*_\text{FSF}, \mathcal{B}^*_m; \mathbb{Z}_2)}= \mathbb{C}^{\mathbb{Z}_2}=\mathbb{C}^{2},
\ee
with the number of logical qubits being
\be
k=\dim H_1(\L_\text{FSF}, \B_e; \ZZ_2)=  \dim H_{2}(\mathcal{L}^*_\text{FSF}, \mathcal{B}^*_m; \mathbb{Z}_2)=1.
\ee
The nontrivial contribution  $\mathbb{Z}_2$ in $H_{2}(\mathcal{L}^*_\text{FSF}, \mathcal{B}^*_m; \mathbb{Z}_2)$ comes from the macroscopic logical brane $\lo{X}$ terminating at the four external $e$-boundaries.  This can be seen from the homology decomposition dual to Eq.~\eqref{eq:FSF_homology}, i.e.,  $H_2(\L^*_\text{FSF}, \B^*_m; \ZZ_2)$$=$$ H_2(\L^*_\text{3DSF}, \B^*_m; \ZZ_2) \oplus H_2(\L^*(S^3), \B^*_m; \ZZ_2)= H_2(\L^*_\text{3DSF}; \ZZ_2) \oplus \mathsf{0}$. In combination with  Eq.~\eqref{eq:FSF_homology}, it shows that the nontrivial homology classes only comes from that of the usual 3D surface code $\L_\text{3DSF}$, and  can hence  only  be  macroscopic. As shown in  Fig.~\ref{fig:connected_sum_3D}(d), we can again obtain the lower bound of the $X$-distance as $d_X \sim \Omega(L)$ or an exact scaling $d_X = \Omega(L^{D_H})$ similar to the case of the 3-torus.
In sum, no nontrivial relative 1-cycle or relative 2-cycle with $O(1)$ distance exists in the presence of $m$-holes in the fractal.

Although we have only proved the case of $\ZZ_2$ topological order, the above proof can be directly adapted to the case of $\ZZ_N$ topological order simply by replacing all the $\ZZ_2$ in the above expressions with $\ZZ_N$. 

\nin Q.E.D.

Besides the above theorem for topological orders and codes, proof 3c is also a rigorous proof for the following mathematical theorem:
\begin{theorem}
	There exists a class of simple fractals $\L$ embedded in three dimension with boundary $\B_m$ on its interior holes, such that both its absolute  1-systole $sys_1(\L)$ and its dual relative 2-systole $sys_2(\L^*, B^*_m)$ are macroscopic, i.e., scale as a power law of the linear size $L$ of the fractals.
\end{theorem}
This theorem is stated for the situation that the background manifold is a 3-torus or more generic closed 3-manifold with non-zero 1st Betti number.  It can of course be extended to the case of relative 1-systole $sys_1(\L, \B_e)$ when there is another boundary type $\B_e$ on the external surface.  Also, this theorem holds for the cases of $\ZZ$-, $\ZZ_2$- and $\ZZ_N$-systoles, and the proof simply follows by replacing all the $\ZZ_2$ in the above expressions with $\ZZ$ or $\ZZ_N$.

\begin{figure}[t]
  \includegraphics[width=1\columnwidth]{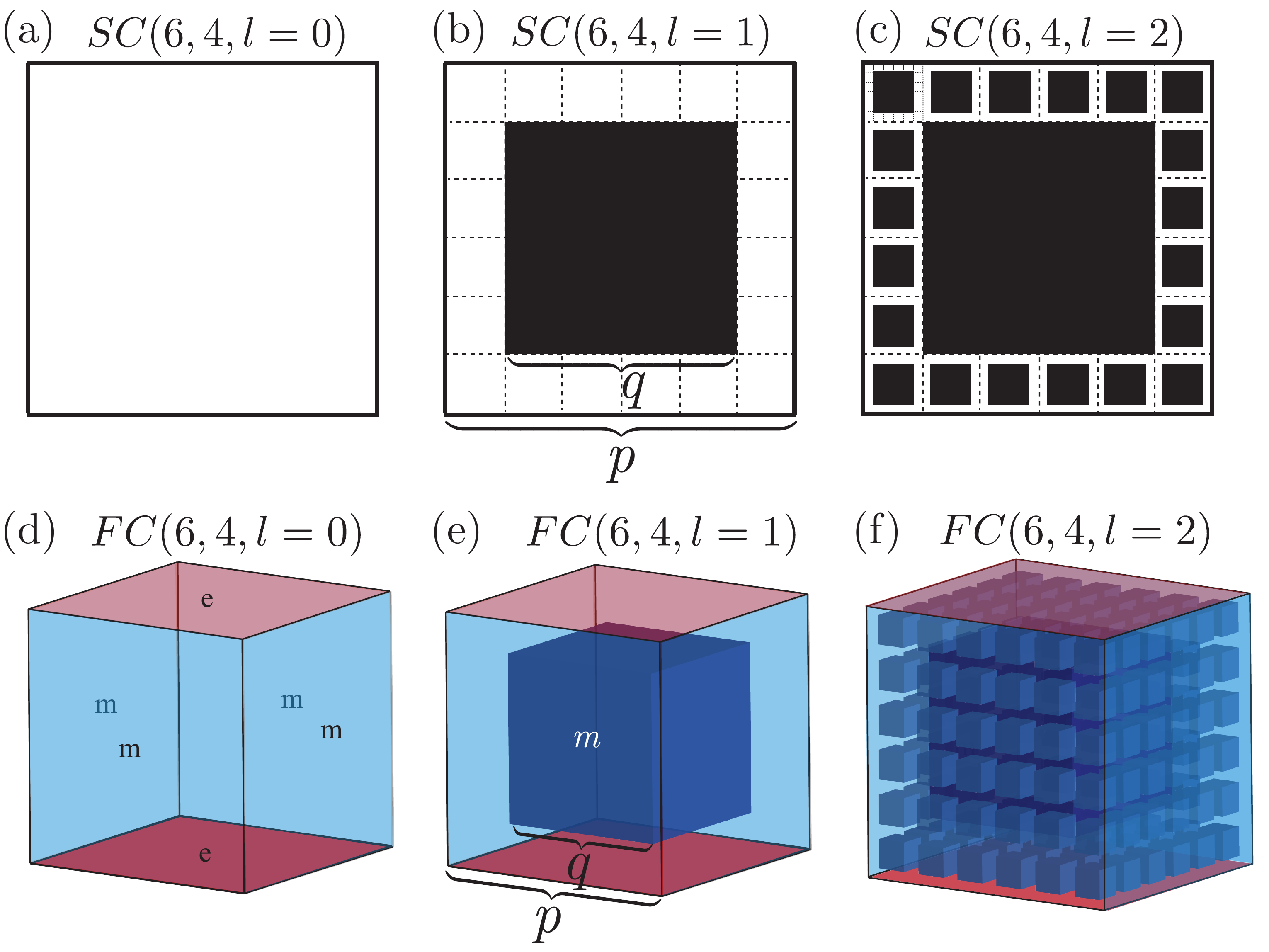}
  \caption{(a-c) Iterative construction of a family of Sierpi\'nski carpet $SC(p,q)$, shown via an illustration of $SC(6,4)$ and Hausdorff dimension $D_H=1.672$. (d-f) Iterative construction of a family of the 3D fractal surface code defined on a fractal cube geometry  $FC(p,q)$ and $m$-holes inside, shown via an illustration of $FC(6,4)$ and Hausdorff dimension $D_H=2.804$.}
  \label{fig:fractal_cube_m_hole_variant}
\end{figure}

\subsection{Construction of a family of fractal codes with Hausdorff dimension $3-\delta$}

In this section, we construct a class of fractal codes defined on simple fractals with $m$-holes and tunable Hausdorff dimension $3-\delta$, where $0<\delta<1$. Most interestingly, the Hausdorff dimension can asymptotically approach $2+\epsilon$ for arbitrary small $\epsilon$.

One can generalize the specific fractal cube geometry $FC(3,1)$ in Fig.~\ref{fig:fractal_cube_m_hole} to a family fractal geometries $FC(p,q)$ as illustrated in Fig.~\ref{fig:fractal_cube_m_hole_variant}. We start by generalizing the Sierpi\'nski carpet fractal $SC(3,1)$ shown in Fig.~\ref{fig:fractal_2D} to $SC(p,q)$ as illustrated in Fig.~\ref{fig:fractal_cube_m_hole}(a-c).  In the first iteration ($l=1$),  one divides the large level-0 ($l=0$) square equally into $p \times p$ small level-1 ($l=1$)  squares, with each one having the linear size $1/p$ of the large level-0 square.  One then punches a level-1 hole region in the center occupying $q \times q$ squares.  Therefore the linear size ratio of the level-1 hole and the level-0 square is $q/p$, where $q=4$ and $p=6$ in the illustration of Fig.~\ref{fig:fractal_cube_m_hole}(a-c). Now we are left with the  remaining $p^2-q^2$ level-1 square, and we denote the present geometry as $SC(p, q,l=1)$. In the next iteration ($l=2$), we repeat the procedure in the previous iteration inside each level-1 square and reach the geometry $SC(p, q,l=2)$, and then further proceed.  In the $l^\text{th}$ iteration, one divides each level-$l$ square equally into $p \times p$ level-$(l+1)$ squares, and punch a hole in the center occupying $q \times q$ squares, and we obtain the geometry $SC(p, q, l)$. The fractal is generated  asymptotically when having an infinite number of iterations, i.e., $SC(p, q) \equiv \lim_{l \rightarrow \infty} SC(p, q, l)$.        

We next generate the general fractal cube geometry $FC(p,q)$ in a similar manner as shown in Fig.~\ref{fig:fractal_cube_m_hole_variant}(d-f), where $p=6$ and $q=4$ is illustrated in this case. In the $l^\text{th}$ iteration, one divides each level-$l$ cube equally into $p \times p \times p$ level-$(l+1)$ cubes with linear size $1/p$ of a level-$l$ cube, and punch an $m$-hole in the center occupying $q \times q \times q$ cubes, and we obtain the geometry $FC(p, q, l)$. The fractal cube geometry is generated  asymptotically, i.e., $FC(p, q) \equiv \lim_{l \rightarrow \infty} FC(p, q, l)$. Note that there is certainly different pairs of $(p,q)$ which represent the same fractal, if one multiply both $p$ and $q$ by the same constant. 
% We can therefore reduce this redundancy in the labeling by requiring $q=p-2$, since we can always just leave a single array of cubes on each side in each iteration.  Therefore, the sequence of fractal obtained in our construction can be represented as $FC(p,p-2)$. 

Now we can calculate the Hausdorff dimension for $FC(p,q)$ and obtain
\be
D_H(FC, p, q) = \frac{\ln(p^3-q^3)}{\ln p}. \ee
For the first example $FC(3,1)$ illustrated in Fig.~\ref{fig:fractal_cube_m_hole}, we have $D_H(FC,3,1)$$=$$\ln(26)/\ln(3)$$=$$2.965$,  For the example $FC(6,4)$ illustrated in Fig.~\ref{fig:fractal_cube_m_hole_variant}(d-f), we have $D_H(FC,6,4)$$=$$2.804$, which is significantly lower than that of $FC(3,1)$.  As we have discussed above, the $Z$-distance is the shortest length of all logical $Z$-string representatives, and is hence $d_Z = O(L)$.  Meanwhile, the $X$-distance for $FC(p,q)$ is the minimal area of all  logical $X$-brane  representatives, which is just the area of the Sierpi\'nski carpet $SC(p,q)$ as illustrated in Fig.~\ref{fig:fractal_logical}, and scales as $d_X$$=$$ O(L^{D_H(SC,p,q)})$, where  $D_H(SC,p,q)$$=$$ \frac{\ln(p^2-q^2)}{\ln p}$ is the Hausdorff dimension of the Sierpi\'nski carpet.  As mentioned before, for $FC(3,1)$ one has $d_X = O(L^{1.893})$, while for $FC(6,4)$ one has $d_X=O(L^{1.672})$. The overall code distance for all $FC(p,q)$ is $d$$=$$\min(d_X,d_Z)$$=$$d_Z$$=$$O(L)$. The space overhead of the fractal surface codes defined on the fractal cube $FC(p,q)$ is hence $O(L^{D_H(FC,p,q)})=O(d^{D_H(FC,p,q)})$. For $FC(3,1)$, one has the space overhead being $O(L^{2.965})$, while for $FC(6,4)$, one has the space overhead of $O(L^{2.804})$. 

\begin{figure}[t]
  \includegraphics[width=1\columnwidth]{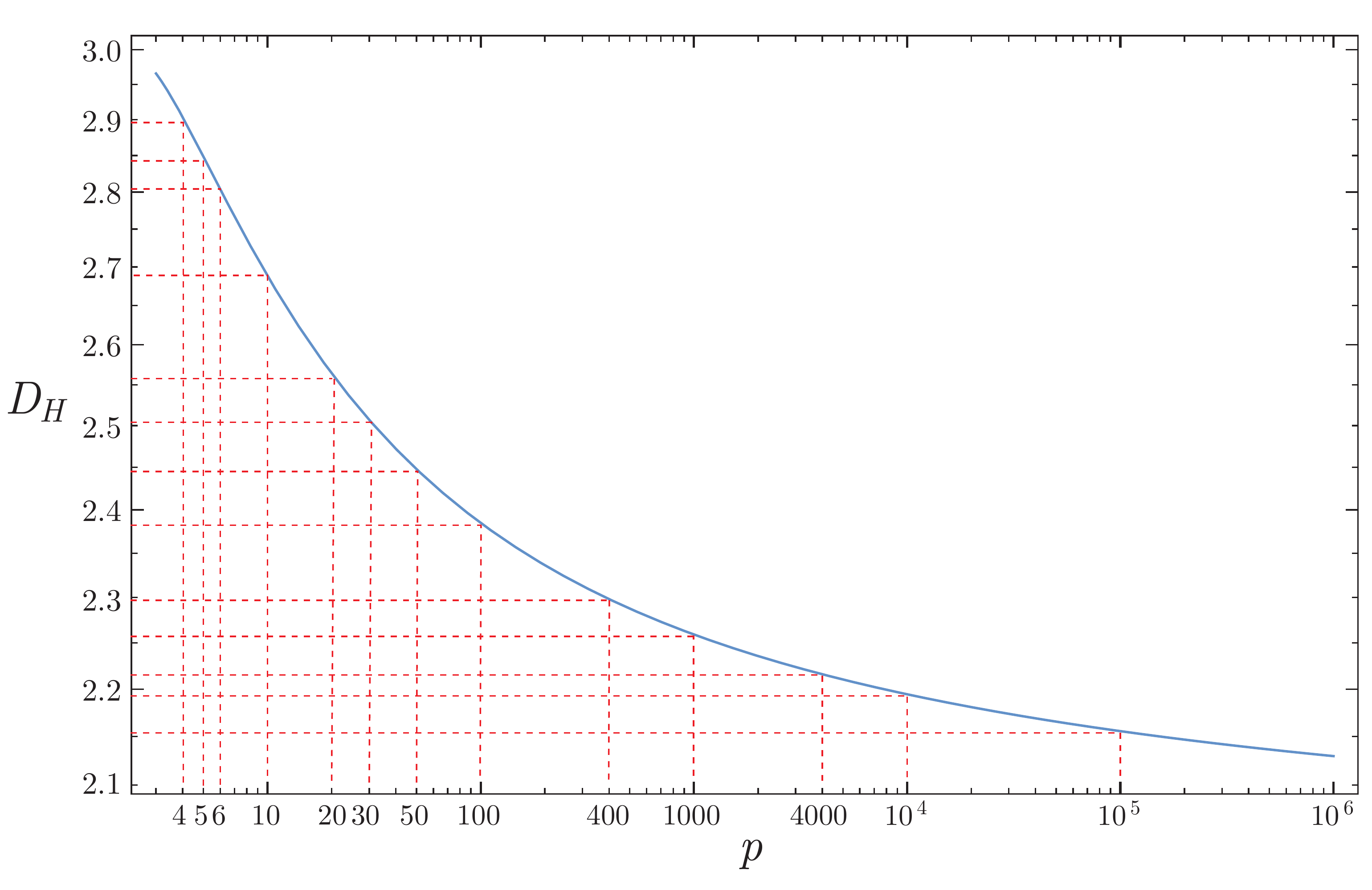}
  \caption{The Hausdorff dimensions $D_H
  $ of a sub-family of fractal cube geometries $FC(p,p-2)$.}
  \label{fig:plot_3D_fractal_dimensions}
\end{figure}

By varying the ratio $p/q$, one can construct fractal codes with Hausdorff dimension $D_H=3-\delta$ where $0<\delta<1$.  In particular, when keeping increasing $p/q$, one can asymptotically approach the Hausdorff dimension $D_H=2+\epsilon$ with arbitrary small $\epsilon$.   For simplicity, we consider a subclass of fractals in this family where we take $q=p-2$, i.e., leave a single array of cubes on each side in each iteration, and keep increasing $p$ to make the ratio $p/q=p/(p-2)$ growing.  The Hausdorff dimension $D_H$ of this subclass of fractal $FC(p,p-2)$ is plotted in Fig.~\ref{fig:plot_3D_fractal_dimensions}. One can see that $D_H$ drastically drops with $p$ above $D_H=2.5$, and the drop slows down around $D_H=2.4$ while $p$ stays reasonably small for $D_H=2.3$.  It then asymptotically approaches $2+\epsilon$, with all the code parameters approaching those of the 2D surface code.  The Hausdorff dimensions and corresponding $X$-distances and the overall code distances for a set of different fractal codes are listed in Table.~\ref{table1}  for detailed comparison. 

% In the inset of Fig.~\ref{fig:plot_3D_fractal_dimensions}, $D_H$ drops below $2.1$ and starts approaching $2$ asymptotically for enormously large $p$.  

\begin{table}
 \begin{center}
 \resizebox{0.6\columnwidth}{!}{%
	\begin{tabular}{|c|c|c|c|}
	\hline
	 	& $D_H$  &  $d_X$ & $d=d_Z$ \\
	 	\hline
	 	3D surface code & 3   & $O(L^2)$  &    \\
	 	%\hline
	 	$FC(3,1)$  & 2.965  & $O(L^{1.893})$   &   \\
	 	%\hline
	 	$FC(4,2)$  & 2.904  & $O(L^{1.792})$   &   \\
	 	%\hline
	 	$FC(5,3)$  & 2.849  & $O(L^{1.723})$   &  \\
	 	%\hline
	 	$FC(6,4)$  & 2.804  & $O(L^{1.672})$   &  \\
	 	%\hline
	 	$FC(7,3)$  & 2.958  & $O(L^{1.896})$   &   \\
	 	%\hline
	 	$FC(7,5)$  & 2.767  & $O(L^{1.633})$   &   \\
	 	%\hline
	 	$FC(10,8)$  & 2.688  & $O(L^{1.556})$   &   \\
	 	%\hline	 
	 	$FC(15,13)$  & 2.611  & $O(L^{1.486})$   & $O(L)$  \\
	 	%\hline	 
	 	$FC(30,28)$  & 2.507  & $O(L^{1.398})$   &   \\
	 	%\hline	 
	 	$FC(100,98)$  & 2.385  & $O(L^{1.299})$   &   \\
	 	%\hline	 
	 	$FC(500,498)$  & 2.288  & $O(L^{1.223})$   &   \\
	 	%\hline	 
	 	$FC(5000,4998)$  & 2.210  & $O(L^{1.163})$   &   \\
	 	%\hline	 
	 	$FC(10^5,10^5-2)$  & 2.156  & $O(L^{1.120})$   &   \\
	 	%\hline	 
	 	$FC(10^{10},10^{10}-2)$  & 2.078  & $O(L^{1.060})$   &  \\
	 	%\hline	 
	 	$FC(10^{20},10^{20}-2)$  & 2.039  & $O(L^{1.030})$   &   \\
        %\hline	 
	 	$FC(10^{80},10^{80}-2)$  & 2.0097  & $O(L^{1.0075})$   &   \\
        %\hline	 
	 	2D surface code  & 2  & $O(L)$   &   \\	 	
	 	\hline	 	 	
	\end{tabular}
	}
\end{center}
  \caption{Hausdorff dimensions and code distances of a set of fractal surface codes defined on the fractal cube geometries $FC(p,q)$, compared with those of the 2D and 3D surface codes. }
  \label{table1}
\end{table}

%\subsubsection{Cylindrical Sierpi\'nski carpet}

%\subsubsection{Cylindrical Apollonian gasket}

%\subsubsection{Apollonian sphere packing: the Apollonian code}

%\subsubsection{Non-simply connected m-holes}

%\subsubsection{More sophisticated cylindrical Sierpi\'nski carpet}

%\subsubsection{Menger sponge}

\section{Topological order on fractals embedded in $n$D}\label{sec:nD}

\subsection{General models in $n$D}
We consider a class of topological orders in $n$ dimensions described by the BF theory:
\be\label{eq:BF_ndimension}
S_\text{BF} = \int  \frac{N}{2\pi} b^{(n-i)} \wedge da^{(i)},
\ee
which generalizes the 3D action Eq.~\eqref{eq:BF}. We have two types of compact U(1) gauge fields coupled with each other, i.e., the $i$-form gauge field $a^{(i)}$ coupled to an $(n-i)$-form gauge field $b^{(n-i)}$. In our convention, we require $i \le  n-i$, such that $a^{(i)}$ always has a lower form than $b^{(n-i)}$.   
% The Wilson operators $W^{(k)}_{a_\nu}=e^{i\int_\Gamma a^{(k)}_\nu} $ and $W^{(n-k)}_{b_\nu}=e^{i\int_{\Gamma'} b^{(n-k)}_\nu} $ are the $k$- and $(n-k)$-dimensional world volume operators representing $(k-1)$- and $(n-k-1)$-dimensional   excitations transported along the paths $\Gamma$ and $\Gamma'$. 
The two types of excitations are called $e$ and $m$ in our definition. Their corresponding Wilson operators $W_e^{(i)}$ and $W_m^{(n-i)}$ are $i$- and $(n-i)$-dimensional branes. 
% In the special cases of $i=1$ and $i=2$, the Wilson operators $W^{(j)}_{a_\nu}$ correspond to the worldline and worldsheet of particle and string respectively.     

We focus on the $N=2$ case, where we can construct  exactly-solvable microscopic models, i.e., the $\mathbb{Z}_2$-toric code models in $n$D. The generalization to $\mathbb{Z}_N$-toric codes is  straightforward and can be found in App.~\ref{append:ZN}. Now we rely on the language of chain complex to describe these models. 

For a general description, we consider the following chain complex:  %in Eq.~\eqref{eq:exact_sequence_nD}.
\be\label{eq:chain_toric}
C_{i+1} \xrightarrow[]{\partial_{i+1}} C_i \xrightarrow[]{\partial_i} C_{i-1}.
\ee
The qubits are placed on the the $i$-cells, with the $X$-type stabilizers associated with the $(i-1)$-cells and the $Z$-type stabilizers with the $(i+1)$-cells.  We call such a model a $(i,n-i)$-toric code, and the corresponding topological phase as a $(i,n-i)$-$\mathbb{Z}_2$ topological order (or more generally $\mathbb{Z}_N$).  In our convention, we require $i < n-i$. 

The generalized $\ZZ_2$ toric code can be described by the following Hamiltonian (see App.~\ref{append:ZN} for the $\ZZ_N$ models):
\begin{align}\label{eq:nDTC}
\nonumber H_\text{TC}=&-\sum_q A^{(i-1)}_q - \sum_r B^{(i+1)}_r, \\
\text{with} \quad  A^{(i-1)}_q=&\Motimes_{j\in \delta e^{(i-1)}_q  }X_j, \quad B^{(i+1)}_r=\Motimes_{j\in \partial e^{(i+1)}_r}Z_j.
\end{align}
Here, $A_q^{(i-1)}$ is the $X$ stabilizer with all the Pauli-$X$ operators supported on the coboundaries of the $(i-1)$-cell labeled by $q$, i.e., $\delta e^{(i-1)}_q$, while $B_r^{(i+1)}$ is the $Z$ stabilizer with all the Pauli-$Z$ operators supported on the boundaries of the $(i+1)$-cell labeled by $r$, i.e., $\partial e^{(i+1)}_r$. In the special case of $i=1$, we have qubits placed on the edges (1-cell), $A_q^{(0)}$ corresponding to the vertex stabilizer (0-cell), and $B_r^{(2)}$ corresponding to the plaquette/face stabilizer (2-cell).

According to the chain complex Eq.~\eqref{eq:chain_toric},  the logical-$Z$ operators, i.e., the Wilson operators corresponding to the world volume of the $e$-excitation, can be associated with the $i^\text{th}$-homology $H_i(\L;\ZZ_2)$$=$$\text{Ker} (\partial_i)/\text{Img}(\partial_{i+1})$ and has the support on a $i$-brane.
 
 To define the logical-$X$ operator, we introduce the dual chain complex of Eq.~\eqref{eq:chain_toric}:
\be\label{eq:dual_chain_toric}
C^*_{n-i-1} \xleftarrow[]{\partial^*_{n-i}} C^*_{n-i} \xleftarrow[]{\partial^*_{n-i+1}} C^*_{n-i+1} .
\ee
  The logical-$X$ operator, i.e., the Wilson operators corresponding to the world volume of the $m$-excitation, is associated with the $(n-i)^\text{th}$-homology on the dual cell complex    $H_{n-i}(\L^*; \ZZ_2)=\text{Ker} (\partial^*_{n-i})/\text{Img}(\partial^*_{n-i+1})$  and has the support on a $(n-i)$-brane.

In the special case of $i$$=$$1$, the logical-$Z$ operators are string (Wilson-line) operators, i.e., corresponding to the worldlines of $e$-particles, and are hence associated with the $1^\text{st}$-homology $H_1(\L;\ZZ_2)$.  The logical-$X$ operators are $(n-1)$-branes corresponding to the world volume of $m$-excitations, and is associated with the $(n-1)^\text{th}$-homology on the dual cell complex $H_{n-1}(\L^*;\ZZ_2)$. In other cases ($i \ge 2$), both the logical-$Z$ and logical-$X$ operators are branes. The $e$-excitation are either loops (for $i=2$) or $(i-1)$-branes.  The $m$-excitations are $(n-i-1)$-branes.

\subsection{Theorems for $n$-dimensional simple fractals}\label{sec:nD_simple_fractal}

\subsubsection{Cases with string-like logical operators $(i=1)$}\label{sec:n_dimensional_no_string}
In the case where the theory contains particle excitations and equivalently string-like logical operators, we have a $\mathbb{Z}_N$ gauge theory and generalized toric-code model with $i=1$ in Eq.~\eqref{eq:nDTC}, i.e., described by the following chain complex:
\be
C_{2} \xrightarrow[]{\partial_{2}} C_1 \xrightarrow[]{\partial_1} C_{0}.
\ee

We can then generalize Theorem 1 into the $n$-dimensional case:
\begin{theorem}\label{theorem1}
	 The $(1,n-1)$-$\mathbb{Z}_N$ topological order exists on a simple fractal embedded in an $n$-dimensional manifold with $m$-holes.
\end{theorem}
Recall that a simple $n$-dimensional fractal consists of $n$-dimensional simply connected holes which are  homeomorphic to $n$-dimensional balls $D^n$, as has been defined previously in Def.~\ref{def:simple_fractal}. The $m$-hole in this case is an  $n$-ball $D^n$, with its boundaries being $m$-boundaries corresponding to a $(n-1)$-sphere ($S^{n-1}=\partial D^n$), where $m$-loop/brane excitations can condense onto while $e$-particle excitations cannot.

As in the 3D case, we consider setups with three different background topologies in our discussions: (1) The $n$-dimensional fractal surface code $\L_\text{FSF}$: we choose the background manifold to be topologically equivalent to an $n$-dimensional hypercube, with two of its $(n-1)$-dimensional hyper-surfaces on the opposite sides being the $e$-boundaries where  the $e$-particles can condense onto while the $m$-loop/brane excitations cannot.  The other $(2n-2)$ of its $(n-1)$-dimensional hyper-surfaces are $m$-boundaries with the opposite condensation property as just introduced above. We also call this code a $(1,n$$-$$1)$-fractal surface code. (2) Punctured $n$-torus  $\L(T^n)$: the background manifold is chosen as an $n$-dimensional torus $T^n$. (3) Punctured $n$-sphere $\L(S^n)$: the background manifold is chosen as an $n$-dimensional sphere $S^n$.        

In contrast to what we have done in 3D, where we have made three different proofs for Theorem 2, here we just use one way based on algebraic topology  generalizing Proof~3c, which is quite concise in the higher-dimensional case.  For clarity, we focus on the case of $\mathbb{Z}_2$ topological order, and the generalization to the  $\mathbb{Z}_N$ case is straightforward.    

\nin \textbf{Proof}:    Similar to the 3D case, the ground-state subspace (code space) of the $\mathbb{Z}_2$ topological order can be written as  
\be
\mathcal{H}_\mathcal{C}= \mathbb{C}^{H_1(\mathcal{L}, \mathcal{B}_e; \mathbb{Z}_2)},
\ee
where $H_1(\mathcal{L}, \mathcal{B}_e; \mathbb{Z}_2)$ represents the 1st relative $\ZZ_2$-homology group.  Here,  the argument $\mathcal{L}$ represents the $n$-dimensional lattice, $\mathcal{B}_e$ represents all the $(n-1)$-dimensional $e$-boundaries. The generalization to the $\ZZ_N$  topological order can be done simply by replacing $H_1$ with the $\mathbb{Z}_N$-homology group $H_1(\mathcal{L}, \mathcal{B}_e; \mathbb{Z}_N)$.

The $e$-particle cannot condense onto the $m$-holes, which corresponds to the fact that its worldline, the $Z$-string,  cannot terminate on the $m$-holes. Mathematically, this means that there is no nontrivial (non-contractible) relative 1-cycle in the bulk region connecting the holes. Meanwhile, the $Z$-loop also cannot enclose an $n$-dimensional hole, similar to the $n=3$ case discussed before.
A $Z$-loop ($S^1$) orbiting around an $m$-hole ($D^n$) can always be shrunk into a trivial point via gauge transformation, for example, by multiplying $Z$-plaquette stabilizers in the exact-solvable case of toric code model, i.e., 
\be
\raisebox{-0.5cm}{\includegraphics[scale=1]{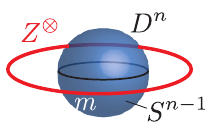}} \quad  =  \raisebox{-0.5cm}{\includegraphics[scale=1]{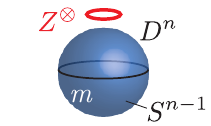}}  = \raisebox{-0.5cm}{\includegraphics[scale=1]{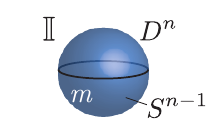}}. 
\ee
This can also be interpreted by the following mathematical  property of the homology groups of a $(n-1)$-sphere, i.e., the boundary of the $m$-hole ($S^{n-1}=\partial D^n$):
\begin{align}\label{eq:sphere_homology_nD}
     H_i(S^{n-1}) = \begin{cases} \mathbb{Z}  \quad  \ &i=0 \ \text{or} \ n-1; \\ 
     \mathsf{0} \quad &\text{otherwise}. 
     \end{cases}
\end{align}
Here, $H_1(S^{n-1})=\mathsf{0}$ corresponds to a loop $S^1$ (absolute 1-cycle) on the sphere $S^{n-1}$ and is trivial (contractible), meaning that a loop cannot enclose a sphere $S^{n-1}$ (the boundary of $m$-holes).  In sum, these properties suggest that there is no nontrivial relative or absolute 1-cycles in the bulk region either encircling or connecting the holes.  A rigorous mathematical proof of this observation requires us to directly compute the relative homology group of the entire complex, i.e., $H_1(\L, \B_e; \ZZ_2)$, which is shown in the following. 

We discuss the setups of  the three types of background topologies mentioned above.  The simplest case is the punctured $n$-sphere $\L(S^n)$. In this case, there is no $e$-boundary in the system at all, i.e., $\mathcal{B}_e = \emptyset$. The geometry is equivalent to an $n$-sphere with a set of interior holes homeomorphic to $n$-balls $D^n$ being cut out, i.e., $\L(S^n) = S^n \setminus (D^n \cup D^n \cup \cdots \cup D^n)=  S^n \setminus \cup_j D^n_{(j)}$.  We can again use a `divide-and-conquer' strategy to compute the homology group.  In particular, we apply the Alexander duality in Eq.~\eqref{eq:Alexander_duality} and get
\begin{align}\label{eq:punctured_n-sphere_homology}
\nonumber &  H_1(S^n \setminus \cup_j D^n_{(j)}; \ZZ_2)
\cong H^1(\cup_j D^n_{(j)}; \ZZ_2) \\
 \cong& \bigoplus_j H^1(D^n_{(j)}; \ZZ_2) = \bigoplus_j \mathsf{0} = \mathsf{0}.
\end{align}
Using the Poincaré-Lefschetz duality, we get the isomorphism $ H^1(\mathcal{L}(S^n); \mathbb{Z}_2) \cong H_{n-1}(\mathcal{L}^*(S^n), \mathcal{B}^*_m; \mathbb{Z}_2)$ as a higher-dimensional generalization of Eq.~\eqref{eq:punctured_sphere_homology}. Combined with Eq.~\eqref{eq:cohomology_eqaul_homology}, we can obtain that the dual $(n$$-$$1)^\text{th}$ relative homology group (on the dual cell complex $\L^*$) corresponding to the $X$-brane is also trivial:
\be\label{eq:sphere_homology_duality_nD}
H_{n-1}(\mathcal{L}^*(S^3), \mathcal{B}^*_m; \mathbb{Z}_2)  \cong H_1(\mathcal{L}(S^3); \mathbb{Z}_2) = \mathsf{0}.
\ee
Therefore, there is no nontrivial logical $Z$-string or $X$-brane in this fractal.  The ground-state subspace (code space) is hence 1-dimensional:
\be
\mathcal{H}_C= \mathbb{C}^{H_1(\mathcal{L}(S^3); \mathbb{Z}_2)} \cong \mathbb{C}^{H_{n-1}(\mathcal{L}^*(S^3), \mathcal{B}^*_m; \mathbb{Z}_2)}=\mathbb{C}^1,
\ee
  meaning that it encodes zero logical qubit and there is a single ground state.  Nevertheless, it is expected that there is still long-range entanglement in the ground-state wavefunction as in the 3D punctured sphere. 
  
  We then consider the $(1, n$$-$$1)$-fractal surface code supported on the complex $\L_\text{FSF}$, which has a background manifold with two $e$-boundaries and $(n$$-$$2)$ $m$-boundaries. Topologically speaking, it is equivalent to a usual $(1, n$$-$$1)$-surface code geometry $\L_\text{SF}$ in $n$ dimensions with a set of $n$-balls being cut out, i.e., $\L_\text{FSF} = \L_\text{SF} \setminus  \cup_j D^n_{(j)}$.  Using  Eq.~\eqref{eq:connected_sum_presentation},  we can present the fractal surface code geometry as $\L_\text{FSF}$$=$$ (\L_\text{SF} \# S^n) \setminus  \cup_j D^n_{(j)}= \L_\text{SF} \# ( S^n \setminus  \cup_j D^n_{(j)})$, namely the connected sum of the $n$D surface code and a punctured $n$-sphere. This connected sum can be interpreted as cutting out a large $n$-ball $D^n$ with radius of $O(L)$ inside the $n$D surface code $\L_\text{SF}$ and gluing a punctured $n$-ball $( S^n \setminus  \cup_j D^n_{(j)})$  into this hollow region, in analogy with the 3D case illustrated in Fig.~\ref{fig:connected_sum_3D}(d). This punctured ball is equivalent to a punctured $n$-sphere with a $n$-ball being cut out, i.e., $( S^n \setminus  \cup_j D^n_{(j)}) \setminus D^n$.   Using the Mayer–Vietoris sequence \cite{Hatcher:2001ut}, we can now compute the 1st relative homology of this connected sum with the following homology decomposition: 
\begin{align}\label{eq:FSF_homology_nD}
\non & H_1(\L_\text{FSF}, \B_e; \ZZ_2) \cong  H_1(\L_\text{FSF}/\B_e; \ZZ_2) \\
\non \cong &H_1(\L_\text{SF}/\B_e \# ( S^n \setminus  \cup_j D^n_{(j)}) ; \ZZ_2) \\
\non \cong & H_1(\L_\text{SF}/\B_e; \ZZ_2) \oplus  H_1( S^n \setminus  \cup_j D^n_{(j)} ; \ZZ_2)\\
 \cong &
H_1(\L_\text{SF}, \B_e; \ZZ_2) \oplus  \mathsf{0} = \ZZ_2. 
\end{align}
From the decomposition, one can see that the only  nontrivial contribution $\ZZ_2$ comes from the $(1, n$$-$$1)$-surface code, corresponding to a single macroscopic logical string $\overline{Z}$ connecting the opposite $e$-boundaries.    According to the Poincaré-Lefschetz duality in Eq.~\eqref{eq:Poincare_duality_cellulation} and the isomorphism in Eq.~\eqref{eq:cohomology_eqaul_homology}, we have the dual $(n$$-$$1)^\text{th}$ relative homology group as
\be
H_{n-1}(\mathcal{L}^*_\text{FSF}, \mathcal{B}^*_m; \mathbb{Z}_2)  \cong H_1(\mathcal{L}^n_\text{FSF}, \mathcal{B}_e; \mathbb{Z}_2) = \mathbb{Z}_2.
\ee
By using the Poincaré-Lefschetz dual of the homology decomposition in Eq.~\eqref{eq:FSF_homology_nD}, we get 
\begin{align}
\non & H_{n-1}(\L^*_\text{FSF}, \B^*_m; \ZZ_2) \\ 
\cong & \non H_{n-1}(\L^*_\text{SF}, \B^*_m; \ZZ_2) \oplus H_{n-1}(\L^*(S^3), \B^*_m; \ZZ_2) \\
=& H_{n-1}(\L^*_\text{SF}, \B^*_m; \ZZ_2) \oplus \mathsf{0},
\end{align}
where the last equality uses Eq.~\eqref{eq:sphere_homology_duality_nD}. We can see that the only nontrivial contribution $\ZZ_2$ comes from the $(1,n-1)$-surface code part $H_{n-1}(\L^*_\text{SF}, \B^*_m; \ZZ_2)$, which corresponds to a single macroscopic logical brane $\overline{X}$ (the world-volume of the $m$-loop/brane)  connecting all the $2n$$-$$2$ external hyper-surfaces with $m$-boundaries and intersects with the macroscopic logical string $\overline{Z}$ at a single point (in terms of algebraic intersection),  leading to the anti-commutation relation $\{\overline{X}, \overline{Z}\}=0$.  The ground-state subspace (code space) of the fractal surface code is $\mathcal{H}_C=\mathbb{C}^{\mathbb{Z}_2}=\mathbb{C}^2$, corresponding to a single logical qubit and 2-fold ground-state degeneracy.  

The discussion of the punctured $n$-torus geometry can be found in App.~\ref{app:n_dimensional_string}.

It is clear from the above discussions  that in both the fractal surface code and the punctured torus geometry (in App.~\ref{app:n_dimensional_string}), there are only macroscopic logical operators. The $Z$-distance corresponding to the minimal length of the macroscopic logical-$Z$ string is $d_Z=O(L)$ ($L$ is the linear size of the system) since the $Z$-string needs to circumvent all the $m$-holes. The $X$-distance corresponding to the minimal volume of the macroscopic logical-$X$ brane has a lower bound determined by the minimal volume of the brane outside the region of the large $n$-ball being cut out and with radius of $O(L)$, i.e., $d_X \sim \Omega(L^{n-2})$ as a generalization of the 3D case illustrated in Fig.~\ref{fig:connected_sum_3D}(c,d).  One can also get an exact scaling of the $X$-distance as     $d_X=O(L^{D_H})$, where $n-2<D_H<n-1$ is the Hausdorff dimension of the minimal-volume fractal logical-$X$ brane, as will be discussed in the next subsection in details. In sum, the distance of this code is the smaller one of the $Z$- and $X$-distance:  $d=\min(d_Z, d_X) = O(L)$.   No  microscopic logical operator with $O(1)$ distance exists.  Therefore, the corresponding code has macroscopic code distance and hence topological order exists in these fractal geometries.   
We note that the proof for the more general case of the $\mathbb{Z}_N$-topological order follows directly from the above proof by simply replacing all the $\mathbb{Z}_2$ coefficients   with $\mathbb{Z}_N$ in the relative/absolute homology and cohomology groups.

\nin Q.E.D.

Similar to the 3D case, the above proof for topological order also provides a rigorous
proof for the following mathematical theorem:
\begin{theorem}
	There exists a class of simple fractals $\L$ embedded in an $n$-dimensional manifold with boundary $\B_m$ on its interior holes, such that both its absolute  1-systole $sys_1(\L)$ and its dual relative $(n-1)$-systole $sys_{n-1}(\L^*, \B^*_m)$ are macroscopic, i.e., scale as a power law of the linear size $L$ of the fractals.
\end{theorem}
This theorem is stated for the situation that the background manifold is a $n$-torus (see App.~\ref{app:n_dimensional_string}) or more generic closed $n$-manifold with non-zero 1st Betti number. As in the 3D case, the theorem holds for the cases of $\ZZ$-, $\ZZ_2$- and $\ZZ_N$-systoles.

\vspace{0.1in}

\begin{figure*}[t]
  \includegraphics[width=1.4\columnwidth]{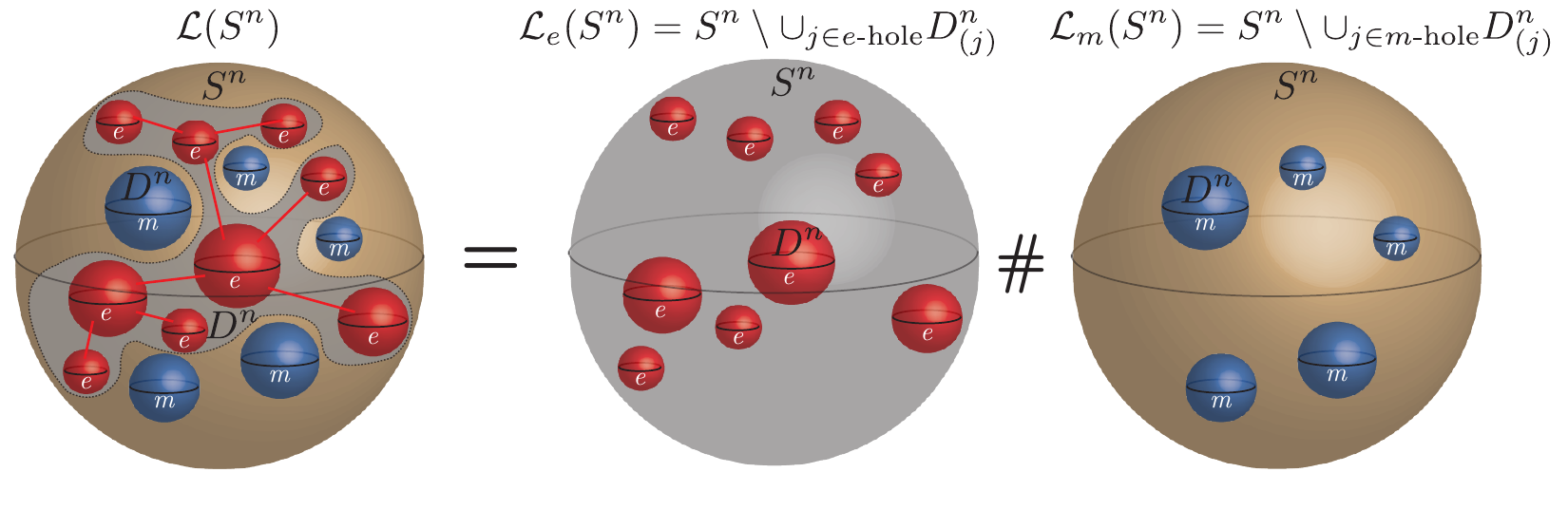}
  \caption{Illustrating the decomposition of a punctured $n$-sphere $S^n$ with both $e$-holes and $m$-holes into a punctured $n$-sphere containing only $e$-holes and another containing only $m$-holes. One first constructs a spanning tree connecting all the $e$-holes, and then thicken the spanning tree into a punctured $n$-ball. The entire punctured $n$-sphere is hence equivalent to the gluing of two punctured $n$-balls with only $e$-holes and $m$-holes respectively, which is in turn equivalent to the connected sum of two punctured $n$-spheres.}
  \label{fig:spanning_tree}
\end{figure*}

\subsubsection{Cases without string-like logical operators ($i \ge 2$): self-correcting quantum memories}\label{sec:general_k}
Now we consider other general situations, i.e., topological orders/codes without particle excitations and string-like logical operators ($i \ge 2$) .  These systems are also expected to be self-correcting quantum memories when supported on a manifold, since the logical operators are all branes.   In these cases, we have the following theorem:

\begin{theorem}
	The $(i,n-i)$-$\mathbb{Z}_N$ topological order exists on a simple fractal embedded in an $n$-dimensional manifold with a certain type of boundary on each hole for $2 \le i \le n-i$, independent of the boundary type on each hole. \footnote{This theorem is expected to be generalizable to the case that each hole can have  mixed boundary types, as will be discuss in future works.}
\end{theorem}

\nin \textbf{Proof}: Recall that in a  simple fractal, all the holes are homeomorphic to an $n$-dimensional ball $D^n$. A key feature in the case without particle excitations ($i \ge 2$) is that either $Z$- or $X$-branes, i.e., world-volumes of the $e$- or $m$-excitations, cannot enclose the holes. In other words, neither $e$-hole nor $m$-hole can encode any logical qubit in this case. This is because the boundary of the holes are $(n-1)$-dimensional spheres: $S^{n-1}=\partial D^n$. For $i \ge 2$, we have $i <n-1$ and $n-i<n-1$. Therefore, both $Z$- and $X$-branes have dimensions lower than $n-1$, the dimension of the hole boundary, and can hence always be shrunk into a single point via gauge transformation, i.e., for $1<i<n-1$, we have 
\be
\raisebox{-0.9cm}{\includegraphics[scale=1]{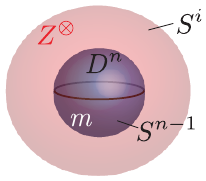}} \quad  =  \raisebox{-0.9cm}{\includegraphics[scale=1]{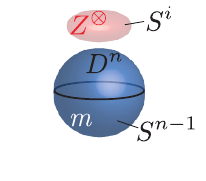}}  = \raisebox{-0.9cm}{\includegraphics[scale=1]{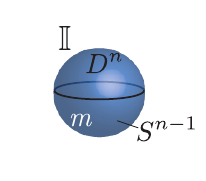}}, 
\ee 
\be
\raisebox{-0.9cm}{\includegraphics[scale=1]{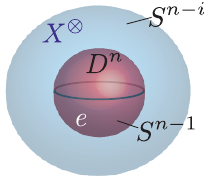}} \quad  =  \raisebox{-0.9cm}{\includegraphics[scale=1]{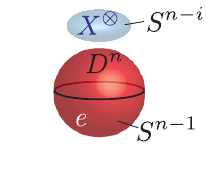}}  = \raisebox{-0.9cm}{\includegraphics[scale=1]{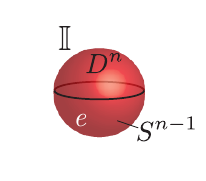}}. 
\ee 
Mathematically, this is a manifestation of the trivial $i^\text{th}$- and $(n-i)^\text{th}$-homology groups on $S^{n-1}$:
\be 
H_i(S^{n-1}) \cong H_{n-i}(S^{n-1})= \mathsf{0} \quad \text{when } 1<i<n-1, 
\ee
as has been summarized in Eq.~\eqref{eq:sphere_homology_nD}.   

Note that there is a significant difference from the $i=1$ case, i.e., the $(1,n$$-$$1)$-$\mathbb{Z}_N$ topological order. In the $i=1$ case, the $X$-brane (world volume of the $m$-excitations) has dimension $n-1$, which can then enclose the hole, since  $H^{n-1}(S^{n-1})= \mathbb{Z}$ as summarized in Eq.~\eqref{eq:sphere_homology_nD}.  To trivialize the $X$-brane enclosing the holes, we have to choose the holes equipped with $m$-boundaries, such that the enclosing   $X$-brane is turned into trivial relative cycle which can be absorbed onto the $m$-hole.  On the other hand, an $e$-hole traps a non-trivial logical-$X$ brane. Similar to the 3D case, there is also a no-go result for the existence of (1,$n-1$)-$\ZZ_N$ topological order when the holes in the fractal have $e$-boundaries, which can be proven  by a straightforward generalization of Theorem \ref{theorem:no_go_3D}.  Nevertheless, in the case of $i \ge 2$, we can choose arbitrary boundary on each holes, either $e$ or $m$, which does not affect the trivial cycles/cocycles around them.   To make a rigorous proof, we need to directly compute the relative homology groups $H_i(\L, B_e; \ZZ_2)$ and $H_{n-i}(\L^*, B^*_m; \ZZ_2)$ as shown below.   
% As an example of a hole with mixed boundaries, we could take an $n$-dimensional hypercubes, with $p$ of its hyper-surfaces being $e$-boundaries, and the other $(2n-p)$ hyper-surfaces being the $m$-boundaries. In sum, there is no non-trivial contributions to the $k^\text{th}$-homology group $H_k$ and the $(n-k)^\text{th}$-homology group $H_{n-k}$ from the holes in the bulk.  

Similar to the $i$$=$$1$ case, we now discuss setups with three types of background manifolds respectively.  We start with the punctured $n$-sphere $\L(S^n)$.
We first decompose $\L(S^n)$ as a connected sum of two $n$-punctured spheres $\L_e(S^n)$ and $\L_m(S^n)$ which only  contain $e$-holes and $m$-holes respectively, i.e., $\L(S^n)= \L_e(S^n) \# \L_m(S^n)$.  This decomposition is enabled by the following construction and is illustrated in Fig.~\ref{fig:spanning_tree}. We first connect all the $e$-holes with a spanning tree, which does not touch any of the $m$-holes.  This spanning tree is simply connected, and by thickening the tree and including the surrounding regions of each $e$-hole we get a subsystem equivalent to a  punctured  $n$-ball with all the interior $e$-holes being cut out, i.e., $D^n \setminus \cup_{j\in e \, \text{hole}} D^n_{(j)}$.  Note that an $n$-ball can be obtained by cutting out an $n$-ball from an $n$-sphere, i.e., $D^n = S^n \setminus D^n$.   Therefore, the subsystem we have just obtained can be viewed as a punctured sphere $\L_e(S^n) = S^n \setminus \cup_{j\in e \text{-hole}} D^n_{(j)}$ with an additional  $n$-ball being cut out.  When cutting out this subsystem, the remaining subsystem can also be viewed as a punctured $n$-sphere (where the $m$-holes are cut out) $\L_m(S^n) = S^n \setminus \cup_{j\in m \text{-hole}} D^n_{(j)}$ with an additional large $n$-ball $D^n$ being cut out.  Therefore, the whole complex $\L(S^n)$ can be obtained from gluing the two punctured $n$-spheres  by cutting out a $n$-ball $D^n$ from each subsystem and identifying the boundaries of the two $n$-balls, i.e.,  two $(n-1)$-spheres $S^{n-1}$, which is literally a connected sum of the two subsystems, i.e., 
\be
\L(S^n)= \big(S^n \setminus \cup_{j\in e \text{-hole}} D^n_{(j)} \big) \# \big(S^n \setminus \cup_{j\in m \text{-hole}} D^n_{(j)} \big).
\ee

In analogy with the $i=1$ case, we choose to compute the absolute homology for the punctured $n$-sphere instead of the relative homology since the former is much easier. Therefore, we aim to first compute $H_i(\L_m(S^n); \ZZ_2)$ and $H_{n-i}(\L^*_e(S^n); \ZZ_2)$, and the rest can be obtained from the Poincaré-Lefschetz duality. We again use the `divide-and-conquer' strategy
through the Alexander duality from Eq.~\eqref{eq:Alexander_duality} and  obtain
\begin{align}\label{eq:punctured_n-sphere_homology_general_mhole}
\nonumber & H_i(\L_m(S^n); \ZZ_2) =  H_i(S^n \setminus \cup_{j\in m \text{-hole}} D^n_{(j)}; \ZZ_2)\\
\non \cong& H^{n-1-i}(\cup_j D^n_{(j)}; \ZZ_2) \cong \bigoplus_{j\in m \text{-hole}} H^{n-1-i}(D^n_{(j)}; \ZZ_2) \\
=& \bigoplus_{j \in m \text{-hole}} \mathsf{0} = \mathsf{0},
\end{align}
where we have used the property $H^{n-1-i}(D^n; \ZZ_2)=H_{n-1-i}(D^n; \ZZ_2)=\mathsf{0}$ for $1$$<$$i$$<$$n-1$ and $n>3$ since any absolute $(n-1-i)$-cocycle or $(n-1-i)$-cycle is contractible in $D^n$.   In a similar fashion, we can obtain
\begin{align}\label{eq:punctured_n-sphere_homology_general_ehole}
\nonumber & H_{n-i}(\L^*_e(S^n); \ZZ_2) =  H_{n-i}(S^n \setminus \cup_{j\in e \text{-hole}} D^n_{(j)}; \ZZ_2)\\
\non \cong& H^{i-1}(\cup_j D^n_{(j)}; \ZZ_2) \cong \bigoplus_{j\in e \text{-hole}} H^{i-1}(D^n_{(j)}; \ZZ_2) \\
=& \bigoplus_{j \in e \text{-hole}} \mathsf{0} = \mathsf{0},
\end{align}
where we have used the property $H^{i-1}(D^n; \ZZ_2)=H_{i-1}(D^n; \ZZ_2)=\mathsf{0}$ for $1$$<$$i$$<$$n-1$ and $n>3$, and dropped the dual `*' after the first equality since the the (co)homology of a manifold is independent of the cellulation.  Now using the Poincaré-Lefschetz duality in Eq.~\eqref{eq:Poincare_duality_cellulation} and the isomorphism in Eq.~\eqref{eq:cohomology_eqaul_homology}, we obtain the dual relative homology groups $H_{n-i}(\L^*_m(S^n),\B^*_m; \ZZ_2) \cong H_i(\L_m(S^n); \ZZ_2)$$=$$\mathsf{0}$ and $H_{i}(\L_e(S^n),\B_e; \ZZ_2) \cong H_{n-i}(\L^*_e(S^n); \ZZ_2)$$=$$ \mathsf{0}$. We can then compute the $i^\text{th}$ relative  homology groups of the entire complex using the homology decomposition of connected sum:
\begin{align}
\non & H_i(\L(S^n), \B_e; \ZZ_2)  \cong   H_i(\L_e(S^n) \# \L_m(S^n), \B_e; \ZZ_2) \\
=&  H_i(\L_e(S^n), \B_e; \ZZ_2)  \oplus  H_i(\L_m(S^n); \ZZ_2) = \mathsf{0}.
\end{align}
By using the Poincaré-Lefschetz duality again, we can get  $ H_{n-i}(\mathcal{L}^*(S^n), \mathcal{B}^*_m; \mathbb{Z}_2) \cong H_i(\mathcal{L}(S^n), \mathcal{B}_e; \mathbb{Z}_2) = \mathsf{0}$, which shows that both the $i^\text{th}$ and $(n-i)^\text{th}$ relative  homology groups in this setup is trivial. In other words, any absolute or relative $i$-cycle and $(n-i)$-cycle in the punctured sphere is trivial. This leads to the 1-dimensional ground-state subspace (code space):
\be
\mathcal{H}_C= \mathbb{C}^{H_i(\mathcal{L}(S^n), \mathcal{B}_e; \mathbb{Z}_2)} \cong \mathbb{C}^{H_{n-i}(\mathcal{L}^*(S^n), \mathcal{B}^*_m; \mathbb{Z}_2)}=\mathbb{C}^1.
\ee
Therefore, no logical qubit or ground-state degeneracy exists in this case, but it is expected that the ground-state wavefunction is long-range entangled as discussed before.

Now we consider the case of the $(i,n-i)$-fractal  surface code geometry $\L_\text{FSF}$, with the background manifold being a hyper-cube having $2i$ hyper-surfaces with $e$-boundaries and $2(n-i)$ hyper-surfaces with $m$-boundaries. One can write the background manifold as $\mathcal{M}_\text{SF}=D^i \times D^{n-i}$, i.e., a manifold product of an  $i$-dimensional disk (ball) and an $(n-i)$-dimensional disk (ball). We first consider the logical operators in the $(i, n-i)$-surface code without holes inside. A single logical $i$-brane $\overline{Z}$ (a relative $i$-cycle) travels through the disk $D^i$ and terminate at all the $2i$ hyper-surfaces with $e$-boundaries
%, while the intersection with these hyper-surfaces are the boundaries of the $i$-disk, i.e., an $(i-1)$-dimensional sphere $S^{i-1}=\partial D^i$.
Similarly, a single logical $(n-i)$-brane $\overline{X}$ (a relative $(n-i)$-cycle) travels through the disk $D^{n-i}$ and terminate at all the $(2n-2i)$ hyper-surfaces with $m$-boundaries.
%, while the intersection with these hyper-surfaces are the boundaries of the $(n-i)$-disk, i.e., a $(n-i-1)$-dimensional sphere $S^{n-i-1}=\partial D^{n-i}$. 

We now start computing the relative homology group of $\L_\text{FSF}$. Note the fractal surface code geometry can be expressed as a connected sum of the $(i, n-i)$-surface code and a punctured $n$-sphere similar to the $i=1$ case, i.e., $\L_\text{FSF}$$=$$  \L_\text{SF} \# ( S^n \setminus  \cup_j D^n_{(j)})$. We then use the following homology decomposition of the connected sum:
\begin{align}\label{eq:FSF_homology_nD_general}
\non & H_i(\L_\text{FSF}, \B_e; \ZZ_2) \cong  H_i(\L_\text{FSF}/\B_e; \ZZ_2) \\
\non \cong &H_i(\L_\text{SF}/\B_e \# ( S^n \setminus  \cup_j D^n_{(j)}) ; \ZZ_2) \\
\non \cong & H_i(\L_\text{SF}/\B_e; \ZZ_2) \oplus  H_i( S^n \setminus  \cup_j D^n_{(j)} ; \ZZ_2)\\
 \cong &
H_i(\L_\text{SF}, \B_e; \ZZ_2) \oplus  \mathsf{0} = \ZZ_2. 
\end{align}
Here, the only nontrivial contribution $\ZZ_2$ comes from the macroscopic logical $i$-brane $\lo{Z}$. By using the the Poincaré-Lefschetz duality,  we get the following expression of the $(n-i)^\text{th}$ relative homology group on the dual complex:
\be\label{eq:homology_equivalence_nD_general}
H_{n-i}(\mathcal{L}^*_\text{FSF}, \mathcal{B}^*_m; \mathbb{Z}_2) \cong H_i(\mathcal{L}_\text{FSF}, \mathcal{B}_e; \mathbb{Z}_2) =\mathbb{Z}_2, 
\ee
leading to the following ground-state subspace (code space): $\mathcal{H}_C$$=$$\mathbb{C}^{\mathbb{Z}_2^{\oplus n}}$$=$$\mathbb{C}^{2^n}$ which corresponds to $n$ logical qubits and a $2^n$-fold ground-state degeneracy.  By using the Poincaré-Lefschetz dual of the homology decomposition in Eq.~\eqref{eq:FSF_homology_nD_general}, we get 
\begin{align}
\non & H_{n-i}(\L^*_\text{FSF}, \B^*_m; \ZZ_2) \\ 
\cong & \non H_{n-i}(\L^*_\text{SF}, \B^*_m; \ZZ_2) \oplus H_{n-i}(\L^*(S^3), \B^*_m; \ZZ_2) \\
=& H_{n-i}(\L^*_\text{SF}, \B^*_m; \ZZ_2) \oplus \mathsf{0},
\end{align}
where the last equality uses Eq.~\eqref{eq:homology_equivalence_nD_general}. We can see that the only nontrivial contribution $\ZZ_2$ comes from the $(i,n-i)$-surface code part $H_{n-i}(\L^*_\text{SF}, \B^*_m; \ZZ_2)$, which corresponds to a single macroscopic logical $(n-i)$-brane $\overline{X}$. 

The discussion of the punctured $n$-torus geometry can be found in App.~\ref{app:general_k}.

% Finally, we consider the simple fractal nested on an $n$-dimensional torus $T^n=(S^1)^{\times n}$. There are $C_n^i$  macroscopic logical $i$-branes $\overline{Z}_j \ (j$$=$$1,2,\cdots, C_n^i)$  going through the $C_n^i$ non-contractible $i$-cycles $(S^1)^{\times i}$ around $T^n$. Similarly, there are $C_n^{n-i}=C_n^{i}$  macroscopic logical $(n-i)$-branes  $\overline{X}_j \ (j$$=$$1,2,\cdots, C_n^i)$  going through the $C_n^i$ non-contractible $(n-i)$-cocycles $(S^1)^{\times (n-i)}$ around $T^n$. We hence get the following relative homology groups:
% \be
% H_i(\mathcal{L}(T^n), \mathcal{B}_e; \mathbb{Z}_2) \cong H_{n-i}(\mathcal{L}(T^n), \mathcal{B}_m; \mathbb{Z}_2)=\mathbb{Z}_2^{\oplus C^i_n}, 
% \ee
% leading to the following ground-state subspace (code space): $\mathcal{H}_\mathcal{C}$$=$$\mathbb{C}^{\mathbb{Z}_2^{\oplus C^i_n}}$$=$$\mathbb{C}^{2^{C^i_n}}$, corresponding to $C^i_n$ logical qubits and a $2^{C^i_n}$-fold ground-state degeneracy. By using the homology decomposition similar to the fractal surface code case, we can see all the nontrivial contribution $\mathbb{Z}_2^{\oplus C^i_n}$ comes from the macroscopic logical  $i$-branes $\lo{Z}_j$ and logical $(n-i)$-branes $\lo{X}_j$.  

\begin{table*}
\centering
\resizebox{2.1\columnwidth}{!}{%
	\begin{tabular}{|c|c|c|c|c|c|c|c|c|c|}
	\hline
	 	Model & 
	 	\begin{tabular}{@{}c@{}}
	 	TQFT \\
	 	action 
	 	\end{tabular} & 
	 	\begin{tabular}{@{}c@{}}
	 	hole \\ 
	 	boundary
	 	\end{tabular}
	 	& logical-$Z$  & logical-$X$  & \begin{tabular}{@{}c@{}} 
	 	  Hausdorff \\
	 	  dimension $D_H$
	 	 \end{tabular} 
	 	  &  $Z$-distance $d_Z$ & $X$-distance $d_X$ &  code distance $d$ & logical gates \\
	 	 %& action &   &   & dimension $D_H$ &  overhead & $d_Z$ & $d_X$ &  $d$ \\
	 	\hline
	 	\begin{tabular}{@{}c@{}}
	 	$(1,n-1)$-fractal  \\
	 	surface code ($n\le 3$)
	 	\end{tabular}
	 	& $\int  \frac{N}{2\pi} b^{(n-1)} \wedge da^{(1)}$   & $m$  & 
	 	\begin{tabular}{@{}c@{}}
	 	string  \\
	 	$H_1(\L, \B_e)$
	 	\end{tabular}
	 	&  \begin{tabular}{@{}c@{}}
	 	$(n-1)$-brane  \\
	 	$H_{n-1}(\L^*, \B^*_m)$
	 	\end{tabular}
	 	&
	 	\multirow{7}{*}{
	 	\begin{tabular}{@{}c@{}}
	 	$\frac{\ln(p^{n}-q^{n})}{\ln p}$ \\
	 	 $ \in (n-1+\epsilon, n)$
	 	 \end{tabular}}
	 	 & $O(L)$ &  \begin{tabular}{@{}c@{}}
	 	$O\bigg(L^{\frac{\ln(p^{n-1}-q^{n-1})}{\ln p}} \bigg)$ 
	 	\end{tabular}
	 	& $O(L)$ & 
	 	\begin{tabular}{@{}c@{}}
	 	$\lo{\text{C}^{k} \text{Z}}$ \\
	 	$1\le k \le n-1$
	 	\end{tabular}
	 	 \\
	 	\cline{1-5}
	 	\cline{7-10}
	 	
		\multirow{3}{*}{
	 	\begin{tabular}{@{}c@{}}
	 	$(i,n-i)$-fractal  \\
	 	surface code, \\
	 	$2 \le i \le n-i$\\
	 	(self-correcting) 
	 	\end{tabular}
	 	}
	 	& 
	 	\multirow{4}{*}{
	 	$\int  \frac{N}{2\pi} b^{(n-i)} \wedge da^{(i)}$   
	 	}
	 	& arbitrary  & 
	 	\multirow{4}{*}{
	 	\begin{tabular}{@{}c@{}}
	 	$i$-brane  \\
	 	$H_i(\L, \B_e)$
	 	\end{tabular}
	 	}
	 	&  
	 	\multirow{4}{*}{
	 	\begin{tabular}{@{}c@{}}
	 	$(n-i)$-brane  \\
	 	$H_{n-i}(\L^*, \B^*_m)$
	 	\end{tabular}
	 	}
	 	&  &  
	 	\begin{tabular}{@{}c@{}}
	 	$O \bigg(L^{\frac{\ln(p^{i}-q^{i})}{\ln p}}\bigg)$   \\
	 	$\le d_Z \le O(L^i)$ 
	 	\end{tabular}
	    & \begin{tabular}{@{}c@{}}
	 	$O \bigg(L^{\frac{\ln(p^{n-i}-q^{n-i})}{\ln p}}\bigg)$   \\
	 	$\le d_X \le O(L^{n-i})$
	 	\end{tabular}  
	 	& 	 	\begin{tabular}{@{}c@{}}
	 	$O \bigg(L^{\frac{\ln(p^{i}-q^{i})}{\ln p}}\bigg)$   \\
	 	$\le d \le O(L^i)$
	 	\end{tabular} 
	 	& 
	 	\multirow{3}{*}{
	 	\begin{tabular}{@{}c@{}}
	 	Within  the  \\
	 	$n^\text{th}$ level of \\
	 	Clifford \\
	 	hierarchy
	 	\end{tabular}
	 	}
	 	 \\ 	
	 	\cline{3-3}  \cline{7-9} 
	 	& & \multirow{3}{*}{$m$} & & & & \multirow{3}{*}{$O(L^i)$} & \multirow{3}{*}{$O \bigg(L^{\frac{\ln(p^{n-i}-q^{n-i})}{\ln p}}\bigg)$}   
	 	& \multirow{3}{*}{$O(L^i)$} & \\
	 	& & & & & & & & & \\
	 	& & & & & & & & & \\
	 	\hline
	 \end{tabular}
	 }
	 \caption{Summary of the main results for   $n$-dimensional fractal surface codes. The logical gates will be discussed in Sec.~\ref{sec:gates}.}
	 \label{n_dimensional_table}
\end{table*}

In sum, we can see that in the case of the fractal surface code and the punctured torus geometry (in App.~\ref{app:general_k}),  no microscopic logical operator with $O(1)$ distance exists, and all the logical operators are macroscopic. Therefore, we have proved the existence of the $(i,n-i)$-$\ZZ_2$ topological order in these fractal geometries.  Again, we can obtain the proof in the general case of $(i,n-i)$-$\ZZ_N$ topological order by simply replacing all the $\mathbb{Z}_2$ coefficients   with $\mathbb{Z}_N$ in the relative/absolute homology and cohomology groups.

\nin Q.E.D.

Similar to the $i=1$ case, the above proof for topological order also provides a rigorous
proof for the following mathematical theorem:
\begin{theorem}
	There exists a class of simple fractals $\L$ embedded in an $n$-dimensional manifold with boundary $\B_e$ or $\B_m$ on each of the interior holes, such that both its relative   $i$-systole $sys_1(\L, \B_e)$ and its dual relative $(n-i)$-systole $sys_{n-1}(\L^*, \B^*_m)$ are macroscopic, i.e., scale as a power law of the linear size $L$ of the fractals.
\end{theorem}
This theorem includes the special case that there is only one boundary type, i.e., $\B_e =\emptyset$ or $\B_m =\emptyset$. As in the previous cases, the theorem holds for the cases of $\ZZ$-, $\ZZ_2$- and $\ZZ_N$-systoles.

\subsection{Construction of a family of fractal codes with Hausdorff dimension $n-\delta$}

Here, we construct a family of fractals and the corresponding codes with Hausdorff dimension $D_H=n-\delta$, where $0<\delta<1$, which approaches $D_H=n-1+\epsilon$ with arbitrary small $\epsilon$ asymptotically.  We consider both the $(1,n-1)-$ and $(i,n-i)$-fractal surface codes supported on these fractals. 

We construct the generalization of the 2D and 3D fractals shown in Fig.~\ref{fig:fractal_cube_m_hole_variant}, and call them fractal hyper-cube geometries embedded in $n$ dimensions, which is denoted by  $FC^{(n)}(p,q)$.  These fractals can be constructed in the following way:  in the $l^\text{th}$ iteration, one divides each level-$l$ $n$-dimensional hyper-cube equally into $p^n$ level-$(l$$+$$1)$ hyper-cubes with linear size $1/p$ of a level-$l$ hyper-cube, and punch a hole in the center occupying $q^n$ cubes, and we obtain the geometry $FC^{n}(p, q, l)$. The fractal hyper-cube geometry is generated  asymptotically, i.e., $FC^{(n)}(p, q) \ $$\equiv$$\ \lim_{l \rightarrow \infty} FC^{(n)}(p, q, l)$.  The Hausdorff dimension of this fractal and the corresponding codes can be expressed as 
\be
D_H(FC^{(n)}, p, q) = \frac{\ln(p^n-q^n)}{\ln p}. \ee

In the case of the $(1,n-1)$-surface code, we put $m$-boundaries on the $(n-1)$-dimensional hyper-surfaces of all the holes.  The external boundaries have been specified above in Sec.~\ref{sec:n_dimensional_no_string}. The $Z$-distance is just the minimal length of all logical $Z$-string representatives and is hence $d_Z = O(L)$. Meanwhile, the $X$-distance of this code is determined by the the minimal volume of all logical $X$-brane representatives, which is just the volume of the fractal hyper-cube embedded in ($n-1$) dimensions, $FC^{n-1}(p,q)$, i.e., $d_X = O(L^{D_H(FC^{(n-1)}, p,q)})$, where $D_H(FC^{(n-1)}, p, q) = \frac{\ln(p^{n-1}-q^{n-1})}{\ln p}$. The overall code distance is chosen from the smaller one of $d_X$ and $d_Z$, and is hence $d=d_Z=O(L)$. The space overhead of this family of codes is hence $O(L^{D_H(FC^{(n-1)}, p,q)})=O(d^{D_H(FC^{(n-1)}, p,q)})$. As in the 3D case, when increasing the ratio of $p/q$, one can keep reducing the Hausdorff dimension $D_H=n-\delta$ (i.e., increase $\delta$) and asymptotically approach $D_H= n-1+\epsilon$ for arbitrary small $\epsilon$.  The space overhead can hence asymptotically approach $O(L^{n-1+\epsilon}) = O(d^{n-1+\epsilon})$.

In the case of the $(i,n-i)$-fractal surface code ($2 \le i \le n-i $), we can choose either $e$- or $m$-boundary on each hole as has been  proved in Sec.~\ref{sec:general_k}.  The $Z$- and $X$-distances are determined by the minimal volume of all logical $Z$-brane (world-volume of $e$ excitations) and $X$-brane (world-volume of $m$ excitations) representatives respectively. We can hence have the range of the $Z$-distance being $L^{D_H(FC^{i}, p, q)}$$\le$$d_Z$$\le$$L^i$. The lower bound is achieved in the case when all the holes have $e$-boundaries such that the $m$-excitations can condense on these boundaries and the $Z$-brane terminate on them. In this case the minimal-volume $Z$-brane are the fractal hyper-cube geometry $FC^{(i)}(p,q)$. The upper bound is achieved when all the holes have $m$-boundaries such that the $Z$-brane cannot terminate on them and the minimal-volume $Z$-brane are just the $i$-dimensional hyper-surface. Similarly the range of the $X$-distance is $L^{D_H(FC^{n-i}, p, q)} \le d_X \le L^{n-i}$, where the lower (upper) bound is achieved when are the holes have $m$-boundaries ($e$-boundaries). 

Since in our convention logical $Z$-brane ($e$-type) has lower or equal dimension than the logical $X$-brane ($m$-type), i.e., $i \le n-i$, we could set all holes having  $m$-boundaries if we aim to maximize the overall distance.   In this case, the $e$-branes cannot terminate on these holes, and we hence have the maximized $Z$-distance $d_Z=O(L^i)$. In the situation that the $X$-brane has larger dimension, i.e., $i<n-i$, we always have $d_Z < d_X$ (since $D_H(FC^{n-i},p,q) <n-i-1 \le i$), and the overall distance is just $d=\min(d_X, d_Z) = d_Z=O(L^i)$, which is the same as the $(i,n-i)$-surface code defined on a manifold but requires lower space overhead, i.e., $O(L^{D_H(FC^n,p,q)})$ (asymptotically approaching $O(L^{n-1+\epsilon})$).  In the situation that the logical $X$-brane and $Z$-brane have the same dimension, i.e., $i=n-i$, we have $d_X<d_Z$,  and the overall distance is $d=d_X=O(L^{D_H(FC^{n-i}, p, q)})$.

A summary of the various types of fractal codes defined on the fractal hyper-cube geometry, including the class with and without string-like logical operators, is given in Table \ref{n_dimensional_table}.

\section{Fault-tolerant quantum computation with fractal codes}\label{sec:gates}

\subsection{Non-Clifford logical gates for 3D fractal codes}

\subsubsection{A brief review of non-Clifford gates in 3D surface  codes}

\begin{figure*}[hbt]
  \includegraphics[width=1.7\columnwidth]{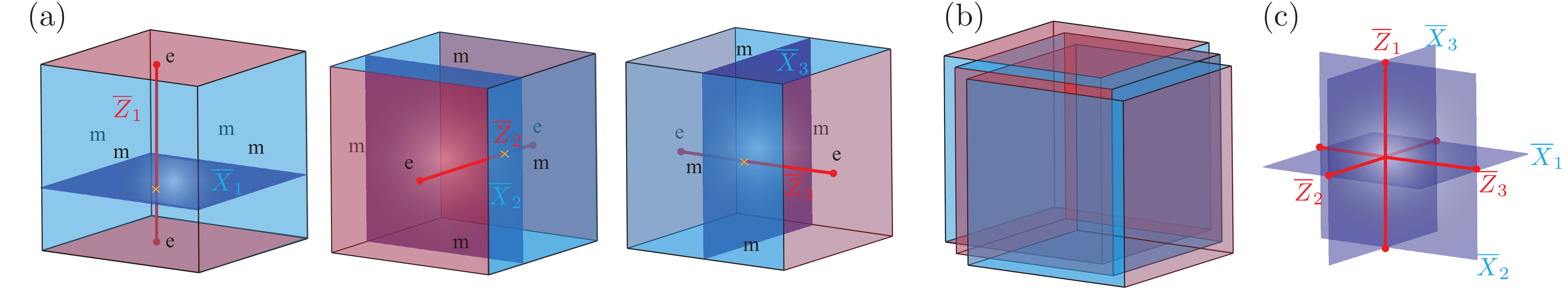}
  \caption{The arrangement of a stack of three 3D surface codes for the application of a transversal logical CCZ gate. (a) The pairs of $e$-boundaries are chosen to be perpendicular to the $z$-, $x$-, and $y$-directions respectively. (b) The three copies are stacked together such that the corresponding qubits acted by the transversal CCZ gate in each copy are aligned with each other.  (c) The alignment of the three pairs of logical operators. All logical $Z$-strings are perpendicular to each other, and the same for the logical $X$-branes.}
  \label{fig:CCZ_arrangement}
\end{figure*}

It has been realized that a non-Clifford logical CCZ (multi-controlled-Z) gate can be applied to a stack of three copies of 3D surface codes with different orientations and cellulations of a 3-manifold~\cite{Kubica:2015br, Vasmer2019}. Similarly, transversal T gate can be applied to a single copy of 3D color code defined on a tetrahedron~\cite{Bombin:2015jk}.   

We begin with a generic picture of the transversal controlled-Z (CZ) and controlled-CZ (CCZ) gates between three copies of  3D surface codes by ignoring the detailed lattice structures.  We consider a stack of 3D surface codes that are oriented in three directions, namely the pair of $e$-boundaries (red) are perpendicular to the $z$-, $x$-, and $y$-directions respectively as shown in Fig.~\ref{fig:CCZ_arrangement}. The logical $Z$-strings (red) terminate on  the two opposite $e$-boundaries. The logical $X$-branes (dark blue) are parallel to the $e$-boundaries and terminate on the $m$-boundaries (light blue). This specific alignment of the three 3D surface codes is carefully chosen such that it satisfies the condition for a transversal CCZ gate \cite{Vasmer2019}. The reason behind the varying choice in orientation of the different copies of the code can be understood from the desired action of the logical CCZ gate. By definition, logical CCZ must map logical~$X$ on one code copy to the product of that same logical~$X$ and the logical~CZ on the other two codes, that is: $\overline{\text{CCZ}}: \overline{X}_a \rightarrow \overline{X}_a \overline{\text{CZ}}_{b,c}$, for any choice of different labels~$a,\ b, \ c$. As such, if for a given codeblock~$a$ the logical~$X$ resides in the $xy$-plane, then there must exist a logical~CZ gate, $\overline{\text{CZ}}_{b,c}$ that can be applied transversally within that plane for the codeblocks $b$ and $c$ and as such their logical~$Z$-strings operators must be parallel to that plane. By iterating this argument for all three codeblocks one concludes that the logical-$Z$ must be in orthogonal directions for all three code blocks, resulting in the orientation outlined above and presented in Fig.~\ref{fig:CCZ_arrangement}.

We outline the requirements for a transversal logical~CCZ gate in much more detail in App.~\ref{app:CCZ_requirements} and discuss the corresponding microscopic construction of the three copies of 3D surface models given by Ref.~\cite{Vasmer2019} in App.~\ref{app:mmm_hole_example}.

\subsubsection{Connection between transversal gates and domain wall sweeping: a TQFT description}\label{sec:domain_wall_picture}

To facilitate later discussions, we now introduce the TQFT picture for transversal gates. It has been realized in Refs.~\cite{Yoshida_gate_SPT_2015, Yoshida_global_symmetry_2016, Yoshida2017387, Webster_gates_2018, Zhu:2017tr} that transversal gates are related to certain global symmetry or subsystem symmetry in the corresponding topological order, and is equivalent to sweeping certain gapped domain walls across the system.  It has been further pointed out in Ref.~\cite{Zhu:2017tr} that in general one can consider transversal gates as onsite topological symmetries in a symmetry-enriched topological (SET) order \cite{barkeshli2014SDG}.

 We consider a generic type of topological order which can be classified as being equivalent to multiple copies of elementary topological orders or equivalently code blocks $\C$, where $\C$ can also be mathematically interpreted as a certain tensor category  representing such topological order\footnote{In 2D, it is understood that $\C$ corresponds to a modular tensor category.}. We also call this setup a stack code. The generic topological order we consider is hence represented as $\C \times \C \cdots \times \C $. We say this type of topological order is a symmetry-enriched topological order with onsite permutation symmetry between different copies $\C$. In the rest of this paper, we consider the elementary topological order to be the toric code\footnote{In the 2D case, one has $\C=\D(\ZZ_2)$. }.  For example, we consider a stack of three copies of 3D toric codes, equivalent to a single copy of 3D color code, which can be classified as a 3D  $\ZZ_2 \times \ZZ_2 \times \ZZ_2$ topological order. %Similarly, we can denote $n$ copies of $n$-dimensional toric code and equivalently a single copy of $n$-dimensional color code as $\ZZ_2^{\times n}$.  

We denote certain transversal gates, or more generally geometrically local constant-depth quantum circuit as $U$.  In the case of transversal gates, we can represent it as $U = \Motimes_j U_j$, where $j$ is the horizontal site label for transversally aligned qubits (in the case of multiple copies of code blocks  $\C$), where $U_j$ only acts on individual qubits or couples aligned qubits in different code blocks with the same label $j$.  Note that $U_j$ does not couple qubits within the same code block $\C$.  If we relax the transversal condition to local constant-depth quantum circuits, $U$ still maps any local operator $\mathcal{O}$ to another local operator $\mathcal{O}'$ in the $O(1)$ neighborhood, i.e., $U \mathcal{O}U^\dag =\mathcal{O}'$.  Now, for $U$ to be a logical gate, it should satisfy the following condition:
\be\label{eq:logical_condition_1}
U:\H_C \rightarrow \H_C,
\ee  
i.e., it preserves the code (ground) space $\H_C$.  For our discussion, we write the parent Hamiltonian of the topological order as:
\be
H = H^\text{bulk} + H^\text{boundary} \equiv - \sum_{i} S_i^\text{bulk} - \sum_{k} S_{k}^\text{boundary}. 
\ee 
Here, $H^\text{bulk}$ and $H^\text{boundary}$ represent the bulk and boundary Hamiltonians, while $S_i^\text{bulk}$ and $S_{k}^\text{boundary}$ represent the local stabilizers or more generally interaction terms (for non-stabilizer models) on the bulk and the boundary respectively. To satisfy condition \eqref{eq:logical_condition_1}, we need the parent Hamiltonian to be preserved under the unitary up to a logical identity, i.e.,
\be\label{eq:Homiltonian_condition}
P_C (UHU^\dag) P_C = P_C H P_C,
\ee
where $P_C$ is the projector onto the code (ground) space $\H_C$. 
In the case that $U$ is a transversal gate, we say $U$ is an onsite topological symmetry for a translationally invariant system if
\begin{align}\label{eq:bulkcondion}
\non  P_C (U H^\text{bulk} U^\dag) P_C &= P_C H^\text{bulk} P_C \\
 \Longleftrightarrow P_C (U S_i^\text{bulk} U^\dag) P_C &= P_C S_i^\text{bulk} P_C.
\end{align}
In the more general case of a  constant-depth circuit, the above conditions can also be generalized accordingly and we say $U$ corresponds to certain type of  topological symmetry for a translationally invariant system. In general, $U$  can be a global symmetry, meaning the constant-depth circuit is applied to the entire system, or a higher-form ($q$-form) symmetry \cite{kapustin2015higher, Gaiotto_2015}, where the circuit is applied to a codimension-$q$  sub-manifold $\M^{n-q}$.  Note that global symmetry is also a $0$-form symmetry according to the above definition.     

Now the presence of the boundary breaks translational invariance, and we have to impose the invariance of the boundary Hamiltonian under the action of $U$ in order to satisfy condition \eqref{eq:logical_condition_1}, i.e., 
\begin{align}\label{eq:boundary_condition}
\non  P_C (U H^\text{boundary} U^\dag) P_C &= P_C H^\text{boundary} P_C \\  \Longleftrightarrow  P_C (U S_i^\text{boundary} U^\dag) P_C &= P_C S_i^\text{boundary} P_C.
\end{align}
We denote the $i^\text{th}$ gapped boundaries of the system as $\B_i$.  The above condition can also be expressed as 
\be\label{eq:domain_wall_condensation_1}
U: \B_i \rightarrow \B_i, \ \text{for} \ \forall i.
\ee

\begin{figure}[t]
  \includegraphics[width=0.9\columnwidth]{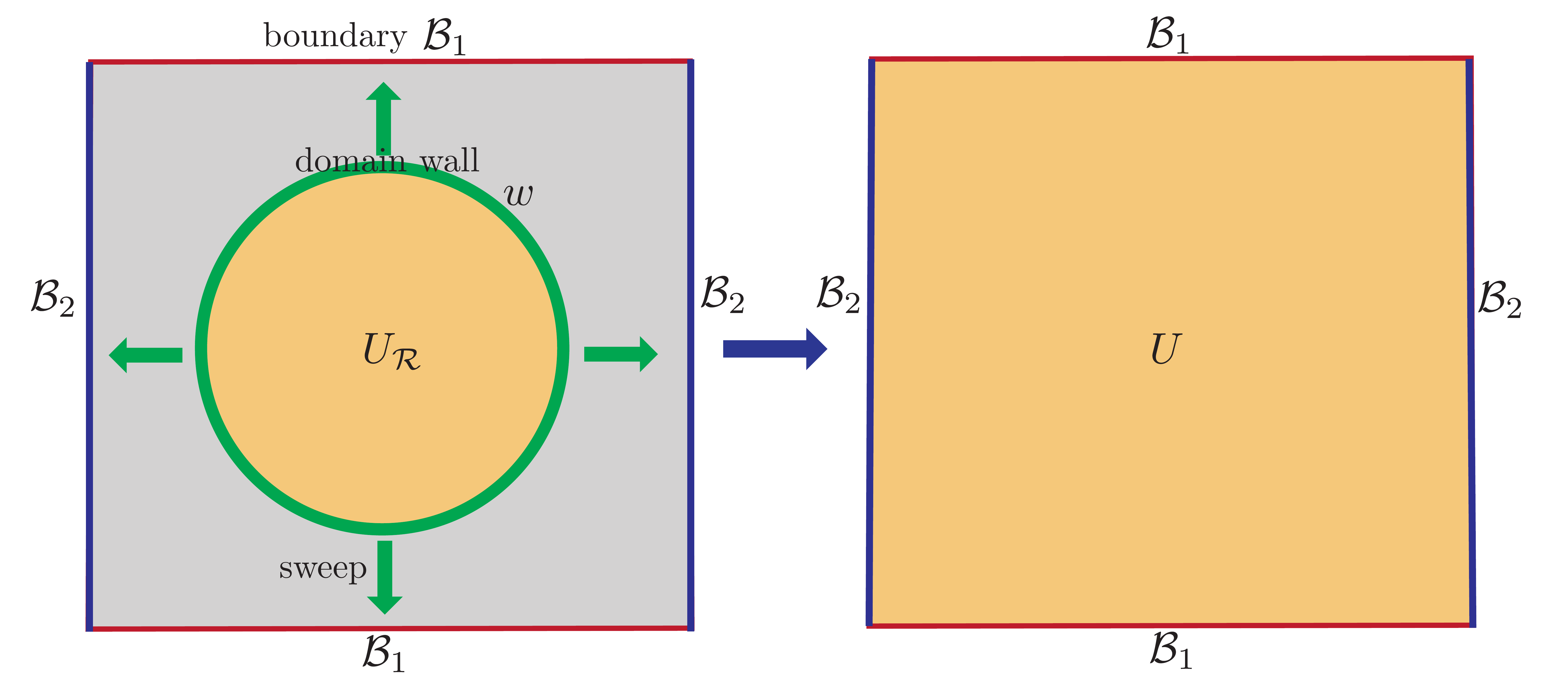}
  \caption{Illustration of the correspondence between transversal logical gate (or a local constant-depth circuit) $U$ and domain-wall sweeping in 2D. Applying $U_\R$ in a region $\R$ is equivalent to sweep the corresponding gapped domain wall $w$ to the boundary of the region $\R$. When sweeping the domain wall $w$ across the entire system, the domain wall condenses on the external boundaries $\B_i$ and one effectively applies the global symmetry $U$ to the entire system corresponding to the transversal logical gate (or constant-depth circuit).  The external boundaries $\B_i$ is required to be invariant under the action of $U$ according to Eq.~\eqref{eq:domain_wall_condensation_1}.}
  \label{fig:domain_wall}
\end{figure}

An alternative physical picture of transversal gates or more  generally constant-depth circuits which corresponds to the topological symmetry is that the application of $U$ is equivalent to sweeping a corresponding gapped domain wall $w$ across the system, as illustrated in Fig.~\ref{fig:domain_wall}. During the domain wall sweeping process, the region $\mathcal{R}$ which has been swept by the wall has been applied a constant-depth circuit  restricted in region $\R$ (denoted by $U_\R$).  When the domain wall $w$ has been swept across the whole system, the entire $U$ is applied, and the domain wall hits on the boundaries $\B_i$. If $U$ is a logical gate, i.e., satisfying condition \eqref{eq:logical_condition_1}, then the domain wall $w$ should condense on all the boundaries, i.e.,
\be\label{eq:domain_wall_condensation_2}
w:\B_i \rightarrow w\B_i w^\dag= \B_i, \ \text{for} \ \forall i,
\ee    
which is equivalent to the condition in Eq.~\eqref{eq:domain_wall_condensation_1}.

\begin{figure}[t]
  \includegraphics[width=1\columnwidth]{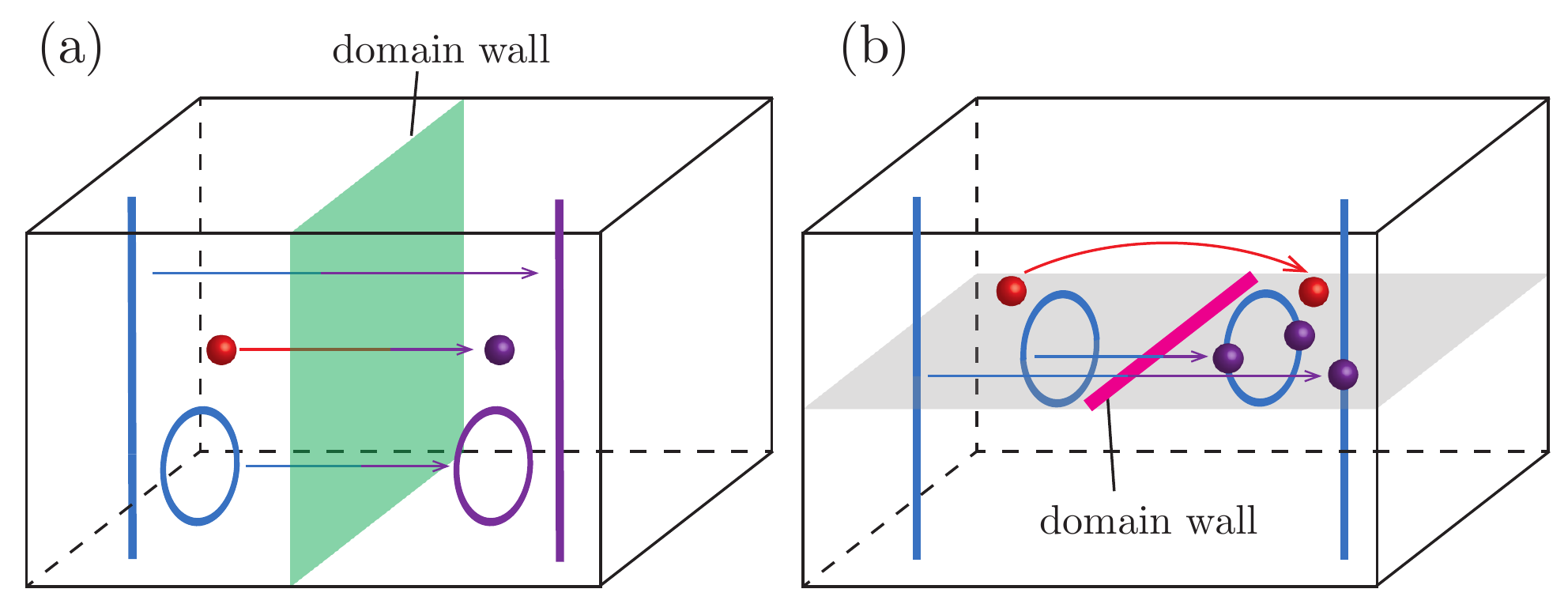}
  \caption{(a) Both the particle and string (loop)  excitations get transformed when passing the  codimension-1 domain wall (green) in 3D. (b) A particle excitation cannot be transformed by the codimension-2 domain wall (pink) in 3D since the particle can generically avoid such a wall. The string (loop) gets transformed by the wall only at the intersection point between the string and the wall, where a new particle (purple) is generated.}
  \label{fig:domain_wall_mapping}
\end{figure}

In 2D topological orders, the domain wall is codimension~1 (dimension~1). Generally, codimension is $d-d_w$, where $d$ is the space dimension of the topological order and $d_w$ is the dimension of the domain wall. In 2D case, one has $d=2$ and $d_w=1$. In 3D topological orders, there are two generic types of gapped domain walls, the codimension-1 
    (dimension-2) wall and the codimension-2 
    (dimension-1), as shown in Fig.~\ref{fig:domain_wall_mapping}. When particle and string excitations go through a codimension-1 domain wall, both types  can get transformed. This is illustrated in Fig.~\ref{fig:domain_wall_mapping}(a), where both a contractible loop (closed string) excitation and a non-contractible string excitation either going along a torus cycle or opposite boundaries (in the case with gapped boundaries) are shown. The worldsheet of the non-contractible loop excitation corresponds to a logical brane  operator. On the other hand, the particle excitation can always circumvent the codimension-2 domain wall as illustrated in Fig.~\ref{fig:domain_wall_mapping}(b). This can be understood via the intersection dimension $d_s+d_w -d=-1$, where $d_s=1$, $d_w=1$, and $d=3$ represent the dimensions of the worldline of a particle (i.e., string operator), domain wall and the space  respectively. The intersection smaller than 0 means no generic interaction can occur between the worldline of a particle and the codimension-2 domain wall, a reflection of the fact that two lines in 3D will generically avoids each other. Even two lines cross each other in 3D, an infinitesimal perturbation can removes that intersection. Due to this topological constraint, particle excitation cannot be transformed by a codimension-2 domain wall. In contrast, a string excitation will generically intersect with a codimension-2 domain wall, as illustrated in Fig.~\ref{fig:domain_wall_mapping}(b). This can be understood via the intersection dimension $d_b+d_w -d=0$, where $d_b=1$ is the dimension of the worldsheet (i.e., brane) of a string excitation. The zero intersection dimension means the intersections are points, which cannot be removed by an infinitesimal perturbation. One can see that when a contractible loop excitation or non-contractible string excitation goes through the domain wall, the intersection point becomes a new particle excitation while the rest of the loop/string remains unchanged. 

\begin{figure}[t]
  \includegraphics[width=1\columnwidth]{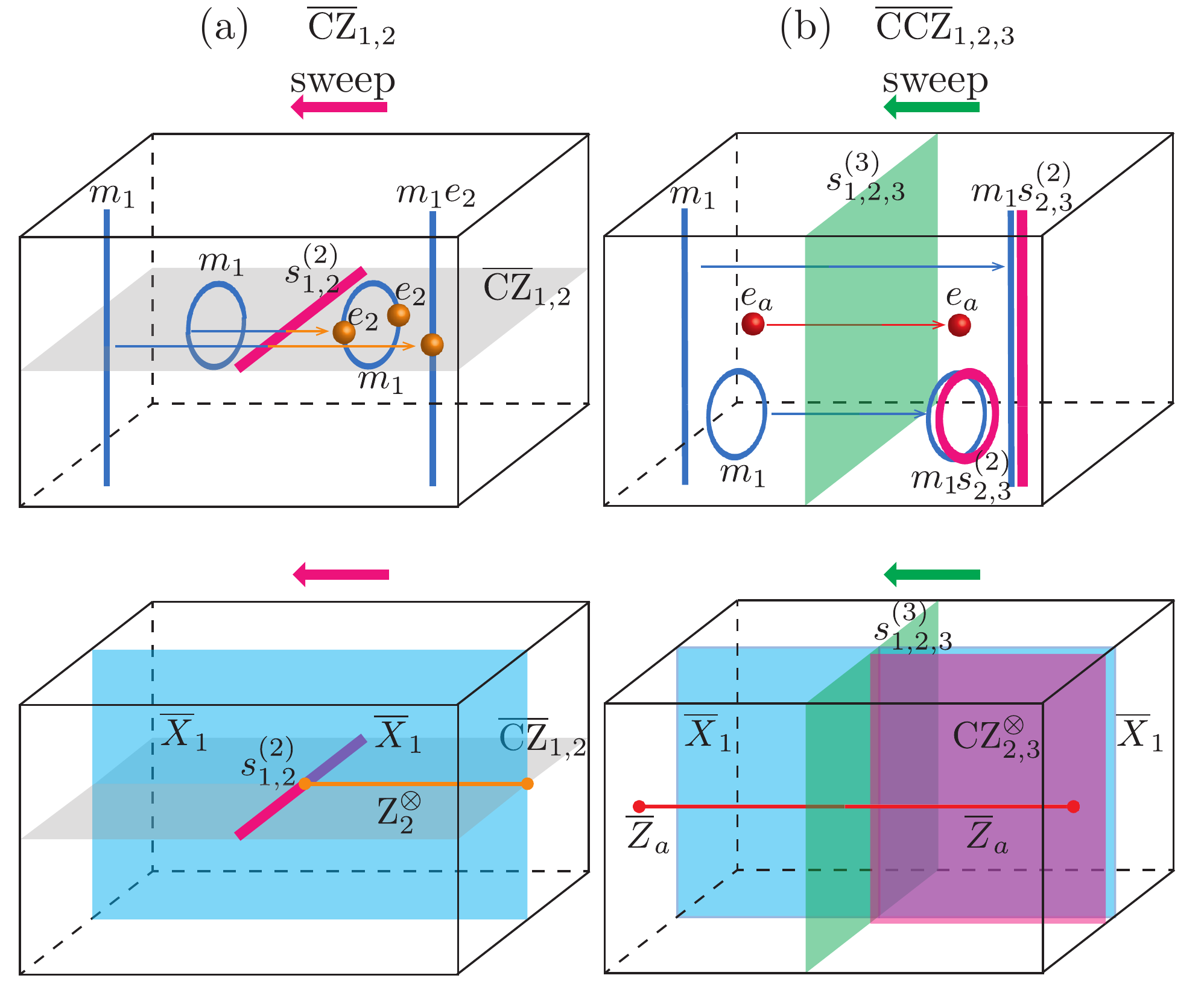}
  \caption{(a) Sweeping the CZ domain wall $s^{(2)}_{1,2}$ (pink) across a non-contractible codimension-1 sub-manifold $\M^2$ (grey) gives rise to a transversal logical gate $\lo{\text{CZ}}_{1,2}$ applied along this sub-manifold. When passing the $m_1$-string (loop) in copy 1 across the wall $s^{(2)}_{1,2}$, a new particle $e_2$ (orange) in copy 2 is generated at the intersection point. Therefore,   a logical brane $\lo{X}_1$ is transformed into the product of itself and an additional string $Z_2^{\otimes}$ (yellow) which becomes the logical string $\lo{Z}_2$ when the sweeping is completed, as illustrated in the lower panel. (b) Sweeping the codimension-1 CCZ domain wall $s^{(3)}_{1,2,3}$ across the system gives rise to the transversal logical gate $\lo{\text{CCZ}}_{1,2,3}$. An $m_1$-string (loop) gets transformed to an $m_1 s^{(2)}_{2,3}$-string (loop) when passing across the wall. Therefore, a logical brane $\lo{X}_1$ is transformed into the product of itself and an additional brane $\text{CZ}^{\otimes}_{2,3}$ (pink), which becomes the logical brane $\lo{\text{CZ}}_{2,3}$ when the sweeping is completed.}
  \label{fig:domain_wall_sweep}
\end{figure}

Now we discuss the symmetry and domain wall sweeping picture of the transversal CZ and CCZ  in a stack of three 3D toric codes, i.e., a 3D $\ZZ_2 \times \ZZ_2 \times \ZZ_2$ topological order. 
% Since the 3D toric code is a CSS code, there exists transversal CNOT gate between any two copies of 3D toric codes.  We denote the domain wall corresponding to $\lo{\text{CNOT}}_{1,2}$ (transversal CNOT on code block 1 and 2) as $c_{1,2}$. It is a codimesnion-1 (dimension-2) domain wall with a membrane shape.  When sweeping it cross the whole system, the domain wall applies the following mapping to the excitations going through it:
% \be
% c_{1,2} : m_1 \rightarrow m_1 m_2,  e_2 \rightarrow e_1 e_2,
% \ee 
% as illustrated in the upper panel of Fig.~\ref{fig:domain_wall_sweep}(a). We note that the mapping is reversible: when  the excitation on the right passes the domain wall, they will become excitation on the left. 
% The above map is closely related to the following mapping on the logical operators (worldlines/worldsheets of the excitations):
% \be
% \lo{\text{CNOT}}_{1,2} : \lo{X}_1 \rightarrow \lo{X}_1 \lo{X}_2, \lo{Z}_2 \rightarrow \lo{Z}_1 \lo{Z}_2,
% \ee
% as illustrated in the lower panel of Fig.~\ref{fig:domain_wall_sweep}(a).
% Note that in a 3D toric code with periodic boundary conditions, there are three logical qubits. For simplicity, we only list one conjugate pair of logical operators in code 1 and the corresponding %     pair in code 2 with the same orientation.  
The CZ wall on code block 1 and 2, denoted by $s^{(2)}_{1,2}$, is a codimension-2 (dimension-1) gapped domain wall with a string shape. When sweeping it across a non-contractible  brane-shaped sub-manifold $\M^2$ wrapping around the 3-torus, it applies a transversal gate $\lo{\text{CZ}}_{1,2}$ supported on the brane which corresponds to a 1-form symmetry. The CZ domain wall applies the following mapping to the excitations going through it:
\be
s^{(2)}_{1,2}:m_1 \rightarrow m_1 e_2, m_2 \rightarrow e_1 m_2,
\ee  
while keeping the point excitation $e_1$ and $e_2$ invariant under the mapping, as illustrated in the upper panel of Fig.~\ref{fig:domain_wall_sweep}(a). It is related to the following mapping on the logical operators
\be
\lo{\text{CZ}}_{1,2}: \lo{X}_1 \rightarrow \lo{X}_1 \lo{Z}_2, \lo{X}_2 \rightarrow \lo{Z}_1 \lo{X}_2,
\ee
as illustrated in the lower panel of Fig.~\ref{fig:domain_wall_sweep}(a).
Note that in a 3D toric code with periodic boundary conditions, there are three logical qubits. For simplicity, we only list one conjugate pair of logical operators in code 1 and the corresponding pair in code 2.

In terms of the microscopic model $H$ (discussed in detail in App.~\ref{app:mmm_hole_example}), we also need it to satisfy the conditions in Eqs.~\eqref{eq:bulkcondion} in order for $\lo{\text{CZ}}_{1,2}$ to be a 1-form symmetry. Since the $Z$ stabilizers is obviously invariant under the action of  $\lo{\text{CZ}}_{1,2}$ (diagonal in the $Z$-basis), we only investigate the action of $\lo{\text{CZ}}_{1,2}$ on the $X$ stabilizers $A_{q;a}$, where $q$ labels the stabilizer and  $a=1,2$ is the label of the toric-code copy. One hence obtains the following mapping: 
\begin{align}\label{eq:CZ_conjugation}
\nonumber \lo{\text{CZ}}_{1,2} A_{q;a} \lo{\text{CZ}}_{1,2}^\dag \equiv& \lo{\text{CZ}}_{1,2} \big(\Motimes_{j \in A_{q;a}} X_j\big) \lo{\text{CZ}}_{1,2}^\dag \\
=&\Motimes_{j \in A_{q;a}} X_{j; a}  \Motimes_{k \in A_{q;a}} Z_{k; b},
\end{align}
where $a,b$ is an arbitrary permutation of $1,2$. Here, $k$ labels the sites on copy $b$ aligned with the sites in stabilizer $A_{q;a}$ in copy $a$. In the microscopic model presented in App.~\ref{app:mmm_hole_example}, the additional term $\Motimes_{k \in A_{q;a}} Z_{k; b}$ is a stabilizer, and is hence a logical identity, i.e.,  $P_C \big(\Motimes_{k \in A_{q;a}} Z_{k; b} \big) P_C = I$,  satisfying the requirement in Eq.~\eqref{eq:bulkcondion}. %Note that since each stabilizer $A_{q;a}$ is an local operator, only the local property of the $\lo{CZ}_{1,2}$ brane is needed to satisfy Eq.~\eqref{eq:CZ_conjugation}.  
One should expect that Eq.~\eqref{eq:CZ_conjugation} also holds if we replace the logical $\lo{\text{CZ}}_{1,2}$ with  transversal CZ gates acted on a contractible closed brane, denoted by $\text{CZ}^{\otimes}_{1,2}$, which is hence also a 1-form symmetry. Moreover, it is also a logical identity, i.e., $P_C\text{CZ}^{\otimes}_{1,2}P_C=I$, since it does not apply any logical gate. We hence get the following TQFT relations in the code (ground) space $\H_C$:
\begin{align}
\raisebox{-0.3cm}{\includegraphics[scale=.60]{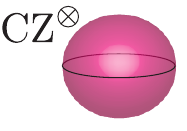}} \ \ \ \ =& \quad \quad \mathbb{I} \label{eq:CZ_sphere} \\
\raisebox{-0.3cm}{\includegraphics[scale=.60]{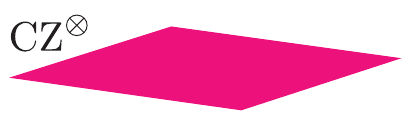}} =& \ \raisebox{-0.3cm}{\includegraphics[scale=.60]{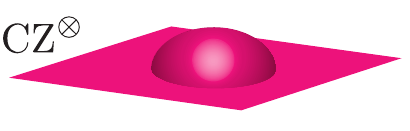}} . \label{eq:CZ_sheet_bending} 
\end{align}
Note that the second relation can be derived from the first relation by multiplying the left CZ brane with a CZ sphere, which suggests that the CZ brane can be freely deformed, including the case of the logical brane $\lo{\text{CZ}}_{1,2}$. This is just a signature of the 1-form symmetry. 

Next, the CCZ wall on all three copies, denoted by $s^{(3)}_{1,2,3}$,  is a codimension-1 (dimension-2) gapped domain wall with a brane shape. When the domain wall is swept across the system, a global  transversal CCZ gate is applied, corresponding to a global symmetry transformation. A translationally invariant 3D  $\ZZ_2 \times \ZZ_2 \times \ZZ_2$ topological order has a global onsite CCZ symmetry, if  the bulk parent Hamiltonian (discussed in detail in App.~\ref{app:mmm_hole_example}) is invariant under such symmetry transformation up to a logical identity.  Since the $Z$ stabilizers are obviously invariant under the action of the transversal CCZ gate (diagonal in the $Z$-basis), we again investigate the mapping of the $X$ stabilizers $A_{q;a}$:
 \begin{align}
\nonumber  \lo{\text{CCZ}}_{1,2,3} A_{q;a}\lo{\text{CCZ}}_{1,2,3}^\dag \equiv&  \lo{\text{CCZ}}_{1,2,3} \big( \Motimes_{j \in A_{q;a}} X_{j;a} \big) \lo{\text{CCZ}}_{1,2,3}^\dag \\
  =& \Motimes_{j \in A_{q;a}} X_{j;a}  \Motimes_{k \in A_{q;a}} \text{CZ}_{k;b,c} , 
  \end{align}  
where $a,b,c$ is an arbitrary permutation of $1,2,3$.  Note that the product of neighboring $X$ stabilizers (including the case of a single $X$ stabilizer) is just a closed $X$-brane in the TQFT picture as illustrated in Fig.~\ref{fig:stabilizer_relations_3D}(e). Therefore, the additional term generated by the CCZ conjugation, $\Motimes_{k \in A_{q;a}} \text{CZ}_{k;b,c}$, is supported on the region in copy $b$ and $c$ aligned with the $X$ stabilizer $A_{q;a}$ and hence the closed $X$-brane. Thus, $\Motimes_{k \in A_{q;a}} \text{CZ}_{k;b,c}$ is just a closed CZ brane, and is hence a logical identity according to Eq.~\eqref{eq:CZ_sphere}.  The condition in Eq.~\eqref{eq:bulkcondion}  is hence satisfied and the transversal CCZ gate is indeed a global symmetry.   

The CCZ domain wall applies  the following mapping to the excitations going through it:
\be
s^{(3)}_{1,2,3}:m_1 \rightarrow m_1 s^{(2)}_{2,3}, m_2 \rightarrow m_2 s^{(2)}_{3,1}, m_3 \rightarrow m_3 s^{(2)}_{1,2},
\ee
while keeping the point excitations $e_1$, $e_2$ and $e_3$ invariant, as illustrated in the upper panel of Fig.~\ref{fig:domain_wall_sweep}(b). It is related to the following mapping on the logical operators:
\be
\lo{\text{CCZ}}_{1,2,3}: \lo{X}_1 \rightarrow \lo{X}_1 \lo{\text{CZ}}_{2,3}, \lo{X}_2 \rightarrow \lo{X}_2 \lo{\text{CZ}}_{3,1}, \lo{X}_3 \rightarrow \lo{X}_3 \lo{\text{CZ}}_{1,2},
\ee
as illustrated in the lower panel of Fig.~\ref{fig:domain_wall_sweep}(b).
Note that only one conjugate pair of logical operators in each copy among three existing pairs is shown. 
% Note that the above discussion and the illustration in Fig.~\ref{fig:domain_wall_sweep} not only show how the logical operators are transformed under domain-wall sweeping, but how the errors (excitations) are transformed.   This helps to understand the error propagation when applying these transversal gates.

\begin{figure}[t]
  \includegraphics[width=0.7\columnwidth]{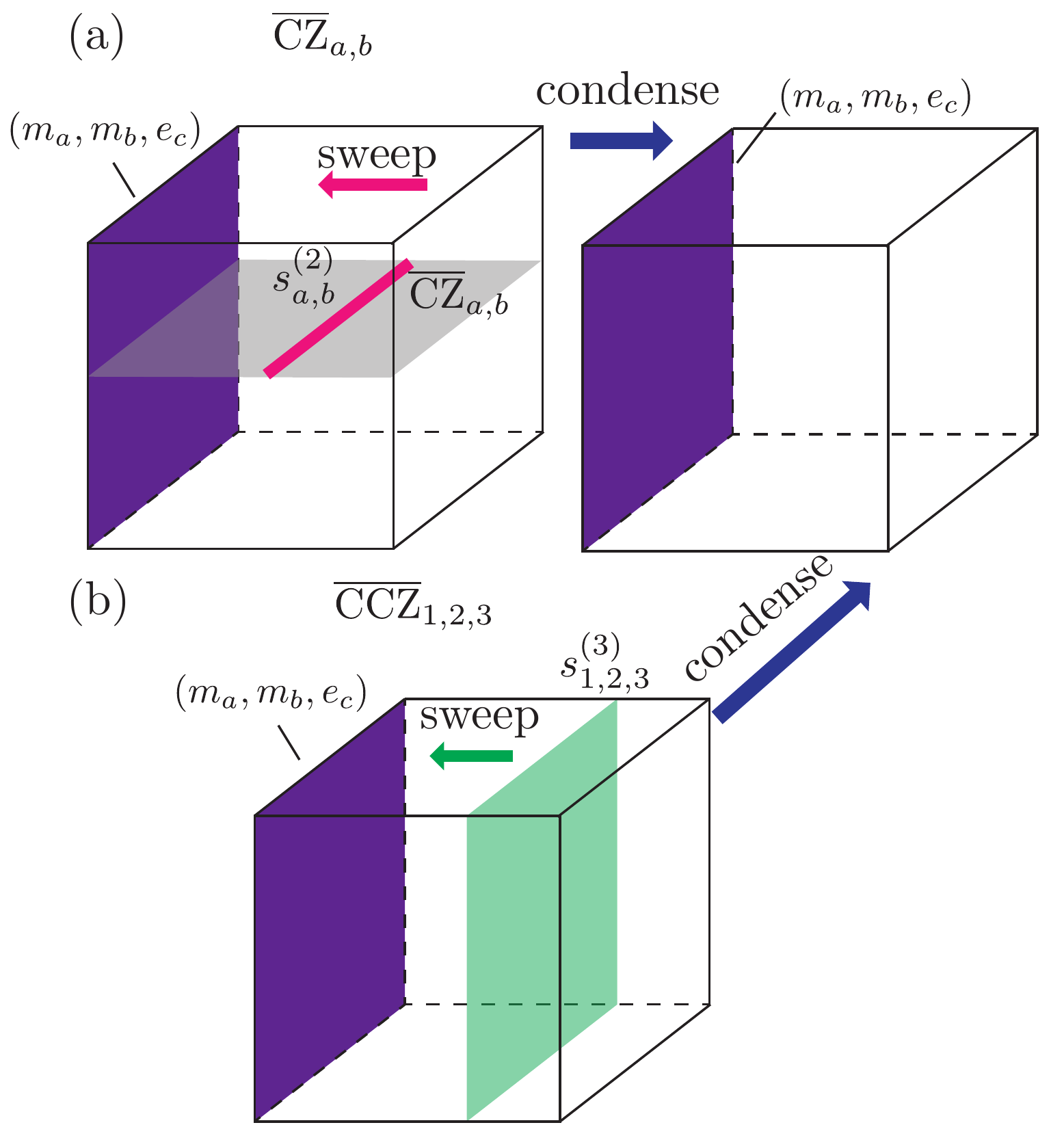}
  \caption{The existence of transversal logical CZ and CCZ gate in a stack of three 3D surface codes is ensured by the condensation of their corresponding domain walls $S^{(2)}_{a,b}$ and $S^{(3)}_{1,2,3}$ on the boundary $(m_a,m_b,e_c)$.}
  \label{fig:domain_wall_condensation}
\end{figure}

Thus far we have been focused on the case of 3D toric codes, i.e., the system is defined on a 3-torus ($T^3$) without any boundary.   Now we consider the situations with boundaries.   When there are gapped boundaries, the domain wall should condense on each boundary, i.e., satisfy the condition in Eq.~\eqref{eq:domain_wall_condensation_2}, such that the transversal gate or more generally the constant-depth circuit correspond to a logical gate.   

% In the case of $\lo{\text{CNOT}}_{1,2}$ , we need to make sure each boundary of the interacting  copies (code blocks) having the same type, i.e., permitting the following boundary types for a stack code with two copies of toric codes: $\B_i$~$=$~$ (e_1, e_2)$~or~$(m_1, m_2)$. This means the corresponding boundary needs to have a consistent condensation of $e$-particles (or $m$-strings) in both copies. One can verify this by checking the mapping of the gapped boundary under the domain wall action:
% \begin{align}
% \nonumber c_{1,2} : \ & (e_1,e_2)\rightarrow (e_1, e_1 e_2) \equiv (e_1, e_2), \\
%  		   &(m_1,m_2)\rightarrow (m_1 m_2, m_2) \equiv (m_1, m_2).   
% \end{align}
% Note that $(e_1, e_1 e_2)$ means the boundary condense both $e_1$ and a composite particle $e_1 e_2$, which is equivalent to the boundary $(e_1, e_1 e_2)$ up to a basis change.  This is because, due to the $\Z_2$ fusion rule $e\times e = \mathbb{I}$, a composite of $e_1$ and $e_1 e_2$ is just $e_2$ ($e_1\times e_1 e_2= e_2$), which should still condense on the the boundary.   Same argument applies to $(m_1 m_2, m_2)~\equiv~(m_1, m_2)$.

\begin{figure*}[hbt]
  \includegraphics[width=1.8\columnwidth]{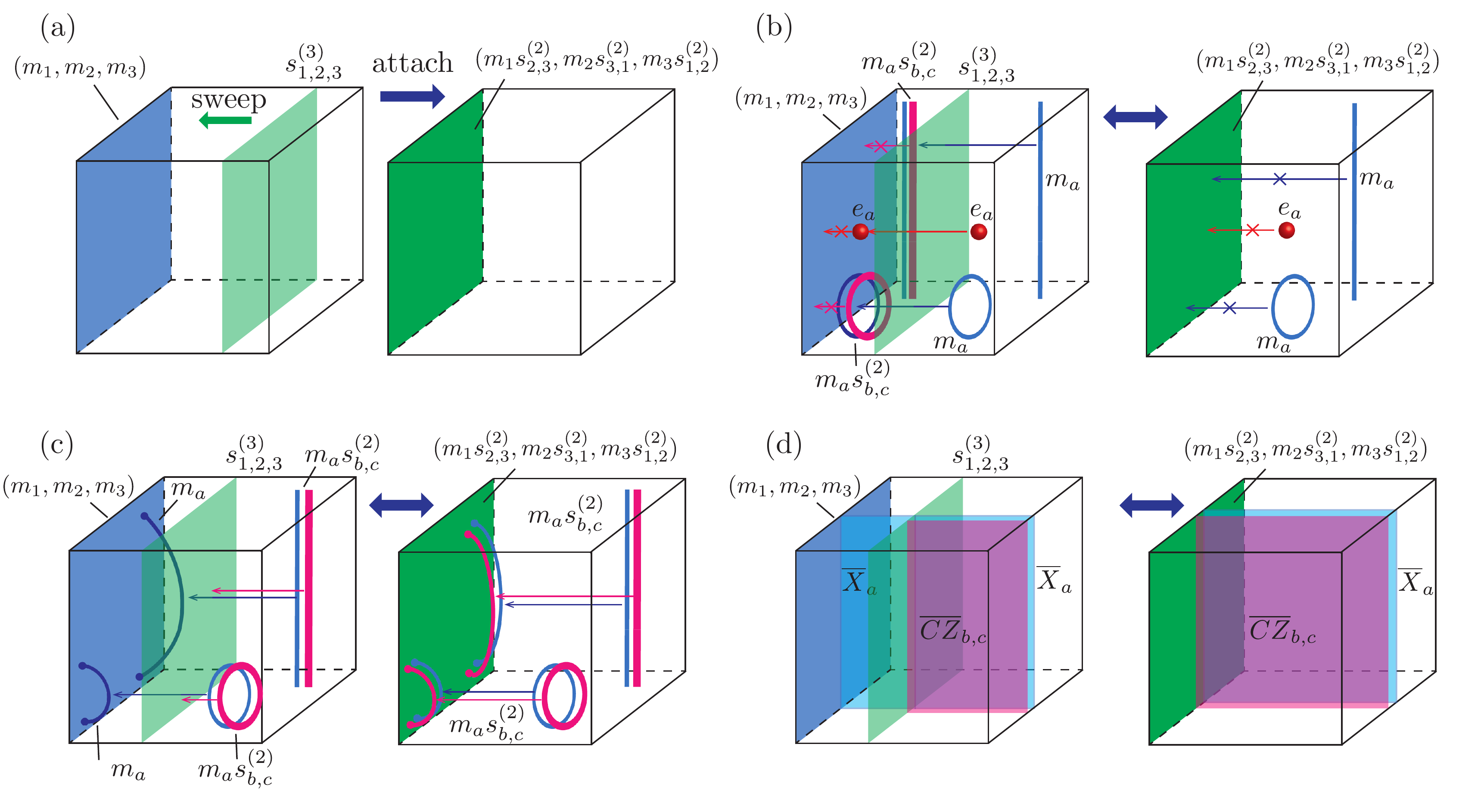}
  \caption{(a) Applying a transversal CCZ effectively sweeps the domain wall $S^{3}_{1,2,3}$ across the system and attaches it onto the $(m_1,m_2,m_3)$-boundary without being condensed, generating an exotic $(m_1 s^{(2)}_{2,3}, m_2 s^{(2)}_{3,1}, m_3 s^{(2)}_{1,2})$-boundary (green).  (b) The $e$-particle and $m$-string alone cannot condense on the $(m_1 s^{(2)}_{2,3}, m_2 s^{(2)}_{3,1}, m_3 s^{(2)}_{1,2})$-boundary (shown on the right),  which can be considered as a composite of the $S^{3}_{1,2,3}$-wall and $(m_1,m_2,m_3)$-boundary (shown on the left). The $e_a$-particle  and $m_a$-string  get transformed into $e_a$ and $m_a s^{2}_{b,c}$ when passing the domain wall first, and hence cannot condense on the remaining $(m_1, m_2, m_3)$-boundary. (c) The $m_a s^{2}_{b,c}$-string can condense on the exotic $(m_1 s^{(2)}_{2,3}, m_2 s^{(2)}_{3,1}, m_3 s^{(2)}_{1,2})$-boundary, since it gets transformed to $m_a$-string when passing the $S^{3}_{1,2,3}$-wall and can then condense on the remaining $(m_1,m_2,m_3)$-boundary. (d) The product of logical branes $\lo{X}_a \lo{\text{CZ}}_{b,c}$ can terminate on the $(m_1 s^{(2)}_{2,3}, m_2 s^{(2)}_{3,1}, m_3 s^{(2)}_{1,2})$-boundary since it gets transformed to $\lo{X}_a$ when passing the $S^{3}_{1,2,3}$-wall and then terminates on the remaining $(m_1, m_2, m_3)$-boundary.  }
  \label{fig:exotic_domain_wall}
\end{figure*}

In the case of $\lo{\text{CZ}}_{1,2}$,  we need to make sure the two interacting copies have opposite $e$ and $m$ boundary types, i.e., permitting the following boundary types for a stack code with two copies of toric codes: $\B_i$~$=$~$ (e_1, m_2)$~or~$(m_1, e_2)$.  The mapping of the gapped boundary under the domain wall action is as follows:
\begin{align}\label{eq:CZ_wall_transform}
\nonumber s^{(2)}_{1,2} : \ & (e_1,m_2)\rightarrow (e_1, e_1 m_2) \equiv (e_1, m_2), \\
 		   &(m_1,e_2)\rightarrow (m_1 e_2, e_2) \equiv (m_1, e_2).   
\end{align}
Note that $(e_1, e_1 m_2)$ in the first line means the boundary condense both $e_1$ and a composite excitation $e_1 m_2$, which is equivalent to the boundary $(e_1, m_2)$ up to a basis change (i.e., different choices of generators).  This is because, due to the $\Z_2$ fusion rule $e\times e = \mathbb{I}$, a composite of $e_1$ and $e_1 m_2$ is just $m_2$ (i.e., $e_1\times e_1 m_2= m_2$), which should still condense on the boundary. The second line in Eq.~\eqref{eq:CZ_wall_transform} is just symmetric to the first line up to  a permutation of the two copies of codes. We note that for the stack of three 3D surface codes considered in Fig.~\ref{fig:CCZ_arrangement} for the purpose of applying CCZ gate, any gapped boundary has the form $(m_a, m_b, e_c)$, where $a,b,c$ is any permutation of $1,2,3$. Also, opposite boundaries have the same form. The CZ domain walls $s^{(2)}_{a,c}$ and $s^{(2)}_{b,c}$ condense on the $(m_a, m_b, e_c)$ boundary according to the above arguments, i.e.,
\begin{align}
\nonumber s^{(2)}_{a,c}:(m_a, m_b, e_c) &\rightarrow (m_a, m_b, e_c) \\
s^{(2)}_{b,c}:(m_a, m_b, e_c) &\rightarrow (m_a, m_b, e_c),
\end{align}
as illustrated in Fig.~\ref{fig:domain_wall_condensation}(a). 
One can also represent this boundary with the redundant notation $(m_a, m_b, e_c)$$\equiv$$ (m_a, m_b, e_c, s^{(2)}_{a,c}, s^{(2)}_{b,c})$, since $s^{(2)}_{a,c}$ and $s^{(2)}_{b,c}$ condense on the boundary just like $m_a$, $m_b$, and $e_c$. 
 A $\lo{\text{CZ}}$ gate is supported on a membrane connecting two opposite boundaries.  Therefore,  both $\lo{\text{CZ}}_{a,c}$ and $\lo{\text{CZ}}_{b,c}$ are allowed.   

Now, in the case of $\lo{\text{CCZ}}_{1,2,3}$ applied to a stack of three 3D surface codes, we need the boundary type to be the form $(m_a, m_b, e_c)$ as mentioned in the above paragraph. The mapping of the boundary under the domain wall action is as follows:
\begin{align}\label{eq:CCZ_condensation}
\nonumber &s^{(3)}_{a,b,c} : \  (m_a, m_b, e_c)\equiv  (m_a, m_b, e_c, s^{(2)}_{a,c}, s^{(2)}_{b,c})  \\
\nonumber 		    \rightarrow & (m_a s^{(2)}_{b,c}, m_b s^{(2)}_{c,a}, e_c, s^{(2)}_{a,c}, s^{(2)}_{b,c}) \equiv  (m_a, m_b, e_c, s^{(2)}_{a,c}, s^{(2)}_{b,c})  \\
 		    \equiv &  (m_a, m_b, e_c), 
\end{align}
implying that $s^{(3)}_{a,b,c}$ also condense onto  the $(m_a, m_b, e_c)$ boundary and hence keeps it invariant, as illustrated in Fig.~\ref{fig:domain_wall_condensation}(b).  Note that the equivalence between $(m_a s^{(2)}_{b,c}, m_b s^{(2)}_{c,a}, e_c, s^{(2)}_{a,c}, s^{(2)}_{b,c})$ and $(m_a, m_b, e_c, s^{(2)}_{a,c}, s^{(2)}_{b,c})$ in the above equation is due to the fact that the composite of  $m_a s^{(2)}_{b,c}$ ($m_b s^{(2)}_{c,a}$) and $s^{(2)}_{a,c}$ ($s^{(2)}_{c,a}$) is $m_a$ ($m_b$), where we have used the fact that two domain walls fuse into vacuum $s^{(2)}\times s^{(2)} = \mathbb{I}$.

\subsubsection{Exotic gapped boundaries generated by the CCZ domain wall}\label{sec:exotic_boundary}

In the above discussion, we require the transversal gate or more generally constant-depth circuits  $U$ to implement a logical gate in the code space $\mathcal{H}_C$, which needs to preserve all the gapped boundaries, equivalently requiring the gapped domain wall $w$ to condense on all boundaries $\B_i$. However, this is not the only interesting situation.  If the domain wall $w$ does not condense on the boundary $\B_i$, it is attached on the boundary, leading to a change of the boundary to some other boundary type: 
\be
\B'_i = w \B_i w^\dag. 
\ee
The transversal gate or the constant-depth circuit hence applies a \textit{transversal logical map} or \textit{locality-preserving logical map}, which not only operates the logical qubits but also maps the original code space to a different code space $\mathcal{H}'$. More interestingly, the generated boundary $\B'_i$ can be some other more exotic boundary type different from the usual $e$- and $m$-type boundaries. This is exactly the situation when applying logical CCZ gate on the 3D fractal codes.  

As have been discussed in Sec.~\ref{sec:3D}, the fractal codes have holes with $m$-boundaries in the bulk. Therefore, for a stack code with three copies of 3D fractal codes, the corresponding hole boundaries can only be $(m_1,m_2,m_3)$. In the following, one can see that neither the CZ or CCZ domain wall condenses  on the  $(m_1,m_2,m_3)$ boundaries.  
We first consider the CZ wall $s^{(2)}_{1,2}$ acting on this boundary:
\be
s^{(2)}_{1,2} : (m_1, m_2, m_3) \rightarrow (m_1 e_2, e_1 m_2, m_3),
\ee
implying that $s^{(2)}_{1,2} $ cannot condense on the boundary.  Similarly, $s^{(2)}_{1,3} $ and $s^{(2)}_{2,3} $ also cannot condense on this boundary.   Now the CCZ domain wall applies the following mapping to this boundary:
\be
s^{(3)}_{1,2,3} : \  (m_1, m_2, m_3) \rightarrow (m_1 s^{(2)}_{2,3}, m_2 s^{(2)}_{3,1}, m_3 s^{(2)}_{1,2}).
\ee
Note that since $s^{(2)}_{1,2}$ and $s^{(2)}_{2,3}$ do not condense on the $(m_1, m_2, m_3)$-boundary, a rewriting similar to Eq.~\eqref{eq:CCZ_condensation} is impossible.   We point out that $(m_1 s^{(2)}_{2,3}, m_2 s^{(2)}_{3,1}, m_3 s^{(2)}_{1,2})$ is a new type of codimension-1 (dimension-2) boundary generated by attaching the CCZ domain wall $s^{(3)}_{1,2,3}$ on the gapped boundary  $(m_1, m_2, m_3)$, since the domain wall does not condense on the boundary. Interestingly, the usual $e$- and $m$-type excitations cannot individually  condense on this new boundary, as opposed to the previously known gapped boundaries in 2D and 3D. Only a combination of the $m$-string in one copy and the CZ domain wall $s^{(2)}$ in the other two copies with the same boundary support as the $m$-string can condense on the  $(m_1 s^{(2)}_{2,3}, m_2 s^{(2)}_{3,1}, m_3 s^{(2)}_{1,2})$-boundary.  This new type of boundary is beyond the standard classification of gapped boundaries using the Lagrange subgroup \cite{Levin:2013tc}.

One can also understand the property of the exotic $(m_1 s^{(2)}_{2,3}, m_2 s^{(2)}_{3,1}, m_3 s^{(2)}_{1,2})$-boundary via an intuitive domain-wall attachment  picture.  This exotic boundary is a composite of a domain wall and a boundary, i.e, the CCZ domain wall $s^{(3)}_{1,2,3}$ is attached on the  $(m_1, m_2, m_3)$-boundary, as illustrated in Fig.~\ref{fig:exotic_domain_wall}(a). One can hence imagine placing the domain wall $s^{(3)}_{1,2,3}$ slightly away from  the $(m_1, m_2, m_3)$-boundary, as shown in Fig.~\ref{fig:exotic_domain_wall}(b).   One can first verify that any $e$- and $m$-type excitations cannot condense on the exotic composite boundary. After penetrating the domain wall $s^{(3)}_{1,2,3}$, $e_a$ ($a=1,2,3$) particle is mapped to $s^{(3)}_{1,2,3} : e_a \rightarrow e_a$, and cannot condense on the $(m_1, m_2, m_3)$-boundary. Similarly,  $m_a$ is mapped to $s^{(3)}_{1,2,3}~:~m_a~\rightarrow~m_a s^{2}_{b,c}$ ($a,b,c$ can be any permutation of $1,2,3$) after penetrating the the domain wall, and also cannot condense on the  $(m_1, m_2, m_3)$-boundary. On the other hand,  $m_a s^{(2)}_{b,c}$ excitation is mapped to $s^{(3)}_{1,2,3}: m_a s^{(2)}_{b,c}~\rightarrow~m_a$ after penetrating the domain wall, which can now condense on the  $(m_1, m_2, m_3)$-boundary, as shown in Fig.~\ref{fig:exotic_domain_wall}(c). When considering the worldsheet of the excitations  $m_a s^{(2)}_{b,c}$, one can conclude that only the composite logical operator $\lo{X}_a \lo{\text{CZ}}_{b,c}$ can terminate at the  $(m_1 s^{(2)}_{2,3}, m_2 s^{(2)}_{3,1}, m_3 s^{(2)}_{1,2})$-boundary, while $\lo{X}_a$ alone cannot terminate, as illustrated in Fig.~\ref{fig:exotic_domain_wall}(d).  This can be understood again via the domain-wall attachment picture:  the  composite logical operator $\lo{X}_a \lo{\text{CZ}}_{b,c}$ penetrating the domain wall is mapped into  $\lo{X}_a$ which can then condense on the $(m_1, m_2, m_3)$-boundary.

\begin{figure}[t]
  \includegraphics[width=0.8\columnwidth]{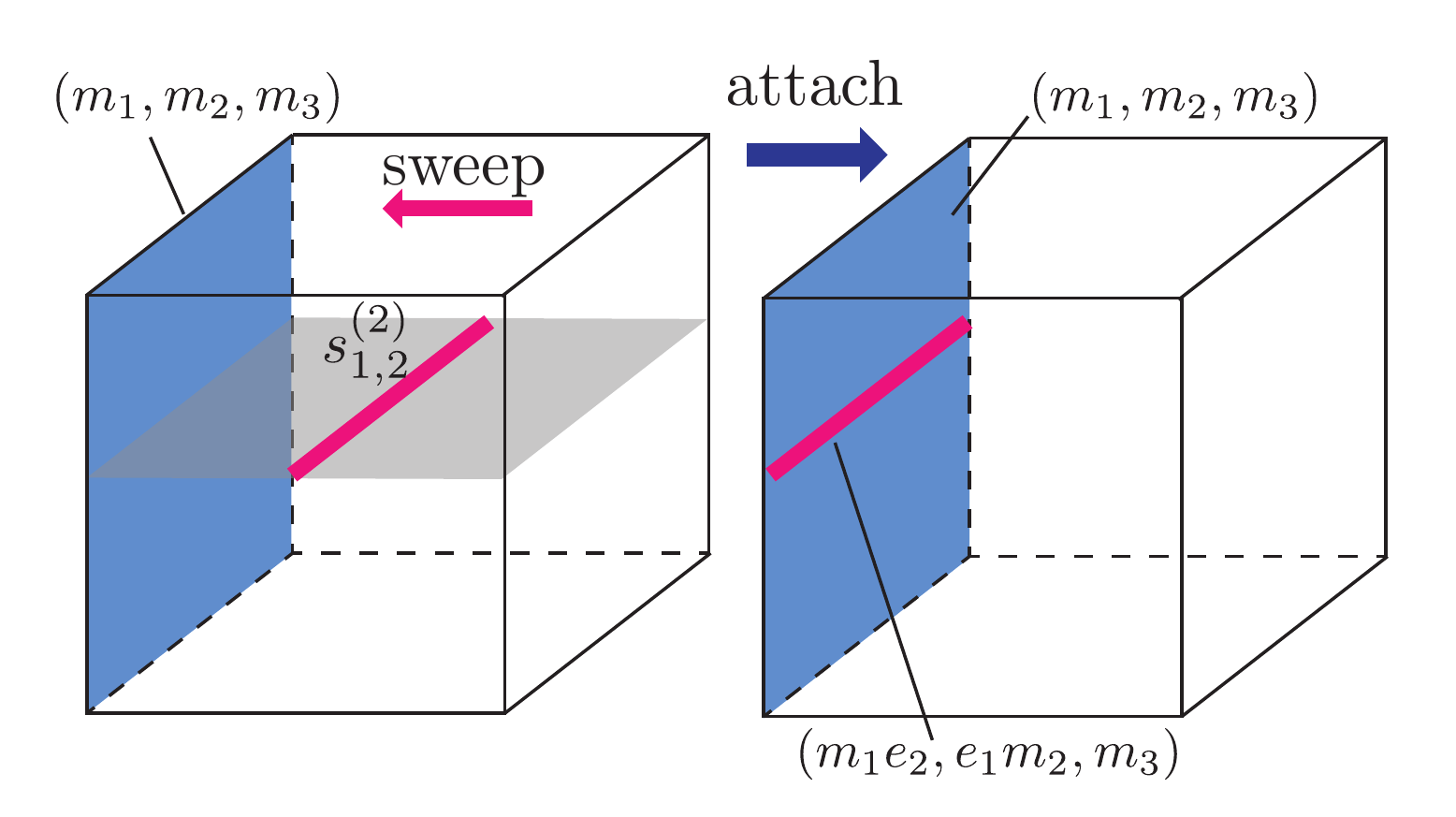}
  \caption{Applying a transversal CZ along a codimension-1 sub-manifold (grey) effectively sweeps the  the $s^{2}_{1,2}$-domain wall across the sub-manifold and attaches the wall onto the $(m_1, m_2, m_3)$-boundary, generating a codimension-2 boundary $(m_1 e_2, e_1 m_2, m_3)$ nested on the $(m_1, m_2, m_3)$-boundary.}
  \label{fig:exotic_subsystem_domain_wall}
\end{figure}

Finally, we note that when attaching the $s^{(2)}_{1,2}$-domain wall to the $(m_1, m_2, m_3)$-boundary, one also generates a new type of codimension-2 (dimension-1) boundary $(m_1 e_2, e_1 m_2, m_3)$ nested on the usual $(m_1, m_2, m_3)$-boundary, as illustrated in Fig.~\ref{fig:exotic_subsystem_domain_wall}.  This nested boundary is exotic since the string excitation $m_1$ ($m_2$) can only condense on the line-shape nested boundary along with the particle excitation $e_2$ ($m_1$). Of course, such exotic codimension-1  boundary also exists in the absence of the third copy, i.e.,  a $(m_1 e_2, e_1 m_2)$ codimension-2 boundary  nested on the usual $(m_1, m_2)$-boundary. 

In summary, when applying global or 1-form topological symmetries $U$ on a system with boundaries,  new types of gapped boundaries are generated if the corresponding gapped domain walls do not condense on the original boundaries, i.e., do not satisfy the conditions in Eqs.~\eqref{eq:domain_wall_condensation_1} and \eqref{eq:domain_wall_condensation_2}.

\begin{figure}[hbt]
  \includegraphics[width=1\columnwidth]{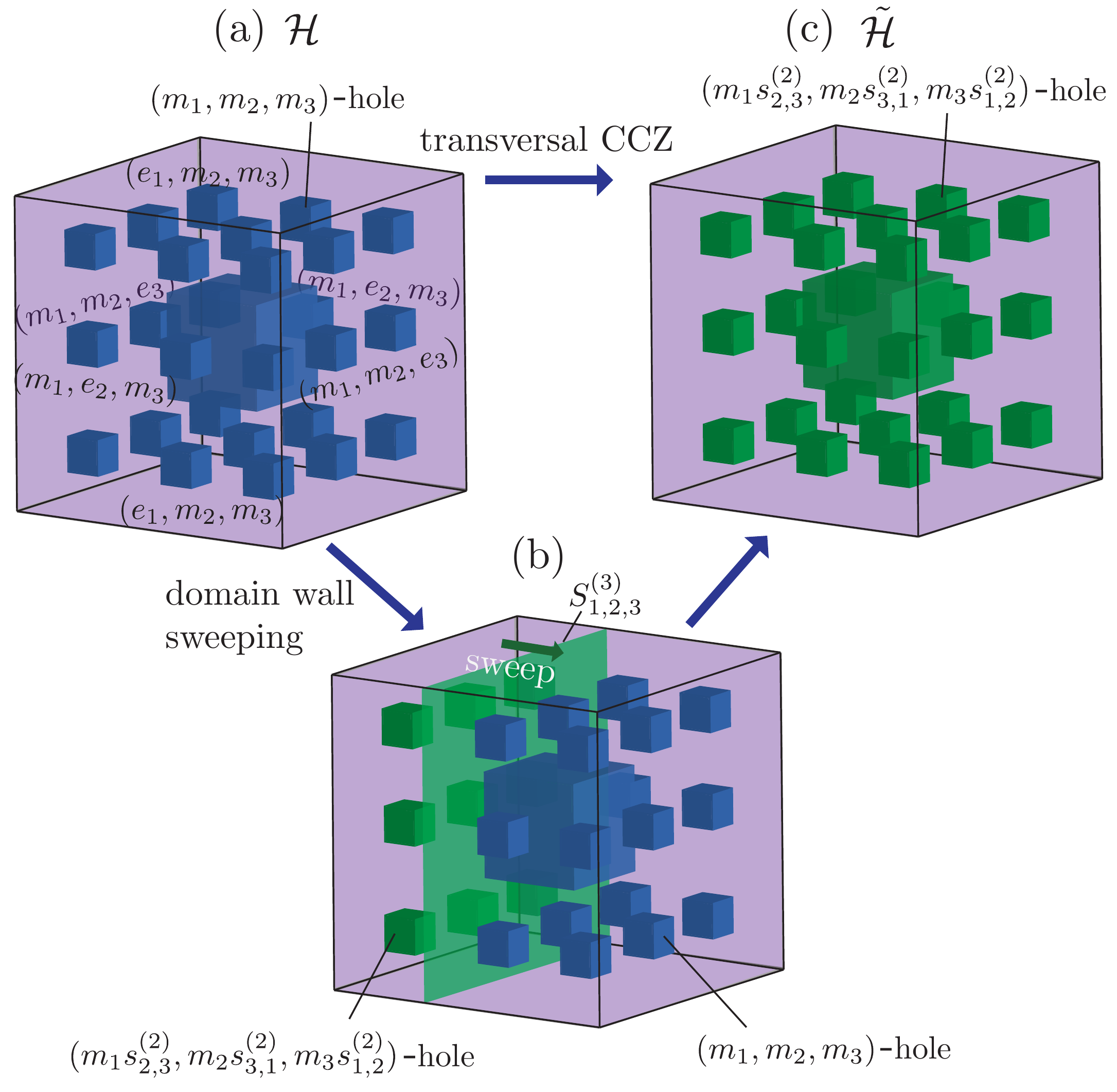}
  \caption{Applying a transversal CCZ gate on a stack of three fractal surface codes effectively sweeps the $s^{3}_{1,2,3}$-domain wall across the system and attach the wall onto all the $(m_1,m_2,m_3)$-holes (blue) in the bulk, producing new  $(m_1 s^{(2)}_{2,3}, m_2 s^{(2)}_{3,1}, m_3 s^{(2)}_{1,2})$-boundaries (green) on these holes which couple the three codes together as previously illustrated in Fig.~\ref{fig:exotic_domain_wall}. }
  \label{fig:fractal_domain_wall_sweeping}
\end{figure}

\begin{figure}[hbt]
  \includegraphics[width=0.8\columnwidth]{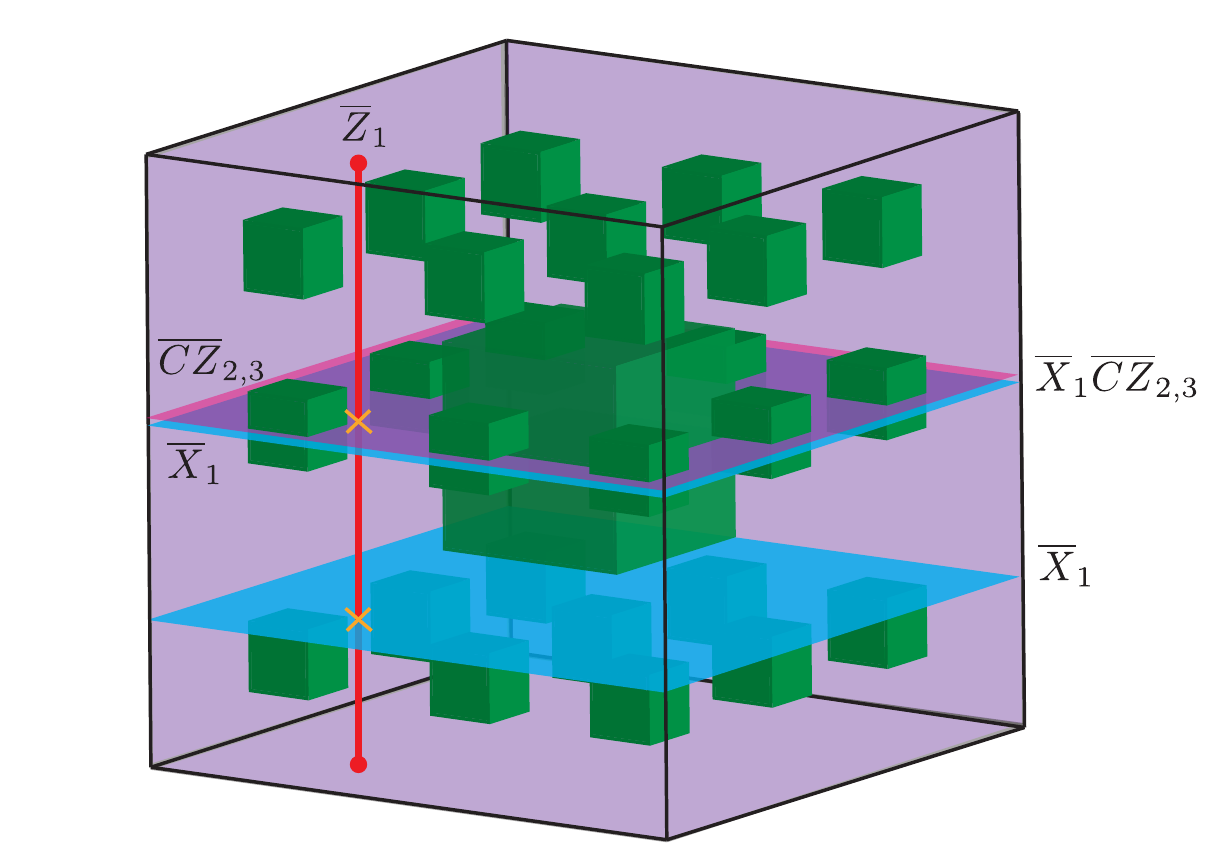}
  \caption{Logical operators in the new stack code with exotic hole boundaries $(m_1 s^{(2)}_{2,3}, m_2 s^{(2)}_{3,1}, m_3 s^{(2)}_{1,2})$ (green). The logical $Z$-strings remain the same as those in the original stack code.  A logical $X$-brane can terminate only on the outer boundaries. The combined logical brane $\lo{X}_1 \lo{\text{CZ}}_{2,3}$ can instead terminate on the exotic hole boundaries and its minimal support forms a Sierpi\'nski carpet as shown.}
  \label{fig:exotic_boundary_logical}
\end{figure}

\subsubsection{Fractal topological orders  and codes with the exotic $(m_1 s^{(2)}_{2,3}, m_2 s^{(2)}_{3,1}, m_3 s^{(2)}_{1,2})$-holes}

In order to implement a logical CCZ gate, we first need to apply a transversal CCZ gate in a stack of 3D fractal codes with the particular gapped boundary configuration on its outer surface: $(m_a, m_b, e_c)$ ($a,b,c$ is an arbitrary permutation of $1,2,3$),  as shown in Fig.~\ref{fig:CCZ_arrangement}, while the hole boundaries are chosen to be $(m_1, m_2, m_3)$. The transversal gate is defined as $\widetilde{\text{CCZ}}=\Motimes_j \text{CCZ}_{j;1,2,3}$.   Here, we use $\widetilde{\text{CCZ}}$ instead of $\lo{\text{CCZ}}$ as before to emphasize that this transversal gate is not a logical gate. The application of $\widetilde{\text{CCZ}}$ corresponds to pushing the CCZ domain wall $s^{(3)}_{1,2,3}$ across the entire system. As we have discussed above, the CCZ domain wall condenses on all the outer boundary $(m_a, m_b, e_c)$ but not on the hole boundary $(m_1, m_2, m_3)$. Therefore, after applying the transversal CCZ gate, the CCZ domain wall $s^{(3)}_{1,2,3}$ is attached to the $(m_1, m_2, m_3)$-boundaries and turn all the hole boundaries into $(m_1 s^{(2)}_{2,3}, m_2 s^{(2)}_{3,1}, m_3 s^{(2)}_{1,2})$, as illustrated in Fig.~\ref{fig:fractal_domain_wall_sweeping}.

Now we discuss the parent Hamiltonian or equivalently the stabilizers of the stack  code and the new code generated by applying $\widetilde{\text{CCZ}}$.  The original  Hamiltonian of the stack code can be divided into three parts: 
\begin{align}
\nonumber H&=H^\text{bulk}+H^\text{o.b.}+H^\text{h.b.} \\
&\equiv - \sum_{i \in \text{bulk}}\sum_a S_{i;a}^\text{bulk} - \sum_{j \in \text{o.b.}} \sum_a S_{j;a}^\text{o.b.}-\sum_{k \in \text{h.b.}}\sum_a S_{k;a}^\text{h.b.}, 
\end{align}
where $H^\text{bulk}$ ($S_{i;a}^\text{bulk}$), $H^\text{o.b.}$ ($S_{j;a}^\text{o.b.}$) and $H^\text{h.b.}$ ($S_{k;a}^\text{h.b.}$) represent the bulk, outer boundary and the hole boundary terms in the parent Hamiltonian $H$ (stabilizer group $\SS=\langle \{S_{i; a}\} \rangle$), and   $a=1,2,3$ labels $a^\text{th}$ copy of the 3D fractal codes.  After applying the transversal CCZ gate, the original Hamiltonian will be conjugated by $\widetilde{\text{CCZ}}$, leading to the parent Hamiltonian of the new code:
\be
\tilde{H} = \widetilde{\text{CCZ}} \ H \ \widetilde{\text{CCZ}}^\dag. 
\ee
Note that $\widetilde{\text{CCZ}}$ is a Hermitian operator, i.e., $\widetilde{\text{CCZ}}^\dag=\widetilde{\text{CCZ}}$. Similarly, the original code space is mapped to a  new code space:
\be
\widetilde{\text{CCZ}} : \H \rightarrow \tilde{\H}.
\ee

 We choose the bulk and outer boundary Hamiltonians ($H^\text{bulk}$ and $H^\text{o.b.}$) and the corresponding stabilizers ($S^\text{bulk}_i$ and $S^\text{o.b.}_i$) to be exactly the same as the case of implementing transversal logical CCZ in a stack of 3D surface codes with the construction in Ref.~\cite{Vasmer2019} (illustrated in Fig.~\ref{fig:CCZ_arrangement}), which satisfies the transversal logical CCZ gate conditions in Eqs.~\eqref{eq:CCZcondition1} and \eqref{eq:CCZcondition2}, and also the generic conditions for transversal logical gates in Eqs.~\eqref{eq:bulkcondion} and \eqref{eq:boundary_condition}.  The later means the bulk and outer boundary Hamiltonian and stabilizers are invariant under the conjugation of $\widetilde{\text{CCZ}}$ up to a logical identity:
\begin{align}
\nonumber P_C (\widetilde{\text{CCZ}} \ H^\text{bulk} \ \widetilde{\text{CCZ}}^\dag)P_C &= P_C H^\text{bulk} P_C \\
\nonumber \Longleftrightarrow P_C (\widetilde{\text{CCZ}} \  S_{i,a}^\text{bulk} \  \widetilde{\text{CCZ}}^\dag) P_C &= P_C S_{i,a}^\text{bulk} P_C, \\
\non P_C(\widetilde{\text{CCZ}} \ H^\text{o.b.} \ \widetilde{\text{CCZ}}^\dag) P_C &= P_C H^\text{o.b.} P_C  \\
\Longleftrightarrow P_C (\widetilde{\text{CCZ}}  S_{j;a}^\text{o.b.} \widetilde{\text{CCZ}}^\dag) P_C &= P_C S_{j;a}^\text{o.b.} P_C.
\end{align}
   In other words, $\widetilde{\text{CCZ}}$ is an onsite topological symmetry in the bulk and outer boundary. Note that $P_C$ here projects onto the original code space $\H$. 
  
  Nevertheless, $\widetilde{\text{CCZ}}$ does not keep the $(m_1,m_2,m_3)$-hole boundary invariant as discussed in the previous subsection. This means the boundary Hamiltonian  $H^{\text{h.b.}}$ and corresponding stabilizers $S^{\text{h.b.}}_{i;a}$ on the hole boundaries (i.e., the first layer on the surface of the hole) are not in general invariant up to logical identity when conjugated by $\widetilde{\text{CCZ}}$. As in all previous cases, we denote the two types of stabilizers, the $X$ and $Z$ stabilizers as $A^{\text{h.b.}}_{q;a}$ and $B^{\text{h.b.}}_{p;a}$ respectively. The $Z$ stabilizers $B_{p;a}^{\text{h.b.}}$ on the hole boundaries remain invariant under the conjugation of $\widetilde{\text{CCZ}}$, since  $\widetilde{\text{CCZ}}$ is diagonal in the $Z$-basis.  
%   \be
%   \widetilde{\text{CCZ}} B_{p; a}^{\text{h.b.}}\widetilde{\text{CCZ}}^\dag \equiv \widetilde{\text{CCZ}}  \big( \Motimes_{j \in B_{p;a}^{\text{h.b.}}} Z_{j;a} \big)\widetilde{\text{CCZ}}^\dag  = B_{p;a}^{\text{h.b.}}. 
%   \ee
  On the other hand the $X$ stabilizers in copy $a$, $A_{q,a}^{\text{h.b.}}$, will be transformed under the  conjugation of transversal CCZ as:
  \begin{align}
\nonumber \tilde{A}_{q;a}^{\text{h.b.}}=  \widetilde{\text{CCZ}} A_{q;a}^{\text{h.b.}}\widetilde{\text{CCZ}}^\dag \equiv&  \widetilde{\text{CCZ}} \big( \Motimes_{j \in A_{q;a}^{\text{h.b.}}} X_{j;a} \big) \widetilde{\text{CCZ}}^\dag \\
  =& \Motimes_{j \in A_{q;a}^{\text{h.b.}}} X_{j;a}  \Motimes_{k \in A_{q;a}^{\text{h.b.}}} \text{CZ}_{k;b,c} , 
  \end{align}  
where $a,b,c$ is an arbitrary permutation of $1,2,3$.  
Note that in the new hole boundary stabilizer $\tilde{A}_{q;a}^{\text{h.b.}}$, the part in copy $a$ is kept invariant as $A_{q;a}^{\text{h.b.}}$, while an additional CZ interaction  is applied between copy $b$ and $c$ with the site label $k \in A_{q;a}^{\text{h.b.}}$ aligned with the corresponding site $j$ in copy $a$ within the support of the boundary $X$ stabilizer $A_{q;a}^{\text{h.b.}}$. Note that although the three copies of 3D fractal codes are independent in the original stack code, these three copies of fractal codes are coupled together along the hole boundaries via the CZ interactions in the new stabilizers $\tilde{A}_{q;a}^{\text{h.b.}}$  after the CCZ conjugation.  See App.~\ref{app:mmm_hole_example} for the detailed illustration with the lattice model. 

Interestingly, this new model $\tilde{H}$ is a non-Pauli stabilizer model with non-Pauli boundary stabilizers. All the stabilizers are contained in the non-Pauli stabilizer group:
\begin{align}
\nonumber \tilde{\SS}=&\langle \{\tilde{S}_{i; a}\}  \rangle \equiv \langle \{S_{i; a}^\text{bulk}, S_{j; a}^\text{o.b.}, \tilde{S}_{k; a}^\text{h.b.}\}  \rangle \\
\equiv & \langle \{A_{q; a}^\text{bulk}, B_{p; a}^\text{bulk}, A_{q'; a}^\text{o.b.}, B_{p'; a}^\text{o.b.}, \tilde{A}_{q"; a}^\text{h.b.}, B_{p"; a}^\text{h.b.} \}  \rangle
\end{align}
Note that since the stabilizers in the original stabilizer group $\SS$ all commute, they remain commute when conjugated by $\widetilde{\text{CCZ}}$.  Therefore, all the stabilizers in $\tilde{\SS}$ commute. 
 
%  Now we show that the non-Pauli stabilizers $\tilde{A}_{q;a}^{\text{h.b.}}$ still commute with each other and all the $Z$-type stabilizers $B_{p;b}^{\text{h.b.}}$.   First, note that $\tilde{A}_{q;a}^{\text{h.b.}}$ and $A_{q;a}^{\text{h.b.}}$ contain exactly the same Pauli~$X$ operators, and we already have the commutation relation $[A_{q;a}^{\text{h.b.}}, B_{p;b}^{\text{h.b.}}]=0$ in the original model. We hence get the commutation relation $[\tilde{A}_{q;a}^{\text{h.b.}}, B_{p;b}^{\text{h.b.}}]~=~0$ in the new model, since the additional operators in $\tilde{A}_{q;a}^{\text{h.b.}}$ are all CZ operators which commute with any Pauli~$Z$ operators in $B_{p;b}^{\text{h.b.}}$.  Secondly, in the original model, any two $X$ stabilizer commute, i.e., $[A_{q;a}^{\text{h.b.}}, A_{r;b}^{\text{h.b.}}]~=~0$. Now under the CCZ conjugation, this commutation is still preserved, i.e., 
% \begin{align}
%  \widetilde{\text{CCZ}}[A_{q;a}^{\text{h.b.}}, A_{r;b}^{\text{h.b.}}]\widetilde{\text{CCZ}}^\dag = [\tilde{A}_{q;a}^{\text{h.b.}}, \tilde{A}_{r;b}^{\text{h.b.}}] =0
% \end{align}
% This shows that any two new stabilizers $\tilde{A}_{q;a}^{\text{h.b.}}$ and $\tilde{A}_{r;b}^{\text{h.b.}}$ also commute. 

Note that in the new stack code with exotic hole boundaries, the centralizer group $\mathsf{C}(\tilde{\SS})$ is generated by the elements in the  non-Pauli stabilizer group $\SS=\langle \{S_{i; a}\} \rangle$ and the logical operators $\lo{Z}_a$ and $\lo{X}_a \lo{\text{CZ}}_{b,c}$ ($a,b,c$ are permutation of $1,2,3$). On the other hand, the standard logical operator $\lo{X}_a$ is no longer in the centralizer group $\mathsf{C}(\SS)$. The ``\textit{canonical}" logical operators of this code which commute with all the stabilizers $S_{i; a}$  are contained in the quotient group $\mathsf{C}(\SS) / \SS$.  As illustrated in Fig.~\ref{fig:exotic_boundary_logical}, the standard logical operator $\lo{X}_1$ can still terminate on the $(m_1,m_2, e_3)$-boundary on the outer surface, but not the exotic $(m_1 s^{(2)}_{2,3}, m_2 s^{(2)}_{3,1}, m_3 s^{(2)}_{1,2})$-hole boundary as explained in the previous subsection.   On the other hand, the combined logical operator $\lo{X}_1 \lo{\text{CZ}}_{2,3}$ can also terminate on the exotic hole boundary and is hence able to vertically move across the whole system. This is because $\lo{X}_a \lo{\text{CZ}}_{b,c}$ commutes with all the non-Pauli stabilizers on the hole boundary, while $\lo{X}_a$ does  not commute with all of them.

\subsubsection{Logical CCZ and CZ gates in 3D fractal codes}\label{sec:CCZ_protocol} 

 \begin{figure*}[hbt]
   \includegraphics[width=2\columnwidth]{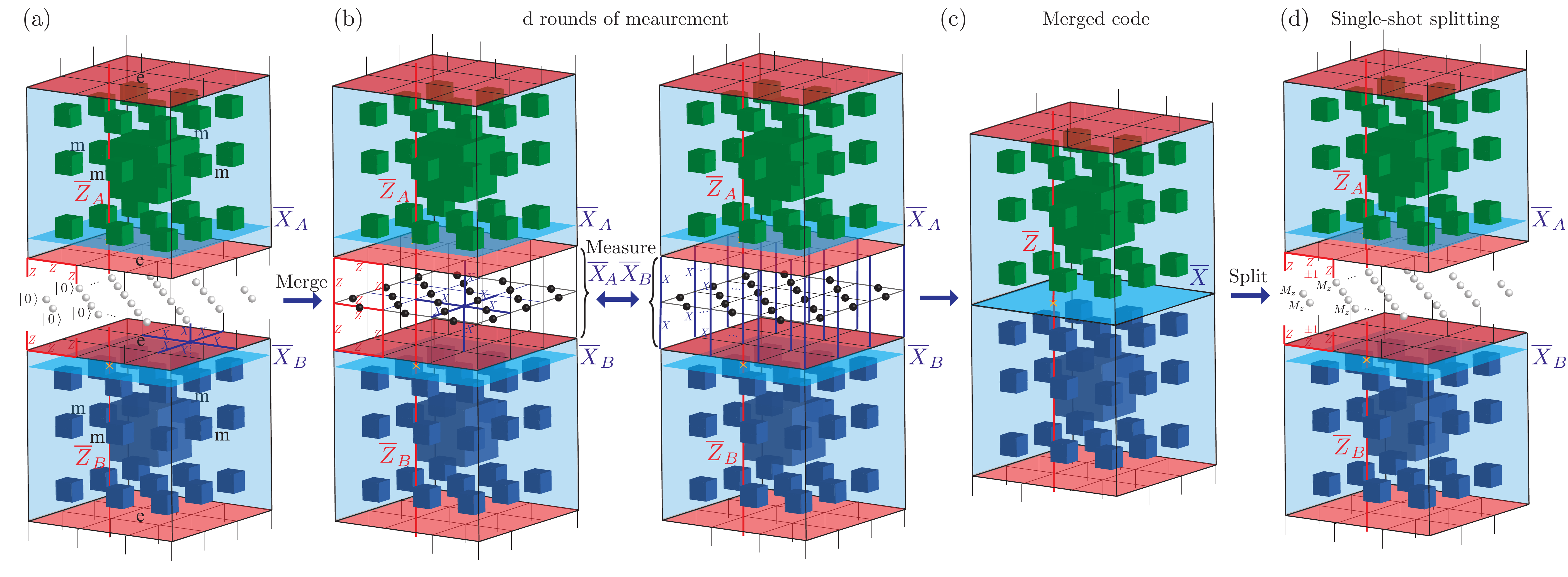}
   \caption{The lattice surgery protocol to transform the logical information inside one of the code copies within the coupled stack fractal code (top) to an ancilla fractal code block (bottom). (a) Initialize the ancilla qubits (white balls) between the two codes in state $\ket{0}$. (b) Lattice merging via $d$ rounds of syndrome measurement and subsequent decoding. Two types of new stabilizers are introduced at the interface: the 4-body $Z$ stabilizers (red) with $+1$ eigenvalues and the 6-body $X$ stabilizers (blue) with unknown eigenvalues and hence random $\pm 1$ measurement outcome. The measurement outcome of the new interface X stabilizers equivalently gives the measurement outcome of the joint parity of logical operators: $\lo{X}_A\lo{X}_B$ (supported on the highlighted blue vertical edges). (c) The merged code after $d$ rounds of measurement. (d) Lattice splitting protocol. The unknown stabilizers at the interface are the 3-body $Z$ stabilizers which have random $\pm 1$ measurement outcome, which can be inferred by the projective measurement of the ancilla qubits in the $Z$-basis ($M_Z$). The $(-1)$-eigenvalue corresponds to the $m$-loop excitations, which can be corrected with a single-shot decoding process even in the presence of measurement error.}
   \label{fig:lattice_surgery}
 \end{figure*}

Now we discuss the entire protocol for implementing the logical CCZ gate as listed below:

\begin{enumerate}
\item
\textbf{Transversal CCZ:} As we mentioned in the last subsection, we first apply a transversal CCZ gate to a stack of three fractal codes as shown in Fig.~\ref{fig:fractal_domain_wall_sweeping}, which maps the original code space $\H$ to the new code space $\tilde{\H}$ with the exotic $(m_1 s^{(2)}_{2,3}, m_2 s^{(2)}_{3,1}, m_3 s^{(2)}_{1,2})$-hole boundaries coupling the three copies of fractal codes together.     

\item
\textbf{Lattice merging along $e$-boundaries:} To complete the logical gate, we must return the quantum information to the original code space $\H$. We propose using lattice surgery to complete this process.   

We will discuss the lattice surgery between one of the code copies within the coupled stack fractal and an ancilla fractal code, as illustrated in Fig~\ref{fig:lattice_surgery}. Although the logical information of each copy in a stack code block is in general entangled with each other and also with other code blocks, we still explain the protocol by denoting logical information in the first copy as $\lo{\ket{\psi}}_A=\alpha \lo{\ket{+}}_A + \beta \lo{\ket{-}}_A$  for simplicity, yet the protocol will also apply in the more general case. The ancilla fractal code is prepared in the logical state~$\lo{\ket{0}}_B$, which is completed by preparing all of the physical qubits of that codeblock in the $\ket{0}$ state and measuring out the $X$~stabilizers fault-tolerantly by repeating their measurements $d$~times (rounds).

 \begin{figure}[hbt]
   \includegraphics[width=0.7\columnwidth]{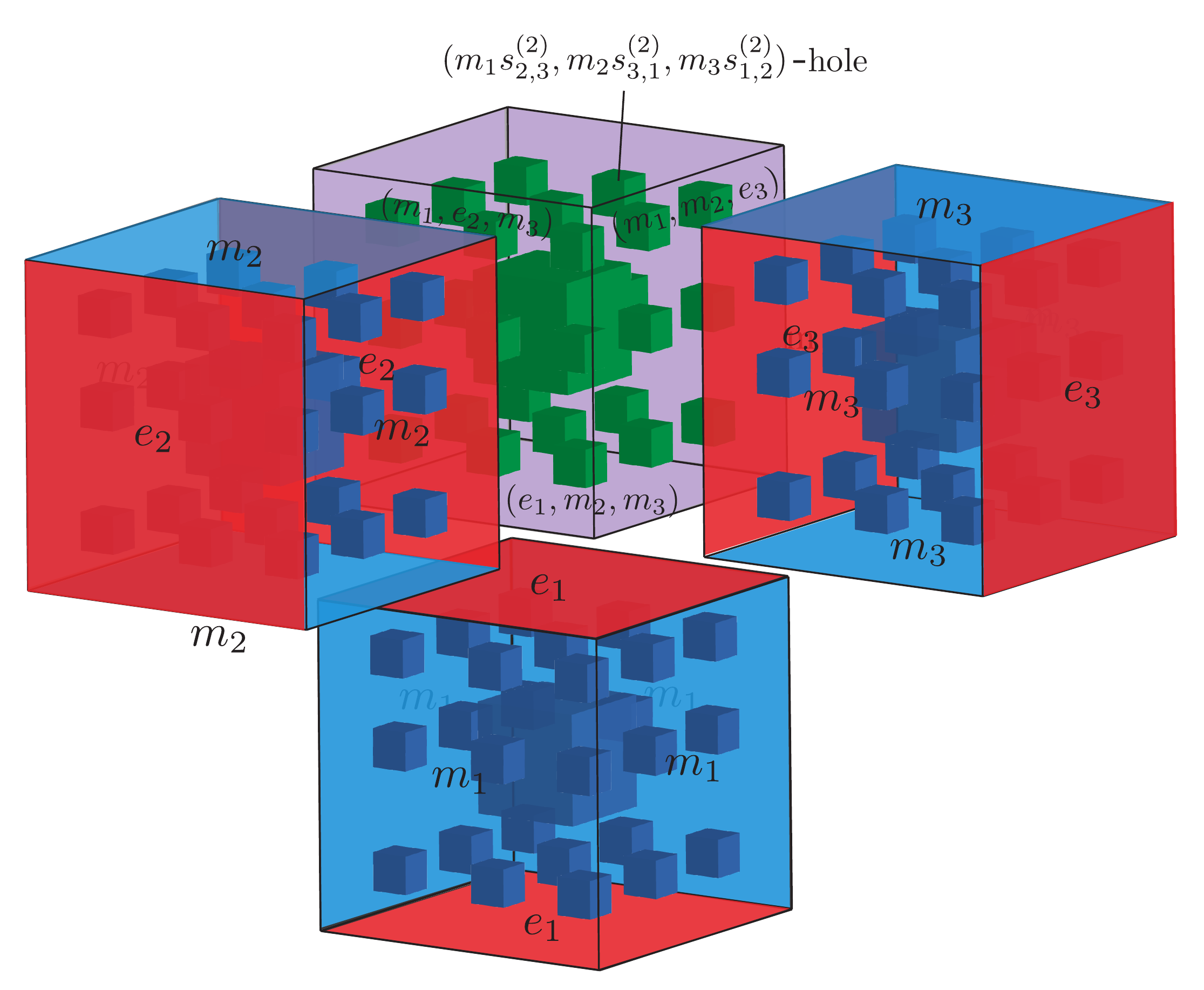}
   \caption{One performs lattice surgery between the new coupled stack fractal code with three ancilla fractal codes along three directions in order to transfer the three logical qubits encoded in the coupled stack fractal code into the three ancilla code blocks.}
   \label{fig:lattice_surgery_stack}
 \end{figure}

The two lattices are merged by introducing an extra set of physical ancilla qubits $\otimes_j \ket{j}$ placed between the two codes, see Fig.~\ref{fig:lattice_surgery}(a). The merging  process involves measuring the $X$~stabilizers between these two code copies.
Besides the stabilizers in the original codes, there two new types of stabilizers, the 4-body $Z$ stabilizers and 6-body $X$ stabilizers as shown in  Fig.~\ref{fig:lattice_surgery}(b).  The eigenvalues of the 4-body $Z$ stabilizers is the product of the eigenvalue of the original 3-body boundary $Z$ stabilizers ($+1$) and the ancilla $Z$ eigenvalue ($+1$), which remains $+1$ and are therefore immediately certain (in the absence of errors).  On the other hand, the newly introduced $X$ stabilizers have completely unknown eigenvalues.  When performing measurements, they will have a random distribution of $\pm 1$ eigenvalues. The $-1$ eigenvalues correspond to generation of $e$-particle excitations.  Measuring all the new interface  $X$ stabilizers equivalently gives the measurement outcome of the joint parity of the $X$-logical operators in these two codes, i.e., $\lo{X}_A \lo{X}_B$, which is a product of all the interface $X$ stabilizers as illustrated in the right panel of Fig.~\ref{fig:lattice_surgery}(b). The parity $\lo{X}_A \lo{X}_B= 1 (-1)$ corresponds to even (odd) number of $e$-particle excitations at the interface. We must perform $d$~rounds of syndrome measurements to ensure the fault tolerance of the protocol in the presence of measurement errors~\footnote{Note that the $d$~rounds of measurement to insure the fault tolerance of the preparation of the ancilla~$\lo{\ket{0}}_B$ and the $d$~rounds required for lattice merging can actually by performed in parallel in a single set of $d$~rounds.}. After~$d$ rounds, one can fault-tolerantly decode and correct the errors and hence clean up  all the $e$-particle excitations at the interface.
Conditioned on the measurement outcome of the joint parity $\lo{X}_A \lo{X}_B$, the merging map can be expressed as follows:
\begin{align}
\nonumber \quad \hat{M}_+ =& \lo{\ket{+}}_M \ \lo{\bra{++}}_{AB} + \lo{\ket{-}}_M \ \lo{\bra{--}}_{AB}, \ \text{if } \lo{X}_A\lo{X}_B=+1 \\
\hat{M}_- =& \lo{\ket{+}}_M \ \lo{\bra{+-}}_{AB} + \lo{\ket{-}}_M \ \lo{\bra{-+}}_{AB}, \  \text{if } \lo{X}_A\lo{X}_B=-1,
\end{align}
where $\lo{\ket{\psi}}_M$ represents the  logical qubit state of the merged code as illustrated in Fig.~\ref{fig:lattice_surgery}(c). 
The logical state is then transformed by the lattice merging map as
\begin{align}
\quad \hat{M}_{\pm}: \ &  (\alpha \lo{\ket{+}}_A + \beta \lo{\ket{-}}_A)\lo{\ket{0}}_B \rightarrow \alpha \lo{\ket{+}}_M + \beta \lo{\ket{-}}_M. 
\end{align}
The merged state is independent of the parity eigenvalue $\lo{X}_A\lo{X}_B=\pm 1$ because of the specific choice of state  $\lo{\ket{0}}_B$ in the ancilla code block.
As we can see from the above expression, the initial logical information $\lo{\ket{\psi}}_A$ is transferred to the merged code block as $\lo{\ket{\psi}}_M$.
 
 The other two copies in the coupled stack code are merged with the other two independent ancilla code blocks along the corresponding $e$-boundaries in the other two directions as illustrated in  Fig.~\ref{fig:lattice_surgery_stack}.

\item 
\textbf{Lattice splitting along $e$-boundaries:}  The next step is to split the merged code again as illustrated in Fig.~\ref{fig:lattice_surgery}(d).  This is achieved by projective measurements of the qubits (white) introduced at the interface in the $Z$ basis.  The measurement results will be a random distribution of $\pm 1$, and hence the 3-body boundary stabilizers will have the corresponding random $\pm 1$ eigenvalues, which corresponds to $m$-loop excitations at the interface. One could also do $d$ rounds of measurement to clean up these excitations, but in this case a single-shot measurement and decoding suffice.  This is because 3D toric code can correct $X$-errors ($m$-loop   excitations) in a single shot~\cite{Bombin:2015hia}.  The details of the single-shot error correction property of the fractal code will be discussed in an upcoming paper~\footnote{Manuscript in preparation.}.

The lattice splitting map can be expressed as:
\begin{align}
\nonumber \quad \hat{S} =& \lo{\ket{++}}_{AB} \ \lo{\bra{+}}_M + \lo{\ket{--}}_{AB} \ \lo{\bra{-}}_M.
\end{align}
After the lattice splitting, the merged logical state is mapped again into a logical state distributed over two codes:
\begin{align}
\nonumber & \hat{S}: \ \alpha\lo{\ket{+}}_M + \beta \lo{\ket{-}}_M  \rightarrow  \alpha\lo{\ket{++}}   + \beta \lo{\ket{--}} \\
= & \frac{1}{\sqrt{2}}\ket{\lo{0}}_A (\alpha\lo{\ket{+}}_B + \beta \lo{\ket{-}}_B) + \frac{1}{\sqrt{2}}\ket{\lo{1}}_A \left(\alpha\lo{\ket{+}}_B - \beta \lo{\ket{-}}_B\right),
\end{align}
where we have represented the logical states in the $X$-basis.

\item
\textbf{Measure out the original code in the $Z$ basis:}. The final step is to measure out the logical qubit information of the original stack code in the $Z$ basis.  In the illustrated example, copy 1 of the original code is measured out, and the logical state of the new code hence becomes 
\be
\alpha\lo{\ket{+}}_B + (-1)^{M_z}\beta \lo{\ket{-}}_B \equiv (\overline{Z}_B)^{M_z} \lo{\ket{\psi}}_B,
\ee
where $M_z$ is the measurement result of copy 1 in the original stack code. In the convention here, $M_z=0$ and~$1$ corresponds to $\lo{\ket{0}}_A$ and $\lo{\ket{1}}_A$, respectively.  Therefore, after the measurement, one can just apply a logical-$Z$ correction $\lo{Z}_B$ to the new code  conditioned on the measurement value $M_z = 1$ \footnote{This correction can always be applied in the software without an actual physical operation.}.  We can see the logical information in the first copy  $\lo{\ket{\psi}}_A$ is successfully transferred to the ancilla copy as $\lo{\ket{\psi}}_B$.   Similarly, the logical information in the other two copies in the coupled stack code is transferred to the other two ancilla code blocks, as illustrated in Fig.~\ref{fig:lattice_surgery_stack}.   Since the three ancilla copies are the same as the original three copies of the uncoupled 3D fractal codes, we have successfully mapped the new code space $\tilde{\H}$ to the original code space $\H$, with a logical CCZ gate being applied.

\end{enumerate}

\begin{figure}[hbt]
  \includegraphics[width=0.4\columnwidth]{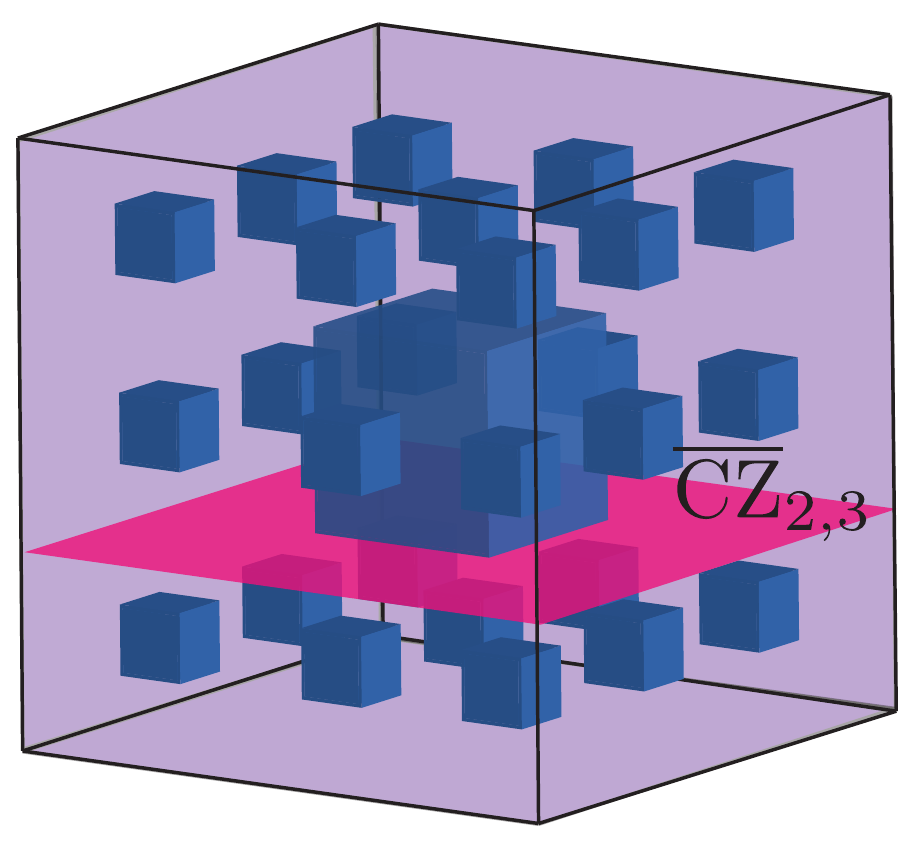}
  \caption{A transversal logical gate $\lo{\text{CZ}}_{2,3}$ between copy $2$ and $3$ in the stack code is acted on a brane perpendicular to the $z$-direction avoiding all the  holes. Similarly, $\lo{\text{CZ}}_{1,2}$ and $\lo{\text{CZ}}_{1,3}$ can be applied on a brane perpendicular to the $y$- and $x$-directions respectively. }
  \label{fig:CZ_logical_gate}
\end{figure}

Finally, we point out that a logical CZ gate between any two copies of 3D fractal surface codes in the stack code can be implemented instantaneously as a traversal CZ gate, without the need of subsequent lattice surgery.  This is because the transversal CZ gate is a 1-form symmetry [see Eq.~\eqref{eq:CZ_sheet_bending}] and can hence be chosen to acted on a brane avoiding all the $(m_1,m_2,m_3)$-holes as illustrated in Fig.~\ref{fig:CZ_logical_gate}, which is surprisingly convenient.

% \begin{figure*}[hbt]
%   \includegraphics[width=1.5\columnwidth]{color_code.pdf}
%   \caption{}
%   \label{fig:color_code}
% \end{figure*}

% \begin{figure}[hbt]
%   \includegraphics[width=1\columnwidth]{color_code_3D.pdf}
%   \caption{}
%   \label{fig:color_code_3D}
% \end{figure}

\subsection{Non-Clifford logical gates for $n$-dimensional fractal codes}\label{sec:logical_gate_nD}

In this subsection, we extend the scheme of realizing non-Clifford logical gates to $n$-dimensional topological codes supported on a simple fractal.  In particular, we consider the case of
$(1, n-1)$-toric codes with particle-like excitations and string-like logical operators, and the case of $(i, n-i)$-toric codes ($i \ge 2$) without particle-like excitations and string-like logical operators.

\subsubsection{Cases with string-like logical operators $(i=1)$}
We begin by generalizing our construction of logical CCZ gate in 3D fractals to logical $\text{C}^{\otimes n-1} \text{Z}$ gate in a stack of $n$ copies of $(1, n-1)$-surface codes supported on $n$-dimensional simple fractals.   We follow the treatment in Ref.~\cite{Kubica:2015br}, where a logical $\text{C}^{\otimes n-1} \text{Z}$ gate corresponding to a local constant-depth circuit is constructed by applying a logical $R_n$ gate in a $(1, n-1)$-color code and then disentangling the color code into $n$ copies of $(1, n-1)$-surface codes.  In the following, we show that such a construction can also be adapted to the case of fractal surface codes.

We start by introducing the construction of $n$-dimensional color codes.  Since any   $n$-dimensional color code belong to the same phase as multiple copies of $n$-dimensional $\Z_2$-toric codes, i.e., $\Z_2^{\times n}$ as stated in Sec.~\ref{sec:domain_wall_picture}, one can also classify color codes with the bi-label $(i,n-i)$, meaning the $Z$ and $X$~logical operators correspond to $i$- and $(n-i)$-dimensional branes, respectively, where we have $1 \le i \le n-i$. Therefore there are $\ceil{\frac{n-1}{2}}$ ways of defining an $n$-dimensional color code, similar to the case of defining an $n$-dimensional toric code. We consider an $n$-dimensional lattice 
$\L$, which is a cell complex and forms a tessellation of an $n$-dimensional manifold. An additional constraint in the case of color code is that $\L$ needs to be ($n+1$)-valent and its $n$-cells are ($n+1$)-colorable.  We always put qubits on the vertices, i.e., 0-cell, as opposed to the toric code case where qubits are put on a $i$-cell.  The $X$- and $Z$ stabilizers are associated with  $(n+1-i)$-cells and $(i+1)$-cells respectively.   

In this subsection, we only focus on the $i=1$ case such that the corresponding $(1,n-1)$-color codes contain string logical operators and particle excitations.  
In this case, the $X$ stabilizers $A_q$ and $Z$ stabilizers $B_f$ are associated with  $n$-cells (volumes) labeled by $q$ and $2$-cells  (faces) labeled by $f$ respectively.  The color code model can be defined with the following parent Hamiltonian:
\begin{align}\label{eq:color_code}
\nonumber H_\text{CC}=&-\sum_q A_q - \sum_f B_f, \\
\text{with} \quad  A_q=&\Motimes_{j\in \{v_q\}  }X_j, \quad B_f=\Motimes_{j\in \{v_f\}}Z_j,
\end{align}
where $\{v_q\}$ represent all vertices in the $n$-cell labeled by $q$, and $\{v_f\}$ represent all vertices in face $f$. We denote the stabilizer group of color code by $CC(\L) \equiv \langle \{A_q, B_p\} \rangle$.  Similarly, we denote the stabilizer toric code on lattice $\L$ as $TC(\L)$.

\begin{figure*}[hbt]
  \includegraphics[width=1.5\columnwidth]{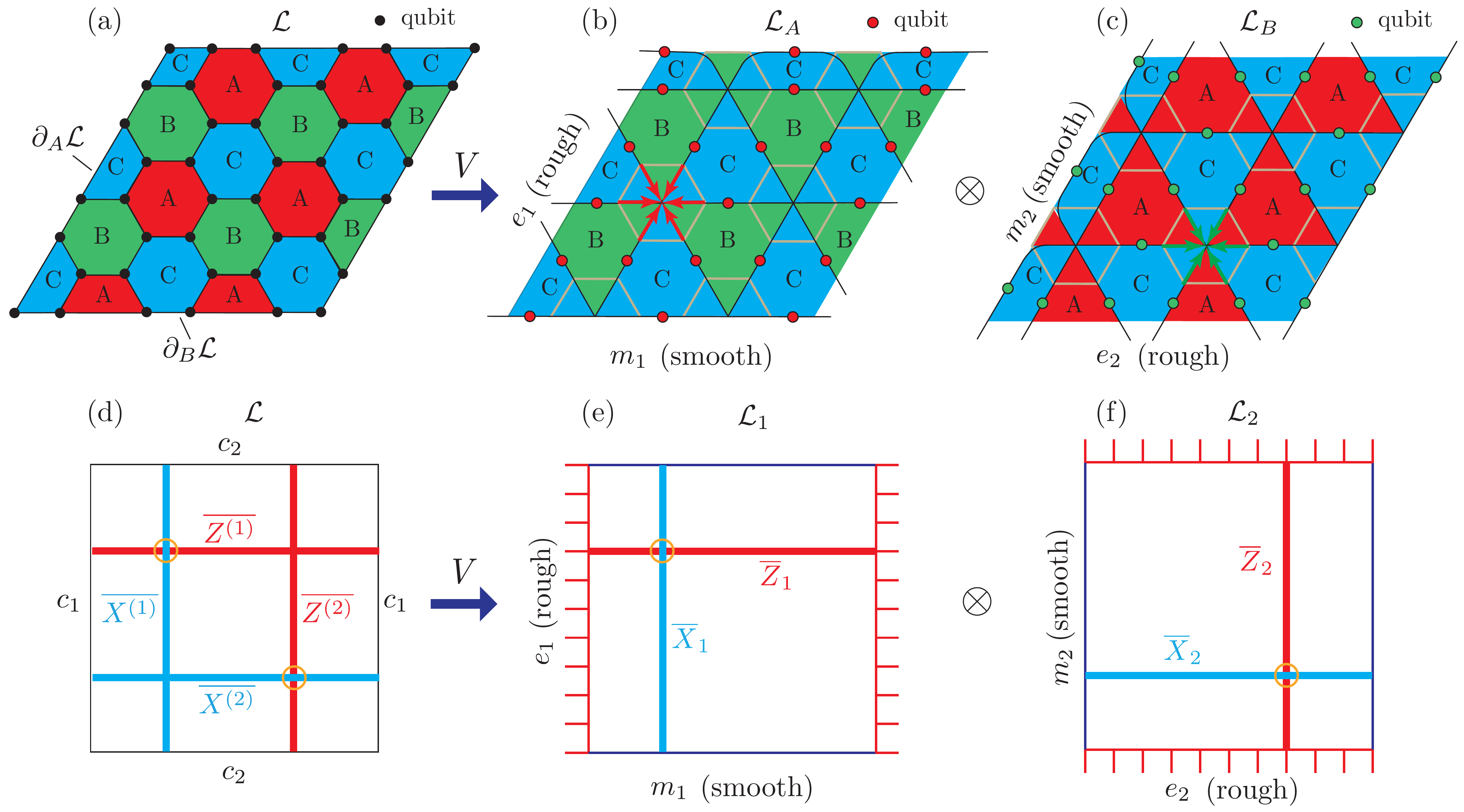}
  \caption{(a-c) Disentangling a 2D color code defined on a hexagonal lattice (qubits located on the vertices) into a tensor product of two surface codes defined on triangular lattices (qubits located on the edges) by shrinking the color $A$ faces and color $B$ faces into a single vertex respectively. The color-code boundary with color $A$ ($B$) is turned into a rough (smooth) boundary condensing $e$-anyons ($m$-anyons) in the first (second) copy of surface code. (d-f) The effective TQFT picture of the lattice model in (a-c), i.e., disentangling a square-patch color code into two square patch surface codes. Two pairs of logical operators in the color code are mapped into the corresponding pair of logical operators within each surface code.}
  \label{fig:color_code_boundary}
\end{figure*}

\begin{figure*}[hbt]
  \includegraphics[width=1.4\columnwidth]{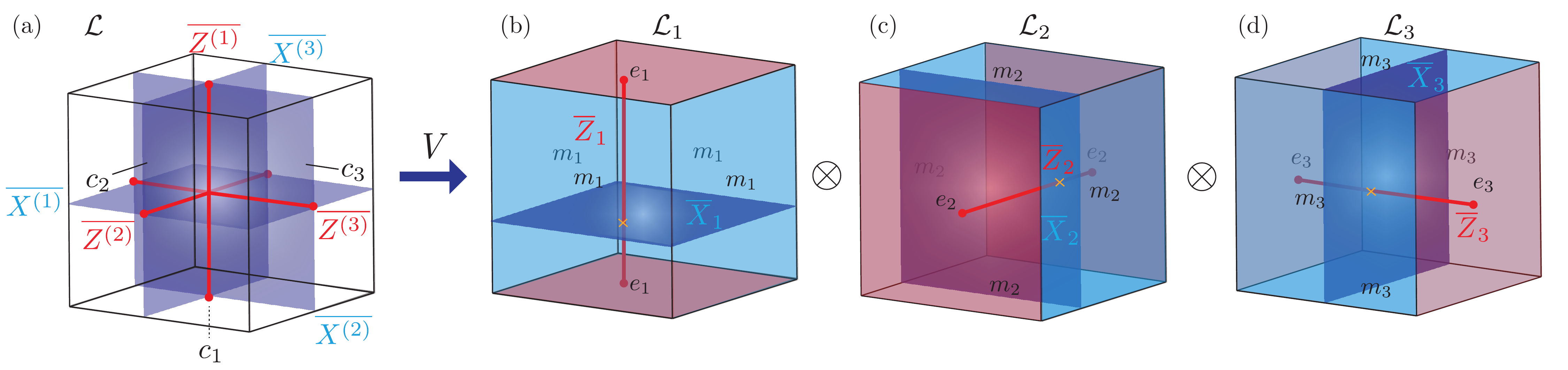}
  \caption{Disentangling a single 3D color code into three copies of 3D surface codes aligned in perpendicular directions. The correspondence of the logical operators in the color code and the three surface codes is shown. }
  \label{fig:color_code_3D_boundary}
\end{figure*}

The equivalence between a color code and multiple copy of toric codes has been established in  Ref.~\cite{Kubica:2015br}, which can be summarized by the following lemma:
\begin{lemma}[\cite{Kubica:2015br}]\label{lemma:nDCC}
Let CC$(\L)$ be a $(1,n-1)$-color code defined on an $n$-dimensional lattice $\L$ without boundaries ($n \ge 2$). The lattice $\L$ is $(n+1)$-valent and colored with $c_0, c_1, \dots, c_n$. Let $X$- and $Z$ stabilizers be supported on $n$-cells and $2$-cells respectively.   There exists a local Clifford unitary $V$ such that
\be
V[CC(\L)\otimes \SS]V^\dag = \Motimes_{i=1}^n TC(\L_i),
\ee
where $\SS$  represents  the stabilizer groups of decoupled ancilla qubits and  $TC(\L_i)$ represents the stabilizer group of the toric code defined on the shrunk lattice $\mathcal{L}_i$ obtained from $\L$ by shrinking $n$-cells of color $c_i$.  The local unitary $V$ can be chosen to be of the form
\be
V = \Motimes_{c \in \mathbf{C}_0} V_c,
\ee
where $\mathbf{C}_0$ is the set of $n$-cells of color $c_0$ in $\L$, and $V_c$ is a Clifford unitary acting only on qubits on vertices of the $n$-cell $c$.
\end{lemma}
We also call $V$ a disentangling unitary, since  a single $n$-dimensional color code $CC(\L)$ can disentangled into $n$ copies of toric code $TC(\L_i)$ by $V$. The qubits in the color code are all placed on the vertices while the qubits in the toric codes are all placed on the edges. Therefore, for each $n$-cell $c$ colored in $c_0$, one needs to add $\mathsf{E}-\mathsf{V}$ ancilla qubits, where $\mathsf{E}$ and $\mathsf{V}$ stand for the number of edges and vertices in $c$. Due to the fact that lattice $\L$ is ($n+1$)-valent, one has $\mathsf{E}= n\mathsf{V}/2$ and $\mathsf{E}-\mathsf{V} \ge 0$ for $n \ge 0$. One can see that, in the 2D case ($n=2$), one has $\mathsf{E}-\mathsf{V} =0$ and so no ancilla qubits are needed.  On the other hand, for three and higher dimensions, one needs to introduce $\mathsf{E}-\mathsf{V} > 0$ ancilla qubits to disentangle the $n$ copies of toric codes. 

% We first illustrate this disentangling procedure in three dimension, as shown in Fig.~\ref{fig:color_code_3D}. The 3D color-code lattice $\L$ has volumes (3-cells) with four colors: $A$, $B$, $C$ and $D$. We can also assign colors to faces (2-cells) with the colors of the two volumes which it belongs to, such as $AD, BD$ and $CD$ as illustrated in Fig.~\ref{fig:color_code_3D}(a).   Similarly, one can assign colors to each edge with the colors of the three volumes it belongs to. We can obtain the three shrunk lattices $\L_A$, $\L_B$, and $L_C$ by shrinking volumes with color $A$, $B$ and $C$ respectively.   In the example in Fig.~\ref{fig:color_code_3D}(a), we show all the faces of a volume (3-cell) with color $D$. We then shrink each volume with color A into a single vertex, and obtain the toric code lattice $\L_A$ in Fig.~\ref{fig:color_code_3D}(b),  where qubits now reside on the edges. As we can see, all the faces belonging to cell are also shrunken, including the face with color $AD$ illustrated in Fig.~\ref{fig:color_code_3D}.  We further list the details of lattice $\L_A$ as follows.  (1) Vertices: centers of $A$-volumes in $\L$; (2) Edges: $BCD$-edges in $\L$; (3) Faces: $BC$-, $BD$- and $CD$-faces in $\L$; (4) Volumes:  $B$-, $C$-, and $D$-volumes in $\L$.

% Now we consider the general $n$-dimensional case. 

The $n+1$ colors of the $n$-cells in the color code lattice $\L$ are labeled as $c_0, c_1, ..., c_n$. One can obtain the toric-code lattice $\L_i$ ($i=1,2, ..., n$) by shrinking $n$-cells with color $c_i$ in $\L$.  For a $k$-cell ($0 \le k \le n$), we can associate it with $n-k+1$ colors corresponding to the $n$-cells it belongs to. In this case, the toric-code lattice $\L_i$ has the structure detailed as follows.  (1)~Vertices: centers of $n$-cells of color $c_i$ in $\L$; (2)~Edges: edges of color $\{c_0, c_1, ..., c_n\} \backslash  \{c_i\}$ in $\L$; (3)~Faces:  faces in $\L$ of color $\{c_0, c_1, ..., c_n\} \backslash  \{c_i, c_j\}$ for all $j \neq i$.  The above lattice information is enough to define the $n$-dimensional toric code by associating qubits with edges, $X$~stabilizers with vertices, and $Z$~stabilizers with faces.  

Up to now, we have discussed how to disentangle a translationally invariant $n$-dimensional color code into $n$~copies of toric codes. Now we investigate the case with boundaries. We say each $n$-dimensional color code boundary $\partial_i \L$ has color $c_i$ if all $n$-cells adjacent to $\partial_i \L$ do not contain color $c_i$. We show an illustration of 2D color code with boundaries in Fig.~\ref{fig:color_code_boundary}(a).  In this example, the 2D color code has two types of boundaries, $\partial \L_A$ and $\partial \L_B$ (can also be labeled as $\partial \L_1$ and $\partial \L_2$) with color $c_1=A$ and $c_2=B$ respectively, and is hence identical to the abstract picture of a square-patch color code with two types of boundaries shown in Fig.~\ref{fig:color_code_boundary}(d). As one can see, the faces (2-cells) adjacent to the boundary $\partial_A \L$ ($\partial_B \L$) only has color B and C (A and C), but does not contain color A (B).  

We first discuss the disentangling unitary $V$ acting on this square-patch 2D color code with boundaries  illustrated in Fig.~\ref{fig:color_code_boundary}.   It has been shown in Ref.~\cite{Kubica:2015br} that there exists a local unitary $V$ to  disentangle the square-patch color code with boundary $CC(\L)$ into two independent copies of surface codes (square-patch toric codes with two types of boundaries), i.e., $SF(\L_A) \otimes SF(\L_B)$, as shown in Fig.~\ref{fig:color_code_boundary}. In particular, $SF(\L_A)$ ($SF(\L_B)$) is obtained by shrinking the faces with color $A$ ($B$). We denote the anyon excitations in $SF(\L_A)$ and $SF(\L_B)$ as $\{e_1, m_1\}$ and $\{e_2, m_2\}$ respectively. The corresponding two pair of logical operators are denoted by $\overline{Z}_{1,2}$ and $\overline{X}_{1,2}$.
As can be seen in Fig.~\ref{fig:color_code_boundary}(b-f), $SF(\L_A)$ ($SF(\L_B)$) has rough boundaries condensing $e_1$ ($m_2$) on the left/right sides, and smooth boundaries condensing $m_1$ ($e_1$) on the upper/lower sides. The logical strings $\overline{Z_{1,2}}$ connect the rough boundaries, while the logical strings $\overline{X_{1,2}}$ connect the smooth  boundaries. We also denote the corresponding logical operators in the original color code as $\overline{Z^{(1,2)}}$ and $\overline{X^{(1,2)}}$.   Since the 2D color code $CC(\L)$ is equivalent to a stack of two surface codes up to a local unitary, one can also label the color code boundary with the anyon condensation of the two surface code layers, i.e., $\partial \L_A \equiv (e_1, m_2)$ and $\partial \L_B \equiv (m_1, e_2)$.

More generally, one considers an $n$-dimensional $(1,n-1)$-color code on an $n$-dimensional hypercubic lattice $\L$ with hyperfaces $\partial\L_i$ perpendicular to the direction $\hat{i}$ ($i=1,2, ..., n$) colored in $c_i$.  Such an $n$-dimensional color code $CC(\L_i)$ can be disentangled by a local unitary $V$ with additional ancilla qubits into $n$ independent copies of $n$-dimensional $(1,n-1)$-surface codes, i.e., $\Motimes_{i=1}^n SF(\L_i)$.  An illustration for the 3D case is shown in Fig.~\ref{fig:color_code_3D_boundary}. In particular, the lattice of  the $i^\text{th}$ copy of surface code $\L_i$ is obtained by shrinking all the cells colored in $c_i$ in the original color code lattice $\L$.   Now the particle and string excitations in the $i^\text{th}$ copy of surface code $SF(\L_i)$ are labeled as $e_i$ and $m_i$ respectively, with the corresponding logical operators denoted by $\overline{Z}_i$ and $\overline{X}_i$. The surface code $SF(\L_i)$ has two rough boundaries condensing $e_i$ particles, which is perpendicular to direction $\hat{i}$.  The logical string $\overline{Z}_i$ is along the direction $\hat{i}$ and connecting the two rough boundaries.  The remaining $n-2$ hypersurfaces are smooth boundaries condensing $m_i$ strings.  The logical brane $\overline{X}_i$ is parallel to the rough boundaries and connecting all the smooth boundaries. We also denote the corresponding logical operators in the original color code $CC(\L)$ as $\overline{Z^{(i)}}$ and $\overline{X^{(i)}}$, which are along the same directions as $\overline{Z}_i$ and $\overline{X}_i$ respectively. Since the color code is equivalent to a stack of surface codes up to a local unitary, we can also label the $2n$ color code boundaries by the condensation of excitations in the surface codes,  i.e., for the $i^\text{th}$ and $(i+n)^\text{th}$ boundaries ($i=1,2, ..., n$),   one has $\partial_i \L = \partial_{i+n} \L = (m_1, m_2, ..., m_{i-1}, e_i, m_{i+1}, ..., m_n)$. 

Now we review the result from Ref.~\cite{Kubica:2015br} that a transversal $R_n$ gate in an $n$-dimensional hypercubic color code is equivalent to logical $\text{C}^{\otimes n-1}\text{Z}$ up to a local unitary.  The generalized phase gate can be defined as $R_n = \text{diag}(1,e^{2\pi i/2^n})$, with the more familiar special cases being $R_2= S$ and $R_3= T$, i.e., the phase (S) gate and the T gate. The $n$-dimensional color code lattice $\L$ is an $(n+1)$-valent and an $(n+1)$-colorable lattice. It is known that the corresponding graph $G=(\mathcal{V}, \mathcal{E})$ containing the vertices and edges of $\L$ is a bipartite graph.  Therefore, the vertices of $\L$ can be divided into two groups, $\mathcal{V}^a$ and $\mathcal{V}^b$, i.e.,  $\mathcal{V}=\mathcal{V}^a \cup  \mathcal{V}^b$.  In particular, vertices in $\mathcal{V}^a$ are only adjacent to vertices in $\mathcal{V}^b$ and vice versa. We then define the transversal $R_n$ gate as:
\be\label{eq:defineR_n}
\widetilde{R}_n = \Motimes_{j\in \mathcal{V}^a} R_n(j) \Motimes_{j\in \mathcal{V}^b} R^{-1}_n(j), 
\ee

We now start addressing what logical gate the  transversal gate $\widetilde{R}_n$ corresponds to in the hypercubic color code and the corresponding stack of surface codes after applying the local disentangling unitary $V$.

Here, we directly consider the general case in $n$~dimensions, while leaving the detailed illustration for the 2D case in App.~\ref{app:color_codes}.  Note that both $R_n$ gate and $\text{C}^{\otimes n-1}\text{Z}$ belong to the $n^\text{th}$ level of Clifford hierarchy, but are outside the $(n-1)^\text{th}$ level of Clifford hierarchy, which is reflected in the following relations:
\begin{align}
\nonumber &K[R_k, X] = e^{-2\pi \textsf{i}/2^k} R_{k-1}, \\
& K[\text{C}^{\otimes k-1}\text{Z}, X\otimes I^{\otimes k-1}] = I \otimes C^{\otimes k-2}\text{Z}.
\end{align}
for $k \ge 2$ and we have used the group commutator $K[A,B]$$=$$ABA^\dag B^\dag$.  

We have defined the transversal $R_n$ operators in Eq.~\eqref{eq:defineR_n}, while the rest of the  transversal operators $\widetilde{R}_p$ ($p=n-1, n-2, ..., 1$) can be define the  recursively:
\begin{align}\label{eq:recursive_R}
\nonumber & \widetilde{R}_{n-1} = K[\widetilde{R}_{n}, \overline{X^{(1)}}],     \\
\nonumber & \widetilde{R}_{n-2} = K[\widetilde{R}_{n-1}, \overline{X^{(2)}}],  \\
\nonumber  & \qquad \cdots   \\
& \widetilde{R}_{1} = K[\widetilde{R}_2, \overline{X^{(d-1)}}] = \overline{Z^{(n)}}.
\end{align}
The disentangling unitary $V$ maps the logical operator in the color code to those in the stack of $n$ surface codes:
\be
V: \overline{X^{(p)}} \rightarrow V\overline{X^{(p)}}V^\dag = \overline{X}_p, \quad \overline{Z^{(p)}} \rightarrow V\overline{Z^{(p)}}V^\dag=\overline{Z}_p.
\ee
By conjugating Eq.~\eqref{eq:recursive_R} with the disentangling unitary $V$, we hence obtain the following recursive relations:
\begin{align}\label{eq:recursive_CZ}
\nonumber & I \otimes  \overline{\text{C}^{\otimes n-2}\text{Z}} = K[I \otimes \overline{\text{C}^{\otimes n-1}\text{Z}}, \overline{X}_1 ],     \\
\nonumber & I^{\otimes 2} \otimes  \overline{\text{C}^{\otimes n-3}\text{Z}} = K[I \otimes \overline{\text{C}^{\otimes n-2}\text{Z}}, \overline{X}_2 ],  \\
\nonumber  & \qquad \cdots   \\
& I^{\otimes n-1} \otimes  \overline{Z}_n = K[ \overline{\text{CZ}}, \overline{X}_{n-1} ],
\end{align}
where we have established the following relation recursively (from bottom to top):
\be
V \widetilde{R}_p V^\dag = \overline{\text{C}^{\otimes p-1} \text{Z}}.
\ee
In particular, we can have the logical gate in the $n^\text{th}$~level of Clifford hierarchy:  $V \widetilde{R}_n V^\dag = \overline{\text{C}^{\otimes n-1} \text{Z}}$.  
% Note that, same as the 2D case, $V \widetilde{R}_p V^\dag$ is a local constant-depth circuit.  
Although $\widetilde{R}_p$ is a transversal gate in the color code, $V\widetilde{R}_p V^\dag$ is not guaranteed to be a transversal gate  in the stack of surface codes.  Still, it is clear that $V\widetilde{R}_p V^\dag$ is a local constant-depth  circuit applied on the stack of $p$ copies of  surface codes which implements the logical $\overline{\text{C}^{\otimes p-1} \text{Z}}$ gate.  

One can understand the process of implementing $V\widetilde{R}_p V^\dag$ as follows:  one first applies the inverse of the disentangling unitary, i.e., $V^\dag$, to entangle the stack of $n$ surface codes $\Motimes_{i=1}^n SF(\L_i)$  into a single copy of $n$D color code $CC(\L)$. Now one applies $\widetilde{R}_p$,  corresponding to a logical gate $\overline{\text{C}^{\otimes p-1}\text{Z}}$ in the code space of the color code.   One then applies the disentangling unitary $V$ again which disentangles the color code back to a stack of $n$ surface codes and maps $\widetilde{R}_p$ into the logical gate $\overline{\text{C}^{\otimes p-1} \text{Z}}$ in the code space of the stack code $\Motimes_{i=1}^n SF(\L_i)$.   
 
Now we consider a stack code consisting of $n$ copies of  $n$-dimensional fractal surface codes, defined on an $n$-dimensional simple fractal with $m$-holes in each copy. These fractal surface codes are  constructed in the following way: (1) we start from the surface code $SF(\L_i)$ ($i=1, 2, ..., n$) obtained from disentangling the $n$-dimensional color code via the local unitary $V$. (2) We then create $m$-holes via removing a subset of qubits and putting smooth ($m$) boundaries around these holes, where only $X$ stabilizers near the boundary differ from the X~stabilizers in the bulk while the $Z$ stabilizers near the boundary are all the same as the bulk stabilizers. 

One can label the hole boundary in the stack code as $\mathcal{B}^\text{h.b.}$$=$$(m_1, m_2, ..., m_n)$. According to our discussion in Sec.~\ref{sec:domain_wall_picture},  $V \widetilde{R}_p V^\dag$ is a topological symmetry which preserves the bulk Hamiltonian or stabilizers up to a logical identity, i.e., 
\be
V \widetilde{R}_p V^\dag : H^\text{bulk} \rightarrow H^\text{bulk}  \Longleftrightarrow  V \widetilde{R}_p V^\dag : S_i^\text{bulk} \rightarrow S_i^\text{bulk}.
\ee
Note that the invariance under the  above mapping refers to the Hamiltonians and stabilizers  projected to the code space $\H_C$. In particular, $V \widetilde{R}_n V^\dag$ corresponds to a global symmetry, and $V \widetilde{R}_p V^\dag$ represents a $(n-p)$-form symmetry ($p$$ \le $$n$), i.e., acting on a codimension-$p$ sub-manifold.  In other words, $V \widetilde{R}_p V^\dag$ is a $p$-dimensional membrane operator. Note that the 0-form symmetry ($p$$=$$n$) is just the global symmetry.   Since the outer boundaries $\mathcal{B}_i^\text{o.b.}$$=$$ \mathcal{B}_{i+n}^\text{o.b.} $$=$$ (m_1, m_2, ..., m_{i-1}, e_i, m_{i+1}, ..., m_n)$ of this stack of fractal codes is the same as the stack of $n$-dimensional surface codes, they are also preserved under the action of this topological symmetry:
\be
V \widetilde{R}_p V^\dag : \mathcal{B}_i^\text{o.b.} \rightarrow \mathcal{B}_i^\text{o.b.}, \text{for} \  i=1,2, ..., 2n.
\ee
On the other hand, $V \widetilde{R}_p V^\dag$ does not preserve the hole boundaries $\mathcal{B}^\text{h.b.}$. 

Now we can again use the domain-wall picture in TQFT to understand the effect of $V \widetilde{R}_p V^\dag$.  Here, $V \widetilde{R}_p V^\dag$ corresponds the $s^{(p)}$-domain wall acting on $p$ copies of toric codes, which is an $n$-dimensional generalization of $s^{(3)}$ and $s^{(2)}$ introduced before and does the following mapping to the excitations:
\be
s^{(p)}_{1,2, ..., p}: m_i \rightarrow m_i s^{(p)}_{1,2, ..., i-1, i+1, ..., p}, \quad   e_i \rightarrow e_i. 
\ee
Therefore, applying $V \widetilde{R}_p V^\dag$ is equivalent to sweeping the $s^{(p)}$-domain wall across the system.  Similar to the 3D case, the $s^{(p)}$-domain walls condense on the outer boundaries, i.e.,
\be
s^{(p)} : \mathcal{B}_i^\text{o.b.} \rightarrow  \mathcal{B}_i^\text{o.b.}.
\ee
However, these domain walls do not condense on the hole boundaries.  

From now on, we focus on the protocol of implementing the logical gate in the highest Clifford hierarchy, i.e., $ \overline{\text{C}^{\otimes n-1}\text{Z}}$. We first apply the constant-depth circuit $V \widetilde{R}_n V^\dag$, and the corresponding domain wall $s^{(n)}$ will be swept across the system and attached on the hole boundaries producing new boundaries as follows:
\begin{align}
\nonumber & s^{(n)}_{1,2, .., n} : \  (m_1, m_2, ..., m_n) \\
 \rightarrow & (m_1 s^{(n-1)}_{2,3, ..., m_n}, m_2 s^{(n-1)}_{1,3, ..., m_n}, ..., m_n s^{(n-1)}_{1,2, ..., m_{n_1}})\equiv \tilde{\mathcal{B}}^\text{h.b.}.
\end{align}
In terms of the microscopic details, only the stabilizers which has $O(1)$ distance away from the hole boundary, denoted by $S^\text{h.b.}_i$ are mapped to the new stabilizers in the $\tilde{S}^\text{h.b.}_i$ new code. We have hence mapped the original code space $\H$ to the new code space $\tilde{\H}$, and at the same time applied a logical $\text{C}^{\otimes n-1}\text{Z}$ gate on the encoded logical state.  

Now to map the code space back to $\H$, we again apply the similar lattice surgery protocol as in the 3D case described in Sec.~\ref{sec:CCZ_protocol}.   In particular, we merge $n$ copies of ancilla fractal surface codes prepared in $\ket{\overline{0}}$, and perform rough lattice merging along $e$-boundaries in each copy in the stack code.  We then perform lattice splitting and measure out the original stack code in the same manner as the 3D case, which now  transfers the logical information to the ancilla stack code which still corresponds to the code space $\tilde{\H}$.  The whole protocol implements the logical gate $ \overline{\text{C}^{\otimes n-1}\text{Z}}$.

Finally, we can also implement the logical gates $\overline{\text{C}^{\otimes p}\text{Z}} $$\equiv$$ V \widetilde{R}_p V^\dag$ ($p \le n-2$) as a constant-depth circuit without the need of subsequent lattice surgery, similar to the CZ gate in the 3D case illustrated in Fig.~\ref{fig:CZ_logical_gate}.  This is because $\overline{\text{C}^{\otimes p}\text{Z}}$ is a $(n-p)$-form symmetry and can hence act on a codimensional-$p$ sub-manifold $\M^{n-p}$ and hence avoid all the holes in the bulk. 

\subsubsection{Cases without string-like logical operators ($i \ge 2$):  
self-correcting quantum memories} 

Now we consider the case of logical gates in $(i,n-i)$-fractal codes in $n$ dimensions for $i\ge 2$, including the situations of a stack of surface codes or a color code.  In this case, no particle excitation or string logical operator exists, and the quantum memory is expected to be self-correcting (needs rigorous proof in future).  

Now for any given transversal logical gate (or more generally constant depth circuit) $U$ in the case of the same type of topological code defined on a lattice which corresponds to the cellulation of an $n$-dimensional manifold (i.e.,  non-fractal),  one can construct a corresponding fractal code such that $U$ is also a logical gate in the following way.  First, $U$ is a topological symmetry which keeps the bulk Hamiltonian or stabilizers  invariant, i.e., $U: H^\text{bulk} \rightarrow H^\text{bulk}$. Furthermore, $U$ also preserves all the outer boundaries of the $n$-dimensional topological code, i.e., $U: \mathcal{B}^\text{o.b}_j \rightarrow \mathcal{B}^\text{o.b}_j$.   Now when we introduce holes in the simple fractal structure, we also introduce new types of hole boundaries $\mathcal{B}^\text{h.b.}_j$.  Now in contrast to the case of $i=1$, where only $m$-holes are allowed in each copy of toric code, i.e., $(m_1, m_2, ..., m_n)$, we can assign any type of boundary $\mathcal{B}^\text{h.b}_j$ to each hole as stated in Sec.~\ref{sec:general_k}.  Therefore, in order for these boundaries to be preserved by the logical gate $U$, we can just choose any hole boundary to be the same as any of the outer boundary, i.e., $\mathcal{B}^\text{h.b.}_j =\mathcal{B}^\text{o.b.}_j$ for $\forall j$. In this way, we always have $U: \mathcal{B}^\text{h.b.}_j \rightarrow \mathcal{B}^\text{h.b.}_j$.  

 Thus, we have shown that $U$ is indeed a logical gate on the fractal code with Hausdorff dimension $n-1+\epsilon$, and perform the same logical operation as the corresponding topological codes defined on $n$-dimensional manifolds.

\section{Discussion and Outlook}

Throughout this paper, we have obtained several surprising results.  First of all, it is remarkable that one can still describe topological orders on the fractal geometries using a TQFT picture, especially with the condensation properties of the gapped boundaries.  Although a fractal geometry is nowhere differentiable and hence significantly different from a usual continuum model,  one can still start from a continuum model defined on a manifold and start punching holes on it at all length scales. The boundary properties of the holes becomes extremely important since the hole boundaries are present everywhere in the system. The survival of the topological order is hence equivalent to the preservation of macroscopic code distance, which is in turn determined by the interplay of the condensation types of the hole boundaries and the corresponding systolic geometry. The key mathematical objects considered here are the relative systoles \cite{Babenko_2002} in the presence of boundaries and the corresponding relative homology theory. In particular, the condensation properties of gapped boundaries in an $n$D Abelian topological order can be captured by relative homologies. 

From the condensed matter physics perspective, our results predict the existence of $\ZZ_N$ spin liquids  and topological order supported on a fractal-like structure embedded in 3D. This could be experimentally realized in porous materials supporting quantum spin liquids \cite{savary2017}, or in a synthetic quantum many-body system such as the Rydberg-atom array where a 2D quantum spin liquid has already been observed recently \cite{Lukin:2021_spin_liquid}.

As mentioned in the introduction, fabrication errors forming islands of corrupted or unusable qubits at all length scales and is hence similar to the situation of a fractal lattice. For the practical purpose of scaling up active error correction in a qubit architecture, such errors remain a significant engineering challenge in the lithography process of making solid-state qubit chips, e.g., in the context of superconducting or semiconducting qubit technology~\cite{Kreikebaum:2020_fabrication, Hertzberg:2020laser, zhang:2020fabrication}. This problem is particularly urgent due to the issues of frequency collision and cross-talks errors between qubits, which makes the yield (success rate) for fabricating a chip with zero-collision (equivalently no corrupted region) exponentially decrease  with the system size~\cite{Chamberland:2020heavyhex, Hertzberg:2020laser}, a daunting challenge for scaling up error correction. On the theory side, it has been realized that the error threshold of 2D surface code decays exponentially when increasing the fabrication errors and vanishes far below the percolation threshold~\cite{Auger_fabrication_error_2017}, which is consistent with our result that $\ZZ_N$ topological order and macroscopic code distance is absent in a setup with holes at all length scales, as can be seen from Fig.~\ref{fig:no_go_2D}.  Therefore,  a natural question arises: whether intrinsic stability against large fabrication errors is possible in certain topological codes, in particular, in the extreme condition near the percolation transition.

In this regard, our result suggests that a 3D architecture is significantly more robust against fabrication errors compared to a 2D architecture.  Given a 3D qubit lattice with clusters of corrupted or unusable regions at all length scales due to various sources of fabrication errors including the dominant problem of frequency collisions, one can simply ignore the qubits within such clusters and  implement an error correcting code with the boundary on these clusters being a (smooth) $m$-boundary, that is a boundary that condenses only loop-like excitations. 
The choice of boundary type is simply a choice of measurement circuits of the boundary stabilizers.  Although our proof for the 3D case is restricted to simply-connected holes, it is not a necessary condition and the theory can be generalized in future work.  This defect-tolerant property at all length scales can significantly improve the yield (success rate) of fabricating a qubit device, and potentially solve the problem of exponential decaying of the device yield with system size \cite{Chamberland:2020heavyhex, Hertzberg:2020laser} at a fundamental level.

It might be tempting to think that the behavior of topological order on a fractal with Hausdorff dimension $D_H=n-\delta$ ($0<\delta<1$) just stands somewhere between $(n-1)$D and $n$D topological orders, which is unfortunately not true at all. In the fractal case, the presence of holes at all length  scales  may just completely destroy the long-range entanglement. In fact, we have shown that if the corresponding $\ZZ_N$ topological order has particle excitations or equivalently string logical  operators, it cannot survive in a fractal geometry with generic hole boundaries embedded in any dimension. In order to preserve topological order, the hole boundaries can only be $m$-boundaries, i.e., those condense loop excitations in 3D and $(n-1)$-brane excitations in $n$D ($n\ge 4$). 
On the other hand, quite surprisingly, $\ZZ_N$ topological order without particle excitations defined in those self-correcting models can generically survive in a large class of fractals independent of the types of hole boundaries. Again, although our current proof is limited to fractals formed with simply-connected holes, it is not a really necessary condition and hence will be generalized in future work.  This property suggests that topological order in the self-correcting models has extreme robustness against spatial disorder or fabrication errors. In analogy with  the stability of the topological order in the self-correcting memory at finite temperature, a similar stability of topological order also exists in the presence of disorder. Note that  it has been realized in Ref.~\cite{Bombin:2016cr} that single-shot error correction is resilient to fabrication errors. Thus, the presented results point to the possible existence of a similar resilience in the case of self-correcting models. Moreover, one can further explore the general time-correlated noise which encompasses the situation of slowly fluctuating fabrication defects in the self-correcting models. Finally, it is also an open question whether the self-correcting properties are still preserved on the fractal geometry, which will be further explored in future. 

Perhaps the most exciting discovery in this paper is that, using a form of \textit{code puncturing} idea, we can significantly reduce the space overhead of implementing fault-tolerant non-Clifford logical gates and hence universal quantum computation. In particular, using the 3D fractal surface code, the logical CCZ gate can be implemented with space overhead $O(d^{2+\epsilon})$ with arbitrary small $\epsilon$, where $d$ is the code distance.  In addition, a code switching protocol to 2D surface code allows implement all Clifford gates and hence forms a universal gate set. For example, this code  switching can be implemented straightforwardly with the lattice surgery between the 3D fractal surface code and the 2D surface code similar to the lattice surgery protocol presented in Sec.~\ref{sec:CCZ_protocol}. The current protocol for logical CCZ gate presented in this paper requires $O(d)$ time overhead during the process of lattice surgery. An important improvement will be to establish whether it is possible to reduce the time overhead to $O(1)$ in order to achieve an overall $O(d^{2+\epsilon})$ space-time overhead for a universal gate set by utilizing the single-shot error correction  properties \cite{Bombin:2015hia} in the 3D fractal surface code. This will be addressed in future works along with the detailed study of decoding and error threshold simulations. Another relevant question is that whether one can also fold the fractal lattice into a lattice occupying less than $L^3$ volume, thus not only reducing the qubit overhead but also the overhead corresponding to the occupied space by the memory in the dilution refrigerator.

A particularly interesting conceptual development in this paper is the exploration of the connection between logical gates and global or higher-form topological symmetries, which is also equivalent to sweeping gapped domain walls.  In particular, the logical gates implemented in the fractal codes relax the previous conditions imposed in Ref.~\cite{Yoshida_gate_SPT_2015, Yoshida_global_symmetry_2016, Yoshida2017387, Webster_gates_2018, Zhu:2017tr} that the gapped boundaries in the system need to remain invariant when applying the symmetries.  From the TQFT perspective, this leads to the discoveries of exotic gapped boundaries which only condense the combination of loop-excitations (or more generally $(n-1)$-brane) and certain gapped domain walls and is hence beyond the standard framework of the Lagrangian subgroup description \cite{Levin:2013tc}.  From the quantum computing perspective, this leads to the use of topological symmetries as a \textit{transversal logical map}, which applies the logical gate but maps the code space to a new one.  Subsequent operations (such as lattice surgery in this work) can be performed to map the new code back to the original code and hence complete the logical gate.  An interesting potential direction will be to formulate and classify the global and higher-form topological symmetries, the  corresponding domain walls, and the exotic gapped boundaries, such as those studied in the present paper, in the language of higher categories. 

\section*{Acknowledgments}
G.Z. thanks Maissam~Barkeshli for pointing out the absence of topological order on a 2D Sierpi\'nski carpet which partially motivated this work. A.D. thanks Dominic~J.~Williamson for a discussion on 2D fractal lattice models. We thank Nikolas~Breuckmann, Nicolas~Delfosse, Sam~Roberts,  Sergey~Bravyi, Jay~Gambetta, Andrew~Cross, Jared~Hertzberg, and Su-Kuan~Chu for helpful discussions and comments. A.D. was supported by IBM during the summer internship and later supported by the Simons Foundation through the collaboration on Ultra-Quantum Matter (651438, XC) and by the Institute for Quantum Information and Matter, an NSF Physics Frontiers Center (PHY-1733907). This paper is a special tribute to Benoit Mandelbrot who developed fractal geometry during his career at IBM in 1980s.

\begin{appendix}

\section{Gapped boundaries and relative homology}\label{append:relative_homology}

In this appendix, we describe the gapped boundaries of the Abelian $\ZZ_N$ topological order using the mathematical language of relative homology.

We first introduce the concept of relative homology group.  We consider a cell complex $\L$ and its subcomplex $\L_0$.  The \textit{relative chain group} is a quotient of the chain groups of $C_i(\L)$ and $C_i(\L_0)$, i.e., $C_i(\L, \L_0)$$=$$ C_i(\L)/C_i(\L_0)$.  Therefore, $i$-chains  supported in $\L_0$ are trivial in $C_i(\L, \L_0)$. In order to distinguish the ordinary chain group and corresponding homology of a complex $\L$ from the ones of a pair $(\L, \L_0)$, we sometimes call the ordinary chain group $C_i(\L)$ the \textit{absolute chain group}, and the corresponding homology the \textit{absolute homology}. The quotient partitions $C_i(\L)$ into cosets $c_i+C_i(\L_0)$ whose $i$-chains can only differ in the $i$-complexes in $\L_0$ but not in those in $\L \setminus \L_0$ (represented by $c_i$). We also call $c_i+C_i(\L_0)$ a \textit{relative $i$-chain}.

Since the boundary map on the absolute chain group $\partial_i$$:$$ C_i(\L) \rightarrow C_{i-1}(\L)$ does the following map  $\partial_i$$:$$C_i(\L_0) \rightarrow C_{i-1}(\L_0)$, it induces a \textit{relative boundary map} on the relative chain group, i.e.,
\be
\partial_i : C_i(\L, \L_0) \rightarrow C_{i-1}(\L, \L_0).
\ee
We hence get the following relation when applying $\partial_i$ to a relative $i$-chain:
\be\label{eq:relative_homology_relation}
\partial_i (c_i + C_i(\L_0)) = \partial_i c_i + C_{i-1}(\L_0).
\ee
We still have the relation $\partial^2=0$ for the relative boundary map since it holds before passing to quotient groups.  

Let $c_i + C_i(\L_0)$ be a relative $i$-chain,  then it is a relative $i$-cycle if it is in the kernel of the relative boundary map, i.e., $\partial_i(c_i + C_i(\L_0))= 0$,  which implies that
\be\label{eq:relative_cycle_condition}
\partial_i c_i \in C_{i-1}(\L_0).
\ee
The above condition also includes the possibility that $\partial c_i =0$, i.e., $c_i$ is an \textit{absolute $i$-cycle}. Note that Eq.~\eqref{eq:relative_cycle_condition} is also the condition for the relative chain $c_i$ being an $i$-cycles by taking a zero contribution from $C_i(\L_0)$. We now define the subgroup of relative $i$-cycles as $Z_i(\L,\L_0):=\text{Ker}(\partial_i) \subset C_i(\L,\L_0)$.
Furthermore, $c_i + C_i(\L_0)$ (and $c_i$) is a \textit{relative $i$-boundary} if there exists a $(i+1)$-chain $c_{i+1}$ of $\L$ such that
\be
c_i + C_i(\L_0)=\partial_{i+1}c_{i+1}.
\ee
This means $c_i + C_i(\L_0)$ is in the image of the relative boundary map $\partial_{i+1}$. The above condition also includes the possibility that $c_i = \partial_{i+1}c_{i+1}$, i.e., $c_i$ is an \textit{absolute $i$-boundary}.  We can now define the subgroup of relative $i$-boundaries as $B_i(\L, \L_0) := \text{Img}(\partial_{i+1}) \subset C_i(\L, \L_0)$.  One can see that the relative boundary $c_{i}+ C_i(\L_0)$ is also an relative $i$-cycle, since $\partial_{i} (c_i+ C_i(\L_0))= (\partial_i \circ \partial_{i+1})c_{i+1} =0 $.
We hence have $B_i(\L,\L_0) \subset Z_i(\L, \L_0)$. We also call the relative $i$-boundary a \textit{trivial relative cycle} since it is contractible.  The trivial relative cycles generate continuous deformation of a particular non-contractible relative cycle which forms a \textit{relative homology class}.  We hence define the $i^\text{th}$ \textit{relative homology group} as
\be
H_i(\L,\L_0) := Z_i(\L,\L_0)/B_i(\L,\L_0) = \text{Ker} (\partial_i)/\text{Img}(\partial_{i+1}).
\ee

% In a similar fashion, we can also define the $i^\text{th}$ \textit{relative cohomology group} as
% \be
% H^i(\L,\L_0) := Z^i(\L,\L_0)/B^i(\L,\L_0) = \text{Ker} (\delta^i)/\text{Img}(\delta^{i-1}),
% \ee
% where $Z^i(\L,\L_0)$ and $B^i(\L,\L_0)$ represent the subgroups of \textit{relative $i$-cocycles} and \textit{$i$-coboundaries} respectively. In addition, $\delta^i$ represents the \textit{relative coboundary} map of $i$-cochain.  We have the following condition for a relative $i$-cochain $c^i$ to be a relative $i$-cocycle:  \be\label{eq:relative_cocycle_condition}
% \delta_i c^i \in C^{i+1}(\L_0),
% \ee
% in analogy with the condition a relative $i$-cycle in Eq.~\eqref{eq:relative_cycle_condition}.  Furthermore, $c_i$ is a relative $i$-coboundary if there exists a $(i-1)$-cochain $c^{i-1}$ of $\L$ such that
% \be
% c^i + C^i(\L_0)=\delta^{i-1}c^{i-1}.
% \ee

Now we use the relative homologies  to describe the gapped boundaries,  particularly with the examples of 2D and 3D surface codes.  When dealing with gapped boundaries, we take the subcomplex $\L_0$ in the relative homology group as a union of certain type of gapped boundaries in the model, i.e., $\L_0=\B_a =\cup_j \B_{a,j}$, where $\B_{a,j}$ denotes the $j^\text{th}$ gapped boundary of type $a$ and $\B_a$ the union of all these gapped boundaries of type $a$. We hence consider the relative homology and cohomology groups $H_i(\L, \B_e)$ and $H^{i}(\L, \B_m)$, where $\B_e$ and $\B_m$ represent the union of all $e$- and $m$-boundaries respectively, and $n$ is the spatial dimension. Note that the relative (co)homology description is valid in the generic $\ZZ_N$ topological order and not restricted just to exact solvable models, although we use the stabilizer models to illustrate the idea.

\begin{figure}[t]
  \includegraphics[width=1\columnwidth]{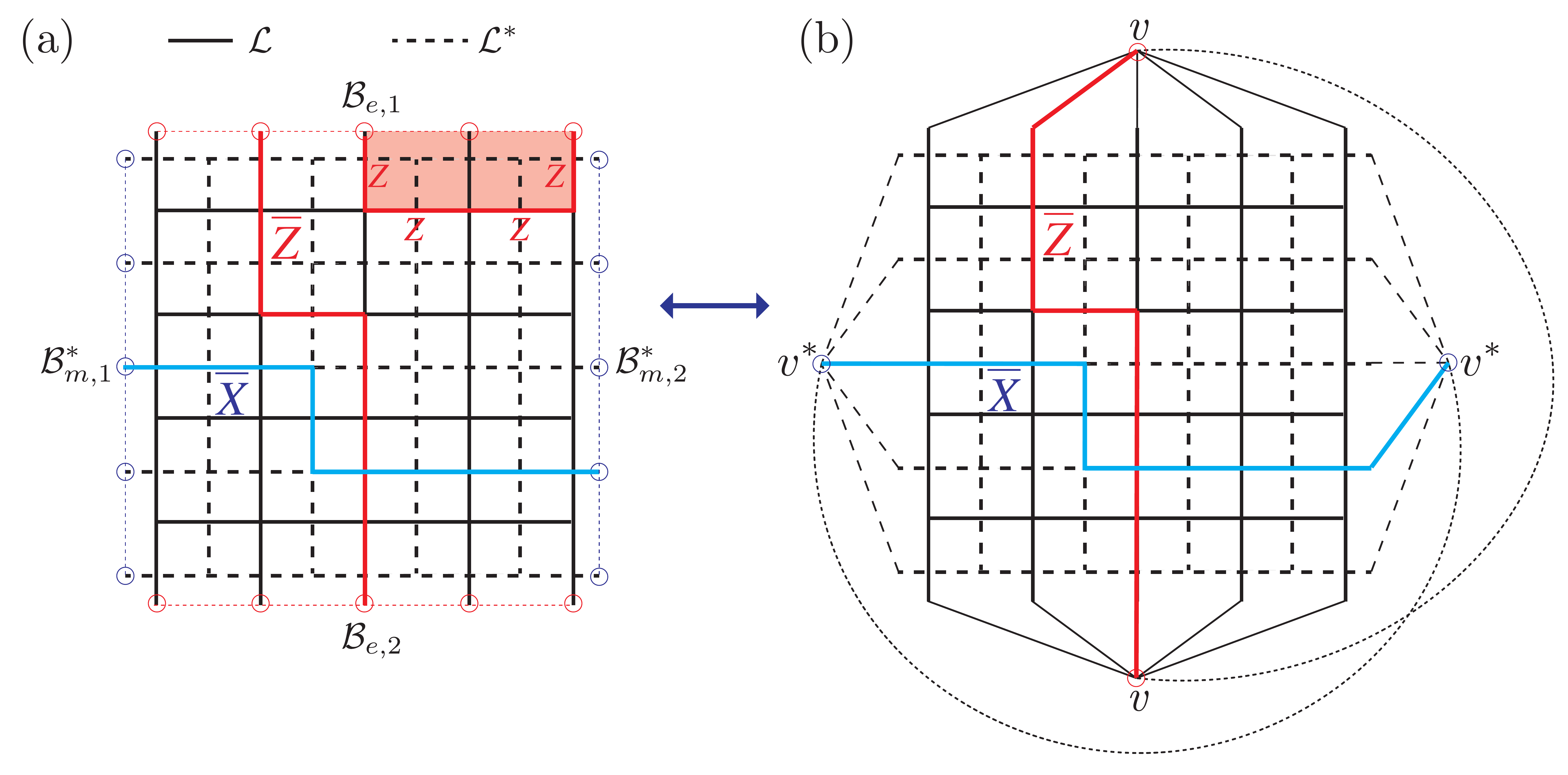}
  \caption{(a) Illustration of the 1st relative $\ZZ_2$-homology $H_1(\L, \B_e; \ZZ_2)$ defined on the cell complex $\L$ (solid lines) of the 2D surface code  and its dual $H_1(\L^*, \B_m^*; \ZZ_2)$ defined on the dual complex $\L^*$ (dashed lines). The free ends on the dangling edges of the original (dual) complex belong to the rough (smooth) boundaries $\B_e$ ($\B_m$). (b) Illustration of the application of the excision theorem and the equivalence between the 1st relative homology $H_1(\L, \B_e; \ZZ_2)$ and the 1st absolute homology $H_1(\L/\B_e; \ZZ_2)$ on the quotient complex $\L/\B_e$, where the rough boundary complex $\B_e$ is identified into a single point (vertex) $v$. In particular, the logical string $\lo{Z}$ corresponding to the relative 1-cycle is equivalent to an absolute 1-cycle on the quotient complex $\L/\B_e$. Similarly, for the dual complex $\L'$ (dashed lines), there is an equivalence between   the 1st absolute homology $H_1(\L^*, \B^*_e; \ZZ_2)$ and the 1st absolute homology $H_1(\L^*/\B^*_m; \ZZ_2)$ on the quotient complex $\L^*/\B^*_m$ where the dual smooth boundary complex $\B^*_m$ is identified into a single point (vertex) $v^*$. In particular, the logical string $\lo{X}$ corresponding to the relative 1-cycle is equivalent to an absolute 1-cycle on the quotient dual complex $\L^*/\B^*_m$. A similar identification procedure was done in the 3D surface code in Fig.~\ref{fig:connected_sum_3D}(d).}
  \label{fig:relative_homology_2D}
\end{figure}

We start with an illustration via the 2D $\Z_2$ surface code, as shown in Fig.~\ref{fig:relative_homology_2D}. The surface code in Fig.~\ref{fig:relative_homology_2D}(a) has two $e$-boundaries (rough boundaries) on the top and bottom sides and two $m$-boundaries (smooth boundaries) on the left and right sides. The code is defined on the chain complex (lattice) $\L$ as shown by the solid lines in Fig.~\ref{fig:relative_homology_2D}(a).   On the rough boundaries, there are free ends (highlighted with red circles) on the vertical dangling edges which are vertices removed from the bulk cell complex. Therefore, there are no vertex stabilizers defined on these free ends. We can define the rough boundary subcomplexes $\B_{e,1}$ and $\B_{e,2}$ as the collection of vertices (0-cells) on the free ends and the edges (1-cells) connecting these vertices (shown as thin red dashed lines).  The union of these two rough boundary subcomplexes are defined as $\B_e := \B_{e,1} \cup \B_{e,2} = \L_0$, and we can hence  substitute all the $\L_0$ in the previous statements for relative homologies with $\B_e$. The logical-$Z$ operator $\lo{Z}([c_1])$ is supported on a class of relative 1-chains $[c_1]$ associated with the 1st relative homology group, i.e.,  $[c_1] \in H_1(\L, \B_e; \ZZ_2)$$=$$\ZZ_2$ based on the relative chain group $C_i(\L, \B_e) $$=$$C_i(\L)/C_i(\B_e)$, where the chain group on the subcomplexes $\B_e$ has been modded out. In particular, any representative of  the logical operator $\lo{Z}(c_1)$ is associated with a relative 1-cycle $c_1$, i.e., $c_1 \in Z_1(\L, \B_e; \ZZ_2)$, which connects $\B_{e,1}$ and $\B_{e,2}$ and is hence non-trivial.  Note that $c_1$ indeed satisfies the condition in Eq.~\eqref{eq:relative_cycle_condition} for a relative cycle since one has 
\be
\partial_1 c_1 \in C_0(\B_e),
\ee
i.e., the boundary of the relative 1-chain $c_1$ belongs to the 0-cells (vertices) of the rough boundaries $\B_e$, i.e., the vertices on the free ends.  While the logical string $\lo{Z}(c_1)$ is associated with a non-trivial relative 1-cycle, the relative 1-cycles $c'_1$ which has its boundary vertices  belonging to the same rough boundary $\B_{e,j}$ are relative 1-boundaries: $c'_1 \in B_1(\L, \B_e; \ZZ_2)$, namely trivial relative 1-cycles, as illustrated by the $Z$-string with four edges on the upper right corner in Fig.~\ref{fig:relative_homology_2D}(a).  Generally speaking, the $Z$-operators corresponding to a relative boundary $Z^\otimes(c'_1)$ are products of stabilizers which contain the boundary stabilizers  particularly (3-body stabilizers on truncated plaquettes on rough boundaries in this case).  On the other hand, all the bulk $Z$ stabilizers (4-body)  are absolute boundaries, and the boundary $Z$ stabilizers (3-body) are relative boundaries.  These absolute and relative 1-boundaries (and their corresponding stabilizers) generate local deformation of the non-trivial 1-cycle $c_i$ and hence all representatives of the logical string $\lo{Z}([c_1])$. 

Now in order to describe the logical-$X$ operator, one can use the relative homology on the dual lattice $\L^*$ (represented by dashed lines) as illustrated in Fig.~\ref{fig:relative_homology_2D}(a). The logical string $\lo{X}([c^*_1])$ is supported on a class of relative dual 1-chain $[c^*_1]$ associated with the 1st relative homology group, i.e., $[c^*_1] \in H_1(\L^*, \B_m^*; \ZZ_2)$$=$$\ZZ_2$. The smooth boundaries $\B_m$ on the left and right sides become again rough boundaries  $\B_m^*$ in the dual lattice. Note that the chain group on the dual boundary subcomplexes $\B_m^*$ consisting of the vertices (blue circles) on the free ends of the dual lattices and the edges connecting them (thin blue dashed lines) is modded out.  This description is basically the same as the case for logical-$Z$ operator by symmetry.  

Now we also show an underlying equivalence between the relative homology description and an absolute homology description already introduced in the main text, which  may hence build up more intuitive understanding.   By construction, relative homology depends only on the part of $\L$ outside the subcomplex $\L_0$ (and in our case the boundary subcomplex $B_a$) and ignores the part inside $\L_0$ (or $B_a$). Hence we can cut out (excise) cells from $\L_0$ (or $B_a$) which also cut out the same set of cells in $\L$ without changing the relative homology.   This property is essentially the \textit{excision theorem} with the following statement:   

\nin \textit{ Let $A \subseteq \L_0 \subseteq \L$ be cell complexes, then there is the following isomorphism between the relative homology group before and after the excision of $A$, i.e.,} 
\be\label{eq:excision_homology}
H_i(\L, \L_0) \cong H_i(\L\setminus A, \L_0 \setminus A), \quad \forall i.
\ee
\nin See Ref.~\cite{Hatcher:2001ut} for the proof of this  theorem. Based on this theorem, we can cut out  the cells in the boundary $\B_a$ as much as possible, leading  to the following isomorphism: 
\be
H_i(\L, \B_a) \cong H_i(\L / \B_a,  \B_a / \B_a) \cong H_i(\L / \B_a,  *) \cong  \tilde{H}_i(\L / \B_a), 
\ee
where `$*$' stands for a single point and $\tilde{H}_i$ represents the $i^\text{th}$ reduced homology as has been introduced in Sec.~\ref{sec:chain_complex} in the main text. Since for the study of topological order and codes, we always just consider $i>0$, therefore the reduced homology always just equals the usual homology.  Therefore, we have the following isomorphism:
\be
H_i(\L, \B_a) \cong H_i(\L / \B_a), \ \text{for} \ i>0, 
\ee
which states that the $i^\text{th}$ relative homology of the pair of complexes $(\L, \B_a)$ is isomorphic to the $i^\text{th}$ absolute homology of the quotient complex  $\L' = \L / \B_a$ obtained from identifying the boundary components $\B_a$ of $\L$ into a single point, as has been stated in the main text.  

Using the 2D surface code in Fig.~\ref{fig:relative_homology_2D}(b) as an illustration, we can identify all the vertices on the free ends of $B_{e,1}$ and $B_{e,2}$ into a single vertex $v$.  Therefore, we get the following isomorphism:  $H_1(\L, \B_e) \cong H_1(\L/\B_e)$, which states that the 1st relative homology of the pair $(\L, \B_e)$ is equivalent to the absolute homology of a modified cell complex $\L'=\L/\B_e$. Here, $\L'$ is equivalent to a complex $\L$ whose  boundary subcomplex $\B_e$ is excised and identified into a single vertex.   This equivalence relation also holds for the dual lattice and the corresponding $X$-logical operator, i.e.,  $H_1(\L^*, \B_m) \cong H_1(\L'^*)$, where $\L'^*$ is a modified complex of $\L^*$ such that its boundary subcomplex $\B_m$ is excised and  identified into a single vertex $v^*$, as shown in Fig.~\ref{fig:relative_homology_2D}(b).

\section{Splitting of ground-state degeneracies on a 2D Sierpi\'nski carpet}
\label{app:degeneracy_splitting}

In this appendix, we discuss in details about the splitting of ground-state degeneracies on a the 2D Sierpi\'nski carpet model illustrated in Fig.~\ref{fig:fractal_surface_code} (Sec.~\ref{sec:2D} in the main text) in the context of passive topological protection.
One can consider the following local  perturbation on top of the parent Hamiltonian $H$:
\be
V_\text{perturb} =g \sum_{A}  O_A, 
\ee
where the sum is over all possible local region $A$ with $O(1)$ support  and $g \ll 1$ being the perturbation strength. In the case of topological order, this local perturbation leads to the following splitting of ground-state degeneracies due to tunnelings of virtual anyons: $\delta \propto \Delta e^{-d/\xi}$ with $d=O(L)$, where $\Delta$ is the many-body gap above the ground-state subspace and one has $\Delta=2J$ for the exact-solvable toric code Hamiltonian Eq.~\eqref{eq:2DTC}. Note that in the presence of a non-trivial local perturbation  $V_\text{perturb}$ which does not commute with $H_\text{2DTC}$ in Eq.~\ref{eq:2DTC}, we always have nonzero correlation length $\xi$. The degeneracy splitting for a generic 2D $\ZZ_2$ topological order on the square-patch geometry (level-0) is shown in Fig.~\ref{fig:degeneracy_splitting} (Sec.~\ref{sec:2D} in the main text). On the other hand, in the 2D fractal case without macroscopic code distance, i.e., $d=O(1)$, this exponential suppression breaks down, and one gets a degeneracy splitting $\delta \propto \Delta$, where $\Delta$ is a constant independent of the overall system size $L$ when $L \gg \xi$. For the toric code model on the Sierpi\'nski carpet with $m$-holes, one can  write down the following effective Hamiltonian for the virtual anyon tunneling induced by the perturbation:
\be\label{eq:effective_tunneling}
H_\text{eff} = -\sum_{i,j} \delta^X_{i,j} \lo{X}_{i,j} - \sum_i\delta^{Z}_i \lo{Z}_i -\delta_L \lo{Z}.
\ee
Here, $\lo{X}_{i,j}$ denotes the $X$-logical string (Wilson lines) connecting the $m$-boundaries labeled by $i$ and $j$ (including the outer and hole boundaries), and $\delta^X_{i,j} \propto \Delta e^{-d^X_{i,j}/\xi}$ represents the tunneling strength, where $d^X_{i,j}$ represents the shortest distance between boundary $i$ and $j$.  Similarly, $\lo{Z}_{i}$ denotes the logical string circulating the $i^\text{th}$ hole, and $\delta^Z_i \propto \Delta e^{-d^Z_i/\xi}$ denotes the tunneling strength where $d^Z_i$ represents the length of the shortest loop circulating the $i^\text{th}$ hole.   Here,    $d^x_{i,j}$ and $d^z_i$ can be considered as the $X$- and $Z$-distance of the corresponding logical operators.  The last term in Eq.~\eqref{eq:effective_tunneling} represents the tunneling corresponding to the macroscopic string $\lo{Z}$, where the corresponding tunneling strength is $\delta_L \propto \Delta e^{-L/\xi}$.  Note the tunneling term of the macroscopic string $\lo{X}$ is automatically included in the first term. For Sierpi\'nski carpet at level-$l$, it encodes $k=3^l+1$ logical qubit, and hence has ground state degeneracy $\text{GSD}=2^{3^l+1}$. From Eq.~\eqref{eq:effective_tunneling}, one can obtain the degeneracy splittings, most of which have a typical size  $O(\Delta)$ and is not exponentially suppressed with the system size $L$.   Such degeneracy splitting is illustrated in Fig.~\ref{fig:degeneracy_splitting}(b), where the split ground state levels can even go above the gap and the topological order is completely destroyed.  In sum, the ground-state degeneracy on the Sierpi\'nski carpet with $m$-holes is not protected.

\section{$\ZZ_N$ toric codes in $n$ dimensions}\label{append:ZN}

\begin{figure}[t]
  \includegraphics[width=0.5\columnwidth]{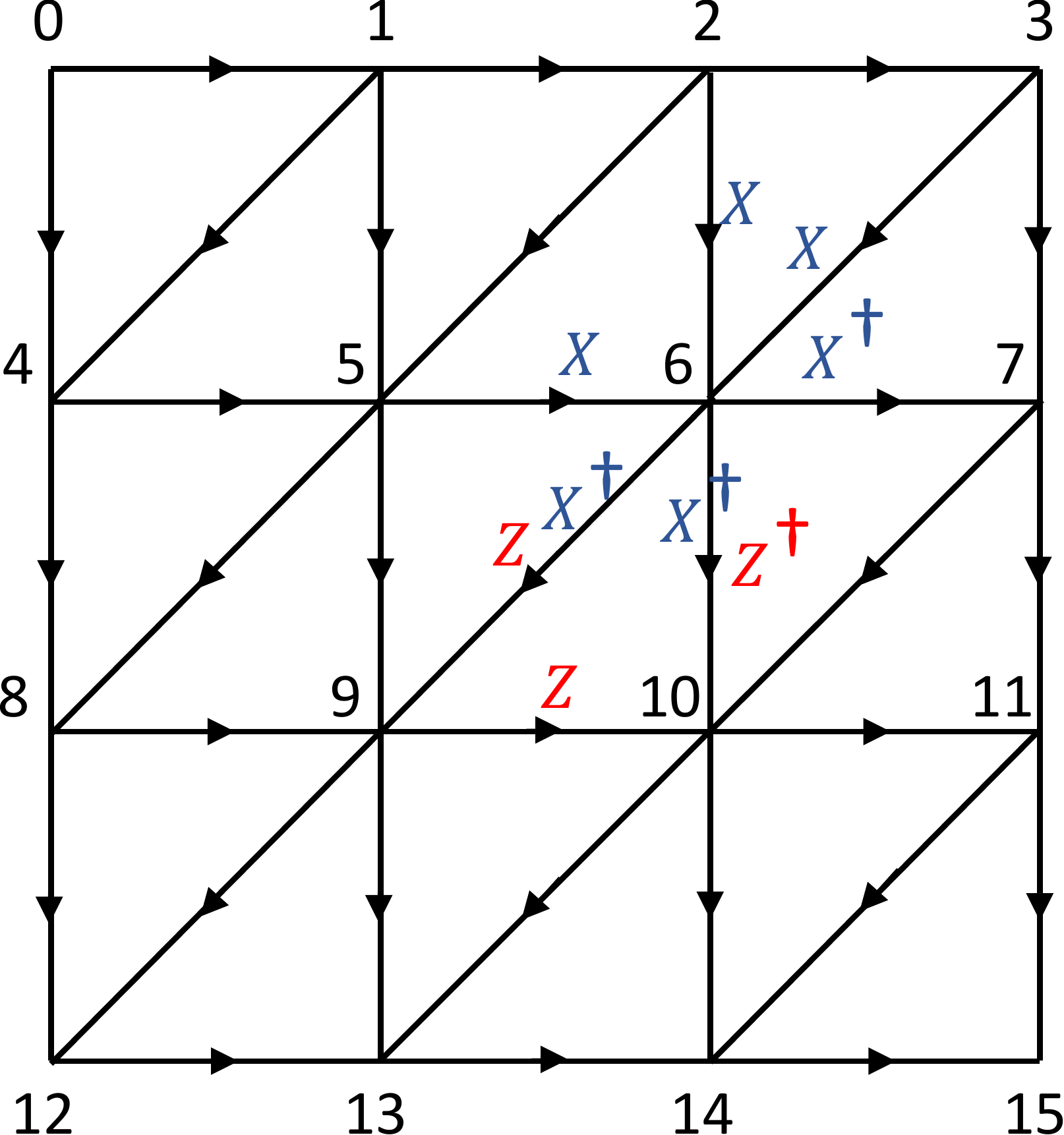}
  \caption{$\mathbb{Z}_N$ toric code in 2D. The vertices on the shown patch are ordered as mentioned. The ordering of vertices provides a branching structure as shown. The $Z$ stabilizer on the 2-simplex has a $Z$ on an edge if the edge orientation is consistent with the ordering of vertices on the 2-simplex up to cyclic permutations and $Z^\dagger$ otherwise. The $X$ stabilizer on a vertex has an $X$ on edges that go into the vertex and $X^\dagger$ otherwise.}
  \label{fig:zntoric}
\end{figure}

As discussed in the main text, we can describe homological codes via
the chain complex 
\[
C_{i+1}\stackrel{\pd_{i+1}}{\rightarrow}C_{i}\stackrel{\pd_{i}}{\rightarrow}C_{i-1}
\]
where the composition $\pd_{i}\cdot\pd_{i+1}$ is $0$. For $\mathbb{Z}_{N}$
toric codes, the chain groups $C_{i}\equiv C_{i}\left({\cal L};\mathbb{Z}_{N}\right)$
are defined for the lattice ${\cal L}$ with coefficients in $\mathbb{Z}_{N}$.
We associate a $\triangle$-complex structure with the lattice
${\cal L}$, i.e., ${\cal L}$ can be defined via a triangulation of
the manifold $\M^{n}$ in $n$ dimensions with a branching structure
which follows from a total ordering of the vertices of the $n$-simplex.
This implies that each element of the chain group $C_{i}$ is formally
a linear combination $\sum_{\alpha}z_{\alpha}e_{\alpha}^{\left(i\right)}$
where $z_{\alpha}\in\mathbb{Z}_{N}$ and $e_{\alpha}^{\left(i\right)}$is
an $i$-simplex, a subsimplex of an $n$-simplex for $i<n$. Given an ordering of the vertices on the $n$-simplices,
we can define the action of the boundary map on an $i$-simplex $\left[v_{0},...,v_{i}\right]$
as 
\[
\pd_{i}\left[v_{0},...,v_{i}\right]=\sum_{j}\left(-1\right)^{j}\left[v_{0},....\hat{v}_{j},....,v_{i}\right]
\]
where $v_{0},...,v_{i}$ represents the vertices of the $i$-simplex
such that their order is consistent with the ordering of the vertices
of the $n$-simplex. The $\hat{v}_{j}$ refers to the vertex that
has been removed from the vertices of the $i$-simplex to give one
face of the $i$-simplex i.e. the ($i-1$)-simplex on its boundary.
Note that the vertices of the ($i<n$)-simplices are a subset of the
vertices of a $n$-simplex and ordered according to the ordering of
vertices in the $n$-simplex. The definition of the boundary map as
defined above leads to the desired property for composition of boundary
maps i.e. $\pd_{i}\cdot\pd_{i+1}=0$. We can define the $\left(i,n-i\right)$-$\mathbb{Z}_{N}$ toric codes on the triangulation of the torus in
$n$-dimensions via the chain complex above such that $N$-level qudits are placed
on the $i$-simplex, $X$-type stabilizers are associated with the
$\left(i-1\right)$-simplices and the $Z$-type stabilizers with the
$\left(i+1\right)$-simplices. For example, in 2D,
we can define the $\left(1,1\right)$-$\mathbb{Z}_{N}$ toric code
with the qudits on edges such that we have the following chain complex
\[
C_{2}\stackrel{\pd_{2}}{\rightarrow}C_{1}\stackrel{\pd_{1}}{\rightarrow}C_{0}
\]
and the $X$ and $Z$ stabilizers can be defined via the maps $\pd_{2}:C_{2}\rightarrow C_{1}$
and $\pd_{1}^{T}:C_{0}\rightarrow C_{1}$. The composition $\pd_{1}\cdot\pd_{2}=0$
ensures that the $\mathbb{Z}_{N}$ stabilizer terms commute; check Fig.~\ref{fig:zntoric}. Similarly,
one can define the $\left(1,2\right)$-$\mathbb{Z}_{N}$ toric code
in three dimensions. 

Alternatively, we can consider a hypercubic lattice instead of a triangulation
and have the natural cellulation associated with the cubic lattice
in $n$-dimensions, i.e., each $n$-cell is an $n$-dimensional hypercube. We can then define the cellulation and boundary maps for the
hypercubic lattice as defined in \cite{chen2021higher}, such that the higher cup products can be defined properly. Given the definition for the cells and boundary maps on the hypercubic lattice, the definition of $(i,n-i)$-$\mathbb{Z}_N$ toric code follows analogously to the triangulated manifold with a $\triangle$-complex structure. 
% In order to define the boundary maps naturally using a simplicial structure, we can consider a triangulation of each hypercube $n$-cell
% and define the maps $\pd_{i+1}$ and $\pd_{i}^{T}$ to give only the $i$-cells associated with the hypercubic lattice, that define the qubits on which the $Z$ and $X$ stabilizers act respectively.

% \section{No-go theorem of generic topological order on 2D punctured fractals}\label{append:no_go}

\section{Higher dimensional fractal topological order defined on a punctured $n$-torus geometry}
\label{app:punctured_torus}

In this appendix, we continue the discussion in Sec.~\ref{sec:nD_simple_fractal} about the  $(i, n-i)$-$\ZZ_N$ topological orders defined on simple fractals with the background manifold chosen as the  $n$-torus.    In the following, we discuss the cases with and without string-like logical operators respectively.

\subsection{Cases with string-like logical operators $(i=1)$}\label{app:n_dimensional_string}

Here, we consider the punctured $n$-torus geometry $\mathcal{L}(T^n)$, where the simple fractal nested on an $n$-dimensional torus $T^n=S^1\times S^1 \times \cdots\times S^1 \equiv (S^1)^{\times n}$, which is a manifold product of $n$ macroscopic  circles $S^1$, i.e., of size $O(L)$. Topologically speaking, $\L(T^n)$ is equivalent to an $n$-torus cutting out a set of $n$-balls, i.e., $\L(T^n)=T^n \setminus \cup_j D^n_{(j)}$. Note that there is no $e$-boundary in the system, i.e., $\mathcal{B}_e=\emptyset$. Using Eq.~\eqref{eq:connected_sum_presentation}, the punctured $n$-torus can also be expressed as a connected sum of an $n$-torus and the punctured $n$-sphere, i.e., $\L(T^n)=T^n \setminus \cup_j D^n_{(j)}= T^n \# S^n \setminus \cup_j D^n_{(j)}$.  Using the Mayer–Vietoris sequence \cite{Hatcher:2001ut}, one can express the 1st-homology of the connected sum of two 3-manifolds as a direct sum of the 1st-homology of each 3-manifold, i.e.,
\begin{align}\label{eq:torus_homology_nD}
\non & H_1(\L(T^n); \ZZ_2) = H_1(T^n \# S^n \setminus \cup_j D^n_{(j)}; \ZZ_2) \\
\non \cong & H_1(T^n; \ZZ_2) \oplus H_1(S^n \setminus \cup_j D^n_{(j)}; \ZZ_2)  \\
= &  (\mathbb{Z}_2 \oplus \mathbb{Z}_2  \oplus \cdots \oplus \mathbb{Z}_2)  \oplus \mathsf{0} = \mathbb{Z}_2^{\oplus n}.
\end{align}
Note that the nontrivial contribution $\mathbb{Z}_2^{\oplus n}$ only  comes from the homology group of the  of the $(1, n$$-$$1)$-toric code, corresponding to the $n$  macroscopic logical strings $\overline{Z}_j \ (j=1,2,\cdots, n)$  going through the $n$ non-contractible 1-cycles $S^1$ around $T^n$.   According to the Poincaré-Lefschetz duality in Eq.~\eqref{eq:Poincare_duality_cellulation} and the isomorphism in Eq.~\eqref{eq:cohomology_eqaul_homology}, we have the dual $(n$$-$$1)^\text{th}$ relative homology group as
\be
H_{n-1}(\mathcal{L}^*(T^n), \mathcal{B}^*_m; \mathbb{Z}_2)  \cong H_1(\mathcal{L}(T^n); \mathbb{Z}_2) = \mathbb{Z}_2^{\oplus n}.
\ee
By using the Poincaré-Lefschetz dual of the homology decomposition in Eq.~\eqref{eq:torus_homology_nD}, we get
\begin{align}
\non & H_{n-1}(\L^*(T^n); \ZZ_2) \\
=& H_{n-1}(T^n; \ZZ_2) \oplus \non H_{n-1}(\L^*(S^n), \B^*_m; \ZZ_2) \\
=& H_{n-1}(T^n; \ZZ_2) \oplus \mathsf{0}.
\end{align}
We can see that the nontrivial contribution $\ZZ_2$ comes from the $(1,n-1)$-surface code part $H_{n-1}(T^n; \ZZ_2)$, which corresponds to the $n$ macroscopic logical branes $\overline{X}_j \ (j=1,2,\cdots, n)$. Each of these $(n$$-$$1)$-dimensional logical branes, $\overline{X}_j$,  travels through the non-contractible  $(n$$-$$1)$-cycle perpendicular to the $1$-cycle experienced by the dual  logical string $\overline{Z}_j$.  The logical brane $\overline{X}_j$ intersects with the logical string $\overline{Z}_j$ at a single point (in terms of algebraic intersection), leading to the anti-commutation relation $\{\overline{X}_j, \overline{Z}_j\}=0$. The ground-state subspace (code space) of this punctured torus fractal geometry is $\mathcal{H}_\mathcal{C}$$=$$\mathbb{C}^{\mathbb{Z}_2^{\oplus n}}$$=$$\mathbb{C}^{2^n}$, corresponding to $n$ logical qubits and a $2^n$-fold ground-state degeneracy.

\subsection{Cases without string-like logical operators ($i \ge 2$)}\label{app:general_k}

Same as above, we also consider the simple fractal nested on an $n$-dimensional torus $T^n=(S^1)^{\times n}$. There are $C_n^i$  macroscopic logical $i$-branes $\overline{Z}_j \ (j$$=$$1,2,\cdots, C_n^i)$  going through the $C_n^i$ non-contractible $i$-cycles $(S^1)^{\times i}$ around $T^n$. Similarly, there are $C_n^{n-i}=C_n^{i}$  macroscopic logical $(n-i)$-branes  $\overline{X}_j \ (j$$=$$1,2,\cdots, C_n^i)$  going through the $C_n^i$ non-contractible $(n-i)$-cocycles $(S^1)^{\times (n-i)}$ around $T^n$. We hence get the following relative homology groups:
\be
H_i(\mathcal{L}(T^n), \mathcal{B}_e; \mathbb{Z}_2) \cong H_{n-i}(\mathcal{L}(T^n), \mathcal{B}_m; \mathbb{Z}_2)=\mathbb{Z}_2^{\oplus C^i_n}, 
\ee
leading to the following ground-state subspace (code space): $\mathcal{H}_\mathcal{C}$$=$$\mathbb{C}^{\mathbb{Z}_2^{\oplus C^i_n}}$$=$$\mathbb{C}^{2^{C^i_n}}$, corresponding to $C^i_n$ logical qubits and a $2^{C^i_n}$-fold ground-state degeneracy. By using the homology decomposition similar to the fractal surface code case, we can see all the nontrivial contribution $\mathbb{Z}_2^{\oplus C^i_n}$ comes from the macroscopic logical  $i$-branes $\lo{Z}_j$ and logical $(n-i)$-branes $\lo{X}_j$.

\section{Requirements for a transversal logical CCZ gate }
\label{app:CCZ_requirements}

We first discuss the condition for the transversal CZ gate as a prerequisite for the discussion of the condition for the transversal CCZ gate in any quantum CSS code. We consider two copies of codes labeled as $a = 1, 2$ stacked on top of each other. The qubits of the two codes are aligned, sharing the same label~$j$, and acted upon by a CZ gate pairwise.  We denote the $i^\text{th}$ $X$ stabilizer of code $a$ by $A_{i;a}$ and the $i^\text{th}$ $Z$ stabilizer of code $a$ by 
$B_{i;a}$. The CZ gate has the following mapping for the Pauli operators: $\text{CZ}_{j; 1,2} : X_{j;a}  \rightarrow X_{j;a} Z_{j;b}$ and $Z_{j;a}  \rightarrow Z_{j;a}$, where $a,b \in \{1,2\}$ such that $a \ne b$. The transversal CZ gate, $\Motimes_j \text{CZ}_{j; 1,2}$, hence leads to the following map on the original stabilizers in  code $1$:
\begin{align}
\nonumber \Motimes_j \text{CZ}_{j; 1,2}: \Motimes_{j \in A_{i;a}} X_{j;a} &\longrightarrow \Motimes_{j \in A_{i;a}} X_{j;a} \Motimes_{k \in A_{i;a}} Z_{k;b}, \\
\Motimes_{j \in B_{i;a}} Z_{j;a}  &\longrightarrow \Motimes_{j \in B_{i;a}} Z_{j;a}.
\end{align}
Here, $Z_{k;b}$ are supported on those qubits in code $b$ which pairwise interact (via the CZ gate) with the aligned qubits supported in the stabilizer $A_{i;a}$ in code $a$. We say that $Z_{k;b}$ are in the image of $A_{i;a}$ on code 2 under $\overline{\text{CZ}}_{1,2}$. This explains the meaning of $k \in A_{i;a}$ in the above expression. Since CZ gate is symmetric with respect to code $1$ and $2$, the above expression obviously holds true for any choice of differing $a$ and $b$. Therefore, the transversal CZ naturally preserves the Z stabilizers of the code but maps the $X$ stabilizers into a product of the $X$ stabilizer and a $Z$ operator on the other code copy.

In order to preserve the stabilizer group  after applying the gates, we require $X$~stabilizer in code~$a$ to be mapped to a product of the original $X$~stabilizer and a $Z$~stabilizer in code~$b$. A necessary and sufficient condition for $\Motimes_{k \in A_{i;a}} Z_{k;b}$ to be a $Z$ stabilizer of codeblock~$b$ is for it to commute with all of the $X$~stabilizers and logical~$X$ operator in that given codeblock. As such, we would like the corresponding support of the operator to have even overlap with all such operators outlined above, which leads to the following set of necessary conditions for the existence of a transversal~CZ gate, i.e. $\forall \ i, j$,
\begin{subequations}
\begin{align}
 \mathbf{supp}(A_{i;a}) \cap \mathbf{supp}(A_{j;b})&=0, \\
\mathbf{supp}(A_{i;a}) \cap \mathbf{supp}(\overline{X}_{b})&=0.
\end{align}
\label{eq:CZcondition1}
\end{subequations}
Here ``$\cap$" represents the number of  overlapping sites of the operator support between codeblocks 1 and 2. It should be noted that these remain separate codeblocks but the overlap is defined in terms of the labeling of the qubits on each codeblock which is determined by their coupling in the transversal CZ application. Since we only consider $\Z_2$-surface codes in this case, the overlap is only counted~$\mod 2$ and $0$ here means even overlap, i.e., $0 \mod 2$. Here, $\overline{X}_b$ denotes logical-$X$ operator of code~$b$.   For simplicity, we assume code~$b$ only has a single logical operator, but it could be generalized to the case with multiple logical-$X$ simply by replacing $\overline{X}_2$ with $\overline{X}_{l,b}$ where $l$ labels the logical qubit. 

The above conditions are sufficient to guarantee that transversal~CZ implements a logical gate, however we also would like it to implement a particular logical gate, that is logical~CZ, $\overline{\text{CZ}}_{1,2}$. Therefore, we additionally require that the logical~$X$ operators from each codeblock have odd overlap with respect to one another, as this would guarantee that they are mapped onto a logical~$Z$ on the other codeblock in a similar fashion to the mappings of the $X$ stabilizers. As such, the resulting last requirement for the transversal~CZ to implement $\overline{\text{CZ}}_{1,2}$ is:
\be\label{eq:CZcondition2}
\mathbf{supp}(\overline{X}_1) \cap \mathbf{supp}(\overline{X}_{2})=1,
\ee
Assuming the underlying codes satisfy equations~\eqref{eq:CZcondition1} and~\eqref{eq:CZcondition2}, then the transversal application of~CZ will result in a logical~CZ, that is $\overline{\text{CZ}}_{1,2} =\Motimes_j \text{CZ}_{j; 1,2}$.

\begin{figure*}[t]
  \includegraphics[scale=0.5]{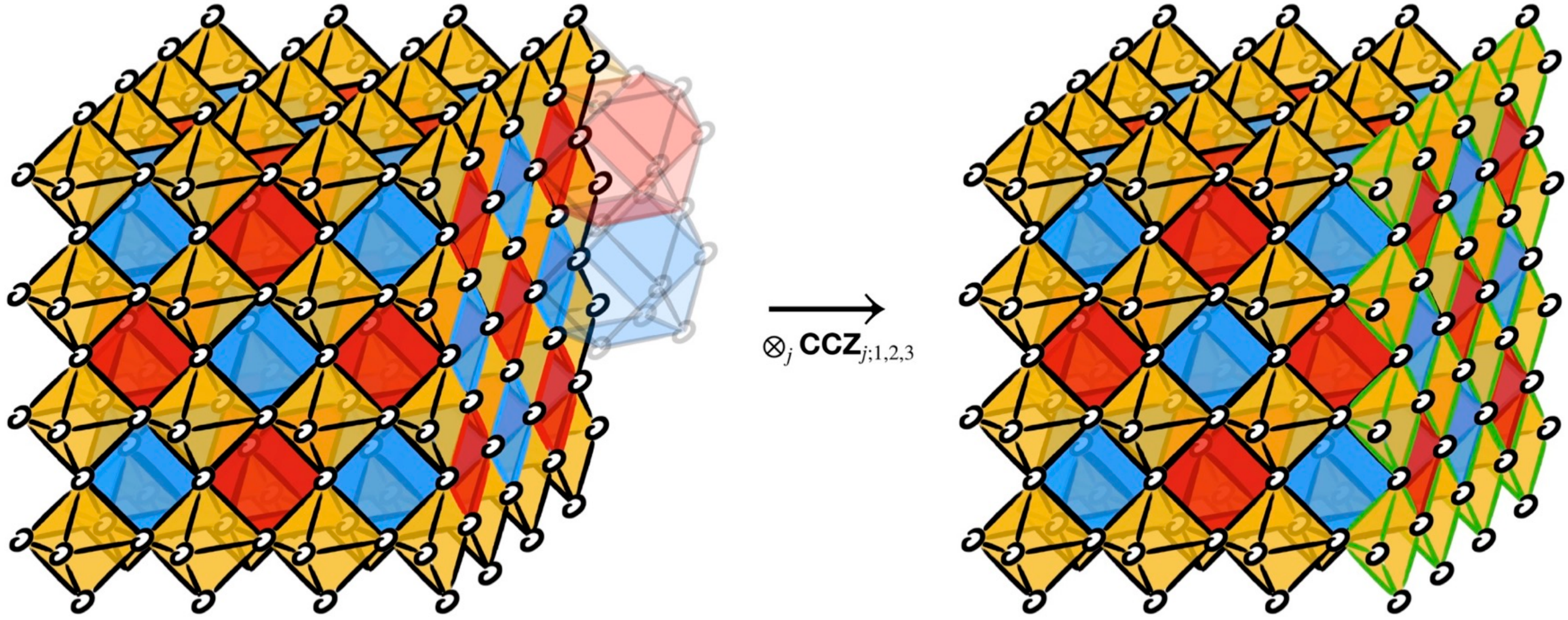}
  \caption{Illustration of an $(m_1,m_2,m_3)$-boundary in a stack of 3 copies of 3D toric codes. The boundary is on the right of the lattice, where the support of the stabilizers is reduced. The code whose $X$~stabilizers are given by the yellow octahedra have the support of their stabilizers reduced to be weight-5. While the red and blue stabilizers, of codes 2 and 3, have weight-4 face stabilizers on the boundary. For illustrative purposes, we show to the right how the stabilizers would normally have larger support, these qubits are faded out. On the right side, after the application of~$\otimes_j \text{CCZ}_{j;1,2,3}$, the bulk $X$~stabilizers are preserved while those whose support were cut at the boundary of the hole are now transformed to be of the form explained in Fig.~\ref{fig:XtransformCCZ}.}
  \label{fig:m_hole3D}
\end{figure*}

We now move on to discussing the conditions for a transversal CCZ~gate.  A CCZ~gate maps a Pauli-$X$ operator on a given qubit to a product of that same Pauli~$X$ and a CZ gate on the other two qubits in the CCZ gate. Given the gate is diagonal in the computational basis it leaves the Pauli-$Z$ operator unchanged, i.e., $\text{CCZ}_{j;a,b,c} : X_{j;a}  \rightarrow X_{j;a} \text{CZ}_{j; b,c}$ and $Z_{j;a}  \rightarrow Z_{j;a}$, for any choice of $a,b,c$. The transversal logical CCZ gate,  $ \Motimes_j \text{CCZ}_{j; 1,2,3}$, hence leads to the following map on the original stabilizers in code~$a$, where $a,b,c \in \{1,2,3\}$ such that $a \ne b \ne c \ne a$:
\begin{align}
\nonumber \Motimes_j \text{CCZ}_{j; 1,2,3}: \Motimes_{j \in A_{i;a}} X_{j;a} &\longrightarrow \Motimes_{j \in A_{i;a}} X_{j;a} \Motimes_{k \in A_{i;a}} \text{CZ}_{k;b,c}, \\
\Motimes_{j \in B_{i;a}} Z_{j;a}  &\longrightarrow \Motimes_{j \in B_{i;a}} Z_{j;a}.
\end{align}
Therefore, in order to preserve the stabilizer group, $\Motimes_{k \in A_{i;a}} \text{CZ}_{k;b,c}$ needs to be logical identity.  This is equivalent to impose Eqs.~\eqref{eq:CZcondition1} and \eqref{eq:CZcondition2} restricted to the image of $A_{i;a}$ on  codes~$b$ and~$c$, i.e., for $\forall i, j, k$, 
\begin{subequations}\label{eq:CCZcondition1}
\begin{align}
 \mathbf{supp}(A_{i;a}) \cap \mathbf{supp}(A_{j;b}) \cap \mathbf{supp}(A_{k;c})&=0, \\
\mathbf{supp}(A_{i;a}) \cap \mathbf{supp}(A_{j;b}) \cap \mathbf{supp}(\overline{X}_{c})&=0. \\
\mathbf{supp}(A_{i;a}) \cap \mathbf{supp}(\overline{X}_b) \cap \mathbf{supp}(\overline{X}_{c})&=0.
\end{align}  
\end{subequations}

Given the above conditions are satisfied, the transversal CCZ gate gives the following map:
\be
\nonumber \Motimes_j \text{CCZ}_{j; 1,2,3}: \Motimes_{j\in \overline{X}_a}X_{j; a} \rightarrow \Motimes_{j\in \overline{X}_a}X_{j; a} \Motimes_{k\in \overline{X}_a} \text{CZ}_{k; b,c},
\ee
i.e., it maps $\overline{X}_a$ to a product of itself and a transversal CZ~gate acted on codes~$b$ and~$c$ restricted to the image of  $\overline{X}_a$.  Therefore,  Eq.~\eqref{eq:CZcondition2} has to be satisfied on codes~$b$ and $c$ restricted to the image of $\overline{X}_a$, i.e., 
\be\label{eq:CCZcondition2}
\mathbf{supp}(\overline{X}_1) \cap \mathbf{supp}(\overline{X}_2)  \cap \mathbf{supp}(\overline{X}_{3})=1.
\ee
In sum, equations~\eqref{eq:CCZcondition1} and \eqref{eq:CCZcondition2} are the necessary and sufficient conditions for a transversal CCZ~gate to implement logical~CCZ, that is $\overline{\text{CCZ}}_{1,2,3} = \Motimes_j \text{CCZ}_{j; 1,2,3}$.

\section{Lattice construction of the stack code and  an example of the $(m_1,m_2,m_3)$-hole}
\label{app:mmm_hole_example}

In this appendix, we provide an explicit example of constructing an $m$-hole in three iterations of the 3D toric code. In order to construct a stack of 3D toric codes that enables the implementation of a transversal~CCZ in the bulk, we can follow the construction of Vasmer and Browne~\cite{Vasmer2019}. In this implementation, shown in Fig.~\ref{fig:m_hole3D} the three copies will be asymmetric in that one of the copies will have weight-6 $X$~stabilizers in the bulk (yellow in the Figure) while the two other copies will have weight-12 $X$~stabilizers (given in red and blue). This choice is made in order to guarantee that when one  takes the mutual intersection of one $X$~stabilizer from each of the three code copies, the intersection is always even. One can simply verify this by checking that a red and blue stabilizer only ever intersect along a weight-4 face and all such faces will intersect their neighboring yellow octahedra at an even number of qubits.

Figure~\ref{fig:m_hole3D} shows the stack code where we have introduced an $(m_1, m_2, m_3)$-hole on the right side of the lattice. That is, all of the $X$~stabilizers have their support cut to the right of the hole. Note that the stabilizers from the yellow block will have boundary stabilizers whose support is cut to now be weight-5. While the weight-12 red and blue stabilizers whose support falls within the hole have their support cut down to just the weight-4 face that lives along the boundary of the hole. At the top right of Fig.~\ref{fig:m_hole3D} we give a faded out versions of what the stabilizers would normally support and all qubits that are transparent can be thought of as bulk qubits that one measures out in the $X$-basis in order to construct the hole (and as such reduce the support of the associated $X$~stabilizers).

\begin{figure}[hbt]
\centering
(a){\includegraphics[scale=0.33]{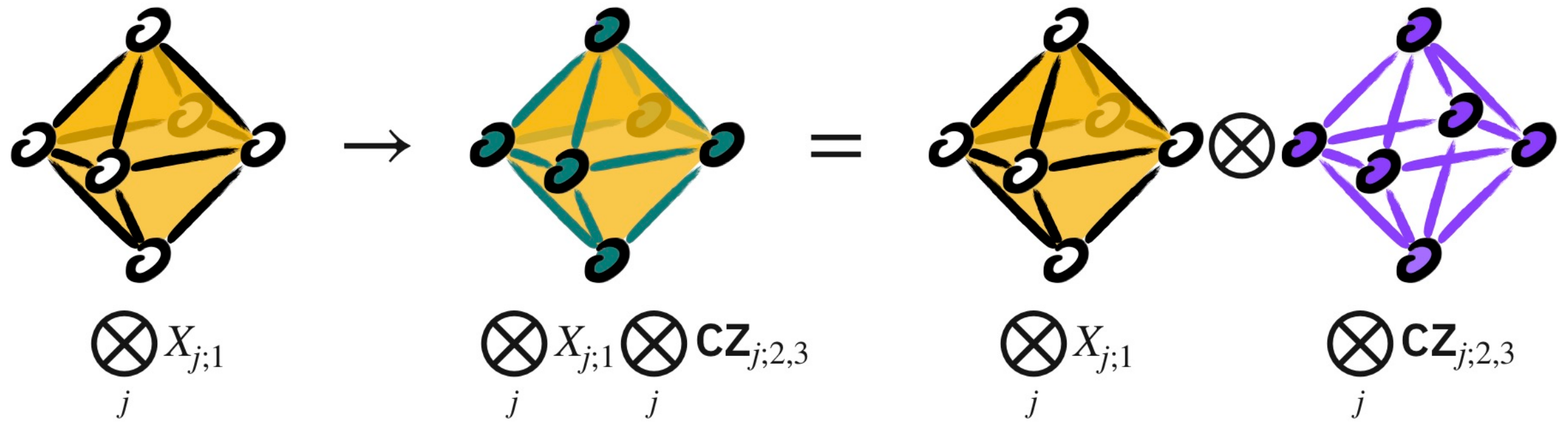}}\hspace{6 mm}
(b){\includegraphics[scale=0.33]{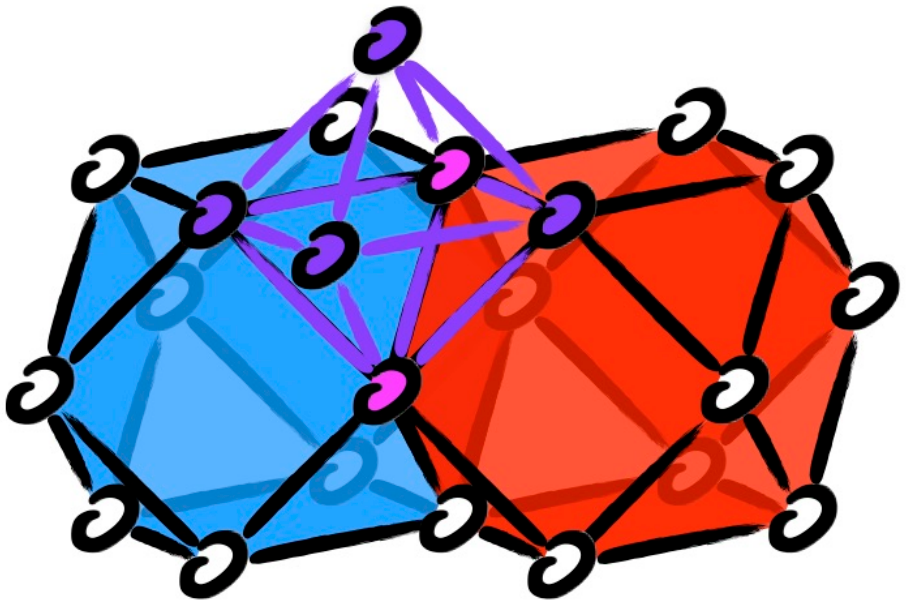}}\hspace{6 mm}
(c){\includegraphics[scale=0.33]{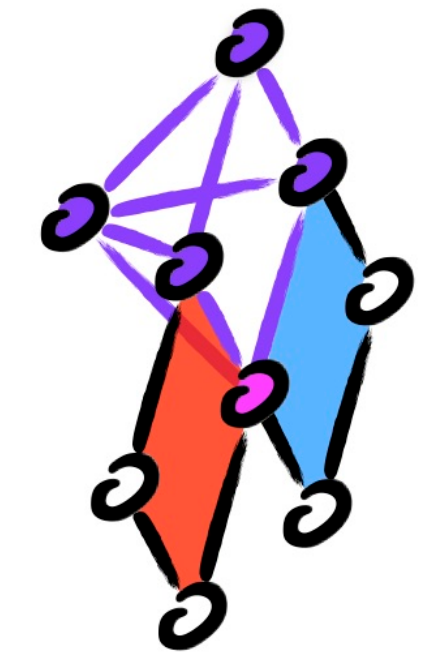}}\hspace{6 mm}
\vspace{6mm}
    \caption{Example of the action of a transversal CCZ gate on the $X$~stabilizers of a code. (a)~We demonstrate the action of a transversal CCZ~gate on the yellow, weight-6 stabilizers in code block~1. Under the action of CCZ, Pauli~$X$ on one qubit is mapped to a product of $X$ and CZ~on the other two code copies. This product is represented in green. We can break this product into two components, one comprised of the original $X$~stabilizer and a resulting tensor product of CZ on the other two codeblocks, which is represented in purple here. (b) The action of CZ on the qubits in the octahedron will result in a logical identity in the bulk as the support of this operator mutually overlaps with two qubits from the intersection of any two red and blue weight-12 stabilizers, these qubits are symbolized in pink. (c)~On the boundary, the overlap of the weight-5 CZ operator will only overlap with a single qubit from the intersection of the red and blue faces, as such this does not satisfy the requirements to be a logical identity operator.}

  \label{fig:XtransformCCZ}
\end{figure}

Finally, it is now simple to verify that the CCZ~gate will no longer preserve the boundary stabilizers. If one considers the weight-4 face stabilizers on the boundary of the hole, they intersect each of the other weight-4 stabilizers on the boundary of different color at a single qubit. Therefore, the requisite condition for the preservation of the $X$~stabilizers will no longer be satisfied as this overlaps with any of the weight-5 yellow stabilizers at a single location. This is visually represented by the purple operator in Fig.~\ref{fig:XtransformCCZ}(c). Upon the completion of the transversal CCZ gate, the resulting stabilizers will remain the same in the bulk however, the boundary stabilizers will undergo the following mapping:
\begin{align}
\nonumber \Motimes_j \text{CCZ}_{j; 1,2,3}: \Motimes_{j \in A_{i;a}^{\text{h.b.}}} X_{j;a} &\longrightarrow \Motimes_{j \in A_{i;a}^{\text{h.b.}}} X_{j;a} \Motimes_{k \in A_{i;a}^{\text{h.b.}}} \text{CZ}_{k;b,c}, \\
\Motimes_{j \in B_{i;a}^{\text{h.b.}}} Z_{j;a}  &\longrightarrow \Motimes_{j \in B_{i;a}^{\text{h.b.}}} Z_{j;a},
\end{align}
where $A_{i,a}^{\text{h.b.}}$ here is a boundary stabilizer of the hole. In the bulk, the term on the right hand side $\Motimes_{k \in A_{i;a}^{\text{bulk}}} \text{CZ}_{k;b,c}$ would be a 1-form symmetry of the code and be equivalent to logical identity, however this is no longer the case on the boundary, as such there is no way to separate this term out from the modified stabilizer. Thus, the stabilizers along the boundary now must include a CZ support on the other two codeblocks, which is the reason behind requiring a more complicated implementation of logical~CCZ in the presence of $(m_1,m_2,m_3)$-holes.

%\section{Construction of fractal stack codes in 3D}

%\section{The $XCZ$ stabilizer formalism}

\begin{figure*}[hbt]
  \includegraphics[width=1.5\columnwidth]{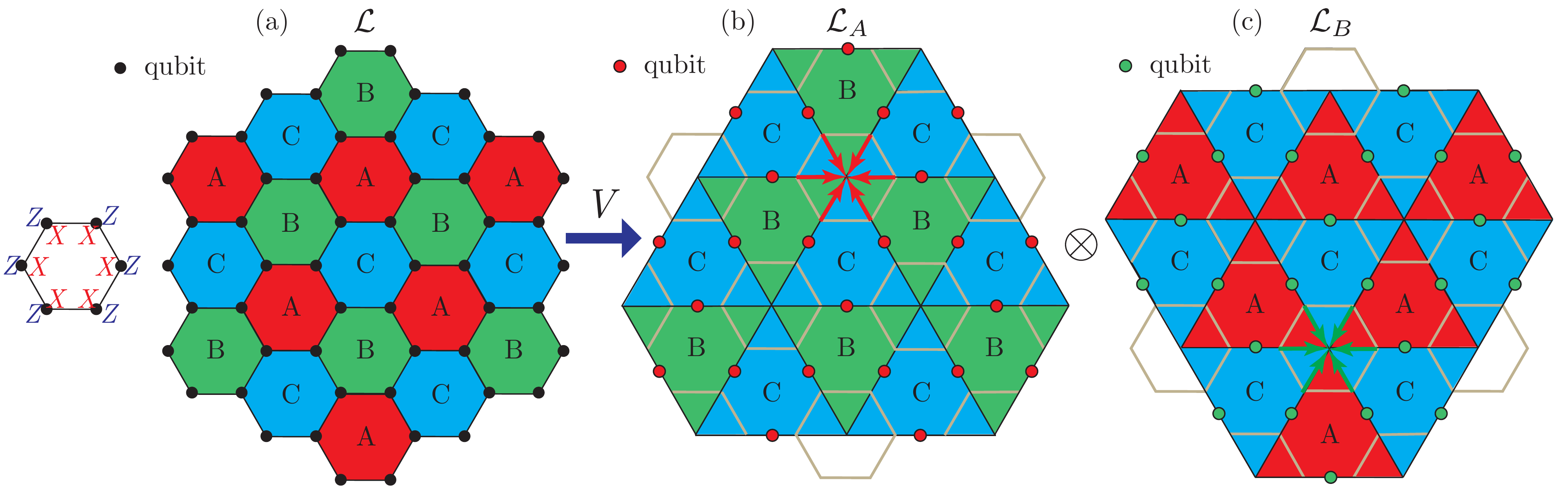}
  \caption{Disentangling a 2D color code supported on  lattice $\L$ shown in (a) into two copies of toric codes supported on the shrunk lattices $\L_A$ and $\L_B$ as shown in (b) and (c) respectively. (a) The qubits (black dots) are located on the vertices of $\L$. The faces have three colors: $A$ (red), $B$ (green) and $C$ (blue). Each face contains two types of stabilizers: the 6-body $X$ stabilizers and $Z$ stabilizers.   (b) The first copy of toric code is supported on the shrunk  lattice $\L_A$, obtained by shrinking all faces in color $A$ (red) of the original lattice $\L$ into vertices (indicated by the red arrows). The qubits (red dots) are located on the edges of $\L_A$. (c) The second copy is supported on $\L_B$, obtained by shrinking all faces in color $B$ (green) of $\L$ into vertices.  The qubits (green dots) are located on the edges of $\L_A$.}
  \label{fig:color_code}
\end{figure*}

\begin{figure}[hbt]
  \includegraphics[width=1\columnwidth]{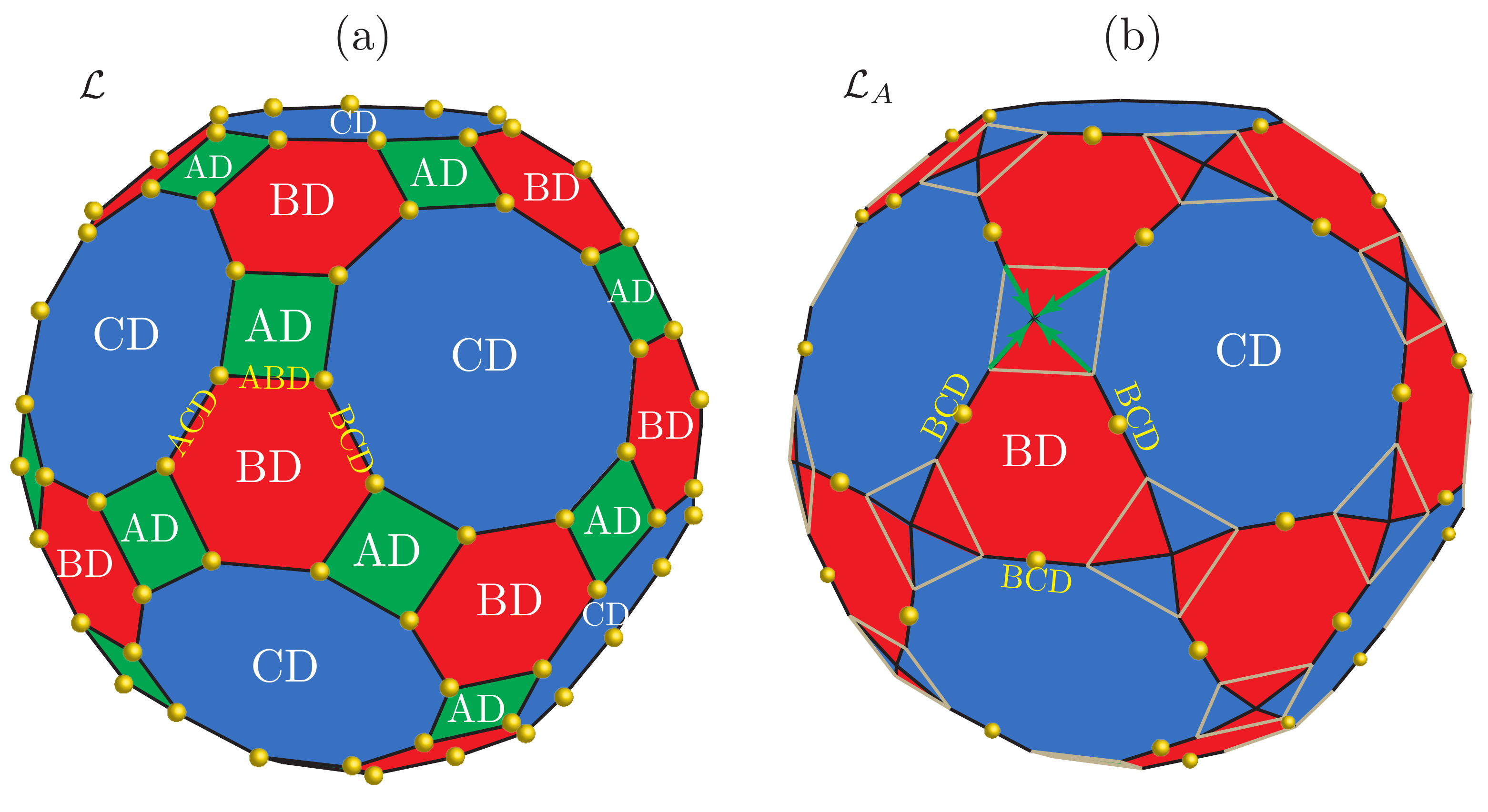}
  \caption{Illustrating the disentangling process for a 3D color code.  (a) The 3D color code is defined on a lattice $\L$ with qubits (yellow balls) located on the vertices (0-cells). All the faces (2-cells) of a volume (3-cell) with color $D$ are shown and are labeled by the colors of the two volumes they belong to: $AD$, $BD$, and $CD$. Similarly, each  edge is labeled by the colors of the three volumes it belongs to: $ABD$, $BCD$, and $ACD$.
  (b) Illustration of one copy of the disentangled  toric code defined on the shrunken lattice $\L_A$. The cells with color $A$ are shrunk   into vertices (indicated by the green arrows and dashed square). The qubits (yellow balls) are located on the edges of the toric-code lattice $\L_A$.}
  \label{fig:color_code_3D}
\end{figure}

\section{Disentangling a color code into toric codes and the corresponding logical gates in 2D and 3D}
\label{app:color_codes}

This appendix aims to illustrate the general statement in  Sec.~\ref{sec:logical_gate_nD}
about the disentangling unitary and the logical gates in $n$ dimensions with more detailed examples in two and three dimensions.   

We first discuss the disentangling unitary in 2D. We consider a trivalent lattice $\L$ with its $2$-cells (faces/plaquettes) being 3-colorable and are labeled by $A$ (green), $B$ (green) and $C$ (blue), as shown in Fig.~\ref{fig:color_code}(a).  In the $n=2$ case, the $n$-cells and $2$-cells coincide, therefore both the $X$- and $Z$ stabilizers are defined on faces labeled by $f$. We hence get the following parent Hamiltonian for the 2D color code:
\begin{align}\label{eq:2Dcolor_code}
\nonumber H_\text{2DCC}=&-\sum_f A_f - \sum_f B_f, \\
\text{with} \quad  A_f=&\Motimes_{j\in \{v_f\}  }X_j, \quad B_f=\Motimes_{j\in \{v_f\}}Z_j,
\end{align}
where $\{v_f\}$ represent all vertices in face  $p$.  Now we introduce the following lemma from Ref.~\cite{Kubica:2015br} for 2D color code:
\begin{lemma}[\cite{Kubica:2015br}]\label{lemma:2DCC}
Let CC$(\L)$ be a 2D color code defined on a lattice $\L$ without boundaries, colored in $A$, $B$ and $C$.  There exists a local Clifford unitary $V=\Motimes_{f \in \mathbf{C}}V_f$, and two lattices $\L_A$ and $\L_B$ obtained from $\L$ by shrinking faces of color $A$ and $B$ respectively, such that  
\be
V [CC(\L)] V^\dag = TC(\L_A) \otimes TC (\L_B),
\ee
where $\mathbf{C}$ represents the set of all faces in $\L$ colored with $C$, and $V_f$ is a Clifford unitary acting only on qubits in face $f$.
\end{lemma}
We also call $V$ a disentangling unitary, since it disentangles two copies of toric codes $TC(\L_A)$ and $TC(\L_B)$ from a single copy of color code $CC(\L)$. The tensor product $TC(\L_A) \otimes TC (\L_B)$ indicates that the stabilizer group is factored into two independent stabilizers groups associated with the two decoupled toric codes.  As shown in Fig.~\ref{fig:color_code}(b,c), the toric code lattice $\L_A$ ($\L_B$) is obtained by shrinking each $A$-face ($B$-face) in the original color code lattice $\L$ into a single vertex.  We hence call $\L_A$ and $\L_B$ as shrunken lattices. The qubits on the color code vertices are disentangled into two groups belonging to the two decoupled toric codes $TC(\L_A)$ and $TC(\L_B)$,  which reside on the edges of the lattice $\L_A$ and $\L_B$ respectively.  The detailed expression of the disentangling unitary $V$ can be found in Ref.~\cite{Kubica:2015br}. 

Following the general disentangling procedure in $n$ dimensions introduced in Lemma~\ref{lemma:nDCC} of Sec.~\ref{sec:logical_gate_nD} in the main text, we now illustrate the disentangling procedure in three dimensions, as shown in Fig.~\ref{fig:color_code_3D}. The 3D color-code lattice $\L$ has volumes (3-cells) with four colors: $A$, $B$, $C$ and $D$. We can also assign colors to faces (2-cells) with the colors of the two volumes which it belongs to, such as $AD, BD$ and $CD$ as illustrated in Fig.~\ref{fig:color_code_3D}(a).   Similarly, one can assign colors to each edge with the colors of the three volumes it belongs to. We can obtain the three shrunk lattices $\L_A$, $\L_B$, and $\L_C$ by shrinking volumes with color $A$, $B$ and $C$ respectively.   The example in Fig.~\ref{fig:color_code_3D}(a) shows all the faces of a volume (3-cell) with color $D$. We then shrink each volume with color A into a single vertex, and obtain the toric code lattice $\L_A$ in Fig.~\ref{fig:color_code_3D}(b),  where qubits now reside on the edges. As we can see, all the faces belonging to cell are also shrunken, including the face with color $AD$ illustrated in Fig.~\ref{fig:color_code_3D}.  We further list the details of lattice $\L_A$ as follows.  (1) Vertices: centers of $A$-volumes in $\L$; (2) Edges: $BCD$-edges in $\L$; (3) Faces: $BC$-, $BD$- and $CD$-faces in $\L$; (4) Volumes:  $B$-, $C$-, and $D$-volumes in $\L$.

In Sec.~\ref{sec:logical_gate_nD} of the main text, we have discussed how to derive the logical $\text{C}^{\otimes p-1} \text{Z}$ gate for a stack of surface codes from the transversal $R_p$ gate in the nD color code and the corresponding disentangling unitary.   
Here, we discuss this procedure in details with the 2D example. In particular, we discuss the transversal $R_2$ gate in the 2D square-patch color code and the corresponding logical $CZ$ gate derived from it.  Note that $R_2$ ($S$) gate does the following transformation to the Pauli operators:  $R_2 : X \rightarrow Y,  \  Z \rightarrow Z$. Therefore, $\widetilde{R}_2$ maps all the $X$- and $Z$ stabilizers on each face into $Y$- and $Z$ stabilizers. Since $Y$-stabilizers on each face is just a multiplication of $X$- and $Z$ stabilizers on the same face,  $\widetilde{R}_2$ keeps the stabilizer group of the color $CC(\L)$ invariant and hence preserves the code space, i.e., 
\be
\widetilde{R}_2 : CC(\L) \rightarrow \widetilde{R}_2 CC(\L) \widetilde{R}_2^\dag = CC(\L).
\ee
The transversal gate $\widetilde{R}_2$ does the following mapping to the logical operators:
\begin{align}\label{eq:transversal_R_mapping}
\nonumber \widetilde{R}_2 : \ &  \overline{X^{(1)}}   \rightarrow   \widetilde{R}_2 \overline{X^{(1)}} \widetilde{R}_2^\dag =  \overline{X^{(1)}} \   \overline{Z^{(2)}}, \qquad \overline{Z^{(1)}} \rightarrow  \overline{Z^{(1)}},  \\ &\overline{X^{(2)}} \rightarrow  \overline{Z^{(1)}} \   \overline{X^{(2)}}, \qquad \overline{Z^{(2)}} \rightarrow  \overline{Z^{(2)}}.
\end{align}
From the above mapping, we know that the transversal gate $\widetilde{R}_2$ is actually a logical CZ gate in the 2D color code, i.e.,
\be
\widetilde{R}_2 = \overline{\text{CZ}^{(1,2)}}.
\ee
Meanwhile, the local disentangling unitary $V$ does the following mapping between the logical operator in the color code and the logical operators in the disentangled surface codes:
\begin{align}
\nonumber V : \ & \overline{X^{(1)}} \rightarrow  V \overline{X^{(1)}} V^\dag = \overline{X}_1 \otimes I, \qquad   \overline{X^{(2)}} \rightarrow  I \otimes \overline{X}_2,  \\ &\overline{Z^{(1)}} \rightarrow  \overline{Z}_1 \otimes I, \  \qquad  \overline{Z^{(2)}} \rightarrow  I \otimes \overline{Z}_2.
\end{align}
We can show that the transversal $R_2$ gate conjugated by the disentangling unitary, i.e.,   $V\widetilde{R}_2 V^\dag$, preserves the stabilizer group and the code space as follows:
\begin{align}
\nonumber & V\widetilde{R}_2 V^\dag [SF(\L_1)\otimes SF(\L_2)] V\widetilde{R}_2^\dag V^\dag = V\widetilde{R}_2 CC(\L) \widetilde{R}_2^\dag V^\dag \\
& = V [CC(\L)] V^\dag = SF(\L_1)\otimes SF(\L_2)1.
\end{align}
We can also derive the mapping of $V\widetilde{R}_2 V^\dag$ on the logical operators $\overline{X}_1 \otimes  I$ in the code space of $SF(\L_1) \otimes SF(\L_2)$:
\begin{align}
\nonumber V\widetilde{R}_2 V^\dag : \ & \overline{X}_1 \otimes  I \rightarrow  V\widetilde{R}_2 V^\dag (\overline{X}_1 \otimes I) V\widetilde{R}_2^\dag V^\dag \\
= &  V\widetilde{R}_2  \overline{X^{(1)}} \widetilde{R}_2^\dag V^\dag  =V \overline{X^{(1)}} \ \overline{Z^{(2)}} V^\dag =  \overline{X}_1 \otimes \overline{Z}_2.
\end{align}
Similarly, we get the mapping of all the  logical operators in the stack code  as: 
\begin{align}
\nonumber V\widetilde{R}_2 V^\dag : \ & \overline{X}_1 \otimes  I \rightarrow  \overline{X}_1 \otimes \overline{Z}_2, \ \overline{Z}_1 \otimes I  \rightarrow  \overline{Z}_1 \otimes I, \\
& I \otimes \overline{X}_2  \rightarrow  \overline{Z}_1 \otimes \overline{X}_2, \  I \otimes \overline{Z}_2 \rightarrow  I \otimes \overline{Z}_2.
\end{align}
From the above mapping, one can see that the operator $V\widetilde{R}_2 V^\dag$ implements a logical CZ gate between the two surface codes $SF(\L_1)$ and $SF(\L_2)$, i.e.,
\be
V\widetilde{R}_2 V^\dag = \overline{ \text{CZ}}_{1,2}. 
\ee
Although $\widetilde{R}_2$ is a transversal gate in the color code, $V\widetilde{R}_2 V^\dag$ is not guaranteed to be a transversal gate (or equivalently local unitary) in the stack of surface codes.  Still, it is clear that $V\widetilde{R}_2 V^\dag$ is a local constant-depth  circuit applied on the stack of two surface codes which implements the logical CZ gate.  

% One can understand the process of implementing $V\widetilde{R}_2 V^\dag$ as follows:  one first applies the inverse of the disentangling unitary, i.e., $V^\dag$, to entangle the stack of two surface codes $SF(\L_1) \otimes SF(\L_2)$  into a single copy of 2D color code $CC(\L)$. Now one applies $\widetilde{R}_2$,  corresponding to a logical gate $\overline{\text{CZ}^{(1,2)}}$ in the code space of the color code.   One then applies the disentangling unitary $V$ again which disentangles the color code back to a stack of two surface codes and maps $\widetilde{R}_2$ into the logical gate $\overline{\text{CZ}}_{1,2}$ in the code space of the stack code $SF(\L_1) \otimes SF(\L_2)$.

\end{appendix}

%\bibliography{mybib_merge.bib}

%apsrev4-2.bst 2019-01-14 (MD) hand-edited version of apsrev4-1.bst
%Control: key (0)
%Control: author (8) initials jnrlst
%Control: editor formatted (1) identically to author
%Control: production of article title (0) allowed
%Control: page (0) single
%Control: year (1) truncated
%Control: production of eprint (0) enabled
%

\end{document}